\documentclass{cernrep}

\usepackage{longtable}
\usepackage{rotating}
\usepackage{csquotes}
\usepackage[style=ieee,citestyle=numeric-comp, backend=biber]{biblatex}
\usepackage[version=4]{mhchem}  
\usepackage{tablefootnote}
\usepackage[hidelinks]{hyperref}
\usepackage{comment}
\usepackage{hyperxmp}
\usepackage{booktabs}
\usepackage{siunitx}

\ifdefined\qty\else
  \ifdefined\NewCommandCopy
    \NewCommandCopy\qty\SI
  \else
    \NewDocumentCommand\qty{O{}mm}{\SI[#1]{#2}{#3}}
  \fi
\fi
\ifdefined\unit\else
  \ifdefined\NewCommandCopy
    \NewCommandCopy\unit\si
  \else
    \NewDocumentCommand\unit{O{}m}{\si[#1]{#2}}
  \fi
\fi

\ifdefined\qtyrange\else
  \ifdefined\NewCommandCopy
    \NewCommandCopy\qtyrange\SIrange
  \else
    \NewDocumentCommand\qtyrange{O{}mmm}{\SIrange[#1]{#2}{#3}{#4}}
  \fi
\fi

\usepackage{makecell}
\usepackage{multicol}  
\usepackage{enumitem}  
\usepackage{adjustbox}  
\usepackage{colortbl}  
\usepackage{svg}
\newcolumntype{\darkgreencolumn}{>{\columncolor[HTML]{7DFA7D}} p{0.22\textwidth}}
\newcolumntype{\lightgreencolumn}{>{\columncolor[HTML]{BBFA7D}} p{0.22\textwidth}}
\newcolumntype{\yellowcolumn}{>{\columncolor[HTML]{FAFA7D}} p{0.22\textwidth}}
\newcolumntype{\orangecolumn}{>{\columncolor[HTML]{FABC7D}} p{0.22\textwidth}}
\newcolumntype{\redcolumn}{>{\columncolor[HTML]{FA7D7D}} p{0.22\textwidth}}
\usepackage{pifont}
\newcommand{\tick}{\ding{51}}%
\newcommand{\cross}{\ding{55}}%
\def\CCC{C$^{3}$~}

\begin{document}
\title{Sustainability Assessment of Future Accelerators}
\author{C. Bloise$^a$, E. Cennini$^b$, J. Gutleber$^b$, W. Kaabi$^c$, A. Klumpp$^d$, P. Koppenburg$^e$, Y. Li$^f$, B. List$^d$, R.~Losito$^b$, B. Mandelli $^b$, E. A. Nanni$^g$, N. Neufeld$^b$, T. Schoerner-Sadenius$^d$, B. Shepherd$^h$, \\ V. Shiltsev$^i$, S.~Stapnes$^b$, M. Titov$^j$, L. Ulrici$^b$, H. Wakeling$^k$}

\institute{\vspace{8mm}
            $^a$LNF-INFN, Frascati, Italy \hspace{5mm} 
            $^b$CERN, Geneva, Switzerland \hspace{5mm} 
              $^c$ IJCLab, Orsay, France \\
              $^d$ DESY, Hamburg, Germany \hspace{3mm}  
             $^e$ NIKHEF, Amsterdam, The Netherlands \\
               $^f$ IHEP, Beijing, China \hspace{12mm} 
                $^g$ SLAC, Menlo Park, CA, USA \\ 
             $^h$ ASTeC, Daresbury Laboratory, STFC, UK \hspace{0.5mm} 
            $^i$ Northern Illinois University, DeKalb, IL, USA \\ 
            $^j$ IRFU, CEA-Saclay, Gif-sur-Yvette, France \\
                $^k$ John Adams Institute, University of Oxford, Oxford, UK\\ \vspace{1.5cm}}

\begin{abstract}
The Large Particle Physics Laboratory Directors Group (LDG) established the Working Group (WG) on the Sustainability Assessment of Future Accelerators in 2024 with the mandate to develop guidelines and a list of key parameters for the assessment of the sustainability of future accelerators in particle physics.
While focused on accelerator projects, much of the work will also be relevant to other present and upcoming research infrastructures.
The development and continuous update  of such a framework aim to enable a coherent communication amongst scientists
and 
adequately convey the information to a broader set of stakeholders.
This document outlines the key findings and recommendations of the LDG Sustainability WG  and provides a summary of current best practices aimed at enabling sustainable accelerator-based research infrastructures.
Not all sustainability topics are addressed at the same level. The assessment process is complex, largely under development and a homogeneous evaluation of all the aspects 
 requires a strategy to be developed and implemented over time.
\end{abstract}


\maketitle
\tableofcontents
\clearpage

\section{Executive Summary}

The sustainability of future accelerator-based research infrastructures is increasingly central to their scientific, economic, and societal viability. 
These large-scale facilities which pursue scientific excellence in advancing fundamental knowledge in particle physics, present complex challenges related to energy consumption, material demand, long-term operation, and environmental impact.

The LDG Sustainability WG report outlines the reference framework of  best practices and standards for assessing and 
guiding a sustainable development of large-scale facilities for future accelerators.  

\noindent
\textbf{Assessment of socio-economic impacts}

Funding landscapes are rapidly changing in Europe and beyond. 
International organizations are working to establish guidelines and sustainability reporting standards in different sectors relevant to accelerator-based RIs. 
From a socio-economic perspective, particle physics  accelerators generate substantial public value beyond their mission of discovering new laws of nature, across several of the 17 Sustainable Development Goals (SDGs) adopted by all United Nations Member States in 2015. 
Accelerator-based RIs act as innovation hubs, foster global collaboration, drive industrial and technological spillovers, and provide education and training that translate into long-term benefits. 
Understanding of these impacts is necessary for public authorities and credit institutions to establish funding priorities and take informed investment decisions.
There are several examples of comprehensive impact assessments conducted by large research infrastructures using different approaches and methodologies - some more quantitative, like Cost-Benefit Analysis (CBA), carried out at CERN to assess the HL-LHC and the FCC project, and others more qualitative, such as narratives and case studies. 
Building on these works, a \textit{comprehensive Cost-Benefit Analysis of HEP projects is recommended.}  It integrates both tangible and intangible contributions across the UN SDGs and is required by several countries for public investment, including science projects.

\noindent
\textbf{Assessment of environmental impacts}

Life Cycle Assessments (LCA) and the use of Environmental Product Declarations (EPDs)  
aligned with international standards, are recognised as key tools for quantifying environmental impacts  informing both design decisions and policy compliance. 
\textit{Future accelerator projects are encouraged to perform LCAs at every project stage (e.g. proposal, CDR, TDR).}
The LCA goals (and target audience) range from identifying the most critical contributions and supporting design optimization (for researchers), to providing inputs for technological choices during project reviews and recommending improvements (for decision-makers), to fulfilling legal requirements for project submission (for regulatory bodies), and communicating global performance (to general public).

LCAs are new for accelerator components and detectors.
\textit{Coordinated efforts among accelerator-based laboratories to create a shared database of accelerator materials and components, and to develop open-source, cost-free LCA tools (e.g., an accelerator-specific extension to ecoinvent), are necessary to improving the accuracy and reliability of environmental assessments.}

The impact of greenhouse gas emissions (GHG) on the climate, quantified as Global Warming Potential (GWP)  
 [\unit{\kg\ce{CO2}e}], is present in all series of LCA's midpoint categories that include amongst others, emissions of particulates, use of freshwater, land use and transformation, and use of natural resources. 
During  \emph{construction phase}, emissions mainly arise from production of the large quantities of materials used in civil engineering works, accelerator components, services, and detector systems.  
For the \emph{operation phase}, electricity consumption represents the most significant contribution in the assumption that direct GHG emissions will be drastically reduced by new generation of detectors and technical services. 
 This contribution depends on the carbon intensity of electricity, which varies substantially across regions and over time.
The \emph{decommissioning phase} requires early analysis of recycling and disposal of used components, radioactive waste, and land reuse. 
Assessments differ substantially when data from generic, current productions are replaced by information from specific EPDs or from projections based on decarbonization pathways outlined by industrial consortia, such as cement and steel producer associations. The use of data referring to specific materials or manufacturing processes -- whether region-specific or technology-specific -- should be supported by a realistic expectation that their production can meet the required quantities at reasonable costs.

\textit{The Sustainability WG has agreed on a common format for sustainability reporting tailored to the current project stage and based on a small set of figures of merit for GHG emissions in the construction and operation phases of future colliders.}
The report presents up-to-date preliminary evaluations of these key parameters, together with the assumptions and simplifications applied by future accelerators projects at CERN (CLIC, FCC-ee, LCF, LHeC), by ILC in Japan, CEPC in China, and C$^3$.

Operation scenarios are defined by center-of-mass,  energy, average luminosity, duty cycles, annual integrated luminosity, and the number of interaction points. Key data on GHG emissions \textit{in the construction phase} cover civil engineering works, accelerator components, and detectors. The emissions evaluated \textit{for the operation phase} are those associated with electricity consumption only, as it is reasonable to expect that future RIs will significantly improve upon the current situation, in which direct emissions from particle detectors and cooling plants are the main sources of GHG emissions. Data on carbon intensity of the electricity are based on official projections of the energy mix at the time the machine would operate - for example, the second half of 2040s for CLIC, FCC-ee, and LCF - and take into account the increase in renewables targeted by the decarbonization plans at the host sites. 
Impacts from civil engineering works rely on full LCAs carried out by external firms for CLIC, FCC-ee, LCF at CERN, and ILC in Japan.
LCAs of accelerator components and detectors have recently been added to the CLIC/ILC/LCF studies. 
LCAs indicate that the carbon footprint is dominated by the production of the large quantities of materials – concrete, steel, iron and copper - required by civil engineering works, accelerator components (beam pipes, supports, magnets, RF systems), and particle detectors (magnet return yokes and absorbers used in calorimeters and muon detection systems). The annual carbon footprint of the operation phase is estimated to be 3-5\% of the civil engineering works at CERN, and 15-25\% in Japan, reflecting primarly the different electricity carbon intensities in the two regions.

At the current stage of the projects, opportunities to reduce CE environmental impacts by 40-50\% have been identified through the use of low-carbon concrete, and by 15-20\% through tunnel design optimization.
 
\textit{The assessment of 
mitigation and compensation measures -- particularly those enabled by innovative technologies specific to accelerator-based infrastructures -- is essential to sustainable future accelerators. }
The report outlines several pathways, including the use of low-carbon-footprint materials and construction practices, the adoption of responsible procurement policies, the implementation of advanced energy management plans and waste-heat recovery solutions, investments in innovative technologies that enhance sustainability performance.
Nature-based interventions and circular economy principles are recognized as complementary strategies to offset residual impacts.
 
\textit{Improving energy efficiency and reducing consumption -- key to minimise carbon emissions -- complements the transition to low-carbon energy sources.}
Notably, any reduction in consumptions makes additional energy resources available to society. 

Among all actions to reduce environmental impact of future colliders, study and development of novel
technologies is the most specific to the accelerator-based RIs.
Medium- and long-term pathways are underway to reach the technological maturity required for future accelerators, often in synergy with other fields of science, and creating opportunities to improve sustainability performance.
\begin{itemize}[nosep]
\item Overall collider design sets beam power and run time to achieve target luminosity.
  The best feasible control and phase-space reduction of colliding particles is essential to ensuring optimal luminosity performance in future colliders. 
\item   RF sources impact the efficiency of all future accelerators. Technological innovations are being pursued through collaborations between industry and accelerator laboratories and have shown significant progress in recent years. In addition to improving efficiency, increasing the mean time between failures (MTBF) enhances accelerator sustainability by reducing material consumption and manufacturing needs.
\item R\&D on high performance SRF cavities - high-gradient/high $Q_0$ -  is needed for many HEP projects.
  From a sustainability perspective, higher $Q_0$ improves cooling efficiency, and the use of SRF coatings significantly reduces the demand for sourcing materials such as bulk niobium.     
\item The development of environmentally friendly large-area gaseous detector systems is necessary to enable sustainable experiment operation.
\end{itemize}

\vspace{2mm}
In summary, sustainability assessment of future accelerators demands comprehensive evaluation of socio-economic impacts, rigorous appraisal of environmental costs, and analysis of impact reductions enabled by innovative technologies. Assessments at each stage of project maturity are essential to inform actions that  
align accelerator development with global climate and development goals.

\clearpage
\section{Introduction}

The aim of this study is to develop guidelines and a set of key indicators
for assessing  the sustainability of the projects of future particle accelerators.
Recognizing the rapidly evolving nature of this field, we do not propose definitive rules or prescriptions at this stage.
Instead, we present examples of good practices, acknowledging that coordinated efforts from particle physics laboratories worldwide, and continuous updates will be essential to developing a shared, coherent framework.

Sustainability, or sustainable development, refers to the ability to maintain an activity over time without compromising the opportunities available to future generations. 
Sustainable growth should be viewed within the context of the entire society. It requires the integration of cultural and ecological aspects in a three pillar (social, economical, and environmental) model \cite{PMR_2019}. 
The development should be evaluated in accordance with the values that a society deems necessary for its welfare~\cite{Pollock_1973}. For development to be truly sustainable, 
the design and implementation of projects must be tailored to the needs and capabilities of the people who are expected 
to benefit from them \cite{Uphoff_1985}. The individual objectives that emerge from the 
pursuit of scientific and economic goals within the environmental and social constraints 
often conflict -- both across different pillars and, 
at times, within a single pillar. Ultimately, it is therefore necessary to conceive a process of trade-offs to achieve a balanced project scenario \cite{Barbier_1987}.

International organizations and fora are working to establish goals and scopes of a sustainable development in different sectors crossing the accelerator-based research infrastructures. UN resolution 66/288 \cite{UN_SDG_2012} provided the basis for developing sustainability goals. The 2030 Agenda for Sustainable Development, adopted by all United Nations Member States in 2015, defines 17 Sustainable Development Goals {\href{https://sdgs.un.org/goals}{(SDGs)}} "aiming to end poverty and other deprivations, improve health and education, reduce inequality, spur economic growth, tackle climate change and work to preserve oceans and forests".

The HEP projects for future accelerators are international initiatives which serve integrated, global scientific communities. They require infrastructures of a size and cost that can only be built through international collaborations. Their sustainability assessment, as well as the reporting of the organizations involved in creating, operating, and maintaining these projects are being developed and 
good practices are being implemented in the RIs all over the world, as with CERN periodic environment reports~\cite{CERN:2023a} that rely on the guidelines set by the Global Reporting Initiative~\cite{GRI}, 
an international organization of independent stakeholders. 

Future accelerators push the frontiers of science and technology, constitute innovation incubators where researchers work with high-tech companies that supply state-of-the-art services and technologies,
and generate spin-offs of interest in a broader socio-economic context. 
At the same time, accelerator-based research infrastructures constitute knowledge hubs that contribute in preparing and training young scientists and technology developers in cooperation with universities and industries. 
Public perception of the value of the investments in projects for future accelerators relies on the efforts and performance for explaining scientific potentialities and the returns for the society in knowledge, technology and education. 
The assessment is complex as the impact of scientific discoveries or novel technologies extend over long time scales, up to decades after finding.
Models for the evaluation are being developed, however consensus on how impacts can be measured and described still needs to be elaborated. 

Conserving and protecting the environment is one of the three pillars of sustainability, and in the public perception possibly the most prominent one. 
In recent years, public awareness of climate change and its negative consequences has increased, and the measures taken in the Paris accord to limit climate warming to \qty{1.5}{\celsius} 
above the pre-industrial baseline have been heavily debated. 
Reporting green house gas (GHG) emissions in terms of \ce{CO2} equivalents has become almost synonymous with a quantification of environmental impact.
A more holistic view of environmental impact recognises that the environment faces numerous threats: in addition to green house gas emissions, the atmosphere is impacted by ozone layer depleting substances, by particulate matter emissions, by substances causing summer smog, and toxic emissions to air in general. Terrestrial and marine ecosystems suffer from eutrophication, acidification, and ecotoxicity. Land use change furthers global warming, mineral and water resources are depleted.
Different stages of a product, system or project typically impact the environment in different ways, and in different locations, from mining of raw material through emissions during fabrication and from the generation of electricity used to wastes disposal at the end of life. 
Selectively reducing the impact of one aspect only often leads to burden shifting, where the impact is increased in other product stages, locations, or impact categories.
Lifecycle Assessment (LCA) is a systematic approach to quantify these impacts, with the goal to reduce and optimise the environment impact across all categories and lifecycle stages. 
The methods and the specific issues arising in conducting and interpreting LCAs in the context of accelerator projects are presented in section~\ref{sec:lca}. 

In summary, the bulk of issues discussed in the following sections pertain to i) socio-economic benefits of accelerator-based research infrastructures in relation to the UN Sustainable Development Goals that refer to society, economy and environment (sections~\ref{sec:basis}-\ref{sec:enablers}); ii) methodology and reporting of Life Cycle Assessment (sections~\ref{sec:lca}-\ref{sec:edp}); iii) evaluation of greenhouse gas (GHG) emissions in construction, operation and decommissioning phases (section~\ref{sec:GHG-emissions}) and iv) assessment of mitigation and compensation strategies (section~\ref{sec:mitigation}).  
In addition, four annexes have been prepared to provide further information: i) on the P5 process in the United States (Annex~\ref{annex:snowmass21}), ii) on the CEPC project in China (Annex~\ref{annex:CEPC}), iii) on legislation, rules, and methodologies used for project appraisal in the European Union, the UK, Australia, and US (Annex~\ref{annex:RI_appraisal}), and iv) on GWP estimates for specific materials commonly used in the construction  of accelerator-based research infrastructures (Annex~\ref{annex:reference-data}). 

\clearpage

\section{Building Strategic Accountability}\label{sec:account}

\begin{figure}[ht]
    \centering
    \includegraphics[width=0.3\textwidth]{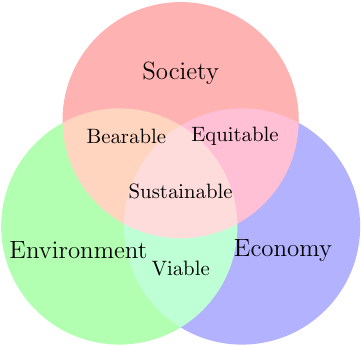}
    \caption{Sustainability dimensions.}
    \label{fig:sustainability_dimensions}
\end{figure}

Sustainability assessment is based on the analysis of three main  pillars: society, economy, and environment (\autoref{fig:sustainability_dimensions}). To achieve sustainable science, the project appraisal integrates financial and socio-economic aspects. 
Socio-economic aspects include both positive externalities -- such as social benefits -- and negative effects, including climate impacts, the consumption of natural resources (e.g., soil, land, and water), degradation of air and water quality, impacts on public health, and various other externalities.\footnote{An externality is a cost or benefit that is caused by one party but financially incurred or received by another. Externalities can be negative or positive. A negative externality is the indirect imposition of a cost by one party onto another. A positive externality, on the other hand, is when one party receives an indirect benefit as a result of actions taken by another.} 

A science programme or project can be considered sustainable if it is able to successfully address and complete its scientific core mission satisfying the requirements of three stakes: it has obtained a `social license' to operate, it maintains the consumption of natural resources within acceptable limits achieving an ecological balance and it is affordable in the long term with well-understood and managed risks.

\begin{figure}[ht]
    \centering
    \hspace{-5mm};
    \includegraphics[ width=0.5\textwidth]{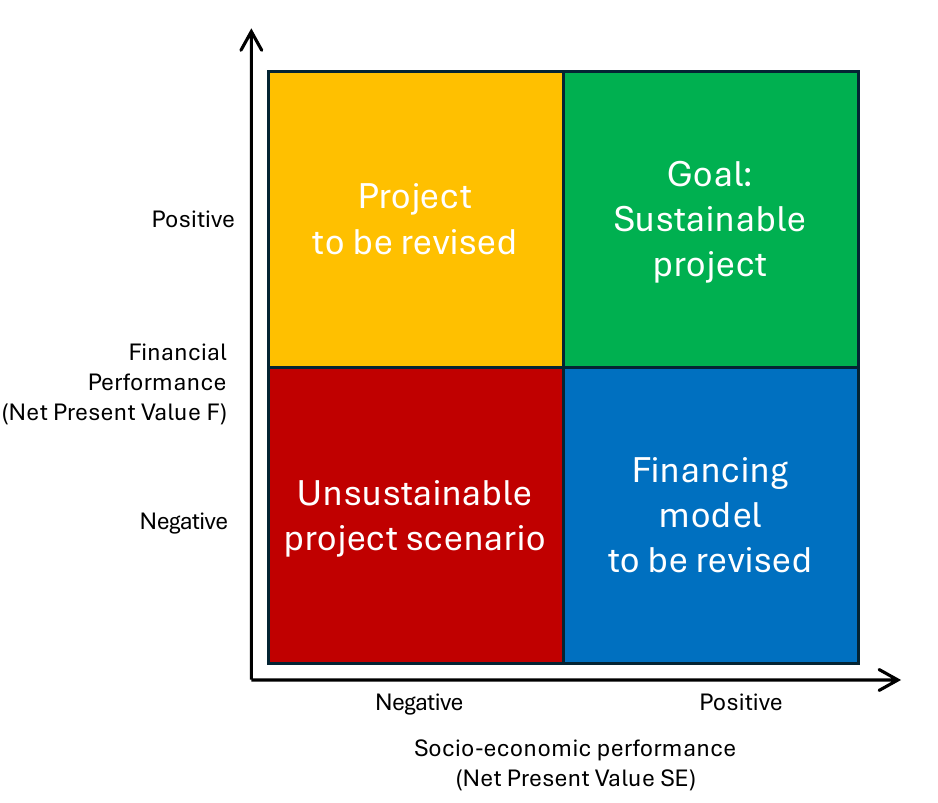}
    \caption{Guidance for the sustainability of public investment projects. From Ref.~\cite{SGPI2018}.}
    \label{fig:guide_sustainability}
\end{figure}

With the recent ISO standard for monetization of environmental impacts (ISO 14008~\cite{iso14008}), and the guidelines for determining environmental costs and benefits (ISO 14007~\cite{iso14007}), there is now also an increasing interest in applying such 
costs in environmental and sustainability
reporting at 
organisation and project levels.
Science projects typically can achieve a positive socio-economic net present value~\cite{FlorioNPV:2019}. 
Projects that are financially and socio-economically positive can be considered sustainable (\autoref{fig:guide_sustainability}).

Early and accompanying socio-economic analysis that integrates all sustainability dimensions helps to benchmark variants and versions of project scenarios and permits planning for long-term sustainability. An RI needs to periodically monitor and track the social, environmental and economic performance against the initial estimations to implement a continuous improvement process.

In line with the `Ecodesign' EU Directive 2009/125/EC~\cite{ECecodesign}, a sustainable development requires proper consideration of all three sustainability elements, including the identification and accounting of positive and negative impacts throughout the entire lifecycle of programme or project. For RIs carrying out science missions, at least the following stakes should be considered:
\begin{itemize}[nosep]
    \item Economy
        \begin{itemize}[nosep]
            \item Scientific excellence
            \item Total costs (capital and operation expenditures)
            \item Risks and residual risks after mitigation
            \item Direct, indirect and induced `value added' and employment
            \item Quantified incremental economic benefit potentials
        \end{itemize}
    \item Society
        \begin{itemize}[nosep]
            \item Quantified incremental social benefit potentials
            \item Common good value (the value of the science mission as perceived by people)
            \item Territorial compatibility
            \item Social license
        \end{itemize}
    \item Environment
        \begin{itemize}[nosep]
            \item Quantified negative externalities
            \item Quantified incremental environmental benefit potentials
        \end{itemize}
\end{itemize}

\subsection{Setting the basis for sustainability}\label{sec:basis}

Future accelerators are international initiatives requiring RIs of a size and cost deliverable only through international collaboration and   represent significant long-term investments for nations. 
To ensure long-term sustainability, an appraisal process from the concept phase onward is typically required for new projects  by funding agencies, public authorities, and credit institutions.

Annex~\ref{annex:RI_appraisal} presents a collection of legal frameworks and appraisal guidelines that exist in selected countries of the European Union -- such as France, and Germany -- as well as  in the UK, Switzerland, Australia, and the United States. It provides an overview of the sustainability requirements that govern the planning and implementation of new research infrastructure projects.
In the countries that are part of the European Research Area, legal frameworks govern the compliance with sustainability aspects. Australia, the UK and the USA are still largely applying panel-based project decision taking, although the UK does carry out ex-post evaluations on a case-by-case basis, and Australia is evaluating the sustainability of research infrastructures periodically. Sustainability and environmental criteria such as climate considerations are increasingly being incorporated into proposal design and assessment. In particular, Switzerland has very recently made sustainability aspects explicit for CERN's strategically important projects, with territorial development in the update of the law~\cite{Swiss_parliament_20240214,Swiss_parliament_20240927} for the 
promotion of research and innovation, which introduces an authorisation process at federal level.

Consideration of quantitative environmental externalities, both positive and negative ones, is still an emerging discipline that is not yet widespread, unless it is explicitly required by national legislations and governed by national guidelines. 
Environmental impact evaluation is already a common practice in  infrastructure project appraisals across many European countries -- including France~\cite{CGEDD:2017}, Germany~\cite{BMBF:2024, LeNa-framework:2023}, Italy and Switzerland -- as well as at the EU level through European Commission (EC)  funding conditions for numerous sectors, e.g. Connecting Europe~\cite{CINEA_CBA:2022}, transport~\cite{UNECE_CBA:2003}, energy~\cite{ENTSOE_CBA:2024}, regional investment projects~\cite{ECCBAGuide2014}, as well as nature preservation and restoring~\cite{Natura200:2013}.

This methodology -- used to assess public investment and welfare-maximising projects across various policy sectors and institutions~\cite{CSILFlorio2023,10143094060}, including environmental impacts~\cite{OECD_CBA_2018} -- is already applied to 
accelerator-based research infrastructure projects 
such as ALBA~\cite{ALBA2023} and SOLEIL~\cite{SOLEIL2020} light sources, the CNAO hadron therapy facility~\cite{BATTISTONI201679}, the LHC~\cite{Florio:2036479}, High Luminosity LHC (HL-LHC)~\cite{Bastianin2021,Bastianin:2319300, doi:10.1080/00036846.2022.2140763}, the Compact Linear Collider (CLIC) study~\cite{Magazinik:2836841}, and the Future Circular Collider (FCC)~\cite{csil_2024_10653396}. It has also been used for other science facilities such as the Einstein Telescope~\cite{EinsteinTelescope2021}, the Paris Saclay heat supply facility~\cite{Saclay_2015}, the Nantes University hospital~\cite{Nantes_2020}, the Cigéo nuclear storage centre~\cite{SGPI2021} and by the Commonwealth Scientific and Industrial Research Organisation (CSIRO)~\cite{CSIRO:2023} to evaluate a number of different science programmes. Ever more RIs, such as the Square Kilometre Array, are considering the approach in view of planning for sustainability~\cite{SKA:2023}. Environmental factors are increasingly being integrated into these assessments~\cite{FS2022,SGPI2018,Enerland2023,STM2022}. 

\subsubsection{Environment and sustainability performance reporting}
\label{par:EMAS}

Different schemes exist for reporting the sustainability and environmental performance of organisations, projects and programmes. It is important to distinguish sustainability performance reporting of organisations from projects. To present a complete picture, the following paragraphs shed light on both topics. Particle accelerator projects are typically developed by research organisations. However, particle accelerators for industrial and medical purposes are also constructed and deployed by other types of publicly or privately funded organisations. During the project development phase, sustainability reporting focuses on the project's potential performance while during the operation phase a research organisation typically includes the performance of active projects and programmes in their overall periodic sustainability performance reporting.

Two prominent sustainability reporting schemes that are used worldwide are the Global Reporting Initiative (GRI) and the European Sustainability Reporting Standards (ESRS). Another simplified scheme that is widely used worldwide, albeit limited to environmental aspects, is the European Union Eco-Management and Audit Scheme (EMAS).

GRI~\cite{GRI} is an international independent standards organisation that aims to understand and communicate a wide range of sustainability-related impacts, including those related to climate change, human rights, and corruption. GRI's voluntary sustainability reporting framework has been adopted by multinational organisations, governments, small and medium-sized enterprises (SMEs), non-governmental organisations (NGOs), and industry groups. Over \num{10000} organisations from more than \num{100} countries use GRI, including CERN. GRI provides the world's most widely-used sustainability reporting standards. 
GRI covers a large and evolving set of indicators that are broken down into two levels (core and additional) in six different areas: economy, environment, human rights, social relations and working conditions, responsible products and society. The full set of indicators are available online~\cite{GRI_indicators,GRI_standards}. 
GRI started to integrate and align with the newly available European Union Corporate Sustainability Reporting Directive in 2021~\cite{EU_Corp_Sus_Reporting}. This European Sustainability Reporting Standard~\cite{ESRS1} came into force on 1 January 2024. It aims at organisations that need to be standards-compliant, i.e. that are subject to mandatory sustainability reporting. The topics covered by the ESRS scheme are shown in \autoref{table:ESRS1}. This development shows that the sustainability reporting landscape is rapidly evolving and EU frameworks are becoming increasingly relevant also at a global scale. Therefore this process needs to be closely followed~\cite{GRI_FAQs}.

\begin{table}
\def\tw{\linewidth-2\tabcolsep}
\centering
\caption{Topics that are covered by the European Sustainability Reporting Standards (ESRS).}
\resizebox{0.8\width}{!}{
\begin{tabular}{ll}
\hline
\textbf{Chapter} & \textbf{Topic} \\
\hline
ESRS 1 & General requirements \\
ESRS 2 & General disclosures \\
ESRS E1 & Climate change \\
ESRS E2 & Pollution \\
ESRS E3 & Water and marine resources \\
ESRS E4 & Biodiversity and ecosystem \\
ESRS E5 & Resource use and circular economy \\
ESRS S1 & Own workforce \\
ESRS S2 & Workers in the value chain \\
ESR S3 & Affected communities \\
ESR S4 & Consumers and end-users \\
ESRS G1 & Business conduct \\
\hline
\end{tabular}
}
\label{table:ESRS1}
\end{table}

The voluntary European Union Eco-Management and Audit Scheme (EMAS)~\cite{EMAS}, also known as the EU EMAS regulation (EC) No.~1221/2009, is a tool for organisations to evaluate, report and improve their environmental performance. The regulation focuses on a standardised environmental statement, which can be certified by an independent environmental verifier, such as for example TÜV SÜD.\footnote{TÜV SÜD is a German service provider for the certification of technical systems and products that emerged out of a supervision society for steam boilers founded in the 19th century.} Thanks to EMAS' Global mechanism~\cite{EMAS_site}, the system is also available worldwide, enabling multinational organisations to register their sites within and outside Europe. Currently more than \num{4600} organisations and \num{7900} sites are EMAS-registered. Certified organisations and sites are entitled to display the EU Eco-Management and Audit Scheme certification logo. The EMAS regulation is entirely integrated into the ISO/EN 14001:2015 standard and norm. A project that follows the ISO/EN 14001 scheme will therefore comply with the requirements that the reporting standard demands. 
EMAS standardises a set of key performance indicators (KPIs) for the environmental efficiency that permits applying it to a project. It covers the direct effects of the project. Compared to GRI, it is a  simplified approach to report environmental aspects of a project or organisation. \autoref{EIB:SCC} outlines relevant indicators for projects to report.

\begin{table}
\def\tw{\linewidth-3\tabcolsep}
\centering
\caption{EMAS key performance indicators for project-oriented environmental reporting.}
\resizebox{0.8\width}{!}{
\begin{tabular}{lp{0.5\textwidth}p{0.2\textwidth}}
\hline
\textbf{KPI} & \textbf{Description} & \textbf{Units} \\
\hline
Energy efficiency & Total annual energy consumption & MWh or GJ \\
Energy use & Percentage of energy from renewable sources used for electricity and heating & Percent \\
Material efficiency & Annual mass flow of different materials used & Tonnes per relevant material \\
Water & Total annual water consumption & Cubic metres \\
Waste & Total annual generation of waste broken down per type & Tonnes \\
Hazardous waste & Total annual generation of hazardous waste broken down per type & Tonnes \\
Biodiversity & Land surface used, consumed or constructed & Square metres \\
Emissions & Total amount of greenhouse gas emissions & Tonnes of \ce{CO2} equivalent \\
Air quality & Total annual emissions into air broken down by type & kg or tonnes \\
\hline
\end{tabular}
}
\label{EIB:SCC}
\end{table}

It is recommended to follow the EU Sustainability Reporting Standards, relying on the support of domain experts for tailoring the reporting to the specific organisation and project. 
GRI and ESRS are relevant for the sustainability reporting at the level of organisations. Generally, organisations with an interest in Europe and who are subject to mandatory sustainability reporting will consider to gradually transit from GRI to ESRS. EMAS, with its simplified approach, aims at encouraging small organisations to report on their environmental performance on a voluntary basis, and permits 
project-oriented environmental performance reporting.

\subsection{Comprehensive sustainability assessment}

New science missions and research infrastructures should carry out a comprehensive quantitative sustainability assessment based on the cost-benefit analysis methodology~\cite{RICBA:2016} by the European Commission, the European Investment Bank, and as described by national guidelines such as the one in France~\cite{SGPI:2023}. 
The goal is to be able to present research infrastructure scenarios that are likely to be long-term sustainable and for which their sustainability performance can be continuously improved with respect to an initial forecast that serves as a baseline. Sustainability analysis can be used as an ingredient to informed decision making. Project and sustainability appraisal of scientific research projects does not capture the opportunity and value of the underlying science mission. The indicators must therefore not be used to compare research projects and to take an investment choice between several projects.

Research infrastructures should foresee accompanying periodic sustainability monitoring and tracking throughout their lifetimes and implement an iterative improvement process following the standard `Plan-Do-Check-Act' principle either at infrastructure level or for new programmes and projects.

Life Cycle Assessment (or analysis) (LCA)~\cite{iso14040:2006,ILCD:2010a}, as discussed in section~\ref{sec:lca}, is a methodology that is suitable for individual project segments. It allows assessment of a set of environmental aspects that are relevant for achieving sustainability. Its goal and scope must be well-defined and its results must be integrated in the overall project and sustainability appraisal. When carrying out an LCA, care must be taken to be as specific as possible and to exhaustively document the assessed scenario variant and version, the assumptions, the input parameters, the data quality and the allocation procedure (algorithms and tools) to assure that the results have a meaningful value for appraisal. 
For a comprehensive sustainability assessment, a quantitative Cost-Benefit Analysis (CBA) method based on state-of-the-art economics knowledge that integrates total costs, negative environmental externalities, industrial, social and environmental benefits is the preferred approach. A description of the approach is provided in Annex~\ref{annex:CBA}.

An example of costs (left column) and benefits or mitigation measures (right column) associated with an accelerator-based facility, which can be considered in a comprehensive sustainability assessment, includes the following: 

\setlength{\columnsep}{-1cm}
\begin{multicols}{2}

\noindent{Direct costs for:}
\begin{itemize}[nosep]
\item Operation
\item Maintenance and repair
\end{itemize}
Shadow cost of carbon for:
\begin{itemize}[nosep]
\item Construction 
\item Equipment 
\item Energy consumed 
\end{itemize}

\columnbreak
\noindent{Revenues from:}
\begin{itemize}[nosep]
\item Research and industrial services
\item Space rental
\end{itemize}

\noindent{Carbon tax paid for:}
\begin{itemize}[nosep]
\item Construction 
\item Equipment 
\item Energy consumed 
\end{itemize}
\end{multicols}
\begin{multicols}{2}
\noindent{Cost of:}
\begin{itemize}[nosep]
\item Land and forest consumed 
\item Environmental pollution 
\item Effects on water bodies 
\item Effects on reduced air quality 
\item Loss of life quality due to nuisances \newline (e.g. noise, dust, vibration, traffic impact) 
\item Loss and impact of natural habitats 
\item Loss of biodiversity 
\end{itemize}
\columnbreak
\noindent{Value of:}
\begin{itemize}[nosep]
\item Scientific products
\item Training and education effects (e.g. lifetime salary premium, increased life quality) 
\item Industrial spillovers from procurement
\item Open data and software
\item On-site tourism
\item Cultural activities
\item Media presence including web and social media 
\item Newly created land (e.g. renaturalised wastelands, agricultural spaces created, parks and recreational areas created)
\item Supplied heat
\item Avoided carbon emissions (e.g. from waste heat and newly created renewable energy sources)
\item Newly created jobs
\end{itemize}
\end{multicols}
\clearpage
\begin{multicols}{2}
\noindent{Cost of:}
\begin{itemize}[nosep]

\item Loss of cultural and natural heritage 
\item Accompanying and compensation measures 
\item Loss of agricultural income
\end{itemize}
\columnbreak
\noindent{Value of:}
\begin{itemize}[nosep]

\item Newly created cultural and natural heritage 
\item Increased life expectancy (e.g. in case of medical use of particle accelerators) 
\item Licenses and patents
\item Clean water produced and made available 
\item New or strengthened community services (e.g. fire fighting, health, emergency and security services) 
\item New or reinforced infrastructures: communication, electrical, water treatment and supply, transport, meeting spaces
\item New or reinforced regional economic activities
\end{itemize}
\end{multicols}

\begin{figure}[htb] 
\centering
  \includegraphics[width=.5\textwidth]{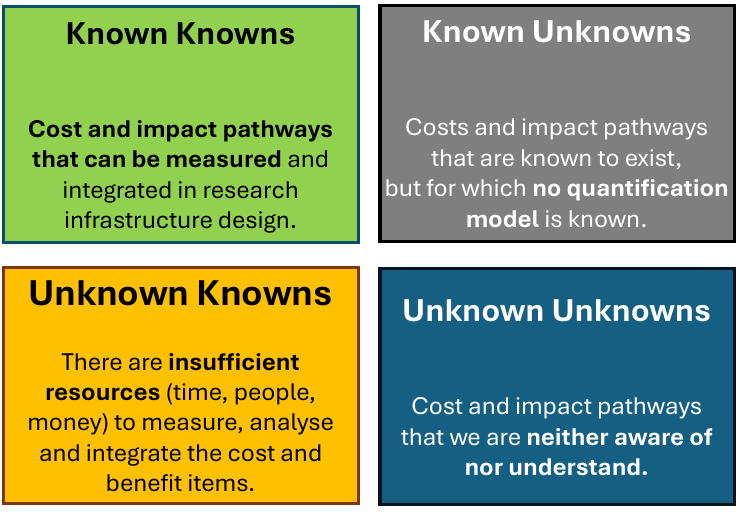}
  \caption{The Rumsfeld matrix~\cite{Krogerus_2012} applied to the identification of costs and benefits of research infrastructures.}  \label{fig:rumsfeld}
\end{figure}

\begin{figure}[ht]
    \centering
    \includegraphics[width=.5\textwidth]{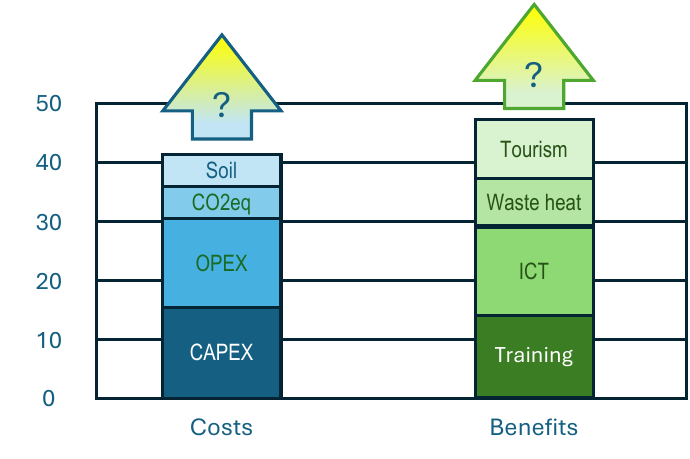}
    \caption{Examples for the accounting of full costs and benefits for an integrated sustainability analysis. Only the identified cost and benefit items that can be converted in monetary terms are considered in the assessment. The question concerning complete coverage of all negative and positive externalities is challenging and incomplete assessment needs to be accepted.}
    \label{fig:fullcoverage}
\end{figure}

A comprehensive sustainability assessment -- as recommended by the Organization for Economic Cooperation and Development (OECD)~\cite{pearce2006cost, oecd2018cost}, the French government~\cite{quinet2014evaluation}, the UK government~\cite{UK_greenbook:2022}, and the European Commission~\cite{sartori2014guide} -- should consider all of the above elements. In practice, not all contributing elements are relevant for a specific investment project, not all elements can always be reliably quantified and not all elements are known or understood (\autoref{fig:rumsfeld}). Both costs and benefits  
are affected by uncertainties (\autoref{fig:fullcoverage}) and a disclaimer or conclusion of any research infrastructure sustainability report should indicate this fact.
It is important to accompany the sustainability assessment with a separate document in which the scope and the considered aspects are described together with the assumptions used for the quantification of the aspects and their conversion in monetary terms.

Rationalising sustainability assessment using standardised cost-benefit analysis~\cite{OMAHONY2021106587} (Annex \ref{annex:CBA}) helps identify both 
limiting and enabling factors, 
and can guide the design and iterative evolution of the research infrastructure to enhance long-term sustainability. The approach accounts for time in a rigorous way, which is essential when considering effects on the environment at large. While  cost items including negative externalities are typically well defined in the existing guidelines, the benefits and positive externalities are not exhaustively captured by a catalogue or taxonomy. They are project specific and need to be identified and captured on a case-by-case basis.

\begin{figure}[htb]
    \centering
    \includegraphics[width=0.8\textwidth]{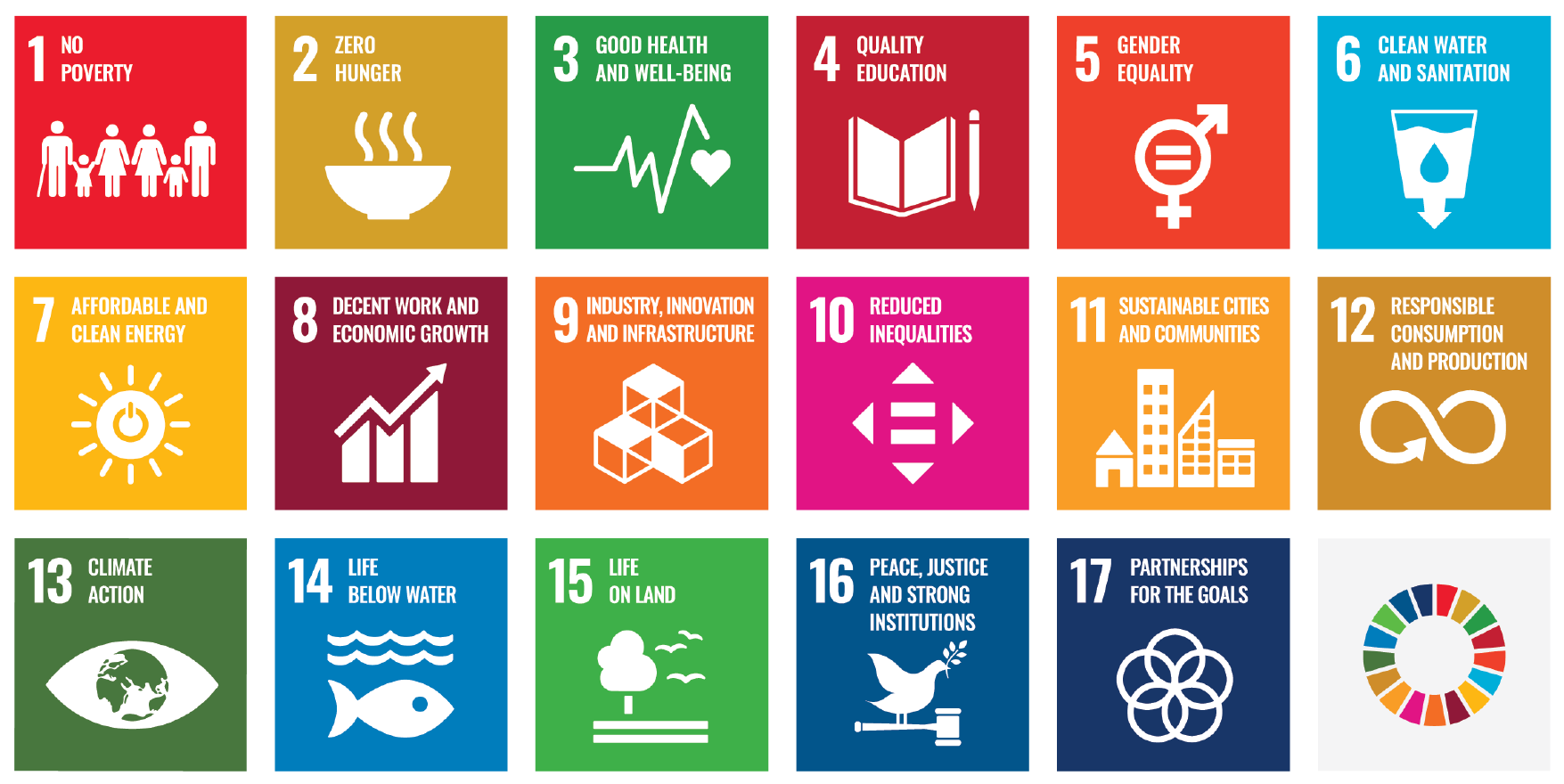}
    \caption{The UN Sustainability Development Goals (SDG)~\cite{UNSDG}.}
    \label{fig:UN_SDG_Goals}
\end{figure}

Sustainability might refer narrowly to the internal sustainability of the project. In reality it refers more broadly to a range of external economic, social or environmental factors  which can be influenced by the project. The 2015 UN Sustainability Development Goals (SDGs) consist of 17 high-level development objectives with more than 160 sub-objectives. \autoref{fig:UN_SDG_Goals} illustrate the breadth of those factors.

Organisations that develop, construct and operate particle accelerator based research infrastructures are encouraged to identify and document quantified positive and negative impacts considering and leaning on the UN Sustainability Goals~\cite{un-sdg-com} as a reference matrix 
for the the objectives to be pursued. 

Although quantitative sustainability analysis cannot provide full coverage, each of the UN SDGs can be regarded as a catalogue of potential negative and positive impacts that can be looked at in terms of causal relationship with the project. The global indicators and monitoring framework for the SDGs~\cite{sdg_transformation_center}  were designed to support countries, not research organisations,  in developing strategies~\cite{UNSDGMonitoringFramework}. They are therefore not directly usable for the reporting of scientific programmes and projects. A research infrastructure can, however, report for each UN goal activity relevant positive and negative impacts in quantitative terms as far as they have been assessed. This approach has for example been implemented with CERN's periodic environment report that relies on the GRI framework~\cite{CERN_env_report_GRI_SDG}. In the absence of established guidelines, the SDG tracker~\cite{owid-sdgs} and the European Commission SDG information site~\cite{ec-sdg} provide good starting points for indicator recommendations based on the UN SDG goal specification. 
It is not expected to give evidence of compliance with specific national or international targets. Rather, the goal is to show how the research infrastructure affects the environment at large in negative ways and how it contributes in very tangible ways to each of the goals. Listed below are some 
non-exhaustive suggestions to be considered during the sustainability  assessment of a research infrastructure and how reporting could be done.

\begin{enumerate}
    \item \textbf{End poverty} in all its forms everywhere.
    \newline \textit{Positive impacts}: Engage people from countries in which more than \qty{30}{\percent} live below the poverty line, \textit{indicated by} number of such people engaged.
    
    \item \textbf{End hunger}, achieve food security and improved nutrition and promote sustainable agriculture.
    \newline \textit{Negative impacts}: Agricultural production decrease by consumption of protected agricultural land.
    \newline \textit{Positive impacts}: Agricultural production increase by new agricultural land created.
    \newline \textit{Indicators}: Amount of land consumed or created.

    \item Ensure \textbf{healthy lives} and promote well-being for all at all ages.
	\newline \textit{Negative impacts}: Catalogue of public health risks (air pollution, water pollution, unintentional poisoning).
	\newline \textit{Positive impacts}: Strengthening of coverage of public health and emergency services.
	\newline \textit{Positive indicators}: Number of personnel and number of equipment contributed to public health and emergency services. 
    
    \item Ensure inclusive and equitable \textbf{quality education}; promote lifelong learning opportunities for all.
	\newline \textit{Positive impacts}: ICT skills improvement, \textit{indicated by} number of people gaining additional skills.
	\newline \textit{Positive indicators}: Number of persons that through the activity gain additional ICT skills.
    
    \item Achieve \textbf{gender equality} and empower all women and girls.
	\newline \textit{Positive indicators}: Number of women and fraction of women vs. men in managerial positions. 
    
    \item \textbf{Clean water} and sanitation
	\newline \textit{Negative indicators}: Total net volume of water consumed and freshwater withdrawal as a proportion of available freshwater resources.
	\newline \textit{Positive impacts}: Improve water quality, wastewater treatment and safe reuse.
	\newline \textit{Positive indicators}: Ratio of the volume of wastewater treated to render it fit to meet environmental or quality standards for recycling or reuse. 
    
    \item \textbf{Affordable and clean energy}
	\newline \textit{Negative indicators}: Amount and fraction of energy consumed that does not come from renewable sources.
	\newline \textit{Positive impacts}: Access and investments in clean energy.
	\newline \textit{Positive indicators}: Direct, indirect and induced financial flows into the creation of new renewable energy sources. 
    
    \item \textbf{Decent work and economic growth}
	\newline \textit{Negative indicators}: Proportion of informal employment in total employment by sector and gender.
	\newline \textit{Positive indicators}: Number of direct, indirect and induced jobs created and quantified cumulative incremental industrial spillover effects. 
    
    \item \textbf{Industry, innovation and infrastructure}
	\newline \textit{Negative impacts}: Carbon intensity, \textit{indicated by} \ce{CO2} emissions per unit of value added.
	\newline \textit{Positive indicators}: Incremental value added indicated by the total industrial spillover generated. 
    
    \item \textbf{Reduce inequality} within and among countries
	\newline \textit{Negative impacts}: Fiscal and social policies that promote equality, \textit{indicated by} the fraction of wages and salaries and social insurance contributions paid with respect to all spendings on engaged personnel.
	\newline \textit{Positive impacts}: Reduce income inequalities, \textit{indicated by} household expenditure growth rate or income growth rate among the bottom \qty{40}{\percent} income of the population. 
    
    \item \textbf{Sustainable cities and communities}
	\newline \textit{Negative impacts}: Urbanisation rates.
	\newline \textit{Positive impacts}: Cultural and natural heritage protection.
	
    \item \textbf{Responsible consumption and production}
	\newline \textit{Negative impacts}: Material footprint, \textit{indicated by} the sum of the weight of the used biomass, fossil fuels, metal ores and non-metal ores.
	\newline \textit{Positive indicators}: Amount of materials re-used. Societal impact estimate of re-use potentials from new technologies developed. 
    
    \item \textbf{Climate action}
	\newline \textit{Negative impacts}: Greenhouse gas emissions
	\newline \textit{Positive impacts}: GHG emission reduction.
	
    \item \textbf{Life below water}
	\newline \textit{Negative indicators}: Surface and type of coastal and marine areas consumed.
	\newline \textit{Positive indicators}: Type and size of coastal or marine ecosystems strengthened or restored. 
    
    \item \textbf{Life on land}
	\newline \textit{Negative indicators}: Type and surface of forest destroyed.
	\newline \textit{Positive indicators}: Ecosystem type and area for biodiversity growth conserved or protected. 
    
    \item \textbf{Peace, justice and strong institutions}
	\newline \textit{Negative indicators}: Percent of public expenditures above approved budget.
	\newline \textit{Positive impacts}: Strengthen the participation in global governance, \textit{indicated by} the proportion of members and voting rights of developing countries in international organisations. 
    
    \item \textbf{Partnerships for the goals}
	\newline \textit{Positive impacts}: Encourage effective partnerships.
	\newline \textit{Positive indicators}: Committed partnerships (types, investments) for public-private infrastructure (energy, ICT, transport, water) partnerships in the frame of the programme or project. 
\end{enumerate}

\subsubsection{Greenhouse gas emissions}\label{sec:carbon}

The recommended method to consider the effect of Greenhouse Gas (GHG) emissions is to identify and quantify all 
project relevant emissions for each year of the observation period, expressed in units of tonnes of carbon equivalent \unit{(\tonne\ce{CO2}e)}. In addition, for each year the project-relevant avoided emissions should be estimated. Then the emission quantities in \unit{\tonne\ce{CO2}e} are converted on a year-by-year basis in monetary terms using the \emph{shadow cost of carbon}~\cite[Sect.4]{EIBCBAGuide2023}, and quantified emissions are discounted using project-specific \emph{social discount rates} (SDR)~\cite[Sect.10]{EIBCBAGuide2023}. 
Finally, the discounted values are summed up, yielding a total ``cost'' of the carbon footprint of the project. 
The entire process is described in ``European Investment Bank Project Carbon Footprint Methodologies''~\cite{EIB_carbon_footprint}.
This document recommends the use of the Intergovernmental Panel of Climate Change (IPCC) 
conversion rates of GHGs based on their 100-year global warming potential into \ce{CO2}e conversion factors. The entire table of the latest 
IPCC conversion factors is available online~\cite{IPCC_GWPs}.

The approach to convert the \ce{CO2}e quantities into monetary terms is based on the least cost to society of meeting the \qty{1.5}{\celsius} temperature goal established in the Paris Agreement~\cite{ParisTreaty}. It is also in line with the empirical evidence that relevant scenarios from large-scale climate-economy models produce.

To help delineate direct and indirect emission sources, three “scopes” (Scope 1, Scope 2, and Scope 3) were defined for GHG accounting and reporting purposes~\cite{world_greenhouse}. Scope 1 direct emissions occur from sources that are owned or controlled by the organisation. Scope 2 accounts for GHG emissions from the generation of purchased energy (electricity, steam, heat and cooling) consumed. Scope 3 is an optional reporting category that catches all other indirect emissions.

For the operation phase of a particle accelerator based research infrastructure only the direct project related (Scope 1) and the indirect emissions (Scope 2 related to energy) are considered. Scope 3 emissions related to the procurement of goods and services are included in the environmental product declarations used in an LCA for the construction and any maintenance and upgrade activities. The Scope 3 emissions during the operation and created by users of the facility are not considered due to the limited possibility to identify and capture them.

Many different carbon shadow cost approaches exist~\cite{oeko_institut_2023}. To our best knowledge, the shadow cost of carbon established by the European Investment Bank is the one that fits best the Paris agreement obligations and it is widely recognised and used in project appraisal. It is the recommended guide to be used for converting \unit{\tonne \ce{CO2}e} in monetary terms (see \autoref{EIB:SCC1} and \cite{EIB_economic_appraisal}).

\begin{table}
\centering
\caption{Shadow cost of carbon under various jurisdictions. Rates in the first row are also published as European Union law 2021/C 280/01 in 2016 Euro values. The second row shows the 2024 time adjusted value to be used for project appraisal today. Note that the \qty{1.5}{\percent} near-term Ramsey Discount Rate is applied to the US rates.}
\label{EIB:SCC1}
\begin{adjustbox}{width=\textwidth}
\begin{tabular}{llccccccccl}
\hline
\textbf{Organisation} & \textbf{Unit per \unit{\tonne\ce{CO2}e}} & \textbf{2020} & \textbf{2025} & \textbf{2030} & \textbf{2035} & \textbf{2040} & \textbf{2045} & \textbf{2050} & \textbf{2060} & \textbf{Source}\\
\hline
EIB and EU & Euro 2016 & €80 & €165 & €250 & €390 & €525 & €660 & €800 & n/a & \cite{EIB_economic_appraisal}\\
EIB and EU & Euro 2024\footnote{Adjusted based on EU-27 GDP deflator 100.0 in 2016 and 126.3 in 2023. It measures the amount to which the real value of an economy's total output is reduced by inflation.} & n/a & €208 & €316 & €492 & €663 & €833 & €1010 & n/a & \\
SGPI France & Euro 2018 & €87 & n/a & €250 & n/a & €500 & n/a & €775 & €1203 & \cite{SGPI:2023}\\
UBA Germany & Euro 2023 & €240 & n/a & €254 & €253 & n/a & n/a & €301 & n/a & \cite{UBA-2024}\\
UK (high) & GBP 2020 & £361 & £390 & £420 & £453 & £489 & £527 & £568 & n/a & \cite{UK_GOV_carbon:2021}\\
EPA USA\footnote{\qty{1.5}{\percent} near-term Ramsey Discount Rate applied} & USD 2020 & \$340 & n/a & \$380 & n/a & \$430 & n/a & \$480 & \$530 & \cite{EPA2023}\\
\hline
\end{tabular}
\end{adjustbox}
\end{table}

Considering that in several countries a carbon tax exists on different products, double counting must be avoided. Therefore, following the recommendations of the World Bank~\cite{World_Bank_carbon_shadow_price}, the project owner needs to subtract the carbon tax paid on the consumed material and products from the shadow cost of carbon used in the CBA-based sustainability assessment on a year by year basis.

No carbon tax is to be subtracted from the avoided amount of \unit{\tonne \ce{CO2}e} and the resulting benefit that is calculated by multiplying the per tonne shadow cost of carbon with the tonnes of \ce{CO2}e avoided.

Carbon emission compensation paid can be considered by deducing the total amount of carbon covered by the compensation from the total amount of carbon emitted. The carbon emission compensation must be entered on cost side of the CBA in the appropriate category (CAPEX or OPEX section). 

The shadow cost of carbon for emitted carbon and for avoided carbon are subject to discounting so that a proper net present value for the 
project can be reported for the set end date of the observation period. The resulting total shadow cost of carbon ($T$) can be calculated by multiplying the residual net value with the per t\ce{CO2}e shadow cost for each year in which the emission takes place. The product-specific and country-specific carbon tax ($C$) paid for any product consumed can then be subtracted from the term $T$. The resulting net shadow cost of carbon is the one to be used in the CBA as negative externality.

\subsubsection{Carbon Footprint Accounting and Reporting}

Projects are encouraged to report their estimated absolute baseline carbon footprint in line with the ``Guideline for a Harmonised Approach to GHG Accounting and reporting''\footnote{\url{https://unfccc.int/documents/268706}} and in line with the ``PCAF Global GHG Standard''\footnote{Partnership for Carbon Accounting Financials (PCAF) web page: \url{https://carbonaccountingfinancials.com/files/downloads/PCAF-Global-GHG-Standard.pdf}} using the following indicators:

\begin{enumerate}[nosep]
\item Description of the baseline scenario boundaries for the carbon footprint accounting considering the physical delineation, the geographical area and the significant sources. 
\item List of the entities who contribute financially and with in-kind contributions to the research infrastructure project, typically countries and 
the population in each of these countries.
\item Absolute emissions: sum of all t\ce{CO2}e emitted for the entire observation period and disaggregated by scope.
\item Absolute emissions per lifecycle phase: sum of all t\ce{CO2}e emitted per individual lifecycle phase and disaggregated by scope.
\item Average absolute emissions: average annual t\ce{CO2}e emitted over the entire observation period and disaggregated by scope.
\item Intensity: average annual t\ce{CO2}e accounted for each contributing country and per capita of the sum of all contributing country, weighted by the sum of the country's financial and in-kind contributions.
\item Net Present Value (NPV) of the total t\ce{CO2}e at the end of the observation period in monetary terms.
\end{enumerate}

Preference should be given to report average annual values rather than a representative year in the observation period.

To obtain the total carbon footprint in t\ce{CO2}e, all identified and estimated emissions are summed up and all identified and estimated avoided emissions are subtracted from the emissions.

The following principles apply:

\begin{enumerate}[nosep]
\item Relevance: the reporting must be relevant for the stakeholders.
\item Completeness: all relevant emissions are accounted and specific exclusions are documented and justified.
\item Consistency: methods used for the accounting must be consistent for all considered emission sources.
\item Transparency: all relevant assumptions, methodological choices, references to the accounting methodologies and data sources are documented.
\item Where data are unavailable, any uncertainty is to be addressed following the principle of conservativeness where it is preferable to over-estimate project emissions and under-estimate baseline emissions.
\end{enumerate}

Scope 3 emissions are only considered where they can be reasonably identified and credibly estimated and where they can be unambiguously accounted to the research infrastructure. 

\subsection{Socio-economic sustainability enablers}
\label{sec:enablers}
\begin{figure}[ht]
    \centering
    \includegraphics[width=0.3\textwidth]{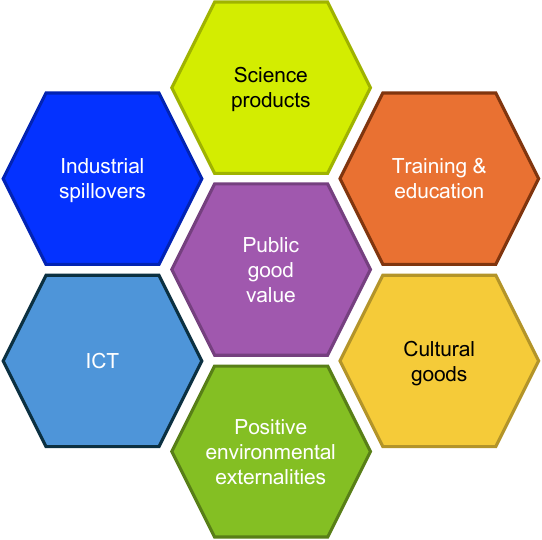}
    \caption{Main socio-economic impact pathways that are sustainability enablers of particle accelerator and particle physics research infrastructures.}
    \label{fig:main_impact_pathways}
\end{figure}

Socio-economic impact pathways are a valuable inventory of sustainability enablers. The RI-PATHS EU funded project has developed a toolkit~\cite{impact_toolkit}, federating research infrastructures across various domains, including also particle accelerator facilities such as ALBA,
CERN and DESY. 
Typical pathways 
for 
accelerator-based facilities are shown in \autoref{fig:main_impact_pathways}. This section outlines how such pathways can directly lead to benefits and positive externalities with specific examples. 

Future projects are strongly encouraged to carry out a comprehensive charting of the impact pathway potentials and develop sustainability-enabling measures around such a systematic exploration and design activity.

\subsubsection{Fundamental Physics Knowledge}

Science, in other words formalised knowledge that is rationally explicable and tested against reality, logic and the scrutiny of peers, is a global public good~\cite{CouncilScience:2021}. Ultimately it is the main output of research infrastructures. 
In our transforming society scientific knowledge is becoming 
the major factor in economic development, 
gradually replacing capital, land and labour. There is evidence that the public sector has and is continuously funding much of the innovative science that stimulates private sector responses~\cite{mazzucato2013entrepreneurial}.
But what is the value of that knowledge that fundamental science projects generate? Are continued investments in new particle accelerator projects for scientific research sustainable in the perception of those people who do not directly use those instruments and who do not directly profit from the generated knowledge? Which means do exist to capture the value of such investments?

Particle accelerators for fundamental science may or may not generate knowledge that can be directly used by society. Other particle accelerators, such as synchrotron light sources, typically serve applied research. This means that a lower bound for the value of the knowledge generated can be quantified in terms of the cost of the publications, the fees charged for the use of the facility and by the resources that the users invest in this research. The approach is based on the assumption that the research results eventually end up in products and services that the society uses. An example is the characterisation of the COVID-19 virus structure at the BESSY-II synchrotron~\cite{Scudellari:2020}. Particle accelerators and colliders that serve curiosity-driven research, such as the LHC or the Electron-Ion Collider, cannot use this model since it is not obvious how the knowledge gain leads to direct societal applications. There exists, however, no doubt that the science outcomes provide benefits for society. This view is shared by the majority of people that constitute society, even if the asset is not directly used.\footnote{The notion for this concept is the `public good' as opposed to the `common good' that is jointly used. Examples for common goods are fish stocks in the ocean and public roads. Examples for public goods are knowledge, natural and cultural heritage, open software and datasets and even tangible assets such as street lighting.}

Several Nobel prizes in economics have been awarded for advances in this domain since it is of fundamental importance for the society and welfare economics.\footnote{J.M.~Buchanan in 1986 for the theory of public choice, R.M.~Solow in 1987 for contributions to the theory of economic growth, A.~Akerlof, A.M.~Spence and J.E.~Stiglitz in 2001 for the work on asymmetric information, E.~Ostrom in 2009 for the economics of common-pool resources, J.~Tirole in 2014 for the analysis of market power and regulation, P.R.~Milgrom and R.B.~Wilson in 2020 for improvements to auction theory and inventions of new auction formats.}

Suitable approaches, such as the contingent valuation also referred to as `stated preference', have been developed for analysing the value of outdoor recreational spaces~\cite{davis1963value}, for assessing levels for environmental protection~\cite{NOAA:1993}, and environmental incident mitigation measures~\cite{ExxonValdez:2003} as well as the investments to protect natural and cultural heritages. The methodology has been transferred to estimating the value of science, first for the LHC project~\cite{RePEc:mil:wpdepa:2016-03}, then for the HL-LHC project~\cite{CERN_Courier:2018} and for the FCC~\cite{GIFFONI2023104627}. The approach has recently been adopted by an atmosphere research infrastructure~\cite{Actris:2023}. Briefly, the value that people associate with a science project can be captured by estimating their willingness to financially participate (WTP) using well-established approaches in economics.

We suggest to carry out a public good value analysis for future large scale research infrastructure projects at a very early stage. 
This will provide a clearer understanding of how the public perceives the scientific vision and help test the validity of the public funding sustainability hypothesis -- namely, what level of periodic investment is considered justified in the public’s view.
 To design, plan and carry out a public good value estimation, domain experts and qualified companies need be engaged. It is 
essential that such analyses are conducted 
independently, free from the 
influence of the project owner and in accordance with 
internationally recognized ethical 
guidelines and the quality standards required for 
survey-based research. 
Examples of this approach can be found in Refs.~\cite{secci_2023_7766949, GIFFONI2023104627}.

Depending on the number of participating countries or funding agencies, 
such a survey can require substantial resource allocation. 
Economists use therefore techniques based on the `Benefit Transfer Method'~\cite{Johnston:2015} to estimate the perceived value of a large population based on the identification of only a few significant parameters that can be derived from a limited number of samples. These parameters may be different for different projects. Therefore such a study requires a pilot phase to determine the significant parameters followed by a mass survey to determine the value that the public associates with the investment.

If the estimated willingness to financially participate (WTP) is larger than the actual or foreseen contribution to the planned new project, the public good value of the funded infrastructure can be considered consistent with the financial participation and can therefore also be considered justified. 

\subsubsection{Sustainability through education and training}

Particle accelerator and experimental physics research infrastructures offer the possibility to engage people at all education and training levels, from apprentices to postdoctoral researchers. If people have the opportunity to actively participate in the design, construction and operation of such projects and programmes they enjoy benefits that translate directly into a \SI{2} to \SI{10}{\percent} lifetime salary premium~\cite{csil_2024_10653396, Camporesi_2017} compared to their peers that are enrolled in conventional training programmes at schools and universities. A dedicated programme to integrate the persons after their active time in a research facility into the labour market in their country or region of origin can significantly improve the return on investment for the participating country or region~\cite{crescenzi_2024_13166167}.

\subsubsection{Sustainability through engagement of international science communities}
Engaging a significantly large and continuously growing community of scientists and engineers in the form of collaborative research projects over long periods of time assures long-term sustainability of a science project or programme. The reasons are that the personnel costs are distributed over many contributing organisations and countries, that a stable generation of scientific products~\cite{csil_2024_13920183, MORRETTA2022121730} (i.e. books, peer-reviewed articles, pre-prints and technical reports, proceedings and presentations) can be assured, that the knowledge can be effectively transferred into lasting education curricula through publications and direct training over generations, and that impact can be generated through the training of early career professionals (see previous section). High value is mainly achieved by assuring that the science products generated directly by the research project or programme are effectively taken up and cited by an even larger second tier community. Open Access 
to trusted (peer reviewed) publications is a prerequisite to enable research infrastructure sustainability through their scientific product generation. Finally, the concept of Open Innovation -- engaging a broad range of knowledge domains around a scientific core mission -- will assure that the 
knowledge generated 
through the challenges of the science mission will eventually spill over into areas of direct societal relevance~\cite{Gutleber2025}. 

\subsubsection{Sustainability through industrial spillovers}

Industrial spillovers from science projects generate benefits directly for industry and society~\cite{Griniece:2020, Castelnovo:2018, Florio:2016}. This impact pathway is most effective if science projects co-construct their instruments and infrastructures with industrial partners~\cite{Florio:2818, Autio:2004, Nordberg:2003}. It is most effective when companies work closely with the research project to develop technologically intensive solutions and deliver non-standard services~\cite{Sirtori:2019}. This approach is more laborious than a conventional client/supplier relationship since it requires exchanges of ideas and knowledge, the development of integrated processes, the mutual adaptation of working methodologies and risk sharing. This,  
however, leads to lasting creation of processes, products and services that the industrial partners can leverage in other markets with earnings multipliers above 3 and effects that can last between five and eight years. Lasting territorial effects have also been reliably documented~\cite{Crescenzi2025, Moretti:2010}.

Sustainability also depends on the design of the science mission: short lived and non-upgradable programmes will lead to lower and shorter lasting industrial spillovers than long-lasting programmes that are characterised by periodic upgrades and operation efficiency improvement measures that engage industrial partners in continuous challenge-based activities.

Industrial spillovers can also lead to lasting positive environmental externalities. The challenge to reduce the carbon footprint and environmental impact of the construction of a new particle accelerator facility creates an opportunity for industrial innovations in numerous construction-related domains. For instance it creates a potential of developing or putting in place low-carbon concrete and other construction materials production facilities. It serves as pilot platform for improved construction techniques, for example using natural resources such as wood and compressed earth. Showcasing the application serves creating market interest. Developing processes and products for the use of excavated materials can generate benefits significantly beyond the needs of the research infrastructure project since the management of construction waste is a challenge that the society faces and for which conventional construction projects typically have insufficient time and budget. Eco-design based industrial architecture is another emerging discipline for which particle accelerator facilities serve as suitable early adopters.
Environmental benefits can also be generated in the area of technical infrastructures that are developed with industrial partners. They range from more efficient refrigeration systems which find their application for example in gas liquefaction and transport over more efficient water cooling systems (adaptive water intake, use of waste water) to improved electrical systems (loss reduction via DC-based systems, high-speed power control management, short and medium term energy storage, adaptive and machine learning infrastructure operations, waste heat buffering and supply). 

\subsubsection{Sustainability through open information and computing technologies}

The development of Information and Communication Technology (ICT) and the generation of widely available data not limited to scientific results (e.g. engineering test data, operation monitoring, system tests, etc.) are essential outputs of particle accelerator based research infrastructures~\cite{FLORIO201638}. Putting in place a global data sharing and processing infrastructure has already led to the creation of numerous, openly accessible software packages, platforms and online services, which are also used in environments outside high-energy physics. They range from scalable data storage and distribution middleware over data analysis and visualisation to data management and workflow systems. Also, openly accessible software has been developed with value that extends beyond the scientific collaborations. Examples include innovative Cloud computing services (e.g. Helix Nebula Science Cloud serving five scientific research domains), meeting and event management software (e.g. Indico), particle/matter interaction modelling and analysis (e.g. Geant4, FLUKA and ActiWiz), and electronic library and information access software (e.g. Invenio and Zenodo). Long term data preservation is another technology domain with high societal relevance that is gaining importance and which is primarily driven by fundamental science research infrastructures.

Particle accelerator and high-energy physics research projects are strongly encouraged to create an inventory of potential ICT tools that can be made available openly and free-of-charge for societal use. The feasibility to quantify the societal impact of those solutions relies strongly on an accounting of the software uptake and use (e.g. number of installations, number of integrations in commercial and other open software packages) over sustained periods of time. Today there is a lack of such accounting, which makes quantitive estimation of the societal impact challenging and labour intensive. However, it is important to put a focus on this topic, since ICT tools developed in the scientific environment have a major impact, with significant and tangible value potentials~\cite{CrespoGarrido:2025a, CrespoGarrido:2025b}.

Open and freely available software developed and maintained by science projects and programmes are sustainability enablers for research infrastructures. Their investment and continuous development and maintenance costs are marginal with respect to the societal benefits they can generate over decades. Future particle accelerator projects and programmes should put a focus on properly managing and promoting ICT developments and make sure they are taken up by the society through a technological competence leveraging process and sustained on-line presence in potential user domains.

\subsubsection{Sustainability through cultural goods}

Creating an interest in science among all citizens is part of the mission of any research infrastructure.
Projects and organisations can develop a broad variety of activities to attract general public, for instance through permanent and travelling exhibitions, open days, guided tours, engagement with schools and teachers in joint workshops, citizen science projects, web sites, social media, engagement with video bloggers, on-line and TV documentaries, art internships, common art projects, feature movies, books, science fairs, presence in radio and TV shows and much more. The limit of cultural good creation is the limit of the imagination of an ever changing and diverse group creative people that are best engaged over sustained periods of time.

Each of these cultural goods have the potential to generate value for the society. Through sustained creation of interest in the society and by de-constructing barriers and fear from the science 
cultural goods also contribute to the increase of the public good value.

It is important for particle accelerator and physics projects to engage lay people in playful and entertaining ways rather than aiming at education and teaching, which is a different socio-economic impact pathway. The two should be kept separate, although effective cultural good creation and engagement raises the possibilities for explaining the underlying science in a second, subsequent step.

The value identification for cultural goods can be very different for each good. Therefore, future projects and programmes will need to focus initially on few cultural pathways. If the estimated value of a pathway turns out to have sufficient potential it should be further developed  with ongoing monitoring to track the evolution of its value. 

The value of on-site visitors represents a cultural good of science infrastructures that can generate substantial tangible economic value in a sustainable way~\cite{CrespoGarrido:2025c, NASA_economic_snapshot:2024}, including the environment and nature discovery~\cite{Weaver:2011, Wyk-Jacobs:2018}. It should therefore be considered for development first.

Social media presence is the most impactful on-line cultural good today and should therefore also be considered with priority~\cite{Bastianin2021}. It is particularly important to create a sustained interest in science and to explain in which ways science impacts the environment and people's everyday lives. Different means of communication are needed for different generational and socio-economic groups~\cite{blekman2024crowdsourcedparticlephysicsstories}.

Finally, citizen science projects in the periphery of the physics science mission are effective tools to engage lay people and to re-assure the environmental and territorial compatibility of the particle accelerator based research infrastructure. Examples are the 
initiatives to create biodiversity inventories of the surface sites, and to improve the quality of habitats on and around sites.

\subsubsection{Sustainability through positive environmental externalities}

Any future particle-accelerator based project has the potential to create positive environmental externalities that can compensate the residual, unavoidable and not further reducible negative environmental effects. Before developing compensation and accompanying measures, negative effects on the environment have to be avoided and if they cannot be avoided be reduced as far as they are compatible with the requirement to achieve the goals and objectives of the infrastructure within accepted cost and schedule constraints.

National and international legal and regulatory frameworks define boundary conditions for the avoidance, reduction and compensation approach. The recently adopted update of the law for research and innovation in Switzerland for instance explicitly requires the consideration of the national and regional climate protection plans and energy-related aspects~\cite{fedlex:24.029}. Other countries, such as France have already encoded the fight against climate change, resource and biodiversity protection, circular economy and sustainable territorial development in the conventional environmental protection laws (L110-1 of~\cite{CodeEnvironnementFrance} and~\cite{Loi_2021_1104}) that govern the authorisation of new projects~\cite{Loi_2021_1104}. Hence, environmental is always to be interpreted in a wide sense. The greenhouse gas emission reduction goals defined by the Paris Agreement are to be integrated in new projects at EU level and due to the translation of that regulation into the national laws~\cite{EU:2018_1999}.

If properly planned, evaluated and monitored, collective compensation can even lead to a net positive effect~\cite{Cerema:2018}. However, some countries, such as for instance Switzerland do not foresee such approach and require 1:1 compensation in the same region~\cite{BAFU:2022, Fedlex:451}. Despite this constraint the overall environmental performance may still be neutral or positive from a socio-economic assessment point of view if the positive environmental externalities are properly evaluated and integrated in the net present value of the project.

Some typical, positive environmental externalities that particle-accelerator facilities can integrate in their designs from the onset are:

\begin{itemize}[nosep]
    \item Re-creation of agricultural spaces by transferring top soil to low quality land plots and wastelands.
    \item Creation of green spaces with covered roads, fertilising wastelands and backfilled quarries, creation of parks, greening of roofs, reforestation around sites, creation and quality improvement of natural habitats and wetlands.
    \item Increase of biodiversity by creation of new habitats on domain of the research infrastructure.
    \item Improvement or creation of new ecological corridors, green and blue continuities.
    \item Creation of forests with climate change adapted trees and plants.
    \item Introduction of forest management in view of protection against wildfires.
    \item Improved water management of existing, but low quality water streams.
    \item Creation of water reservoirs.
    \item Creation of raised hedges to fight soil erosion and create new habitats.
    \item Creation of soft or multimodal mobility concepts for use beyond the research facility.
    \item Creation and improvement of natural habitats and nature protection zones on the research infrastructure sites.
    \item Carbon footprint reduction by helping to avoid fossil fuel use through waste heat supply.
    \item Supply of raw water for non-drinking purposes by supplying purified waste-water when not used by the research facility.
    \item Increase of renewable energy resource capacities through long-term power purchase agreements and energy purchase communities.
    \item Development of products and processes that leverage circular economy principles and low environmental impact technologies that spill over into the industrial domain (e.g. in the areas of electrical substations, construction materials, architecture solutions, power transmission and buffering, industrial cooling).
    \item De-construction of no longer used infrastructures from previous projects and programmes for the purpose of creating environmental and societal value.
\end{itemize}

\subsubsection{Innovation and R\&D}
New accelerators and the experimental detectors operation at the collision points are projects spanning several decade while using cutting-edge technology. As a consequence, they tend to be based on technology that does not yet exist at the time of design. This process continues for the duration of the operation as detectors are upgraded. The required technical advances may be designed in-house and/or in partner institutions (such as universities) in conjunction with external companies. The development can take the form of joint design, or procurement of technologies to be developed by the company.  
In both cases the industries profit from a technological knowledge transfer which may lead to technological breakthrough, patents or new business opportunities~\cite{bastianin2019technologicallearninginnovationgestation,Sirtori:2670056}. Such processes are encouraged and supported by dedicated knowledge transfer (KT) centres.\footnote{{\it e.g.} \url{https://kt.cern/} at CERN, \url{https://innovation.desy.de} at DESY, \url{https://partnerships.fnal.gov/} at Fermilab, \url{https://web.infn.it/TechTransfer/} at INFN, {\it etc}.} While not being the primary goal of the research programme, such created knowledge is often seen as the most valuable output by decision-makers, which significantly affects the social sustainability of a project.

\subsubsection{International collaboration}

If, like humans, accelerators have soft skills, international cooperation and its diplomatic potential are prime examples. 
They are accountable and inclusive enterprises with a unique scientific mission and thus encourage countries to create common strategies at a ministerial level to pursue joint objectives. 
Developing a coordination network among different geopolitical actors entails agreements not limited to science and technology.
As recently seen with the ceremonies of CERN's 70th anniversary\footnote{\url{https://cern70.cern/cerns-70th-anniversary-ceremony/}.}, heads of state and ministers come to visit the site and have bilateral discussions that may or may not be related to the scientific mission of the project, or even to science. National news outlets have reported on the event both focusing on the common project, and on these side aspects. Often who discusses with whom is seen as more important as the event itself. Collaboration leading to more collaboration, there is a societal interest to bring decision makers together to discuss issues of the would in an informal setting. CERN achieves this for Europe and beyond. The SESAME project in Jordan\footnote{\url{https://www.sesame.org.jo/}} was set up with international collaboration in mind, as much or even more than the proposed scientific project.

\subsection{Life Cycle Assessment}
\label{sec:lca}

Future accelerator projects are encouraged to perform Life Cycle Assessments (LCA) at each stage of design and construction. They should be performed to understand the project's baseline environmental impact factors, to identify areas of improvement of the impacts, and to prevent those impacts through optimised design.
This section introduces the measurement of environmental impact factors through an LCA and discusses the LCA landscape. 
Recommendations are made in the context of assessing the sustainability of the projects of future particle accelerators with consideration of observations and highlights.

\subsubsection{What is an LCA?}

A Life Cycle Assessment (LCA) is a methodical approach to quantify the environmental impact of a system, product or organisation.
The 14040 suite of ISO standards~\cite{iso14040:2006}, in particular ISO 14044~\cite{iso14044:2006}, defines how such an assessment is conducted. 
It consists of four stages:
\begin{itemize}[nosep]
    \item Definition of goal and scope,
    \item Life Cycle Inventory (LCI),
    \item Life Cycle Impact Assessment (LCIA), and
    \item Life Cycle Interpretation.
\end{itemize}
In the following subsections, each of these stages are discussed briefly in the context of future accelerator facilities.
Notably, an LCA is an iterative process, which means that new findings can be implemented in previous stages of the LCA at any point in the process and evaluated again.

\subsubsection{Goals}

The motivation and goals for a full or partial LCA depend on the phase of an accelerator project, ranging from an early concept phase or engineering design phase to the construction, operation phase or decommissioning stage.
Accelerator projects are often large and require substantial resources, which in turn come with substantial environmental impacts.
Many existing and planned accelerator laboratories worldwide are already committed to the goal 
of conducting their scientific mission in a responsible and sustainable manner~\cite{CERN:2023a, GreenILC, CCC, ISIS-II, ESS}. 
Performing LCAs for proposed or operating facilities contributes towards this goal as they provide a solid basis to reduce environmental impact through optimising the design and operation.

In the planning phase, LCAs permit comparison and selection of design alternatives.
During this phase, the potential for carbon and environmental impact reduction is at a maximum, as design decisions are still cheap to make or change~\cite{pas2080:2023}.
In the engineering phase, LCAs support decision making on a more detailed engineering level, and are part of the preparation for project approval, where nowadays a quantification and justification of carbon emissions is a legal requirement. 
More generally, LCAs help to demonstrate a laboratory's commitment to sustainability to the state and the public. Approval of large publicly-funded infrastructure projects such as accelerators is conferred directly from funding bodies, but also indirectly via a ``license to operate'': we need to ensure that the public considers its taxes to be well-spent. This is particularly important for particle physics, where the public benefit may not be immediately obvious, and the general public may legitimately ask existential questions about Big Science projects that have a large environmental footprint.

Thus the goals of performing an LCA may include a desire to quantify and optimise environmental impact at an early project stage, a wish to reduce environmental impacts where feasible, the provision of input to a selection and decision process, the fulfilment of legal requirements for the approval of a project, the improvement of public approval of project proposals. 
These goals would have an impact on the methodology and metrics adopted for the LCA, including the scope and the impact categories studied.

For instance, projects may require an assessment according to EN~17372~\cite{en17472:2022} as part of a wider Environmental Impact Assessment for the authorisation process. EN~17372 considers only greenhouse gas (GHG) emissions measured as \ce{CO2} equivalent, no other impact categories. It still adheres to the requirements of ISO~14040/14044, so it constitutes a full LCA with just global warming potential (GWP) as the sole impact category.

\subsubsection{Functional units}

The scope definition of an LCA addresses three aspects:
\begin{itemize}[nosep]
    \item the functional unit, i.e. the system or project under consideration,
    \item the system boundary, i.e. the lifecycle stages considered, and
    \item the methodology, e.g. which impact categories are considered or whether global or regional impact factors are to be used.
\end{itemize}

For an accelerator or similar project, the functional unit would be the accelerator, 
including detectors and supporting infrastructure. 
Ideally, to compare project proposals with similar physics goals on an equal footing, a compatible functional unit definition should be adopted. It should specify:
\begin{itemize}[nosep]
    \item which accelerator systems are considered and how to treat pre--existing systems, e.g. only the main accelerator or injection systems and pre--accelerators,
    \item which program phases are included, e.g. only considering the first run phase, or subsequent upgrades (which may or may not be part of a baseline project proposal),
    \item which supporting infrastructure is considered, e.g. surface buildings, cryogenic plants, computing centres, power stations, water treatment plants, or roads.
\end{itemize}

Other functional units have been proposed, defined in terms of physics outcome, such as the production of a single Higgs boson~\cite{Janot:2022jtn}. 
While a functional unit that is related to a desired physics result could be attractive, such a narrow functional unit definition neglects the broader spectrum of physics topics that motivate the construction of an accelerator facility. It neglects the definition of the actual scope of the project, especially if it does not specify the assumed or intended operation time, and it only makes sense in a very limited analysis that concentrates solely on Scope 2 (electric power) emissions. In addition, presenting a functional unit based on a physics result should be avoided due to the risk of `greenwashing' or `science-washing'. Therefore, we do not recommend the sole use of such functional units. Absolute environmental impacts are recommended as a minimum in an environmental reporting structure.

\subsubsection{Boundaries}

\begin{figure}[ht]
    \centering
    \includegraphics[width=\textwidth]{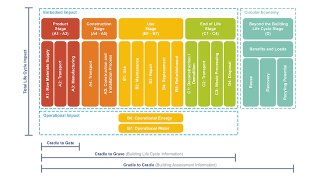}
    \caption{Lifecycle stages according to EN~15978~\cite{en15978:2011}, from D. Overbey~\cite{Overbey:2021a}, by author's permission}
    \label{fig:lca-stages}
\end{figure}

System boundaries are defined in relation to a lifecycle model.
For instance, the European standard EN~15978~\cite{en15978:2011} defines a lifecycle model of buildings (\autoref{fig:lca-stages}) in a manner that is also suitable to other large scale projects such as accelerators, with four main phases, i.e.
\begin{itemize}[nosep] 
    \item product stage, the production of raw materials and components,
    \item construction stage, transport to site and installation,
    \item usage stage, and
    \item end of life.
\end{itemize}
Depending on the phases covered by an LCA, we speak of ``cradle-to-gate'' (raw material production), ``cradle-to-grave'' (full life including operation and demolition) or ``cradle-to-cradle'' (including recycling into the same use) analyses.

Analyses that consider particular processes that are part of a larger production chain are sometimes referred to as ``gate-to-gate''. 
These may constitute an important contribution to a larger database of LCA data and help to get a more complete picture of accelerator relevant processes. 
Relevant examples in the context of accelerators are assembly of cryomodules including clean room operations, special production techniques such as electron beam welding, or treatments such as electropolishing of cavities. 

For large accelerator projects, including the decommissioning or demolition stage in an LCA poses particular challenges: important systems that are constructed for one project may be reused in subsequent projects, possibly after refurbishment. 
Prominent examples are the Proton Synchrotron (PS) at CERN that has been in continuous operation since 1959, undergoing a number of refurbishments, or the LEP tunnel that was reused to host the LHC.
Similarly, the FCC proposal foresees the installation of FCChh after the completion of the FCCee program, reusing much of the civil infrastructure.
Linear collider projects typically foresee building and operating an initial stage, and extending the facility later on to achieve higher luminosity or energy, reusing most of the initial civil infrastructure as well as accelerator and detectors.

\subsubsection{Scope}

A related, though slightly different way to categorise LCA scope has been introduced in section~\ref{sec:carbon}. It is defined in the Greenhouse Gas Protocol~\cite[Sect. 4]{Ranganathan:2004a}. This definition is widely used for reporting emissions during operational stages, in particular of an organisation or company. They are referred to as Scopes 1, 2 and 3:
\begin{itemize}[nosep]
    \item \textbf{Scope 1} refers to emissions from sources under the direct control of an organisation (e.g. from combustion of fossil fuels for transport or heating, or emissions from leaky gas systems).
    \item \textbf{Scope 2} covers emissions from bought-in energy - this almost exclusively refers to grid electricity consumption.
    \item \textbf{Scope 3} covers all other indirect emissions that are a consequence of an organisation's activities, but occur from sources outside of its direct control. For instance: embodied emissions in manufactured goods, commuting, business travel, data transfer and processing.
\end{itemize}
The biannual CERN environmental report~\cite{CERN:2023a} 
has been extended over the years to also cover Scope 3 in its reporting. 

In the context of an accelerator project, Scope 2 and Scope 3 are usually the largest contributors; Scope 2 is dominated by the electricity needs during operation, while Scope 3 covers the emissions caused by the manufacturing of accelerator components and raw material production for the construction of buildings, tunnels and shafts.

\subsubsubsection{Quantifying impacts \label{sec:quantifying_impacts}}

Defining the methodology entails a choice of the impact categories to be evaluated, and the specifics of how the impact is calculated, for instance whether global averages or local impact factors are used for electricity or raw materials.

Emissions of substances influence the environment in different ways: greenhouse gas emissions change the energy imbalance of the earth, causing global warming; emission of fluorocarbons causes ozone layer depletion; emission of phosphate into water causes eutrophication (over-enrichment of nutrients).
These different pathways of environmental impact are referred to as impact categories.   

The impact of emissions on climate change, quantified as global warming potential (GWP), is by far the most prominent consequence of emissions in public debate and legislation. GWP is one of the impact categories used in LCA. There are several atmospheric gases known to influence global warming, of which \ce{CO2} is the best known.
The overall warming caused by a specific substance depends on its  radiation absorption profile and how long it remains in the atmosphere, which may in turn depend on the location of the emission. 
To make emissions of different substances comparable, weighting factors are used that describe the impact effect of a specific substance in relation to a standard substance. 
Thus, the GWP over a 100 year period ($\mathrm{GWP}_{100}$) is given in relation to the emission of \ce{CO2} and thus measured in kilograms of equivalent \ce{CO2} emission, \unit{\kg\ce{CO2}e}. 

A number of methods and standards exist that specify how GWP should be evaluated and reported, among them:
\begin{itemize}[nosep]
    \item The Greenhouse Gas Protocol GHP~\cite{Ranganathan:2004a}
    \item European standards EN 15804 ~\cite{en15804:2022} and others
    \item British Standards Institution (BSI) Publicly Available Specification PAS 2080~\cite{pas2080:2023}
\end{itemize}

These methods are all based on the approach defined by the IPCC in its fifth assessment report~\cite[Tab.~8.A.1]{IPCC:AR5:WG1:Ch8}, using $\mathrm{GWP}_{100}$, taking into account the effect of feedbacks.
There exist other metrics  to quantify the climate impact of greenhouse gases, in particular the concept of Global Temperature-change Potential GTP\footnote{ 
GWP is defined in terms of integrated radiative forcing, while GTP considers temperature change at the end of a reference period. 
For green house gases with lifetimes that are small compared to the reference period, the temperature impact also decays and thus GTP de-emphasises the impact of those gases compared to GWP.}, that are discussed in the fifth~\cite{IPCC:AR5:WG1:Ch8} and sixth IPCC assessment reports~\cite[Sec. 7.6.1]{IPCC:AR6:WG7:Ch7}. However, in the context of LCA, GWP is by far the most widely used metric.

\begin{figure}[ht]
    \centering
    \includegraphics[width=0.6\textwidth]{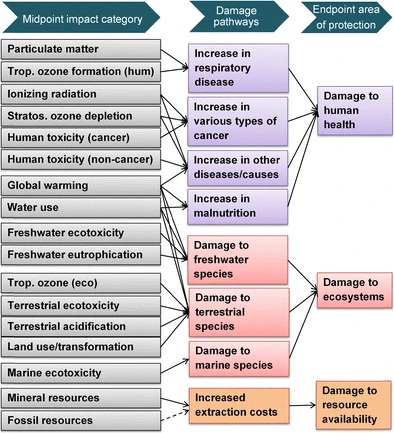} 
    \caption{ReCiPe 2016 impact categories, from~\cite{Huijbregts:2017a} (CC-BY-4.0).}
    \label{fig:ReCiPe2016}
\end{figure}

Beyond GWP, the so-called midpoint impact categories (\autoref{tab:lca-categories}, \autoref{fig:ReCiPe2016}) provide a quantitative way to compare emissions of different substances with respect to the same impact pathway, indicating for instance, that the emission of \qty{1}{\kg} of methane causes 25 times more warming than the emission of \qty{1}{\kg} of \ce{CO2} over a hundred year period (comparing it over 25 years leads to a factor of 80, because methane does not remain as long in the atmosphere as \ce{CO2} does). 
The weighting factors for midpoint impact categories are typically well defined and can be determined to reasonable accuracy. 

Midpoint categories, however, do not answer the question whether it is preferable to reduce emissions that cause harm via one impact category (say, global warming) at the expense of harm via a different pathway (say, eutrophication). 
For instance, replacing resistive electromagnets by permanent magnets poses such a dilemma: the permanent magnet option reduces the GWP, even when the higher carbon intensity of the magnet material is taken into account~\cite{shepherd:2023a}. 
However, the use of materials with a high content of rare earth elements such as samarium or neodymium leads to increased impacts such as ecotoxicity, as well as social effects that are not classified as environmental impact, such as modern slavery or the impact of mining on local communities.

Endpoint categories address this by attempting to quantify the consequences to humans (measured in disability adjusted life years lost, DALY, in the human population), damage to ecosystems (measured in number of species lost) and damage to resource availability -- see \autoref{fig:ReCiPe2016}).
One of the first such methods was Eco--Indicator 99~\cite{Goedkoop:2001a}. 
These endpoint categories are calculated from the midpoint categories by the application of weighting factors.

A number of LCA methodologies exist that define sets of impact categories that go beyond GWP, which is present in all methodologies.
Amongst the most popular ones\footnote{The CML-IA method~\cite{deBruijn2002} was developed at the Institute of Environmental Sciences Leiden (Centrum voor Milieuwetenschappen Leiden CML) as one of the first midpoint LCA methods, defining eleven impact categories in the baseline.
Combining midpoint categories from CML-IA with endpoint categories from Eco-Indicator 99~\cite{Goedkoop:2001a}, a team from the Swiss Federal Institute of Technology Lausanne (EPFL) developed the IMPACT method, with 15 midpoint and four endpoint indicators.
In a similar fashion, ReCiPe ~\cite{Huijbregts:2017a,Huijbregts:2020a} was developed by a collaboration of RIVM and Radboud University, CML, and PRé Consultants, based on CML-IA and Eco-Indicator 99, first in 2008, updated in 2016.
The International Life Cycle Data (ILCD)~\cite{ILCD:web} system was developed by the European Joint Research Centre (JRC) and the European Directorate for Environment of the European Commission. It is regularly updated, the latest update of the default list was made in March 2022, referred to as Version 2.0. 
It has adopted the core environmental impact indicators of EN 15804:2012+A2:2019/AC:2021 as mandatory indicators. The EN 15804 ``reference package'' provided by EU JRC is available at 
\url{https://eplca.jrc.ec.europa.eu/LCDN/EN15804.html}, 
last update February 2023.} 
~\cite{Rybaczewska-Bazejowska:2024aa, Wahl:2018a} are:
\begin{itemize}[nosep]
\item ReCiPe 2016~\cite{Huijbregts:2017a,Huijbregts:2020a}
\item ILCD 2011~\cite{ILCD:web,ILCD:2010a,ILCD:2010b,ILCD:2012a}
\item CML--IA 2012~\cite{deBruijn2002}
\item IMPACT 2002+~\cite{Jolliet:2003aa, Pennington:2005aa}
\end{itemize}

These methods have been compared in a recent publication~\cite{Rybaczewska-Bazejowska:2024aa}.
Owing in part to their common ancestry, the CML-IA method, they share many of the midpoint categories -- see \autoref{tab:lca-categories}.
The EC regulation No 1221/2009 establishes the eco-management and audit scheme (EMAS) introduced in section \ref{par:EMAS}, which is a voluntary premium environmental management tool for all kinds of organisations that are willing to evaluate, report on and improve their environmental performance. While it does not define an LCA method, it is an established framework for impact reporting at global scale. The environmental review, management system, audit procedure and statement must be approved by an accredited environmental verifier. The validated statement is registered and made publicly available.

A survey of LCA practitioners~\cite{Rybaczewska-Bazejowska:2024aa, Wahl:2018a} has shown that of these rather similar LCA approaches, ReCiPe 2016 is the most widely used.

\begin{table}[htp]
\caption{Midpoint impact categories in four popular LCA methods~\cite{Rybaczewska-Bazejowska:2024aa}: ReCiPe 2016a, ILCD+EPD, CML--IA, and IMPACT 2002+.}

\begin{center}
\scriptsize{}
\begin{tabular}{*{5}{>{\raggedright\arraybackslash}p{2.8cm}}}
Main category & ReCiPe 2016a & ILCD+EPD & CML--IA & IMPACT 2002+ \\
\hline

Climate change 
 	& Global warming [\unit{\kg}~\ce{CO2} eq] 
	& Climate change [\unit{\kg}~\ce{CO2} eq] 
	& Glob. warm. (GWP100a) [\unit{\kg}~\ce{CO2} eq] 
 	& Global warming [\unit{\kg}~\ce{CO2} eq]  \\ \hline
Ozone depletion
 	& Stratospheric ozone dep. [\unit{\kg} CFC-11 eq]
 	& Ozone depletion [\unit{\kg} CFC-11 eq]
 	& Ozone layer dep. (ODP) [\unit{\kg} CFC-11 eq]
 	& Ozone layer dep. (ODP) [\unit{\kg} CFC-11 eq] \\  \\ \hline
    &
 	& Ionising radiation HH [\unit{\kilo\becquerel}~\ce{^{235}U}~eq~to~air]
	& 
 	& \\  \cline{3-3}
\multirow{-2}{\hsize}{Ionising radiation}
 	& \multirow{-2}{\hsize}{Ionising radiation [\unit{\kilo\becquerel}~\ce{^{60}Co}~eq]}
	& Ionising radiation E (interim) $[\mathrm{CTU_e}]$ 
    & \multirow{-2}{\hsize}{--}
    & \multirow{-2}{\hsize}{Ionising radiation [\unit{\becquerel}~\ce{^{14}C}~eq]} \\  \hline
	& Ozone form., Human health [\unit{\kg}~\ce{NO_x}~eq]
	& & & \\  \cline{2-2}
 /*
\multirow{-2}{\hsize}{Ozone formation / Respiratory organics}
& Ozone form., Terrestrial ecosyst. [\unit{\kg}~\ce{NO_x}~eq]
    & \multirow{-2}{\hsize}{Photochemical ozone form. [\unit{\kg}~NMVOC~eq]}
	& \multirow{-2}{\hsize}{Photochemical oxidation [\unit{\kg}~\ce{C2H4}~eq]}
	& \multirow{-2}{\hsize}{Respiratory organics [\unit{\kg}~\ce{C2H4}~eq]} \\  \hline
Particulate matter / Respiratory inorganics 
	& Fine particulate matter form. [\unit{\kg}~PM\textsubscript{2.5}~eq]
	& Particulate matter  [\unit{\kg}~PM\textsubscript{2.5}~eq]
	& --
	& Respiratory inorganics [\unit{\kg}~\ce{SO2}~eq]  \\  \hline
	& & & 
	& Terrestrial acid/nutri [\unit{\kg}~\ce{C2H4}~eq]  \\  \cline{5-5}
\multirow{-2}{\hsize}{Acidification}
	& \multirow{-2}{\hsize}{Terrestrial acidification [\unit{\kg}~\ce{C2H4}~eq]}
	& \multirow{-2}{\hsize}{Acidification [\unit{\kg}~\ce{H^+}~eq]}
	& \multirow{-2}{\hsize}{Acidification [\unit{\kg}~\ce{C2H4}~eq]}
    & Aquatic acidification [\unit{\kg}~\ce{SO2}~eq]  \\ \hline
	& Freshwater eutrophication [\unit{\kg}~P~eq]
	& Freshwater eutrophication [\unit{\kg}~P~eq]
	& 
	&  \\  \cline{2-3}
	& Marine eutrophication [\unit{\kg}~N~eq]
	& Marine eutrophication [\unit{\kg}~N~eq]
    & 
    & \multirow{-2}{\hsize}{Aquatic eutrophication [\unit{\kg}~\ce{PO4}~eq,~P-lim]} \\  \cline{2-3} \cline{5-5}
\multirow{-3}{\hsize}{Eutrophication}
    & --
	& Terrestrial eutroph. [\unit{\mole}~N~eq]
    & \multirow{-3}{\hsize}{Eutrophication [\unit{\kg}~\ce{PO4^3-}~eq]}
    & -- \\  \hline
	& Terrestrial ecotoxicity [kg~1,4-DCB~eq]
    & 
	& Terrestrial ecotoxicity [kg~1,4-DCB~eq]
	& Terrestrial ecotoxicity [kg~TEG~soil] \\ \cline{2-5}
 	& Freshwater ecotoxicity [kg~1,4-DCB~eq]
 	& Freshwater ecotoxicity [CTU\textsubscript{e}]
 	& Freshwater aq. ecotox. [kg~1,4-DCB~eq]
    &  \\ \cline{2-4}
\multirow{-3}{\hsize}{Ecotoxicity}    
 	& Marine ecotoxicity [kg~1,4-DCB~eq]
    &
   	& Marine aq. ecotoxicity [kg~1,4-DCB~eq]
    & \multirow{-2}{\hsize}{Aquatic ecotoxicity [kg~TEG~water]} \\  \hline
    & Human toxicity: cancer [kg~1,4-DCB~eq]
    & 
    & 
    & 
    \\ 
    \cline{2-2}
\multirow{-2}{\hsize}{Human toxicity}
    & Human tox.: non-cancer [kg~1,4-DCB~eq]
    & \multirow{-2}{\hsize}{Cancer and non-cancer effects [CTU\textsubscript{h}]}
    & \multirow{-2}{\hsize}{--}
    & \multirow{-2}{\hsize}{--}
    \\ 
    \hline
Land use
    & Land use [\unit{\square\meter yr}~crop~eq]
    & Land use [kg~C~deficit]
    & --
    & Land occupation [\unit{\square\meter}~org.~arable] \\  \hline
    & Mineral resource scarcity [kg~Cu~eq]
    & 
    & Abiotic depletion [kg~Sb~eq]
    & Mineral extraction  [MJ~surplus] \\  \cline{2-2}  \cline{4-5}
\multirow{-2}{\hsize}{Resource scarcity}
    & Fossil resource scarcity [kg~oil~eq]
    & \multirow{-2}{\hsize}{Mineral, fossil \& renewables depletion  [kg~Sb~eq]}
    & Abiotic depletion (fossil fuels) [kg~Sb~eq]
    & Non--renewable energy  [MJ~surplus] \\  \hline
Water consumption
    & Water consumption [\unit{\cubic\meter}]
    & Water resource depletion [\unit{\cubic\meter}~water~eq]
    & -- & -- \\ \hline

\end{tabular}
\end{center}

\label{tab:lca-categories}
\end{table}

An important part of the methodology is the choice whether specific or generic impact factors are used in the evaluation.
Impact factors can vary wildly, depending on the actual production process used, the country of origin of raw materials and location of a processing site, or a specific company, and on the time when a specific product has been or will be produced -- see \autoref{tab:LCA_materials}.
The choice of impact factors depends on the project stage and the goal of the LCA.
During a prospective LCA of a project that is still in a conceptual phase, the specifics of materials and suppliers will generally not be known, in the case of international projects with world-wide contributions possibly not even the country of origin of certain goods. 
In such a case, globally averaged values will generally be appropriate for some materials; however, materials for civil construction and electricity will typically be drawn from the host country, or a known region within the host country, and thus should be taken into account. 
For projects that are envisaged to be realised far in the future, projections of impact factor reductions may also be taken into account, however, reliable projections are hard to find and will typically be limited to GWP reductions.
Usage of specific products such as \ce{CO2}-reduced concrete, is difficult to take into account when purchasing policies are not known; such reductions can be identified as possible savings, or they might be taken if there exists a plan or even pledge by the project to utilise such materials, possibly 
at a higher upfront cost.

LCA evaluations in the context of project approval, or in preparation of a possible approval, will be based on more specific information about available suppliers and time lines, and may come with an obligation from the approval authorities to adhere to the values given. 
In such a case, actual product certificates may be available from potential suppliers and may be used, 
also constraining the LCA method to be chosen if certificates are available only for a limited set of impact categories (possibly only for GWP). 
Environmental product declarations are presented in more detail in section \ref{sec:edp}.

\subsubsection{Life Cycle Inventory}

The Life Cycle Inventory (LCI) is a model of the product or service under consideration (the functional unit) that summarises the flows of products and materials contributing to -- and emanating from -- the functional unit during its lifecycle.
Commercial and open-source LCA databases, such as ecoinvent~\cite{Ecoinvent} (maintained by the Paul Scherrer Institute and partners) and GLAD~\cite{GLAD}, contain impact data for a large number of product flows and processes, which may range from the production of $1\,\mathrm{kg}$ of aluminium to the complete lifecycle of a pair of jeans. 

These data come from dedicated LCAs, the number of publications of which have grown exponentially in recent years. For commonly used materials and components, such as simple resistors or capacitors, the data is very comprehensive. However, for accelerator-specific components and materials such as high-purity niobium or magnetic materials such as samarium-cobalt, publications are not only rare, but the calculated values for these materials vary widely. The details of these LCAs are rarely published in peer reviewed journal articles; a large part of LCA results contained in the databases comes from undisclosed analyses, because manufacturers often agree to only provide data on the condition of anonymity to protect proprietary business information. As discussed in section \ref{sec:quantifying_impacts}, these data are also often highly temporally and geographically dependent.

\subsubsection{Assessment}

In the Life Cycle Impact Assessment (LCIA) step of an LCA, the environmental impact is quantified in the impact categories chosen during the scope and goals step of the LCA, and is based on the inventory gathered in the LCI step.
During the Life Cycle Inventory Assessment (LCIA) stage of an LCA, the system is decomposed into smaller units, product flows are analysed and modelled, and data on the individual process steps are collected. 
The data gathered is structured and collected in a model that is typically processed with a dedicated software tool, such as Simapro\textregistered ~\cite{Simapro}, OpenLCA~\cite{OpenLCA} or One Click LCA\textregistered~\cite{OneClickLCA}.

An LCA is a time-consuming process; as many components and devices as possible need to be included. 
The origin of the materials, the production processes, 
and end of lifecycle must be investigated.
Although the materials and equipment in accelerator facilities are often similar or identical throughout the community, they are very specific to the research sector. A prominent example is niobium for vacuum and superconducting RF technology.
In order to reduce the research effort, but also to facilitate comparability between facilities, it would be necessary to establish a database of materials for accelerator technology as well as standard equipment such as ion getter pumps or power supplies. 
If possible, this database should contain more than one LCA for a material/device so that differences in place of origin or production can also be reflected.

Variations in calculated LCA values can be relatively obvious, such as varying geographical location (transport routes and power generation) or fabrication processes. 
However, the use of different LCIA methodologies can also lead to significant differences in results, which would in turn make comparison between accelerator projects or components difficult.
Annex~\ref{annex:reference-data} contains a table of materials that are widely used in the context of accelerator projects, with indicative values for GWP. It exhibits the potential difficulties accelerator facilities will face in performing LCAs through lack of readily available, bespoke material data to ranges of reported data that the researchers will have to sift through. It also shows how it is important to use the same methodologies throughout the evaluation of accelerator facilities.

\subsubsection{Interpretation}

Interpretation is an integral part of the LCA process.
It is a cross-cutting activity that is performed not just at the end of the LCA, but throughout the procedure, to guide the process and improve results. 
As part of the interpretation, a variety of analyses are typically performed to gather insights and improve the results.
Examples of each type of analysis listed here are taken from a CLIC/ILC LCA study, focusing on tunnel construction~\cite{clic_ilc_lca_arup}.

\textit{Sensitivity Analysis} is typically performed early on, when first or partial results of the LCA and LCIA activities are available.
By varying specific inputs, the impact of particular assumptions or choices of materials and processes on the final result is evaluated, with the goal to identify points where additional work is required or warranted to improve the result.
Sensitivity analysis is important to reduce uncertainties in the final result, which is further discussed in section~\ref{sect:lca:uncertainties}.
For example, amounts of steel and concrete in tunnel construction can be compared to their relative GWP impact. Material qualities can be varied (low \ce{CO2} concrete, steel with large recycling fraction) and results compared to the baseline result.

\textit{Hotspot Analysis}~\cite[section 3]{Schau:2016a} is closely connected to sensitivity analysis.
It aims to identify the most important elementary (material) flows, processes, impact categories or lifecycle phases that dominate the overall impact and therefore warrant particular attention, in the evaluation of impacts as well as in measures targeted to reduce the overall impact.
In our example, GWP contributions of various subprocesses to tunnel construction were compared, showing that in-situ permanent lining and shield wall construction are the largest GWP contributors to the ILC main linac tunnel.

\textit{Contribution Analysis} is another method to identify to which impact categories various components, processes, flows or lifecycle stages contribute in which manner. 
Typically, contributions to all impact categories under consideration are visualized in stacked bar charts, divided by contributions from contributing elements or processes, normalised to the total impact.
The contributions analysis in our example showed that raw materials (phase A1-A3) were the dominant source of all impact categories. Electricity use in construction (A5) was the dominant source for the category of ionising radiation, due to the use of nuclear power plants.
    
\textit{Benchmarking}. LCA results are compared to results from other publications for comparable processes or products, which helps to gauge the credibility of the results or how the product under consideration scores in comparison to similar products.
In our example, \ce{CO2} intensities per km of tunnel length were compared between accelerator tunnels under investigation and tunnelling projects for road and rail traffic for which LCA data was available. 

\begin{figure}[ht]
    \centering
    \includegraphics[width=0.8\textwidth]{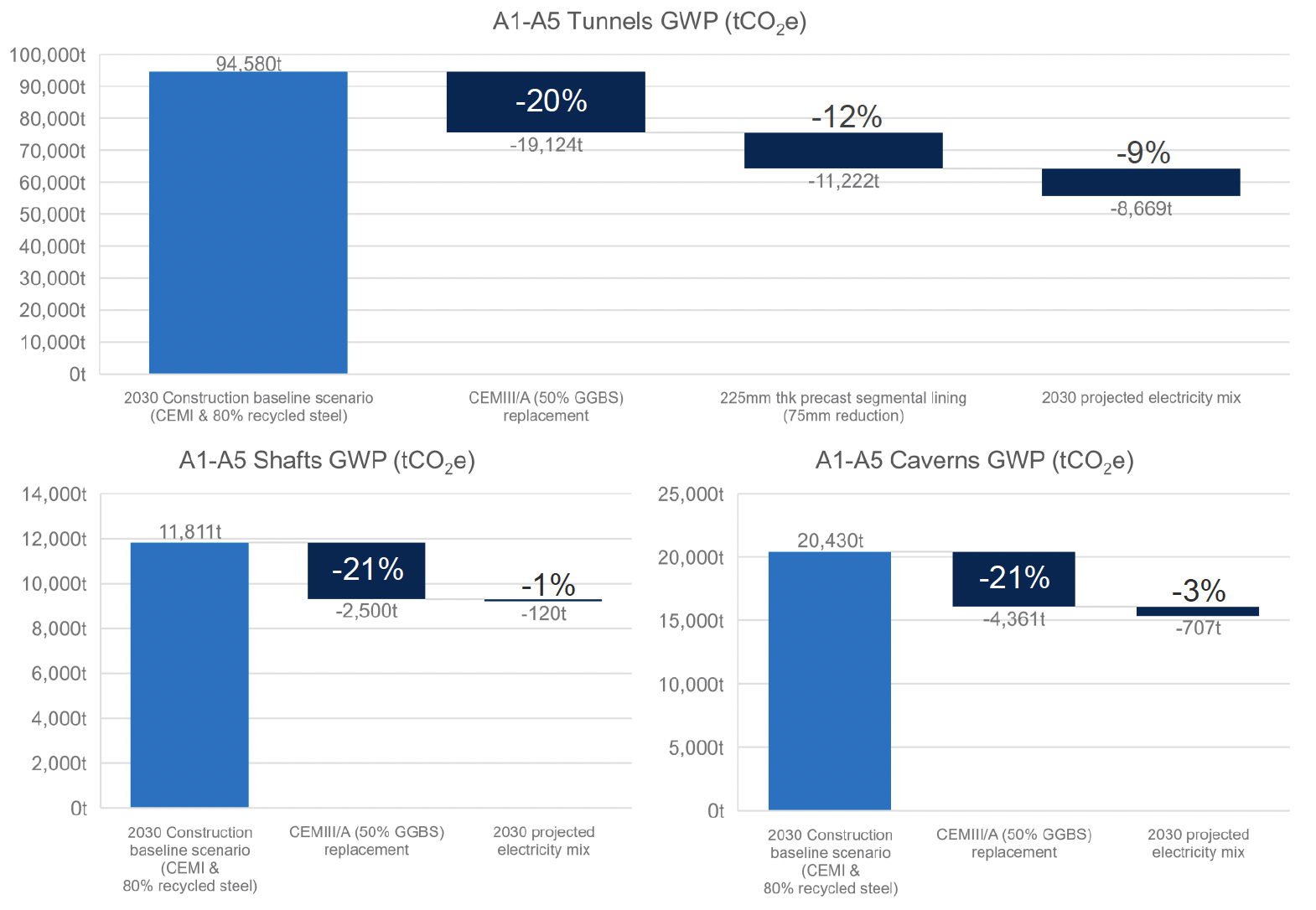}  
    \caption{Examples of possible impact reductions, from~\cite{clic_ilc_lca_arup}.
    }
    \label{fig:reduction}
\end{figure}

The most important outcome of the interpretation process of an LCA is the identification and quantification of measures to improve the project or product under consideration by reducing the environmental impacts measured in the LCA (\autoref{fig:reduction}).
The quantitative LCA provides the data necessary to gauge which of a range of possible measures have the largest impact,  to evaluate possible trade-offs and burden shifting (for instance, increased impact on resource scarcity or environmental toxicity by use of permanent magnet materials versus decreased GWP).

\subsubsection{Evaluation of Uncertainties}
\label{sect:lca:uncertainties}

There are many sources of uncertainty within an LCA. These can materialise from sources such as:
\begin{itemize}[nosep]
    \item the initial assumptions of the LCA, e.g. early design decisions
    \item the chosen LCA database
    \item input data and data quality including missing or unaccounted  materials/components/etc.
    \item the LCIA method, e.g. the relation to the database of certain materials and whether all impact factors of the material are accounted for
    \item material (`product') accuracy, e.g. whether a material substitution has been made, whether a local or global estimation is used
    \item accuracy of extrapolated data, e.g. energy production, end-of-life choices and impact scenarios
\end{itemize}

Uncertainty is one of the key factors in the reliability of the final LCA and a comprehensive uncertainty analysis should be included~\cite{LCAuncertainty}. It must be ensured that sources of uncertainty are documented extensively and transparently with assumptions, uncertainties and what is in and out of scope included. Ultimately, the LCA must be robust under external examination and against criticism.

There are many methodologies for evaluation of LCA uncertainties; approaches such as Monte Carlo simulation could be a familiar method for accelerator researchers to use. 
Other options exist such as data quality systems that could be used to evaluate uncertainties from the LCA database used in the inventory analysis. There are a few data quality systems available. As an example, the widely-used Ecoinvent Data Quality System~\cite{Ecoinvent} considers 5 uncertainty contributions for a product.
Each of the 5 uncertainty categories can be evaluated using a 5-point system as displayed in \autoref{tab:EcoinventDataQuality}. The software (e.g. OpenLCA) can then amalgamate these uncertainties to calculate a variance, which can then be used in a Monte Carlo simulation with a distribution, e.g. a log-normal distribution, to evaluate the uncertainty of the LCIA results. 

In addition, other uncertainty evaluations could be considered or created e.g., a simplified uncertainty evaluation using a form of `traffic light system'.

\begin{table}
\centering
\caption{The ecoinvent data quality system~\cite{Ecoinvent, OpenLCA}}
\label{tab:EcoinventDataQuality}
\begin{adjustbox}{width=\textwidth}
\begin{tabular}{>{\raggedright}p{0.16\textwidth}\darkgreencolumn \lightgreencolumn \yellowcolumn \orangecolumn \redcolumn}
\hline
& \textbf{1: best} & \textbf{2} & \textbf{3} & \textbf{4} & \textbf{5: worst, default} \\
\hline
Reliability & Verified data based on measurements & Verified data partly based on assumptions or non-verified data based on measurements & Non-verified data partly based on qualified estimates & Qualified estimate (e.g. by industrial expert) & Non-qualified estimate \\
\hline
Completeness & Representative data from all sites relevant for the market considered, over an adequate period to even out normal fluctuations & Representative data from >\qty{50}{\percent} of the sites relevant for the market considered, over an adequate period to even out normal fluctuations & Representative data from only some sites (<<\qty{50}{\percent}) relevant for the market considered or >\qty{50}{\percent} of sites but from shorter periods & Representative data from only one site relevant for the market considered or some sites but from shorter periods & Representativeness unknown or data from a small number of sites and from shorter periods \\
\hline
Temporal correlation & Less than 3 years of difference to the time period of the dataset & Less than 6 years of difference to the time period of the dataset & Less than 10 years of difference to the time period of the dataset & Less than 15 years of difference to the time period of the dataset & Age of data unknown or more than 15 years of difference to the time period of the dataset \\
\hline
Geographical correlation & Data from area under study & Average data from larger area in which the area under study is included & Data from area with similar production conditions & Data from area with slightly similar production conditions & Data from unknown or distinctly different area (North America instead of Middle East, OECD-Europe instead of Russia) \\
\hline
Further technological correlation & Data from enterprises, processes and materials under study & Data from processes and materials under study (i.e. identical technology) but from different enterprises & Data from processes and materials under study but from different technology & Data on related processes or materials & Data on related processes on laboratory scale or from different technology \\
\hline
\end{tabular}
\end{adjustbox}
\end{table}

\subsubsection{Recommendations for conducting LCAs}

This section provides recommendations on what is needed in the accelerator field for future accelerator projects to accurately, efficiently and consistently perform LCAs at various stages of planning and construction method.

\paragraph{Perform LCAs throughout the various stages of design of future accelerator facility projects.}

LCAs are essential for future accelerator-based facilities due to their evaluation of a large range of environmental impact factors and their reporting standard qualities. Facilities should perform LCAs to set Key Performance Indicators (KPI) from which to improve their environmental impact through optimisation of design. 

There are many different stages of design for particle accelerator facilities, which have many layers, perturbations and possibilities. The earlier the facility implements positive sustainability actions (like performing LCAs) the larger the ability to influence its whole life environmental impact~\cite{pas2080:2023}. 

For future accelerator facilities at the early conception phase, efforts should be spent on performing a simplified LCA that uses generic data and focuses only on essential environmental aspects (e.g. GWP [\unit{\kg\ce{CO2}e}]), setting up the basis for future LCAs at later design stages. Environmental impacts of whole facilities -- including construction, operation and decommissioning impacts -- should be evaluated. Estimations of the development of power generation methods and policy involving chemicals that may be phased out should be taken into account. Extrapolations in time will be necessary and care should be taken to accurately estimate and take into account various scenarios (e.g. electricity supply decarbonisation policies), assumptions and uncertainties. 

In the optioneering stage, simplified LCAs should be performed for considered design options, to identify the differences of environmental impact. Sustainability could be considered a showstopper for high impact design options when alternatives are available.  

In the design stage, simplified LCAs should be performed for major components, and for accelerator design options to identify whether environmental impacts could be reduced. LCAs that have already been performed and are available for use e.g., sub-systems and components, should be incorporated into these LCAs where applicable. Any assumptions or alterations made e.g., the scaling of the LCA results of a magnet, should be documented.

Finally, the LCA methodology should be followed stringently. For example, alongside the large electricity consumption of a particle accelerator, it may be tempting to consider the other operational resources as negligible; this is not the case. All operational resource consumption (water, helium, etc.) should be included in LCAs. Operational water (B7) encompasses the total amount of water utilised in the operation of the accelerator, which is likely dominated by cooling water use.
CERN's environment report contains comprehensive data on the withdrawal and discharge of water as part of the laboratory's operation. 

\paragraph{Utilise LCAs to reduce environmental impact.}
Throughout the design of the facility, utilise the LCAs to identify areas of potential for reduction in environmental impact and take action to realise these reductions where feasible.
Performing LCAs without using them in this manner removes their potential for making a real difference in a facility's environmental impact.

\paragraph{Evaluate LCAs using common methods.}
As discussed, it is paramount that accelerator facilities use comparable and common methods, databases, tools and assessment methods to enable comparison of components/facilities and facilitation of future work. 
A common, widely-used and accessible Life Cycle Impact Assessment method (LCIA) should be used with a range of indicators such as the ReCiPe 2016 midpoint (H) or the ILCD method.

In addition, a functional unit that is compatible throughout accelerator facilities should be agreed upon and implemented. 
Accelerator components, experimental apparatus, and all necessary infrastructures should be included. 
We do not recommend functional units that have been based on physics results to be the sole reported functional unit of an LCA.

\paragraph{Perform LCAs of typical accelerator components.}

Conduct dedicated LCAs of components that are used widely in accelerator projects such as superconducting cavities, superconducting magnets, NEG coated vacuum chambers.
LCAs should be performed and published to facilitate future facilities adopting the same or similar components. 

Based on estimated average component lifetimes, include the impact of the production of replacement components and the impacts of downtime.

\paragraph{Collect and publish reference data on materials and fabrication processes used in accelerator facilities.} 
Many materials used in accelerator construction have specifications that require special production processes, which incur additional environmental impacts.
This is particularly true for materials used in vacuum, RF and cryogenic applications, such as oxygen-free copper for vacuum chambers and RF equipment, ultra-pure (${\rm RRR}>300$) niobium, aluminium for vacuum chambers, magnetic steel with high cobalt content, permanent magnet materials etc. For some of these materials -- for instance the niobium RRR300 -- commercial LCA databases such as ecoinvent~\cite{Ecoinvent} contain insufficient-to-no data. In these cases, the accelerator field may require LCAs to be performed and collated into a readily available database.

While for some components the environmental impact is determined from the raw material used (accounting for scrap and the material quality), the fabrication process itself can also add considerably to the impact.
Some examples of fabrication processes relevant in the context of accelerator projects that are energy- or otherwise resource-intensive or lead to emissions to be taken into account are:
\begin{itemize}[nosep]
    \item electron beam welding (large energy use)
    \item electropolishing (e.g. of superconducting cavities), requiring sizeable amounts of chemicals such as phosphoric acid and leading to substantial amounts of problematic waste
    \item NEG coating of vacuum chambers
    \item conditioning of RF cavities (large energy use)
    \item cleaning of vacuum components
    \item leak testing of vacuum components with helium
    \item cold tests of super conducting components such as RF cavities or magnets (energy use, helium losses)
    \item assembly under clean room conditions (energy use).
\end{itemize}

Transport can also contribute substantially to the impact of accelerator components. 
Many accelerator components are produced by only a few, highly specialised companies, and their procurement may thus require intercontinental transport. 

Collecting these data in a common database or performing LCAs on these materials and processes will facilitate and expedite LCAs for all accelerator facilities. 
Data could be used to benchmark and compare results 
of different projects.
Ultimately, LCAs and their interpretations will indicate the path to, and facilitate the reduction of, accelerator environmental impacts.

Accelerator laboratories and research institutions should collaborate to provide open-source and cost-free LCA tools, data and software (e.g., an accelerator specific extension to ecoinvent), 
in a similar manner to open-access publishing, to align with wider sustainability goals and UN SDGs. This would not only enable LCAs and environmental impact reduction efforts, it would create availability to those with minimal or no funding; the people, organisations and countries who need it most. 
The released free software and databases will be available for evaluating products, manufacturing processes, and services of general interest --
representing a case where research infrastructures generate goods of direct benefit to society.

\subsection{Environmental Product Declarations}
\label{sec:edp}

Environmental Product Declaration (EPD) reports are standardized, third-party verified documents that transparently present credible information about a product’s impact on the environment. An EPD is a so-called type III environmental declaration that is compliant with the ISO 14025 standard. Following the standard, EPDs are useful for communication, comparison, and decision-making. EPDs are based on the information from a Life Cycle Assessment (LCA) study. The EPD report consists of two key documents: first, the underlying LCA report, a systematic and comprehensive summary of the LCA project to support the third-party verifier when verifying the EPD. This report is not part of the public communication. Second, a public EPD document that provides the LCA results and other EPD content. The public EPD document does not disclose sensitive details that may impact the commercial activity of the company that supplies the product. For example the process or raw materials used to achieve the results may not be reported. Only the results are disclosed to help informed decision making about comparable information within the same product group.

The EPDs are in turn used in Lifecycle analysis studies of further project scenarios to deliver credible results on the environmental effects of the project under specific construction conditions.

EPDs adhere to strict regulations and standards and are valid for a limited period of time, typically five years. The standards that govern them vary according to the specific application sector. In practice, the more specific the guidelines, the more comparable the results are. For instance, every EPD is created according to a specific set of Product Category Rules (PCRs). PCRs offer calculation rules and guidelines to ensure comparability between EPDs within the same product category.

EPDs are administered and supervised by independent agencies called EPD Program Operators (EPD POs). EPD POs are responsible for identifying and creating Product Category Rules (PCRs) for EPDs. PCRs ensure that EPDs within the same product category report comparable information by following the same calculation methods and reporting guidelines (e.g., what environmental indicators to report on). Different countries have their own EPD POs and rules, which can make the selection process complicated. 

EPDs must undergo verification by independent experts before they are published. These independent expert verifiers must also be approved by the EPD PO. Most EPD POs in Europe are linked to ECO Platform, an EPD umbrella organization that helps make EPDs more comparable since 2013.

The mandatory environmental impact indicators of EPDs\footnote{See also a brief online summary at \url{https://www.environdec.com/resources/indicators}} are regulated by the European standard EN 15804+A2. Six indicators are mandatory and 11 are optional for non-construction products. All inventory indicators of EN 15804 are mandatory for construction products. The indicators and their units in the SI framework are detailed in the International Reference Life Cycle Data system (ILCD) handbook~\cite{ILCD_handbook} issued by the Joint Research Centre, Institute of Environment and Sustainability of the European Commission:

\begin{enumerate}[nosep]
\item Climate change (kg \ce{CO2} eq.)
\item Ozone depletion (kg CFC-11 eq.)
\item Photochemical ozone formation, human health (kg NMVOC eq.) \footnote{Non methane volatile organic compounds}
\item Particulate matter / respiratory inorganics (kg $PM_{2.5}$ eq.)
\item Ionising radiation, human health (kBq \ce{^{235}U} eq. to air)
\item Acidification (mol \ce{H+} eq.)
\item Eutrophication: terrestrial (mol N eq.), freshwater (kg P eq.), marine (kg N eq.)
\item Human toxicity: cancer and non-cancer effects ($CTU_h$) \footnote{Comparative Toxic Unit for human (CTUh) expressing the estimated increase in morbidity in the total human population per unit mass of a chemical emitted (disease cases per kilogram emitted).}
\item Freshwater ecotoxicity ($CTU_e$) \footnote{Comparative Toxic Unit for the ecosystem (CTUe), an estimate of the Potentially Affected Fraction of species (PAF) integrated over time and volume, per unit mass of a chemical emitted; [$CTU_e$ per kg emitted] = [PAF × \unit{\cubic\meter} × day per kg emitted]}
\item Land use (kg C deficit) \footnote{Change of soil organic carbon in kg, depending on the soil quality before and after the transformation.}
\item Resource depletion: water (\unit{\cubic\meter} water eq.), mineral, fossil and renewables (kg Sb eq.)
\end{enumerate}

These technical factors are intended for use in the frame of impact assessments with Life Cycle Assessments by practitioners and 
their interpretation requires typically expert knowledge. Based on the interpretation guidelines~\cite{JRC104415}, the individual practitioner will still be left with a range of choices that can change the results and conclusions of an assessment. The ILCD handbook covers guidance and recommendations for methods for each impact category at both midpoint and endpoint. The indicators provide governments and businesses with a basis for assuring quality and consistency of lifecycle data, methods and assessments.

Reporting to stakeholders is typically achieved with simplifying approaches. Indicators like EMAS and contributions to UN SDGs can help communicate the effects. The comprehensive socio-economic evaluation, based on reporting a Net Present Value and a Benefit/Cost ratio, allow stakeholders to immediately grasp the sustainability level of a project. The intensity of the environmental effects can also be reported for a per capita value of the countries that contribute to the project (both financially and in-kind). If required, the per capita value can be calculated by weighing the financial and in-kind contribution level of each participating country. For this purpose the environmental effects need to be converted into monetary terms. Conversion factors for the \ce{CO2}e are for instance provided by the European Investment Bank. Monetary valuation coefficients (MVC) for other environmental effects vary significantly across impact categories and among models and countries~\cite{AMADEI2021129668,SHEN2024100133}.

A project needs to select and cite a specific national MVC source for relevant impact categories. Guidelines and tables exist for instance for France~\cite{Quinet2019, Annexes-giude-france-strategie}, Switzerland~\cite{BAFU-MVC-2020} and Germany~\cite{UBA-MVC-2020}. At European level, the Joint Research Centre, Institute of Environment and Sustainability of the European Commission issued EU-wide data~\cite{JRC67216}.

\clearpage

\section{Environmental Impacts of Accelerator-based Facilities}  
\label{sec:GHG-emissions}
This chapter presents the application of Life Cycle Assessment (LCA), as outlined in section \ref{sec:lca}, to the design, operation  and decommissioning of accelerator-based RIs. Mitigation strategies mentioned in this section are discussed in further detail in section \ref{sec:mitigation}.
The methods and relevant issues for the LCA process will be split between project phases in accordance with the EN~15978~\cite{en15978:2011} definitions, illustrated in \autoref{fig:lca-stages}: A1-A5 (product and construction stages), B1-B5 (operation stage), and C1-C4 (decommissioning and end-of-life stages).

\subsection{Machine and Infrastructure construction}

The construction phase (stages A1 to A5 of the LCA process, see \autoref{fig:lca-stages}) covers the environmental impact up to the start of operation of an accelerator project; it covers the impacts of the processes necessary to build all civil engineering infrastructure and components, from raw material supply (stage A1), transport (A2) to manufacturing (A3), as well as the assembly of the project ready for operation, from transport to site (A4) and construction and installation at the site (A5).

Large accelerator projects are mega projects, with a capital expenditure of hundreds of millions or billions of dollars, and hundreds to thousands of people involved in the design and execution stage.
As a consequence, the amount and complexity of items that need to be considered in an accelerator project LCA is enormous.

The laboratories that build and operate these accelerators have a lot of experience in the design, planning and costing of these projects.
The methods employed in the cost evaluation form a solid basis for the LCI data gathering, 
in particular experience in defining 
consistent Product and Work Breakdown Structures (PBS and WBS) that form a complete, hierarchical decomposition of all product parts and the work required to deliver the final product.

Thus, the LCI process for the construction phase starts with the definition of a PBS of a suitable granularity, such that at each level of decomposition the resulting product is the sum of all its parts. 
Parts or subsystems that are reused across the project can be defined and analysed as separate functional units such that the data gathering and analysis is done only once in the LCI process.

It is important to consider that the lifecycle inventory entails more than a tally of raw material masses.
Depending on the subsystems and components under consideration, resources for and emissions during fabrication and transport processes are relevant and should be considered, as well as the amount of scrap material.

For example, for the fabrication process of the yoke of a resistive electromagnet, the amount of scrap material varies considerably in the fabrication process: production by laminations from stamped metal sheets may produce \qty{30}{\percent} or more of scrap metal, while assembly from forged and machined bars (e.g. the CEPC dipole design~\cite[Section 4.3.3.2]{CEPCStudyGroup:2023quu}) may be much more material-efficient.
During the conceptual design phase of an accelerator, these details are often unknown or too expensive or time consuming to evaluate for all components involved.
In such cases, estimates of overall fractions of scrap production may give at least an indication of the efficiency of the production process. 
Dedicated LCA studies of typical accelerator components such as magnets, power supplies, vacuum chambers or RF accelerating equipment would provide valuable input for more accurate project assessments.  

Once the design has been finalized, it should be possible to report expected construction emissions with high accuracy; however, due to the inflexibility of design at this stage opportunities for emission reductions are likely to be marginal.
 Consequently, 
 to ensure effective mitigation, 
 sustainability considerations should be 
  addressed as early as possible 
  in the design process of a new or upgraded accelerator. 
  Even though the design is inherently uncertain and subject to change, making it unsuitable for a comprehensive LCA approach as outlined in the previous section, 
  we still recommend making estimates of quantities of materials required for the build, as well as approximate power consumption levels, as early as possible, and refining these estimates as the design progresses. This will enable sustainability to be embedded into the design mindset, facilitate hotspot analyses and impact reductions, and allow environmental concerns to be integral to the decision-making process. Trying to add sustainability as an extra concern midway through the design process runs the risk of requiring high effort and financial cost for little impact reduction, and could result in it being seen as an optional -- rather than an essential -- requirement; this could mean that sustainability could be value-engineered out of the project.

On a similar note, project costing must be done on a project lifetime basis, rather than keeping capital and operating costs separate. There are many examples of sustainability interventions that come with a higher initial price tag, but over the multi-decade life expectancy of a typical accelerator will pay for themselves several times over. Cost optimisation for large facilities is often weighted heavily in favour of upfront costs, which will reduce the chances of funding being available for environmental improvements.

The approach used 
early in the design process 
should be similar to that used in a full LCA; clearly, at the start of the project, many open questions will remain. However, even very rough ideas of quantities of materials required will give a good start to the process, and these can be further refined as the design progresses. \autoref{tab:project_phases} shows the typical phases of a large project, with expected levels of detail for the environmental impact assessment at each phase.

\begin{enumerate}[nosep]
    \item \textit{Project definition phase.} At this point, the accelerator specification will be very vague, and we will have to make very rough estimates for quantities of materials required. For some parts of the project, technology choices will not have been made yet; where there are 2-3 possible options, materials estimates should be made for all of them in order to make a more informed decision. At this point, we are looking for headline figures: what are the likely largest single sources of carbon emissions, and are any particular technologies likely to lead to significantly greater emissions than alternatives?
    
    \textit{Examples}: warm vs cold RF systems; permanent magnet vs electromagnet systems; housing a facility above or below ground.

    \item \textit{Conceptual design phase.} This is where the design will be pinned down to a greater amount of detail. During this phase, technology choices will be made for several parts of the facility, and quantities of components should be easier to estimate with a greater level of detail. At this point, the emphasis should be on providing semi-quantitative estimates of embodied emissions. Where there are choices to be made, we should be in a good position to provide detailed data on the environmental impact of each option.
    
    \textit{Examples}: linac module length and frequency; RF power sources.
    
    \item \textit{Technical design phase.} The detailed facility design is produced during this phase. Technology choices are typically `locked in' by this point, and switching to lower-carbon alternatives will probably not be easy. The sustainability team should focus instead on finding the lowest-impact route to delivery for each subsystem, whilst maintaining the baseline scientific merit established during the conceptual design phase.
    
    \textit{Examples}: choice of shielding material; in-vacuum vs out-of-vacuum undulators.
    
    \item \textit{Authorisation-process phase. 
    } The focus of the project shifts from design to finding suppliers for each subsystem. The role of the sustainability team at this point is to find the lowest-emission route to facility delivery, ensuring environmental impact is considered alongside other key project metrics such as quality, value for money, and timely delivery. Setting out a standardised procurement template at the start of this phase will greatly assist in this.
    
    \textit{Examples}: querying magnet manufacturers to find their sources of supply for steel and copper, and comparing expected emission factors.
\end{enumerate}

\begin{table}
\centering
\small
\caption{Expected level of detail of environmental impact assessments 
for each phase of a large RI project.}
\label{tab:project_phases}
\begin{adjustbox}{width=\textwidth}
\begin{tabular}{p{2cm} lcccc p{3cm} l p{6cm}}
\hline
\textbf{Phase} & \textbf{Inventory} & \multicolumn{4}{c}{\textbf{Include in impact assessment}} & \textbf{Impact categories} & \textbf{Scope for} & \textbf{Goals of project} \\
& \textbf{detail level} & \textbf{Examine} & \textbf{Mass of} & \textbf{Factory} & \textbf{Transport} & & \textbf{mitigation} & \textbf{sustainability team} \\
& & \textbf{options} & \textbf{materials} & \textbf{processing} &  &  & & \\

\hline
Definition & Low & \tick & \tick & \cross & \cross & GWP only & High & Find likely largest-contributing areas and potential mitigation strategies \\
Conceptual design & Medium & \tick & \tick & \cross & \cross & GWP and others where data is readily available & High & Evaluate technology choices, including environmental impact as a factor \\
Technical design & Medium & \cross & \tick & \tick & \cross & All: decide on assessment methodology at this point & Medium & Fine-tune details of design; look for further reductions (e.g. alternative sourcing for raw materials) \\ 
Authorisation-process & High & \cross & \tick & \tick & \tick & All & Low & Evaluate suppliers, including environmental impact; 
report 
to funding bodies \\
\hline
\end{tabular}
\end{adjustbox}
\end{table}

Civil engineering works 
give a substantial contribution to the environmental impact of 
a project. Excavation materials and concrete shielding constitute core subjects of the sustainability assessment. 
The materials extracted during the excavation of the underground structures are today undeniably an important circular economy issue. In most cases, pathways for these materials have been found in attempts to obtain a balance between the requirements of spoil and fill materials. In other cases, extracted materials have been placed in final storage with no real thought given in advance to possible uses. However, given the resource preservation challenges and the considerable volumes generated by ambitious projects, a scheme of this kind is no longer satisfactory or sufficient. The use of these materials must be investigated adequately and at a sufficiently early stage of the projects to facilitate access to mineral resources and to meet the demands of society and the regulatory objectives. 
The materials extracted during the excavation of underground structures are essentially inert materials. They can be reused on the same site from which they are extracted or can be recovered. Reuse and recovery are two ways of preserving natural resources and, directly and indirectly, facilitating the achievement of the regulatory objectives.
Owner’s undertaking to recover the excavated materials may result in an economic benefit for the project insofar as these materials will not be accounted for as materials to be sent to a landfill. 
A non-exhaustive list of potential reuse is: 
\begin{enumerate}[nosep]
\item	use of specific excavated materials (e.g. limestone) in making cements and in stabilising constructions;
\item	reuse of the excavated materials in backfilling quarries and mines;
\item	development of new construction materials containing some specific constituents of the excavated materials (for example materials used in sandwich construction and tunnel stabilisation, insulating foams, etc.), for use within the project, where technically appropriate, or outside the project, if there is a potential market for them;
\item	use of the excavated materials in innovative pathways, for example processing of it into fertile soil for use in the renaturation of wasteland and quarries, agricultural and forestry applications, and other areas. 
\end{enumerate}

The environmental impact of the excavated material management scenarios are linked to three main items:
\begin{enumerate}[nosep]
\item	impacts of the excavated material treatment processes used to ensure that the material characteristics align with the specifications of a given recovery pathway (washing, crushing, screening, liming, etc.); 
\item	impacts relating to the streams and uses: impacts of the backfilling processes (e.g. unloading, handling, compacting, etc.);
\item	impacts relating to transport (modes and distances).
Excavated materials are a source of mineral resources for the region in their own right, which must be managed in terms of their quantity and quality, their use locally and for the benefit of the local economic substrate, and to limit adverse effects and the carbon dioxide emissions generated by their transport.
\end{enumerate}

From a certain perspective, recovering the excavated materials may create a better material distribution network by artificially providing new sources of geological materials or by supplying recycling platforms. In this case it will be a question of structuring the supply and demand in such as way as to optimise the use of resources within a given area.

\subsection{Machine and Infrastructure operation}
\label{sec:RIoperation}

The operation phase (stages B1 to B7 of the LCIA process, see \autoref{fig:lca-stages}) covers the environmental impact of the operational lifetime of an accelerator project; it covers the impacts of all processes necessary to operate (B1) and maintain (B2) the accelerator and experiments, including  
repairs (B3), replacement of components (B4) and refurbishments (B5). 
Of particular interest are the impact of energy (B6) and water (B7) required for operation.

Currently, assessments of 
accelerator projects concentrate mostly on the electricity consumptions, assuming that those constitute the dominant impact, at least in terms of green house gas emissions.
This is, however, only valid assuming that future RIs overcome the current situation, in which direct emissions from particle detector and detector cooling plants are the most abundant sources of GHG emissions, as outlined in the CERN environmental report for 
2021-22~\cite{CERN:2023a} and discussed in section \ref{sec:detector}. 

After building an accelerator, it will be operated for an expected lifetime of 10-40 years, depending on the maturity of the technology, the local user base, and the anticipated science programme. Many accelerators have lifetimes significantly in excess of what was originally planned. During the operation phase, an accelerator will usually have a set number of operating periods in any given year, interspersed with planned downtime for maintenance and improvements. The running time varies between 100-300 days per year, 
with a typical duration of around 250 days. During this time, most accelerators run for 24 hours a day, 7 days a week, in order to minimise issues with startup and shutdown of critical systems. In the downtime phases, most systems are either 
switched off or set to a low-power state. However, certain energy-intensive systems -- particularly air conditioning -- must remain operational at all times.

The amount of the emissions produced in operating an accelerator, arising from electricity production, depends on the grid intensity of the host country, 
which in turn is determined by its 
energy generation mix 
(coal, gas, nuclear, wind, solar, etc.). 
Table~\ref{ElectricityCarbonFootprint}, in  section~\ref{sec:mitigation}, presents examples of grid intensity [\unit{\kg\ce{CO2}e\per\mega\watt\hour}] for several representative countries.

An accelerator using \qty{3}{\mega\watt} of power during 250 operating days per year, and \qty{0.3}{\mega\watt} for the rest of the time, would use a total of \qty{19}{\giga\watt\hour} per year. It will therefore generate between \qtyrange{1}{11}{\kilo\tonne\ce{CO2}e} per year depending on its location; even in the best-case scenario, the operations emissions will dwarf those of construction after a relatively short operation period.

This report focuses on 
future accelerator facilities. Grid electricity is produced using a variety of different generation methods, and many countries are transitioning from fossil-based sources to renewables. This means that the present value for grid intensity is likely to be quite different to the value expected in 10-20 years' time. National governments publish future projections of their energy mixes, which depend on various factors including: installation of new generation capacity, retirement of older power plants, construction of international interconnectors, and shifting patterns of domestic and industrial demand.

There are two other significant sources of emissions to consider. The first is direct emissions of greenhouse gases (Scope 1). For most facilities, the most significant of these to consider will be sulphur hexafluoride, \ce{SF6}, with a global warming potential \num{23500} times greater than that of \ce{CO2}. \ce{SF6} is used in RF waveguides and high-voltage power supplies. Given its high GWP value, strong consideration should be given to using alternatives; if it is deemed necessary to use it, care must be taken to minimise release into the atmosphere. 
Another source which should be considered is emissions due to travel of users to the accelerator facility, which falls under Scope 3. Depending on the scale of the facility (national or international), the majority of users will either travel from the host country or overseas, which obviously influences the balance of transport modes (air, rail, road). 
Decisions about siting the accelerator should take into account the proximity of public transport hubs to a proposed site. Rail travel is preferred, since it is the least carbon-intensive form of transport, followed by road (with clear routes to decarbonisation), and finally air travel, where low-carbon technology is still in the early stages of development. Accelerators such as synchrotrons and FELs, with a large scientific user base, often operate a 'sample by post' service, where users submit samples for standardised tests in lieu of travelling to the facility. Increasing the breadth and flexibility of this type of service has clear benefits for the environment as well as potentially increasing the throughput and efficiency of a facility.

Environmental impacts from maintenance (B2), repair (B3), and replacement (B4) have 
been rarely considered in the past in accelerator LCAs.
Expenditures for the procurement of spare parts for those processes is an important contribution to the operating costs of an accelerator, and empirical data exists on the typical lifetime of key components such as klystrons that have to be exchanged on a regular basis.
This data can be used to quantify the impact of maintenance, repair and replacement. 
Evaluation of maintenance also includes the environmental impact of materials used for operation, such as helium or liquid nitrogen, which need to be replenished on a regular basis.

\subsubsection{Electricity use in operation}

Here we consider some of the largest subsystems by power usage for many accelerators.

One common area is air conditioning and water cooling. Accelerators are housed in temperature- and humidity-controlled rooms in order to ensure smooth and reliable running. Large volumes of air must be processed, using refrigeration systems to keep conditions stable even when the outside temperature varies by tens of degrees over a day-night cycle and seasonally. In addition, all the electrical systems that comprise an accelerator eventually convert their electrical input energy into heat, which must be taken away, often by water cooling since this is an efficient way to rapidly carry away heat. A rule of thumb is that water cooling adds an overhead of around \qty{35}{\percent} to the 
electrical power requirements.

For most future high energy physics machines being explored, the RF power is the dominant contributor to 
wall-plug power consumption, either to accelerate the beam to collision energy or replace synchrotron losses. However, RF technology is at present rather inefficient as power is lost at each stage from the modulators, RF amplifiers (including their magnets and cooling), transmission losses, reflected power from the cavity and cooling/cryogenic power to keep the RF at a constant temperature. In most machines the wall-plug to beam efficiency is only \SIrange{5}{20}{\percent} depending on the accelerator. However new technologies are being developed that can substantially improve this efficiency when used together.

\begin{figure}[ht]
    \centering
    \includegraphics[width=0.8\textwidth]{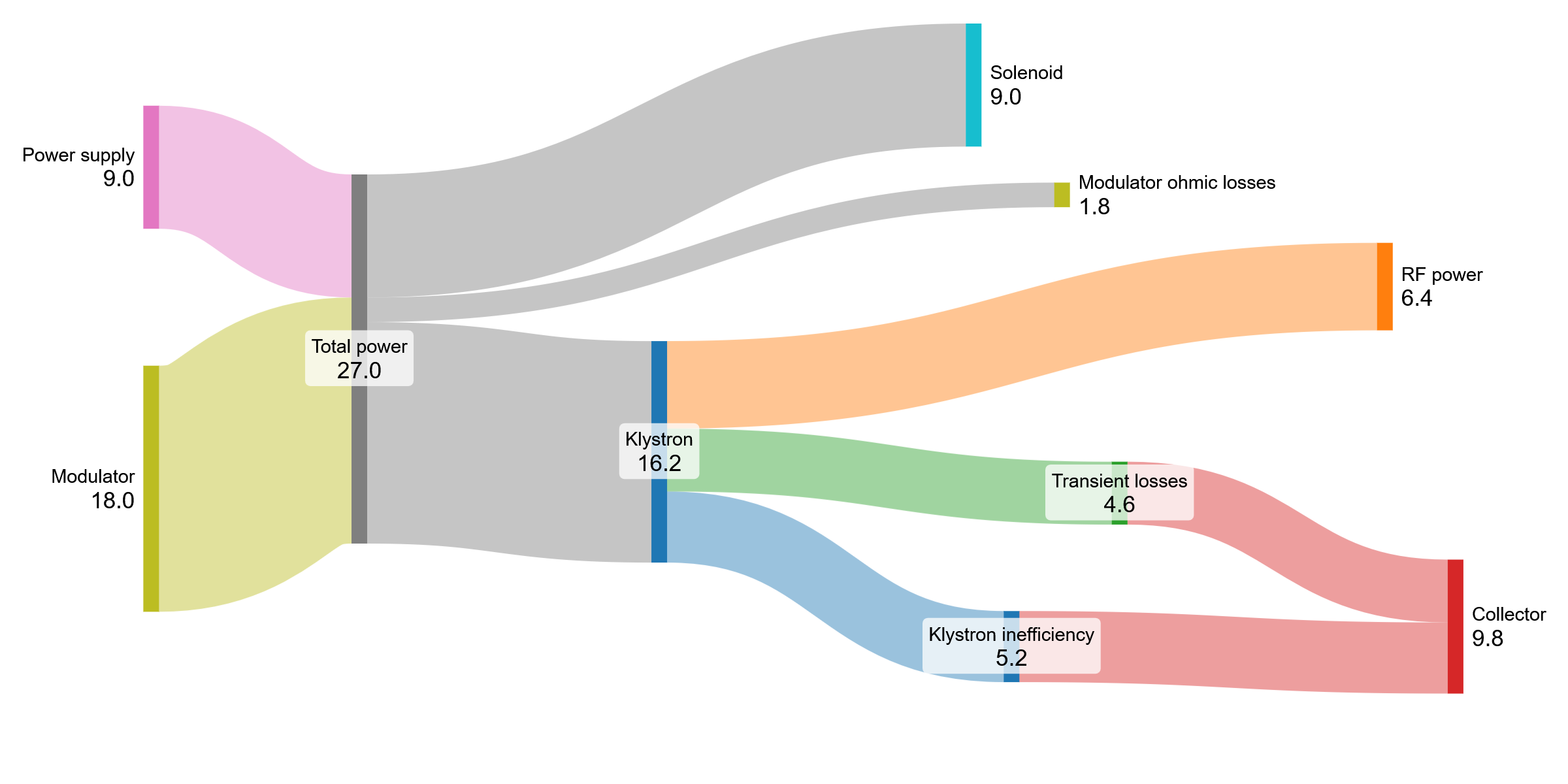}  
    \caption{Energy flows for a high-efficiency klystron operating in pulsed mode~\cite{Catalan_Lasheras_klystrons}.}
    \label{fig:klystron_sankey}
\end{figure}

The first major loss mechanism we encounter moving from the wall-plug is the RF amplifiers. Typically a klystron has a saturated efficiency of \SIrange{40}{60}{\percent}, which is the efficiency when running at maximum power. However, no accelerator is able to run at maximum power, since an overhead is required to allow the Low-Level RF (LLRF) system to deal with transients and hence the amplifier is ``backed-off''. An inherent issue with klystrons is that the efficiency is linearly proportional to the RF power so backing-off by \qty{20}{\percent} reduces the efficiency by the same amount. At present, research is ongoing to develop high efficiency klystrons (HEKs), which have saturated efficiencies between \SIrange{70}{85}{\percent}. A prototype of an HEK for LHC is being tested at CERN, which uses higher harmonic cavities to improve the klystron efficiency: this is known as the core stabilization method. Another concept is the two-stage klystron, which has the beam initially at low voltage to use the increased space-charge forces to improve the efficiency, before using a post-acceleration stage to increase the beam energy to achieve high RF powers. Two-stage klystrons are under investigation at present for FCC-ee, CLIC and Muon Colliders. Another approach is to deal with the backing-off rather than the saturated efficiency. DESY is looking at the use of AI LLRF systems that require less overhead, and Cockcroft/CERN have proposed the use of a smart modulator to control the lower voltage in a two-stage klystron to create fast changes in the saturated output power. Other amplifier types, such as solid-state power amplifiers (SSPAs) and Inductive Output Tubes (IOTs), maintain 
their efficiency when backed off; however,  they currently lack 
the power output  needed to replace klystrons. CERN recently installed SSPAs in the SPS accelerator using a novel cavity combiner which 
links hundreds of transistors to achieve 
high power, but scaling to the MW power scale will need higher power transistors (potentially SiC or GaN) or larger scale combining. Multi-beam IOTs have been proposed to increase their power and CERN has recently launched a study of a two-cavity variant for FCC called 
Tristron. \autoref{fig:klystron_sankey} shows typical energy flows for a high-efficiency pulsed klystron.

Another major loss of power is the signal reflected from the RF cavity, arising from transient changes in frequency, either due to beam loading or mechanical vibrations (microphonics). This results in cavities being designed with significantly lower Q factors that is ideal in order to widen the bandwidth, but this causes an impedance mismatch and hence reflections. A recent proposal to reduce this is the fast reactive tuner, where power flows to and from a reactive load in order to alter the cavity frequency by varying the reactance. The two proposed methods use either ferroelectric or ferromagnetic tuners. The faster response of these is the ferroelectric tuners which are being investigated at CERN, HZB and Cockcroft. Here a DC voltage is applied across a dielectric that is part of an RF circuit, changing its capacitance and hence reactance. Such tuners can in theory respond in \qty{100}{\nano\second} to changes in the beam or cavity, allowing Q factors to be an order of magnitude higher, significantly reducing reflections. It should be noted that this is only useful if the beam loading is reactive or, if it is resistive but small in SRF cavities, otherwise the beam or ohmic losses widens the bandwidth naturally and hence no reflections occur.

The final major loss in efficiency comes from ohmic losses in the RF cavities themselves, which needs to be replaced by the RF system. Superconducting cavities can reduce this loss substantially but cryogenic cooling to keep SRF cavities at the desired frequency is very poor due to the Carnot efficiency, which limits any heat engine to $T/300$ where $T$ is the operating temperature in Kelvin. This efficiency could be significantly increased by operating at a higher temperature; however, the resistance of superconducting materials increases substantially with temperature, meaning typical operation for Niobium is around \qty{2}{\kelvin}. New materials such as \ce{Nb3Sn} have transition temperatures twice that of niobium and hence will double the efficiency; however, they are brittle and can only be applied as thin layers on niobium or copper. Significant international effort is underway to develop and demonstrate reliable methods of creating \ce{Nb3Sn} thin films that achieve similar or better performance than bulk Nb. It should also be noted that mining Nb is very energy-intensive and the amount of \ce{CO2} released due to the production of a kg of Niobium is significantly higher than copper. So only using a thin deposited layer of \ce{Nb3Sn} on the inside of a bulk copper cavity also reduces \ce{CO2} production in manufacturing.

Combining all these technologies is required to reduce the energy usage of the RF system for future accelerators. Significant international R\&D is required to get each technology mature enough to be applied, but it is vital to the environmental sustainability of our field that such developments happen swiftly and in a coordinated manner.

\subsection{Particle Detector operation}
\label{sec:detector}

Detectors have recently been included in the LCAs for the construction phase of future accelerators~\cite{clic_ilc_lca-accel_arup}. Their environmental impact has mostly been evaluated based on their mass, with major contributions coming from the iron used in the return yokes of large magnets for particle tracking, and from absorbers in muon detection systems and calorimeters. This impact is substantial -- comparable to that of civil engineering works and accelerator components, as discussed in section~\ref{par:LCA_acce_det} and shown in table~\ref{tab:ghg-project-FCC}, and tables~\ref{tab:ghg-project-CLIC}-\ref{tab:ghg-project-ILC}.

During operation, indirect emissions from detectors -- due to power consumption -- account for only a small share of total emissions, which are largely dominated by the accelerator and related infrastructure. 

One area that still raises concerns is the direct emissions from cryogenic systems and gaseous detectors used in tracking chambers and muon detectors, which will need to be addressed and mitigated in  experimental apparatus of future accelerators.

The performance and long-term operation of gaseous detectors rely primarily on the use of the optimal gas mixture, which is the detector’s active medium where the primary ionisation happens. As introduced in section~\ref{sec:RIoperation}, several gaseous detector technologies make use of gas mixtures containing expensive or greenhouse gases, which have specific properties that allow optimal detector performance and avoid ageing effects. The GHGs most in use are the \ce{C2H2F4} (known as R134a, GWP of 1430) and \ce{SF6} (GWP of 23900) for the Resistive Plate Chambers, the \ce{C4F10} (GWP of 8860) for Cherenkov detectors and the \ce{CF4} (GWP of 7390) for wire chambers, Cherenkov detectors and micro-pattern gaseous detectors (MPGDs). These gases are necessary to mitigate ageing phenomena, as Cherenkov radiator and to contain charge development (thanks to their electronegative properties) or to improve time resolution.

In the large experiments, like the CERN LHC experiments, the detector volumes range from a few~\unit{\meter\cubed} to hundreds of \unit{\meter\cubed} making the use of gas recirculation systems compulsory to reduce operational costs and GHG emissions. Even with the implementation of these systems, in some cases emissions can be present mostly due to detector requirements or presence of leaks. In the specific case of CERN, the annual emission during Run 2 was around \qty{180}{\kilo\tonne\ce{CO2}e} (in 2022)~\cite{CERN:2023a}, \qty{50}{\percent} of which came from particle detection of the LHC experiments. It is worth mentioning that these GHG emissions would have been about ten times higher without the implementation of gas recirculation systems. 
The remaining emissions were due for around \qty{20}{\percent} to specific cases where it was not possible to recirculate \qty{100}{\percent} of the gas due to detector constraints, and for around \qty{80}{\percent} to the presence of leaks at the detector level in the ATLAS and CMS RPC systems. These leaks are mainly due to the breakage of plastic pipes and connectors, which break due to built-in fragility and mechanical stress. These leaks are not accessible during LHC run periods and, in some cases, also during end of the year technical stops. Big leak search and repair campaigns usually take place during long shutdown periods, but leaks keep developing.

One of CERN objectives is to reduce its {\textit{Scope 1}} emissions in the coming years. To fulfil this objective a strategy based on three action lines has been elaborated:
\begin{itemize}[nosep]
    \item \textit{Gas Recirculation.} The gas mixture is taken at the output of the detectors, purified and sent back to the detectors. It is technically possible to recycle \qty{100}{\percent} of the gas mixture.
    \item \textit{Gas Recuperation.} The gas mixture is sent to a recuperation plant where the GHG is extracted, stored and re-used. This system is used always in combination with a gas recirculation system to allow a further GHG reduction.
    \item \textit{Alternative gases.} Search of alternative gas mixtures suitable for particle detectors that do not contain or limit the use of GHGs.
\end{itemize}

The reduction of the use of F-gases is fundamental for the next LHC runs and future particle detector applications because of the implementation in Europe of the F-gas regulation~\cite{EU:2018_1999}. 

This regulation establishes the total elimination of hydrofluorocarbons by 2050. A phasing down policy for the reduction of national quota of F-gas refrigerants is in progress and the F-gas availability in 2030 will be less than \qty{25}{\percent} with respect to 2014 data. This has already an impact on prices~\cite{oei_2869}, which are increasing, obviously leading to an increase in the operational costs of detector systems. Furthermore the availability of these gases for the time to come is not known and this aspect has to be taken into account seriously for future detector operation. It has also to be mentioned that in 2023  the European Chemicals Agency (ECHA) released a proposal regarding restriction on PFAS, i.e. per- and polyfluoroalkyl substances, which contains at least one fully fluorinated methyl (\ce{CF3}-) or methylene (-\ce{CF2}-) carbon atom (without any H/Cl/Br/I attached to it). The proposal envisages to cover over \num{10000} different PFAS, which are considered environmental pollutants with links to harmful health effects. Most of the so-called `eco-friendly' gases belong to the PFAS family.

\subsubsection{Recommendations concerning gaseous detectors}
With climate change a growing concern and implementation of F-gas regulations, it is fundamental for existing and future particle detector applications to reduce or eliminate the use of GHGs.

Most of today’s emissions at CERN are due to leaks on detectors that are impossible to repair during LHC Runs and, in some cases, also difficult to access during the long shutdown periods. The first point to consider for future experiments is to be sure to have leaks-free detectors. A specific attention has to be addressed in the engineering of the detectors, in particular the use of plastic connectors and pipes have to be absolutely avoided. Reliability tests have to be conducted thinking that the detectors need to work flawlessly for a long time (tens of years) in a harsh environment with limited or no access for repairing. Quality control and assurance tests have to be extensively performed. Another weak point in the construction of the detector is the use of materials that can produce outgassing or that are permeable to air or humidity. A database of allowed materials should be made available for the detector community. It has to be considered that a leak-free detector is essential not only in case of usage of GHGs but also when expensive gases are used and large volumes are involved.

For future experiments, the better way to proceed is for sure the complete elimination of GHGs. Considering also the new concerns on PFAS, these gases should be avoided (this family includes the nowadays tested HFOs). The future detectors will have to be designed with already in mind a well-defined gas mixture and not, as it is usually done, by choosing the gas mixture at the very end of the design process following the physics requirements.

In case it will not be possible to avoid the use of GHGs, the best way to reduce the emissions is the implementation of gas recirculation systems. In case of no constrains at detector level and no presence of leaks, it is technically possible to reach almost \qty{100}{\percent} recirculation, i.e. no emissions. If it is not possible to recycle \qty{100}{\percent} of the gas mixture, recuperation systems can be envisaged. They are complex and custom-made systems permitting to extract the GHG, store and reuse it.

\subsection{Data Centers operation}

All present and future large Research Infrastructures produce enormous amounts of data. The required data-processing is traditionally split into “near-source’” or “edge” processing and "offline" computing. The first is all the data-processing and storage which is most efficiently done in close proximity or even directly connected to the detectors. 
Because at this stage often the data are processed in real-time, i.e.  as they are produced by the instruments, this is also often called “online” processing. In most cases the goal of this step will be significant data-reduction by means of filtering, feature-extraction and compression. 

Since this data-processing is co-located and an integral part of the research infrastructures, it is considered in this report in view of its sustainability impact. 

There is a second crucial phase of the data-processing which has as goal the extraction of the ultimate scientific results, usually in the form of published papers. Data are often processed long after their initial creation, and 
the same data-set can be reprocessed multiple times -- for example, in light of new theoretical models or improved statistical methods. 
This processing is performed on data stored 
in mass-storage systems. The computing and storage infrastructure required for this stage differs in no significant way from other types of data-processing common in the commercial world, which is dominated by cloud computing. Like cloud computing the physical location of the processing and/or the distribution of the compute elements has no technical constraints, but is dictated more by cost considerations\footnote{Privacy considerations, very important in most commercial applications, are usually not critical for data in the physical sciences, but can of course be very relevant in the life-sciences.}. In the world of particle physics a model for federating compute infrastructures from many stakeholders has been developed more than 20 years ago, which is the Worldwide LHC Computing Grid, WLCG (``the grid”). 
Given the ongoing initiatives aimed at studying and improving the sustainability of large-scale computing infrastructures (see, e.g., Ref~\cite{greenDigit_sipos_2025}), and their similarity to widely used commercial systems -- whose environmental footprints are documented -- we do not address the sustainability assessment of these facilities. 
Guidelines and key parameters for their assessment 
are readily available and continuously updated.

The world of on detector processing is full of jargon. While we try to avoid this as much as possible in this section, it is necessary to introduce a few key-concepts here: we refer to \textit{online} computing to all processing on data which is not (yet) permanently stored on some mass-storage device. Very often the data is \textit{in-flight} i.e. existing only in some ephemeral buffer-memory. \textit{Real-time} refers to what is more specifically called \textit{hard real-time}, which is any computing operation which is guaranteed to not take more than a definite, maximum time. This is crucial for hardware trigger systems with limited buffers. It should not be confused with \textit{soft real-time} systems, which is a more vague term referring to all kinds of "fast" compute attempting to minimise latency.

\subsubsection{The landscape for computing for particle detectors}

The LHC experiments are not only dealing with unprecedented volumes of data, they are also requiring the largest amount of on-site computing to process and select interesting data. They are therefore a good starting point to think about the needs of future large detectors.

\subsubsection{Driving factors for increased compute resources}
Complex experimental signatures make it more and more difficult to do efficient selections using simplified or local information only. Often a large fraction of 
available data from a bunch-crossing are required. 

\subsubsection{Energy efficiency, capital expenses and operational expenses}
The resource consumption (energy and others) for the \emph{production} a microchip -- of the same type (e.g.:~an x86 processor) and generation -- is essentially independent of its "inner" features, such as maximum clock-speed or number of compute-cores. The capital expenses are on the contrary very different as the customer essentially buys into the yield-curve. The best chips are often disproportionally expensive. 
This means 
that, for a given total required amount of compute power (in our community often measured in HEPScore/HEPSpec), the most cost effective approach in terms of initial capital expenditure, 
is to deploy 
a larger number of cheaper CPUs rather than the minimal number of most powerful ones. 

The current incentives set by limited budgets are such that minimisation of capital expenses is the paramount goal. This however has a double negative impact on the environment: i) more CPUs and servers mean more environmental resources used to create this hardware than would be necessary if more powerful devices would have been bought, and ii) more servers, even if equipped with weaker CPUs, will use more energy in the aggregate over time. This logic is true ceteris paribus also for other types of compute elements and also for storage (GPUs, FPGAs, SSDs...). 

The adoption of sustainable and responsible procurement criteria -- as discussed in section~\ref{sec:resp_procur} -- which include assessments of the environmental impact of an experiment’s computing infrastructure, not only during production but also throughout operation and disposal, should contribute to reducing \textit{overall} costs. The higher upfront cost of purchasing fewer, more energy-efficient components should be considered an integral part of the resources required to implement responsible procurement practices within the RI.

\subsubsection{Resource usage for compute}
The production of silicon chips uses a lot of resources: manufacturing a single 8-inch wafer causes consumption of \qty{2.58}{\meter\cubed} of water and \qty{361.3}{\kilo\watt\hour} of energy, and emissions of \qty{263.9}{\kg\ce{CO2}e}~\cite{WANG202347}.

There is also resource consumption in the production of PCBs and necessary mechanics, but the production of the microchips dominates by far.
In the operation of the hardware there is the consumption of electrical energy to switch the transistors, which is converted into heat.
The cooling of microchips also requires energy, either by the need of moving air to cool and the treatment of that air, or also of moving some other refrigerant and cooling it in its turn. 
Depending on the location and technology of the data centre infrastructure this requires more or less electrical energy for pumps, fans, ventilators, compressors, water-towers and so on.

\subsubsection*{A sample calculation on the power-savings potential by upgrading}

Typically server chips are ``refreshed'' every 18 to 24 months. While generation-to-generation performance improvements are varying and depend also on the specific workload, let us assume a modest \qty{10}{\percent} improvement in power efficiency. Let us take a modern CPU with a \qty{200}{\watt} maximum power consumption and let us assume, unrealistically, that the CPU is running flat out all the time. Then taking the numbers for chip production above, the typical die size of a server CPU, and some simplifying assumptions, the power consumption for manufacturing a CPU is about \qty{7.2}{\kilo\watt\hour}. That means that after only 360 hours one will have started to actually save on power. In practice other constant overheads in the full-server production and so on will make the time to break-even longer, but still, from a purely energy saving point of view, frequent upgrades are beneficial. 

\subsubsection{The influence of AI}
AI is and will be a pervasive aspects of all compute infrastructures. For future online processing installations one can currently see three of particular importance. 
\begin{enumerate}[nosep]
    \item AI on the front-end can lead to data-reduction at the source, which can reduce resource needs all the way to final storage. This is particularly attractive, when radiation is of no concern and industry standard devices can be used, when power and space permit it. Using commercially available chips made for the ``edge-computing'' will allow profiting from the developments of a very dynamic and resourceful industry, far beyond what ever can be hoped to be developed within the HEP community. 
    \item AI in the later stages of online data processing, in the so-called software triggers is interesting because it can shift processing power to different locations and times. This is because, broadly speaking, AI-models are trained offline and, more often than not, on completely different hardware than is used later for \emph{applying} the trained models to new data. This can allow reduction of the power consumption and also the amount of hardware at the experiment site. Since training is not done as frequently, this could also allow for sharing resources with other fields of research and also common facilities such as supercomputing sites. This again corresponds to a very common scheme applied also in the world of commercial computing and is therefore guaranteed to profit from the developments there.
    \item Lastly, AI can also help directly in the optimisation of the resource usage in the compute facilities by optimising cooling and other operational parameters. Again this is an active area of research, where cross-fertilisation with other fields and industry is very likely. 
\end{enumerate}

\subsubsection{Server cooling and heat-reuse}
In the current computing industry three main methods for cooling computer hardware have been established. They are in order of decreasing popularity:
\begin{enumerate}[nosep]
    \item Cooling with air
    \item Direct to chip liquid cooling (DLC, DCLC)
    \item Immersion cooling
\end{enumerate}

Immersion cooling is plagued with compatibility issues and it cannot use standard rack-infrastructure. There is currently no sign that it will widely displace the other two methods so it will be ignored here.

Cooling with air is very cheap to deploy and well-known but is coming more and more to a limit as the power-dissipation of modern chips tends to go higher and higher. In direct chip liquid cooling normally only the most dissipating chips are cooled with cold-plates cooled by a refrigerant (often just water with glycol) and the rest of the server infrastructure continues to be cooled with air.

In a pure air-cooling system the inlet air is typically heated up by about \qtyrange{20}{30}{\degreeCelsius}. It is possible to capture this heat and use it to warm up water. The efficiency of such a system will not be very high and it will depend on the proximity to a potential use of the waste heat, for instance in a building heating system. 

Direct liquid cooling warms up the refrigerant passed directly through the heat sinks on top of the chips. These loops can run at much higher temperature and the required liquid-to-liquid heat-exchange is quite efficient. Overall efficacy will depend again on the ease of access to a reuse facility. Another advantage of DLC is the possibility to use water directly from cooling towers as primary coolant, 
with no need of pre-cooling. 

The efficiency of the cooling system is characterised by the so called power-usage efficiency (PUE) defined as
\[\mathrm{PUE} = \frac{\mathrm{Total~power~used}}{\mathrm{Power~used~by~the~IT~equipment}}\]
Despite various methodological shortcomings\footnote{For instance: the considerable power consumption of the internal fans of a computer is counted in the numerator, i.e. as useful power, while it should be included in the overhead. The obvious reason for this, is the easy definition of the measurement point.} this number gives a simple scalar figure of merit. Without heat reuse the best theoretical value is $1.0$ and modern, good systems are normally achieving values $< 1.1$. Heat reuse can improve this beyond that limit. Heat reuse is easier with DLC, 
since  the starting point is a warm liquid (up to \qty{50}{\degreeCelsius}). This and the increasing power densities of modern microchips are good arguments to invest in DLC for future compute infrastructures.
          
It should be borne in mind that one important issue with PUE is that it only takes into account electrical energy. In a system with adiabatic assist for instance, the power consumption can to some extent be reduced, and hence the PUE improved, by using more water for adiabatic cooling. The water consumption, though also an important environmental impact, is not showing up in PUE however. 

\subsubsection{Proximity of data centres to facilities}

Modern high-speed networks are very energy-efficient. In 2024, sending \qty{400}{\giga\bit\per\second} requires about \SIrange{200}{300}{\watt} for a range of about \qty{80}{\km}~\cite{10474160}, and this value is likely to decrease in the future. That means transporting all of the data of a large experiment with \qty{100}{\tera\bit\per\second} sustained acquisition will only require about \SIrange{45}{70}{\kW}. So the energy cost of transporting data off-site is almost negligible. Of course, depending on the distance, such an transmission network can have significant capital cost. This interesting fact is mentioned here to make the point that most of the data processing for an experiment, from an energy consumption point of view, can easily be relocated wherever it is most suitable: for instance in a location where the inevitable waste heat can be used meaningfully to heat something. Since longer-distance connections will inevitably be more expensive to put in place, this is another illustration of the conflict between (initial) capital expenses and (long-term) energy efficiency. 

\subsubsection{The size of future HEP data centres}
Given the energy needs of future detectors and accelerators, it is interesting to make an educated guess of the power requirements of the computing infrastructure of the future. Looking at the capital cost of recent large experiments (excluding labour), the budget for compute infrastructure is below \qty{10}{\percent}. Assuming 500~MUSD for a large detector, that means that at most 50~MUSD will be spent on IT infrastructure. High-end CPUs and GPUs today consume around \qty{300}{\watt} and cost somewhere between 3000 and 10000 USD. Neglecting costs for memory, power-supply, chassis and storage, all of which contribute comparatively little to the power consumption, 50~MUSD will buy between 10000 and 15000 chips at most, so the maximum power will be between \SIrange{3}{4.5}{\mega\watt}. This is about two times larger than what is found today in the LHC experiments, but is very small when compared to today's large commercial data-centres.

\subsubsection{Recommendations regarding local data centers}
From the considerations in this section the following recommendations can be derived to minimise the energy consumption and maximise the overall sustainability of future HEP data centres. For the construction of data-centres serving future detectors in the 15 to 20 year time frame:
\begin{enumerate}[nosep]
    \item Since the transport of data can be expected to continue to decrease in cost and energy consumption and larger facilities will almost always be more efficient in cost and energy terms to operate, the first consideration before building an on-site data-centre should always be if the experiment-specific compute cannot be hosted off-site.
    \item If a data-centre at the experiment site needs to be built, prefabricated, modular structures will be more cost-effective, have a lower ecological footprint than traditional brick-and-mortar facilities and allow for easier customisation to leave out unnecessary features.
    \item In moderate climate regions, a year-averaged power efficiency better than $1.1$ 
    should be the minimum target, and energy/heat-reuse should be considered from the start of the design. Direct to chip liquid cooling facilitates heat-reuse and is future-proof for increasingly power-hungry chips, which can justify its higher initial cost. 
    \item Most optimisations of the energy consumption come with a higher initial capital cost for hardware. If one is serious about reducing energy consumption in future IT facilities for experiments, this fact should reflect in the funding schemes, for instance energetically well justified extra cost over the cheapest possible solution should be granted, as discussed in general in section~\ref{sec:resp_procur}. 
\end{enumerate}
For the operation of these data-centres:
\begin{enumerate}[nosep]
    \item When selecting 
    compute equipment, the power-consumption and effective usage over the full lifecycle of the experiment should be considered. Specialised technology (e.g. FPGA, IPU etc\ldots) might have a lower power consumption during operation but in all likelihood will have a significantly less effective usage over the full lifetime and thus a worse ecological footprint than a more generically useful technology (e.g. CPU, GPU). 
    \item The best performing and most expensive chips are the most power-efficient. The energy consumption can be reduced by the willingness to invest more upfront for equivalent computing power. 
    \item When the equipment is not used, idle power consumption should be minimised by appropriate configuration or by selective powering. 
    \item Power-efficiency must be part of the operational principles and the guiding policy for the operations team.
\end{enumerate}

\subsection{Decommissioning}
\label{sec:decomissioning}

The decommissioning phase (stages C1 to C4 of the LCIA process, see \autoref{fig:lca-stages}) of an accelerator covers the environmental impacts after the end of operation until all components and materials no longer needed are properly disposed, namely the demolition stage (C1), transport of material (C2), processing of waste (C3), and final disposal (C4).

A number of difficulties arise in the evaluation of the impact 
of the decommissioning phase.
First, accelerators are often very long-lived, and the actual end of life is often unknown at the time of the project planning.
For example, CERN's proton synchrotron, originally commissioned in 1959, today, 65 years later, is still an integral part of the LHC's injection chain.
Other accelerators, for example HERA at DESY, have been decommissioned, but still await final demolition and disposal of components, many of which are still usable.
The ALPS-II experiment~\cite{Bahre:2013ywa,Hallal:2020ibe} at DESY has demonstrated that this approach is well motivated:
ALPS-II uses 24 superconducting dipoles of the HERA accelerator for an experiment that searches for axion-like feebly interacting particles, utilising the magnets in a way that was never envisaged in the original design.

Some accelerators, such as LEP at CERN, have in fact been demolished completely. 
A report~\cite{international2020iaea} commissioned by the IAEA has collected valuable information on this process. 
Recently, the procedures to reclassify radioactive materials from accelerators at CERN have been published~\cite{Svihrova:2024tzy}.

An issue particular to the disposal phase of accelerator projects is the treatment of activated material.
In adherence to the strict legal regulations concerning activated materials, all equipment removed from the tunnels and caverns of a decommissioned accelerator has to be tested for the presence of activated material. 
All components need to be treated in a controlled way according to the applicable standards, which vary considerably between countries (see e.g.~\cite{Svihrova:2024tzy} for a discussion of the situation at CERN, with two relevant host countries). 

From an inventory collection point of view this poses challenges that are currently unresolved. 
The disposal lifecycle of material that has been subjected to radiation depends on regulations that vary between countries and may be finalised as part of the project approval process.
Therefore, defining material categories such as ``stainless steel, 
induced activity below a threshold in  \unit{\micro\sievert\per\hour}'' or ``stainless steel, total irradiation below a threshold in \unit{\mega\gray}'' that correspond to a specific disposal path or allow an estimation of subsequent environmental impact of the disposal process is difficult and has yet to be done for an LCA.
Also, predicting the amount of radiation materials are subjected to in the course of accelerator operation generally requires dedicated simulations of the radiation field, material composition and operating scenarios, which is typically done only for certain parts of the accelerator, and typically only in the engineering design phase of a project.

Radioactive waste at accelerator installations is mainly produced in activities of preventive and corrective maintenance and during decommissioning. The latter can be motivated by the need to upgrade or completely replace an experiment or installation. Typical waste from maintenance and repair worksites are: small metallic components, cables, light units, ventilation units, massive equipment (e.g. magnets, dumps), technological waste (e.g. gloves, overalls). Experiments can generate specific waste, like irradiated targets, part of detectors and irradiated electronics. Renewal of used or obsolete equipment in consolidation and upgrade projects generates mainly metallic waste (components, supports, etc.). Decommissioning activities can generate considerable volumes of waste (e.g. concrete from civil engineering modifications, infrastructures, massive objects, important quantities of cables and metallic supports like cable trays).   

Most of the radioactive waste that is generated in particle accelerators can be classified as waste with very-low-level (VLL) ~\cite{international2020iaea} activity, which can also be candidate for clearance from regulatory control (also called free-release) in those countries where this procedure is defined. Targets, beam-dumps and any other accelerator components that are directly hit by the beam can reach low- to intermediate-levels of activity. The disposal of activated waste towards final repositories requires accurate radiological characterisation to ensure that the activity they store falls below the activity limits for which the repository was designed.

The activation varies with the type of accelerated particles, the irradiation history, the chemical composition of the material and its geometrical characteristics (e.g. size). Many details on the type of activation with relation to the type of installation are available in a publication on the decommissioning of particle accelerators issued by the International Atomic Energy Agency (IAEA)~\cite{international2020iaea}. 

The possible disposal pathways depend on the national regulations. In Europe, the European Commission (EC) provides directives~\cite{euratom} to be adopted by its member states. Each country is allowed to introduce further regulations and laws as compared to the EC requirements. For instance, while Switzerland has clearance limits corresponding to those recommended by the EU, the French legislation, until February 2022 ~\cite{JO-2022-174, JO-2022-175}, did not provide clearance limits below which waste coming from a radiological area can be treated as non- radioactive and cleared from the regulatory control. Release of such waste was only allowed after a detailed theoretical study supported by extensive experimental measurements to produce a detailed “zoning” of the accelerator and experimental areas, i.e., a classification of areas where material could or could not have been activated. This approach was adopted at CERN at the time of the Large Electron-Positron Collider (LEP) decommissioning in year 2000~\cite{international2020iaea}.

Radioactive waste management starts at the moment of the conception of a facility or an experiment and comes to an end by the disposal of radioactive waste to the final repositories. In the last few years CERN developed and implemented several tools to identify the chemical composition and the radioactive inventory of elements or compounds activated in any CERN facility or experiment (ActiWiz~\cite{Vincke:2636327}). The early consideration of the waste disposal pathways when designing an experiment or a facility not only contributes to the minimisation of the radioactive waste, but also facilitates the entire industrial disposal process and the evaluation of the LCA. 

From an LCA perspective, the important factors are:
\begin{enumerate}[nosep]
\item	The minimisation of the production of radioactive material and waste. This encompasses:
\begin{enumerate}[nosep]
\item Optimisation of the design of machine components (e.g. design adapted to an easy separation of activated parts, emphasis on low maintenance and long lifetime) in order to minimise waste without jeopardising technical or scientific performance;
\item Choice of the best composition of the material (i.e. whenever several different chemical compositions or alloys are equally suitable, identification of those less prone to activation);
\item Optimisation of the beam parameters in order to minimise the beam losses;
\item Radiological zoning and sorting at the source to avoid unnecessary production of radioactive waste.
\end{enumerate}
\end{enumerate}
Further factors that influence indirectly the quantity and quality of radioactive waste to be managed, and therefore the total `inventory' that should be considered in the LCA evaluation, are: 
\begin{enumerate}[nosep]
\item The possible reuse of components within the same installation, contributing to the minimisation of the total quantity of raw material that is used and potentially activated. 
\item The availability of centralised temporary storage in order to profit from natural decay and optimise processing for disposal by campaigns of waste with similar characteristics.
\item The treatment of waste in a centre with state-of-the-art equipment dedicated to the processing (dismantling, sorting, size/volume reduction and packaging) of radioactive waste, eventually disposed  towards the final repositories, in order to minimise the total volumes needing transport and final storage.
\item	The development of further innovative scientific techniques to characterise (i.e. define the radionuclide inventory) installed equipment as well as already stored radioactive waste;
\item	The assessment of the production of future radioactive waste. This estimation and knowledge on flux and radionuclide inventory are key parameters for an optimal radioactive waste management both for the installation and the final repository receiving the waste. 
\item	The communication to the many actors involved (e.g. equipment groups, experiments, etc.) to raise awareness of the elements of the radioactive waste programme and its evolution, such as the reception rules or waste forecast.  
\end{enumerate}

\subsubsection{Recommendations concerning radioactive waste}
A global approach, which takes into consideration the entire lifecycle of waste from the design phase of the facilities to the disposal of waste is essential. 
In general it is recommended that the management of waste is structured backwards, starting from the requirements of the final repositories and from the prescriptions provided by the regulatory framework of the country in which disposal will be performed. These requirements are transferred in internal acceptance criteria, to be complied with by the waste producers when delivering radioactive waste to the centralised radioactive waste treatment and storage facility. Experience at CERN shows that if these acceptance criteria are pragmatic, well defined and regularly updated, they are also rated as very useful during the planning and execution of the decommissioning projects. 

A reliable traceability system is essential during operation (for the recording of the operational beam parameters and for keeping trace of the movement of the components) as well as  during dismantling and decommissioning in order to ensure that sufficient information on waste origin and irradiation conditions are available for the characterisation step.

There is a limitation to the generalisation of recommendations on selection of materials that would reduce the environmental impact of activated materials linked to the existence of country specific regulations for the management of potentially activated materials. In general, tools for the evaluation of the radionuclide inventory as well as for the appropriate choice of component materials are nowadays available (e.g. ActiWiz) and should be used. 

 There is an environmental and economic benefits in performing the release from regulatory control whenever this procedure can be applied and is in line with the local regulatory framework. Instead of being stored for decades in a repository, released material is recycled in the metal industry, thus minimising total raw material required. 

\subsection{Data on Future Accelerator Projects}

A synthesis of data from sustainability assessment of the accelerator projects is requested in public debates,
within research communities and science management panels.

Proposals of future accelerators all contain running scenarios identified by center-of-mass (CoM) energy, average luminosity, duty cycles for the operation, integrated luminosity per year, number of interaction regions (IPs), and include upgrades of different kinds. 

Data from several projects of potential future circular colliders (FCC-ee: \autoref{tab:ghg-project-FCC}, CEPC: \autoref{tab:ghg-project-CEPC}, and LHeC: \autoref{tab:ghg-project-LHeC}) and linear colliders (CLIC: \autoref{tab:ghg-project-CLIC}, ILC: \autoref{tab:ghg-project-ILC}, LCF: \autoref{tab:ghg-project-ILC}, and C$^3$: \autoref{tab:ghg-project-CCC}) are presented in the following sections. 
Because LCAs are new to accelerator components and particle detectors, section~\ref{par:LCA_acce_det} includes, as an example, a description of the studies conducted in close collaboration by CLIC, ILC, and LCF projects.

\subsubsection{Evaluation of GHG emissions for the FCC-ee project}

The Future Circular Collider (FCC) programme~\cite{Benedikt:2928193, Benedikt:2928793,Benedikt:2928194} proposes a new infrastructure connected to CERN permitting the operation of at least two subsequently installed particle colliders: first, a lepton collider (FCC-ee) produces high luminosity beams at different centre of mass energy levels spanning the Z pole, the WW pair threshold, the ZH production peak and beyond the top-pair production threshold. Second, a hadron collider (FCC-hh) addresses the energy frontier with proton and heavy ion beams, running beyond the end of the 21st century. The injector complex for the lepton collider is entirely located on today's fenced CERN Prévessin site in France. The injector of the hadron collider would leverage the existing pre-accelerator complex at CERN fully. 
The FCC foresees a quasi-circular subsurface structure with a circumference of \qty{90.7}{\kilo\meter} and eight surface sites that occupy a total surface of 40 ha. The concept foresees four large experiments that are operated at the same time, permitting a sustained engagement of a user community of more than \num{12000} scientists. 
One experiment site exploits a synergy with the existing LHC surface sit P8 in France. Access shaft depths range between 180 and 400 m. One surface site would be located in Switzerland and 7 in France: 2 in the department of Ain and 5 in the Haute-Savoie department.

The FCC study has adopted the international lifecycle assessment standard EN ISO 14040/44 for GHG emission assessments. For calculations, the EN ISO 15804+A2 was used, the foundational basis for EN 17472. Compliant with the national regulations for environmental evaluations required to obtain project authorisations, this assures that all results are based on actual products available on the market today, thus increasing trust in the accuracy of the results.

Recent tunnel projects such as the construction of the Brenner base tunnel between Austria and Italy employ very low carbon cement and concrete in large quantities already today~\cite{spaun2021path,sauer_dissertation_2017}. The GrandParis project explicitly reports the use of C40/50 concrete with a carbon footprint of 294 kg/$m^3$ in 2023~\cite{SGP2023_GPE_GES}. A range of 250-300 kg CO2e per $m^3$ of C45/50 concrete based on 430 kg cement/$m^3$ with a CO2e emission of 0.54 tCO2e per tonne of cement should therefore be considered reliably available. Hence, the figures presented in Table \ref{tab:ghg-project-FCC} below are to be interpreted as highly conservative and no longer based on state-of-the-art for projects with an envisaged construction phase beyond 2030.

The applicable environmental regulations in the host-states France and Switzerland called for developing the implementation scenario according to the so called "Avoid-Reduce-Compensate" iterative improvement methodology. Independently developed and verified results are made publicly available by the FCC study after review by host state appointed committees. This includes not only the approaches~\cite{kpamegan_2024_14336970}, but also for instance the LCA results for the construction phase~\cite{mauree_2024_13899160}.

Concerning climate footprint of associated Scope 2 emissions, the official values of the Swiss federal office for environment (OFEV) and the French ADEME Agency for the Ecological Transition were used. Although the GHG emissions of existing nuclear energy in France exhibits the lowest carbon footprint, a market-based mix of renewable energy sources complemented with a small fraction of nuclear energy has been selected as the basis for the estimations to account for the engaged energy transition~\cite{Benedikt:2928194}. This choice results in a rather conservative reporting of the electricity consumption induced carbon footprint during construction and operation. 

Considering that the study is at the stage of advanced concepts and feasibility, technology choices for the technical infrastructures and accelerator components have not been taken. Experiment detector concepts are to be developed by international collaborations at a later stage. Therefore, GHG emission estimates for technical infrastructures, particle accelerators and experiment detectors are today excluded from the reporting. However, replacement of detector and cooling gases with alternatives are being studied, investigations of sourcing of low carbon emission materials, manufacturing and transport of equipment has begun and will be included in the environmental impact assessment reports. If needed, estimates of the GHG emissions for the four detectors can be based on the LCA for the CLIC detectors~\cite{clic_ilc_lca-accel_arup}, as similar detector concepts can be expected for all electron-positron  collider proposals.

The presented GHG emissions for FCC-ee are to be considered upper bounds. Further reduction potentials of the carbon footprint of civil construction have been identified. They for example can be achieved with a refined design, use of materials with higher performance and reduced carbon footprint and requiring less steel, further increase of local production of construction materials, use of recycled materials and optimisation of the construction processes. Based on today's knowledge, further improvement potentials are at least 5 \% for the surface sites and 16 \% for the underground construction works. However, implications on cost and schedule have not been assessed. In addition, the carbon footprint avoidance potential due to substitution by waste heat supply has been calculated~\cite{ginger_burgeap_2025_14719832, Benedikt:2928194}, but has not yet been integrated in the overall lifecycle analysis calculation, despite its significant benefit value.

R\&D is ongoing to improve the efficiency of the RF sources with high-efficiency klystrons and tristrons with potentials to further increase their efficiency from the presently assumed value of 80\% to 90\%.

A comprehensive socio-economic impact assessment has been carried out that comprised in addition to the classical terms also the positive and negative environmental externalities~\cite{csil_2025_15421975}. Those have been quantified using international and national standards for monetisation of environmental impacts (ISO 14008) and the guidelines 
for determining environmental costs and benefits (ISO 14007). All terms have been incorporated into a social cost-benefit analysis that led to an overall positive Benefit-Cost-Ratio of 1.20.
Table \ref{tab:ghg-project-FCC} presents in a synthetic form the key parameters of the first FCC stage, the FCC-ee together with the most relevant GHG emissions related to construction and operation and the underlying assumptions.

\begin{table}
 \small
 \centering
 \caption{Data on GHG emissions for the FCC-ee project}
 \label{tab:ghg-project-FCC}
 \small\addtolength{\tabcolsep}{-2pt}
 \resizebox{1.0\textwidth}{!}{
 \begin{tabular}{|l|c|c|c|c|}
 \hline 
   \textbf{ FCC Mode } & Z & W & ZH & ${\rm t}\bar{\rm t}$ \\
 \hline
    CoM energy [\unit{\GeV}] & 91.2 & 160 & 240 & 365\\ \hline
    Luminosity/IP $[10^{34} \unit{\cm^{-2}}\unit{\s^{-1}}~]$ & 140 & 20 & 7.5 & 1.4\\ \hline
    Number of IPs & \multicolumn{4}{c|}{4} \\ \hline
    Operation time for physics/yr  $[10^7 \unit{s}/ \unit{yr}]$ & \multicolumn{4}{c|}{1.2\footnotemark[1]} \\ \hline
    Integrated luminosity/ \unit{yr}  [1/\unit{fb}/ \unit{yr}]  & 68 000  & 9 600  & 3 600  & 670 \\ \hline
    Host countries & 
           \multicolumn{4}{c|}{France and Switzerland } \\
 \hline 
 \multicolumn{5}{|l|}{\textbf{GHG emissions from construction, stage A1-A5}} \\
 \hline 
    Subsurface tunnels, caverns, shafts [\unit{kt}~\ce{CO2}e] & \multicolumn{4}{c|}{\makecell{480 (according to EN ISO 15804+A2)\\  \textasciitilde 1000 (with generic LCA data)} }\\ \hline 
    Surface sites and constructions [\unit{kt}~\ce{CO2}e] & \multicolumn{4}{c|}{\makecell{50 (according to EN ISO 15804+A2)\\ \textasciitilde 184 (with generic LCA data)}}\\ \hline 
    All constructions [\unit{kt}~\ce{CO2}e] & \multicolumn{4}{c|}{\makecell{530 (according to EN ISO 15804+A2)\\ \textasciitilde 1184 (with generic LCA data)}}\\ \hline 
    Accelerator (coll.) [\unit{kt}~\ce{CO2}e] & \multicolumn{4}{c|}{Not estimated: subject to equipment design \& procurement}\\
 \hline 
    Accelerator (inj.) [\unit{kt}~\ce{CO2}e] & \multicolumn{4}{c|}{Not estimated: subject to equipment design \& procurement}\\
 \hline 
    Detectors (A1 - A3, scaled from linear colliders) [\unit{kt}~\ce{CO2}e]  & \multicolumn{4}{c|} {142 - 186}\\
 \hline 
    \textbf{Total [kt~\ce{CO2}e]} & \multicolumn{4}{c|} 
    {Not estimated}\\ \hline 
    Collider tunnel length  [\unit{km}] & \multicolumn{4}{c|}{90.658}\\ \hline
    Collider tunnel diameter [\unit{m}] & \multicolumn{4}{c|}{5.5}\\ \hline
    Concrete GWP used for LCA  [\unit{kg}~\ce{CO2}e/ m$^3$] & \multicolumn{4}{c|}{220 (C12/15), 340 - 422 (C35/45), 400 - 540 (C40/50)}\\ \hline 
    Low-carbon concrete GWP (for comparison)  [\unit{kg}~\ce{CO2}e/ m$^3$] & \multicolumn{4}{c|}{150 (C12/15), 170 (C35/45), 175 (C40/50)}\\ \hline
    Accelerator  GWP / m  [\unit{t}~\ce{CO2}e/ \unit{m}] & \multicolumn{4}{c|}{Not estimated: subject to equipment design \& procurement}\\
 \hline 
 \multicolumn{5}{|l|}{\textbf{GHG emissions from operation}} \\
 \hline 
    Maximum power in operation [\unit{MW}]   & 222 & 247 & 273 & 357 \\ \hline
    Average power in operation [\unit{MW}]   & 122 & 138 & 152 & 202 \\ \hline
    Electricity consumption / yr  [\unit{TWh}/ \unit{yr}] & 1.2 & 1.3 & 1.4 & 1.9 \\ \hline
    Years of operation & 4 & 2 & 3 & 5 \\ \hline
    Carbon intensity of electricity [\unit{g}~\ce{CO2}e/ \unit{kWh}] & \multicolumn{4}{c|}{\makecell{14 - 18, conservative assumption based on actual data: \\12.625\footnotemark[3] in France and 15.7\footnotemark[2] in Switzerland}} \\ \hline
    GHG emissions/year of physics operation [k\unit{\tonne}~\ce{CO2}e/yr] & 18 - 26 & 20 - 29 & 22 - 31 & 29 - 40 \\ 
 \hline 
 \end{tabular}
}
\end{table}
    \footnotetext[1]{185 operation days per year (1.6e7) with an efficiency of 75\% for physics.}
    \footnotetext[2]{Switzerland, Umweltblianz Strommixe Schweiz, BAFU/OFEV, 2018, \url{https://www.bafu.admin.ch/bafu/de/home/themen/klima/fragen-antworten.html}}
    \footnotetext[3]{France, Base empreinte, ADEME, 2023: 75\% off shore wind and 25\% nuclear, \url{https://base-empreinte.ademe.fr}}
    
\clearpage

\subsubsection{Evaluation of GHG emissions for the CEPC  project}

The CEPC is designed~\cite{CEPCTDRDec24} as an advanced electron-positron collider Higgs factory, operating at a center-of-mass energy of 240 GeV. This energy can be upgraded to 360 GeV for $t \bar{t} $ production. To optimize the performance-cost ratio, the collider will feature a 100 km tunnel with two interaction points. Given the 30 MW SR power for Higgs operation and 50 MW SR power for $t\bar{t}$, the expected luminosities are $5.0\times 10^{34}$ cm$^{-2}$s$^{-1}$ for Higgs operation and $0.8\times 10^{34}$ cm$^{-2}$s$^{-1}$ for $t\bar{t}$  production. The machine is projected to operate for 3,600 hours per year ($1.3\times10^7$ s/yr) at full power for electron-positron collisions and data acquisition. Consequently, the annual integrated luminosities are estimated at 1,300 fb$^{-1}$/yr for Higgs studies and 217 fb$^{-1}$/yr for $t\bar{t}$ production.

The CEPC tunnel is situated approximately 100 meters underground. Two cross-sectional designs are under comparison: a circular tunnel excavated using a Tunnel Boring Machine (TBM) and a gate-shaped tunnel constructed via drill-and-blast methods. The circular tunnel has a diameter of 6.5 meters. For the main tunnel, the estimated carbon footprint is 7 metric tons of \ce{CO2} per meter. Accounting for an additional 25\% in emissions from material transportation and on-site construction of auxiliary structures, this figure rises to 8.75 metric tons of \ce{CO2} per meter. A further 30\% emissions increase applies to auxiliary tunnels, including those for the klystron gallery, access shafts, caverns, and experimental halls. In total, the 100 km CEPC tunnel is projected to generate 1137 kilotons of \ce{CO2} emissions. The tunnel will use C25/C30-grade concrete, which has a GWP of 0.16 \unit{\kilogram {\ce{CO2}e} \per \kilogram}.

In line with the projected SR powers and luminosities, the maximum operational electricity demand is expected to reach 260 MW for Higgs operation and 433 MW for $t\bar{t}$ operation. The total annual electricity consumption accounts for two operational phases: a 3,600-hour period of accelerator complex operation at full power, followed by 2,160 hours dedicated to machine studies at 50\% power. This results in an annual electricity consumption of 1.3 TWh/yr for Higgs operation and 2.2 TWh/yr for $t\bar{t}$ operation. Under China’s national dual-carbon strategy, the carbon intensity of electricity generation is steadily decreasing. Assuming the CEPC begins operation in 2035 under the most optimistic scenario, and given that its location has not yet been finalized, the projected electricity carbon intensity could range from 12.5 \unit{g \ce{CO2}e \per kWh} (Qinghai province, lowest carbon intensity) to 673 \unit{g \ce{CO2}e \per kWh} (Neimenggu province, highest carbon intensity). Based on these estimates, the average annual operational emissions would vary from 16.4 to 883.0 \unit{kt \ce{CO2}e \per yr} for Higgs operation and from 27.3 to 1469.8 \unit{kt\ce{CO2}e\per yr} for $t \bar{t}$ operation.
\clearpage

\begin{table}[htbp]
 \centering
  \caption{Data on GHG emissions for the CEPC project}
 \label{tab:ghg-project-CEPC}
 \begin{tabular}{|l|c|c|}
 \hline 
   \textbf{CEPC} & ZH & ${\rm t}\bar{\rm t}$ \\
 \hline
    CoM energy [\unit{\GeV}] & 240 & 360\\ \hline
    Luminosity/IP $[10^{34} \unit{\cm^{-2}}\unit{\s^{-1}}~]$ & 5.0 & 0.8\\ \hline
    Number of IPs & \multicolumn{2}{c|}{2} \\ \hline
    Operation time for physics/yr  $[10^7 \unit{s}/ \unit{yr}]$ & \multicolumn{2}{c|}{1.3\footnotemark[1]} \\ \hline
    Integrated luminosity/ \unit{yr}  [1/\unit{fb}/ \unit{yr}]  & 1300   & 217 \\ \hline
    Host countries & 
           \multicolumn{2}{c|}{China} \\
 \hline 
 \multicolumn{3}{|l|}{\textbf{GHG emissions from construction, stage A1-A5}} \\
 \hline 
    Subsurface tunnels, caverns, shafts [\unit{kt}~\ce{CO2}e] & \multicolumn{2}{c|}{1137\footnotemark[2]}\\ \hline 
    Surface sites and constructions [\unit{kt}~\ce{CO2}e] & \multicolumn{2}{c|}{Not estimated}\\ \hline 
    Accelerator (coll.) [\unit{kt}~\ce{CO2}e] & \multicolumn{2}{c|}{Not estimated}\\
 \hline 
    Accelerator (inj.) [\unit{kt}~\ce{CO2}e] & \multicolumn{2}{c|}{Not estimated}\\
 \hline 
    Detectors [\unit{kt}~\ce{CO2}e] & \multicolumn{2}{c|}{Not estimated}\\
 \hline 
    \textbf{Total [kt~\ce{CO2}e]} & \multicolumn{2}{c|}{\textbf{1137}}\\ \hline 
    Collider tunnel length  [\unit{km}] & \multicolumn{2}{c|}{100}\\ \hline
    Collider tunnel diameter [\unit{m}] & \multicolumn{2}{c|}{6.5}\\ \hline
    Concrete GWP  [\unit{kg}~\ce{CO2}e/ kg] & \multicolumn{2}{c|}{0.16 (C25/C30)}\\
 \hline 
    Accelerator  GWP / m  [\unit{t}~\ce{CO2}e/ \unit{m}] & \multicolumn{2}{c|}{Not estimated}\\
 \hline 
 \multicolumn{3}{|l|}{\textbf{GHG emissions from operation}} \\
 \hline 
    Maximum power in operation [\unit{MW}]   & 260 & 433 \\ \hline
    Average power in operation [\unit{MW}]   &  &\\ \hline
    Electricity consumption\footnotemark[4] / yr  [\unit{TWh}/ \unit{yr}] & 1.3 & 2.2 \\ \hline
    Year of operation & 2035 & 2035 \\ \hline
    Carbon intensity of electricity\footnotemark[5] [\unit{g}~\ce{CO2}e/ \unit{kWh}] & 12.5 & 12.5 \\ \hline
    Average Scope 2 GHG emissions / yr [\unit{kt}~\ce{CO2}e] & 16.4 & 27.3\\
 \hline 
 \end{tabular}
\end{table}

\footnotetext[1]{Effective run-time for physics encompasses 3600 hours of data acquisition, with detectors operating at full capacity.} 
\footnotetext[2]{7 t/m \ce{CO2}e emissions for main tunnel; total footprint increases by 30\% due to auxiliary civil construction of klystron gallery, access shafts, caverns, and experimental halls, and by another 25\% for cargo transport.}
\footnotetext[3]{A Carbon footprint of 8.75 t \ce{CO2e}/m has been calculated by applying a factor of 1.3 to the base-value of 7 t \ce{CO2}e/m which accounts for the additional 25\% carbon emissions from transportation and auxiliary buildings .}
\footnotetext[4]{Electricity consumption accounts for 2 operation phases: a 3600 hours period of accelerator complex at full power (260 MW for Higgs and 433 MW for $t \bar{t}$), and additional 2160 hours dedicated to machine studies, at 50\% of full power.}
\footnotetext[5]{The electricity intensity utilize data from Qinghai province with clean energy sources - predominantly hydroelectric power, photovoltaic and wind energy.}
\clearpage

\subsubsection{Evaluation of GHG emissions for the LHeC project}

Achieving high-luminosity $ep$ collisions with the LHeC is made possible by the Energy Recovery Linac (ERL) accelerator technology (see Section~\ref{sub-sec:ERL} and \cite{LHeC2025Input}). This innovative approach enhances the energy sustainability of modern accelerators by recirculating the beam power rather than the particles themselves. The LHeC conceptual design incorporates 802 MHz superconducting radio frequency (SRF)
technology with an already reachable gradient of around 20 MV/m in CW mode with $Q_0$ values exceeding $10^{10}$.
Additional reductions of the power  requirements can be achieved by further developments of SRF technology including
high-temperature (4.5K) cavities with high $Q_0$, Fast Reactive Tuners (FRT) for microphonics mitigation, and high-efficiency
klystrons – in particular the very compact (2.8~m) two-stage klystron design for the FCC. 
Collectively, these innovations address the broader sustainability challenges facing future particle colliders. The LHeC serves as a pathfinder in demonstrating how overall energy consumption during operation can be significantly reduced.

Table~\ref{tab:ghg-project-LHeC} summarizes the data on the GHG emissions for the LHeC project. Compared to other projects discussed in this report, LHeC has the shortest length (8.9 km) of the 5.5~m diameter collider tunnel. The expected type of the concrete has the same GWP [kg CO$_2$e/kg] as for all other CERN projects. Thought not yet finalized, the estimated GHG emissions from the construction (stages A1-A5) can be prorated from, e.g., CLIC or FCCee numbers by simply taking into account the ratio of the collider circumferences (i.e., about 8.9/90.7 $\approx$ 10\% of the FCCee estimates). The estimates of the GHG emissions from operations, also not yet finalized, are expected to be close to a similar type (SRF) facility project with similar power consumption, e.g., the FCCee at $Z$-pole operation - see table~\ref{tab:ghg-project-FCC} above. Of course, the LHeC assumes availability and operation of the HL-LHC, which will provide 7 TeV protons to collide with 50/60 GeV electrons circulating in the ERL race-track facility, and accounts the added power of 60MW when comparing to HL-LHC. 

\begin{table}[htbp]
 \small
 \centering 
 \caption{Data on GHG emissions for the LHeC project.}
 \label{tab:ghg-project-LHeC}
 \begin{tabular}{|l|c|}
 \hline 
   \textbf{ LHeC } & \\
 \hline
    CoM energy [\unit{\GeV}] & 1180\\ \hline
    Luminosity/IP $[10^{34} \unit{\cm^{-2}}\unit{\s^{-1}}~]$ & 2.4\\ \hline
    Number of IPs & 1\\ \hline
    Operation time for physics/yr  $[10^7 \unit{s}/ \unit{yr}]$ & 1.6 \\ \hline 
    Integrated luminosity/ \unit{yr}  [1/\unit{fb}/ \unit{yr}] & 180 \\ \hline
    Host countries & 
           France and Switzerland  \\
 \hline 
 \multicolumn{2}{|l|}{\textbf{GHG emissions from construction, stage A1-A5}} \\
 \hline 
    Subsurface tunnels, caverns, shafts [\unit{kt}~\ce{CO2}e] & Not estimated \\ \hline 
    Surface sites and constructions [\unit{kt}~\ce{CO2}e] & Not estimated\\ \hline 
    Accelerator (coll.) [\unit{kt}~\ce{CO2}e] & {Not estimated}\\
 \hline 
    Accelerator (inj.) [\unit{kt}~\ce{CO2}e] & {Not estimated}\\
 \hline 
    Detectors [\unit{kt}~\ce{CO2}e] & {Not estimated}\\
 \hline 
    \textbf{Total [kt~\ce{CO2}e]} & { }\\ \hline 
    Collider tunnel length  [\unit{km}] & 
    8.9 \\ \hline
    Collider tunnel diameter [\unit{m}] & 5.5\\ \hline
    Collider tunnel GWP / m  [\unit{t}~\ce{CO2}e/ \unit{m}] & 8.0 \\ \hline
    Concrete GWP  [\unit{kg}~\ce{CO2}e/ kg] & 0.16 (C30)\\
 \hline 
    Accelerator  GWP / m  [\unit{t}~\ce{CO2}e/ \unit{m}] & Not estimated\\
 \hline 
 \multicolumn{2}{|l|}{\textbf{GHG emissions from operation}} \\
 \hline 
    Maximum power in operation [\unit{MW}]   & 
    220 \\ \hline
    Electricity consumption / yr  [\unit{TWh}/ \unit{yr}] & 
    1.06 \\ \hline
    Years of operation & 2041 -  \\ \hline
    Carbon intensity of electricity [\unit{g}~\ce{CO2}e/ \unit{kWh}] &  \\ \hline
    Average Scope 2 GHG emissions / yr [\unit{kt}~\ce{CO2}e] &  \\
 \hline 

 \end{tabular}
\end{table}

\subsubsection{Evaluation of GHG emissions for the CLIC and ILC projects} 

The CLIC and ILC projects have been working closely together at the evaluation of green house gas emissions from construction and operation, and the potentials to reduce those. 
In particular, they have collaborated on a common lifecycle assessment of the underground civil engineering structures (tunnels, caverns, and shafts)~\cite{clic_ilc_lca_arup} and on an upcoming study of the accelerators and detectors at large~\cite{clic_ilc_lca-accel_arup}.
The available data is briefly discussed in the following.

\paragraph{LCA of underground civil engineering structures}

\begin{figure}[ht]
    \centering
    \includegraphics[width=0.58\textwidth]{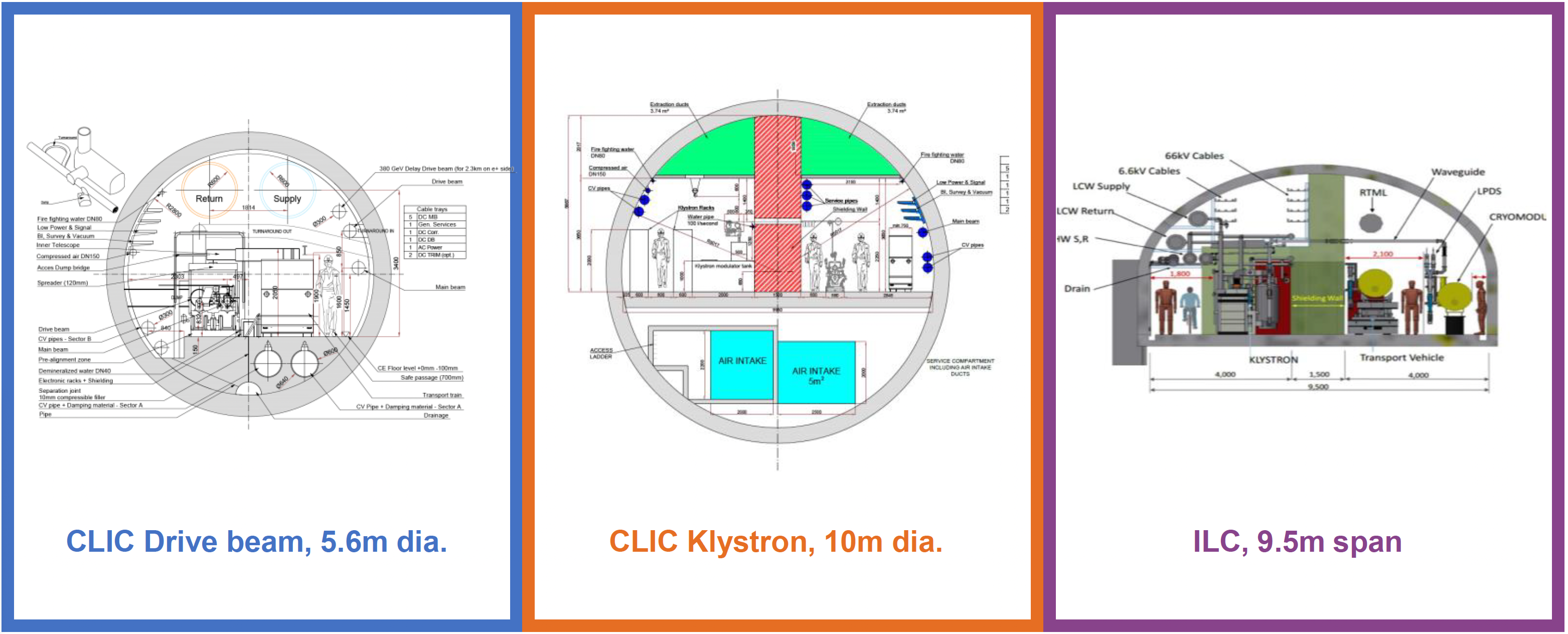}  
  \centering
    \includegraphics[width=0.38\textwidth]{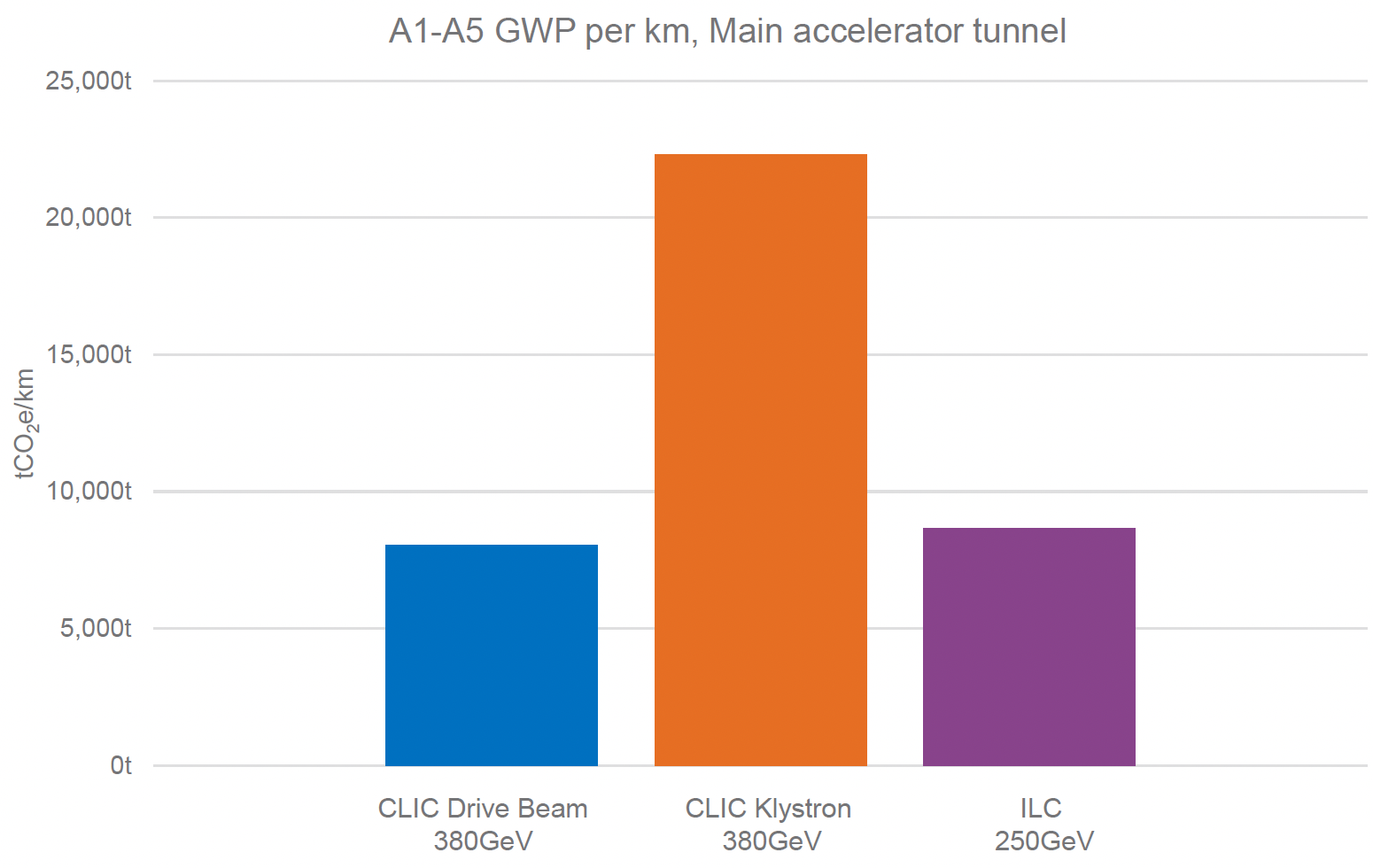}  
    \caption{Left: Main linac tunnel cross sections of the CLIC and ILC configurations considered in the lifecycle assessment study~\cite{clic_ilc_lca_arup}, right: result for the embodied carbon. 
}
    \label{fig:tunnel-crosssections}
\end{figure}

In a first study~\cite{clic_ilc_lca_arup} completed in 2023 by the international consultancy company ARUP, a  lifecycle assessment of the underground civil engineering structures (tunnels, caverns, and shafts) was conducted for the CLIC and ILC projects.
For CLIC, three energy stages were considered: $380\,\unit{GeV}$, $1.5$ and $3\,\unit{TeV}$; 
for the $380\,\unit{GeV}$, a klystron driven version was studied as an alternative configuration in addition to the baseline drive-beam option.
For the ILC, an Higgs factory configuration operating at $250\,\unit{GeV}$ was considered.
Fig.~\ref{fig:tunnel-crosssections} shows the main linac tunnel cross sections of these configurations and the corresponding result for the GWP per km of tunnel.
The functional unit considered were the full underground civil engineering structure, ready for installation of components (excluding cable trays, lightning etc), and, for comparison purposes, a $1\,\unit{km}$ long main linac tunnel section.
The LCA was performed with the goal to identify hotspots and reduction opportunities. 
It covered lifecycle modules A1 to A5 (before use stage, raw material supply to construction process) as defined in EN~17472~\cite{en17472:2022}.
The LCA followed the ISO~14040~\cite{iso14040:2006} methodology, using Simapro~\cite{Simapro} software and the Ecoinvent 3.8 database~\cite{Ecoinvent}.

\begin{figure}[ht]
    \centering
    \includegraphics[width=0.48\textwidth]{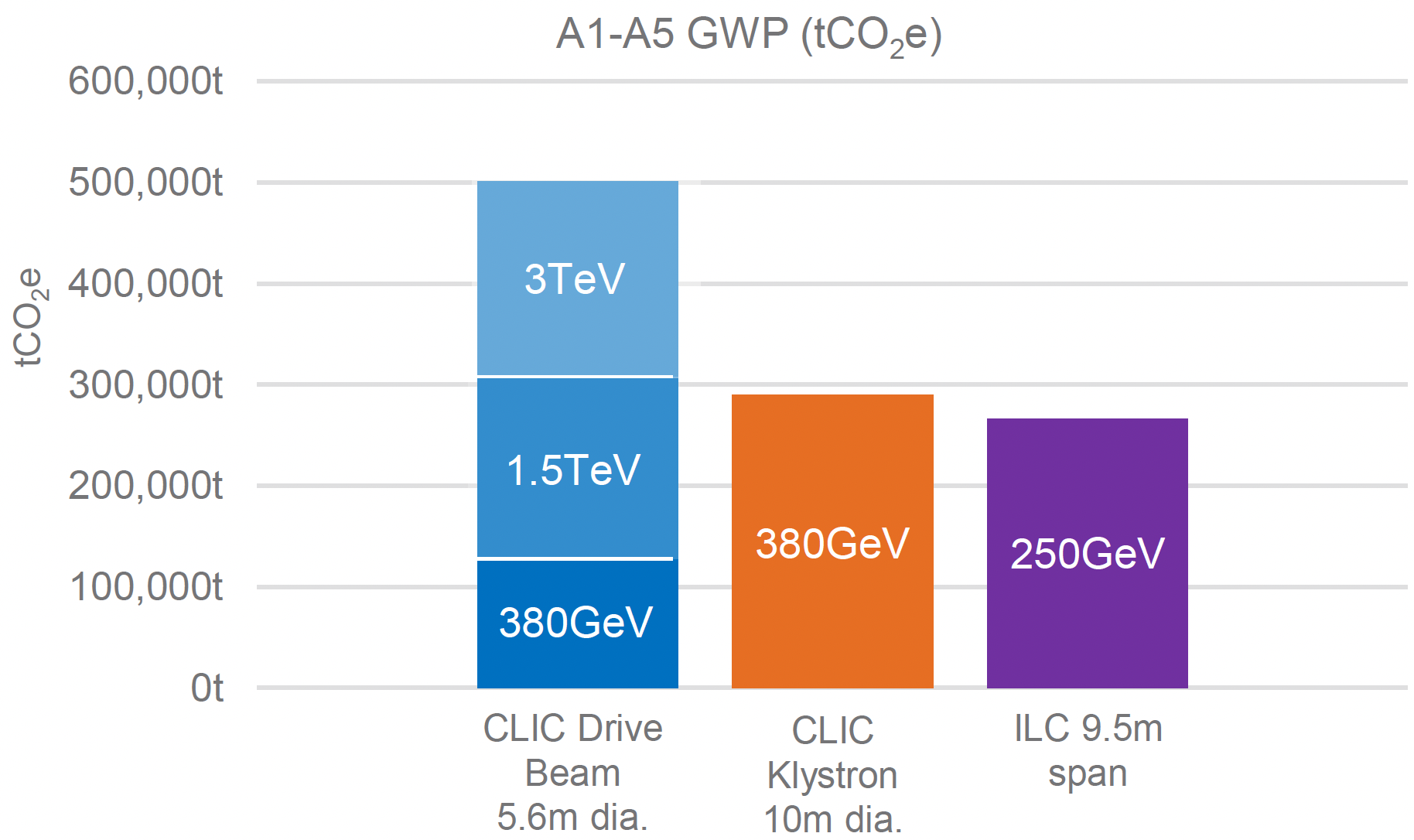}  
  \centering
    \includegraphics[width=0.48\textwidth]{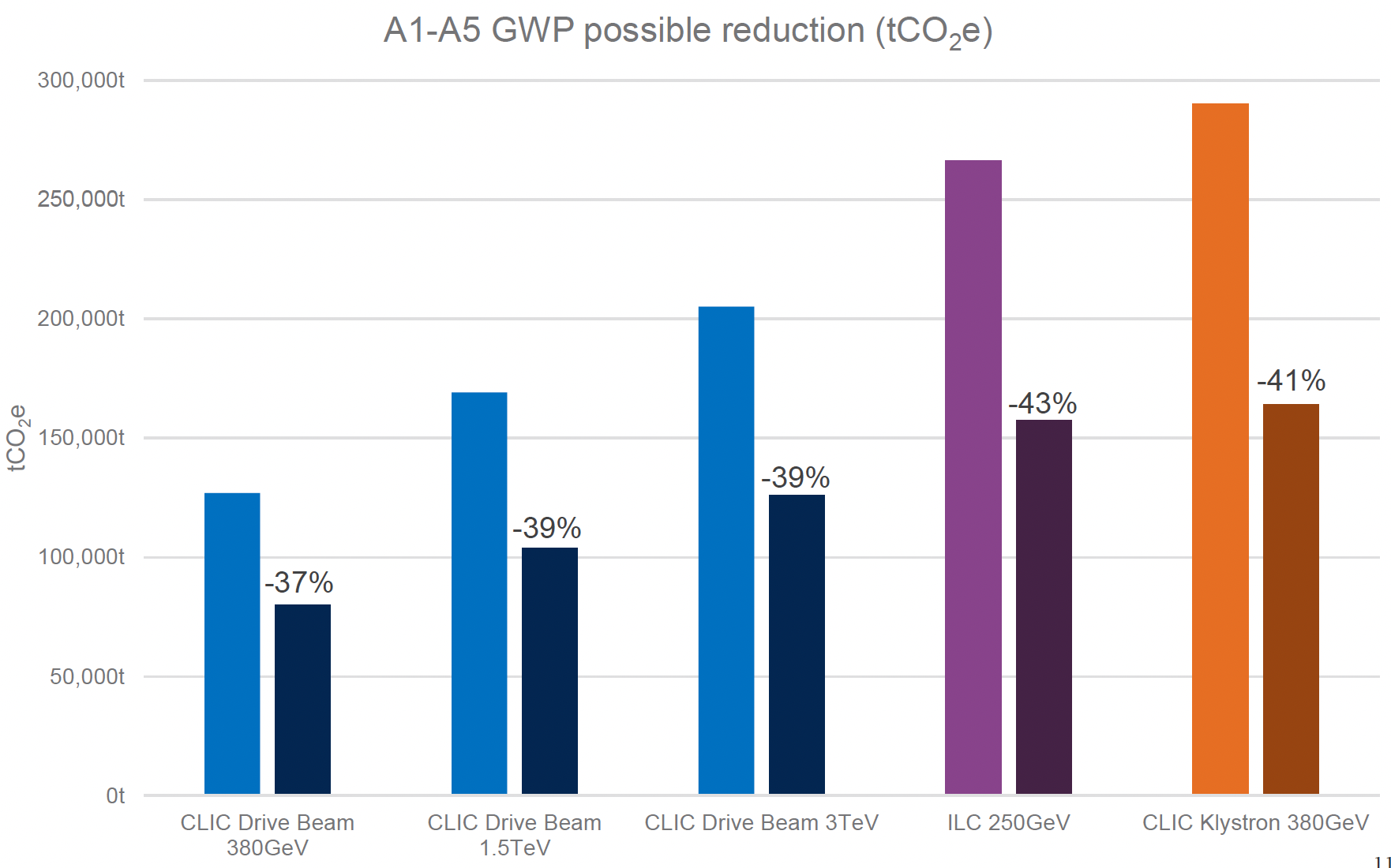}  
    \caption{Left: total embodied carbon for the CLIC and ILC underground civil engineering structured~\cite{clic_ilc_lca_arup}, right: reduction potentials identified.
}
    \label{fig:tunnel-results}
\end{figure}

The main results of the analysis are shown in Fig.~\ref{fig:tunnel-results}:
for the CLIC drive-beam option, the total embodied carbon corresponds to $127 / 296 / 501$ \unit{kt \ce{CO2}e} for the three energies considered; 
for the $380\,\unit{GeV}$ klystron driven option it amounts to $296$ \unit{kt \ce{CO2}e}.
The difference between klystron and beam driven configurations arises 
from the significantly larger ($10\,\unit{m}$ vs. $5.6\,\unit{m}$) tunnel diameter. 
However, the result is exaggerated by the fact that the drive-beam complex adds a significant amount of surface buildings to the design, which was out of scope in this analysis. 
These buildings are considered in a subsequent analysis~\cite{clic_ilc_lca-accel_arup}.

For the ILC, the total GWP amounts to $266$ \unit{kt \ce{CO2}e}.

The report also identified significant reduction potentials, between 37\% for the CLIC drive-beam $380$ \unit{GeV} option, and  up to 43\%  for the ILC. 
The most important reduction opportunitiy is the use of carbon-reduced cement, namely CEMIII/A with 50\% GGBS
\footnote{GGBS stands for Ground Granulated Blast Furnace slag, which is a by-product from iron production that can replace part of the portland cement in concrete to reduce the \ce{CO2} intensity.} 
content. 
Additional reduction opportunities arise from a possible reduction of the lining thickness, material changes in the shielding walls, and a projected reduction of the  \ce{CO2} intensity of the electricity consumed in the construction process.

\paragraph{LCA of accelerators and detectors}
\label{par:LCA_acce_det}
An additional study \cite{clic_ilc_lca-accel_arup} has been conducted to assess the lifecycle impact of the accelerator and detector installations, and the surface buildings.
Owing to the complexity of accelerators, such an LCA is far more complex than for civil construction alone.

The goal of the study was to deliver a comprehensive LCA of the complete facility, including accelerator, detectors and civil infrastructure, over the lifetime of the experimental program, from cradle to grave, as input to comparative studies in the context of the European Strategy for Particle Physics Update (ESPPU) 2026 \cite{Arduini:2025a}, and to identify critical items and hotspots early in the design phase in order to reduce the impact 
as much as possible.
Therefore, the scope definition encompassed the full accelerator, the detectors, and civil construction and infrastructure directly associated with the project, from raw material extraction and construction (modules A1 to A5), operation (modules B, in particular B4, B6, B7) and disposal (modules C), to the extent possible within the time and resources available for this work.

Due to the multitude of components and materials necessary to construct an accelerator and detector, a number of exclusions were necessary.
This concerns individual items such as sources, dumps, electronics racks, or detector components such as trackers that have a very complex composition and production process but contribute little to the overall inventory in terms of mass. 
In the operation stage, replacement of components are incorporated in a very coarse manner by estimating a replacement rate based on typical lifetimes of components and an assumed overall operation time. 
For the use stage the analysis focused on operational water use (module B7) and, most prominently, on the electricity use (module B7).
Estimations of the disposal stage are currently hindered by a lack of quantitative data from previous projects, and also by the inherent uncertainty how an experimental program evolves.
Reuse before recycling has long been a maxim for accelerators, as witnessed by facilities such as the Proton Synchrotron at CERN or the DESY accelerator in Hamburg, which have been in operation since more than half a century.

To compile the Lifecycle Inventory data, a comprehensive Product Breakdown structure (PBS) was defined for both projects, ILC in the \SI{250}{GeV}, \SI{20.5}{km} configuration for the Japanese site in Kitakami, and CLIC in two configurations for \SI{380}{GeV}, driven by klystrons or with two beam acceleration. 
The PBS had two levels, the subsystem level corresponding to different accelerator areas (sources, damping rings, main linac etc) and the component level corresponding to technical systems (magnets, rf systems etc), in line with the structure used in the cost estimation of both projects.

\begin{figure}[ht]
    \centering
    \includegraphics[width=0.45\textwidth]{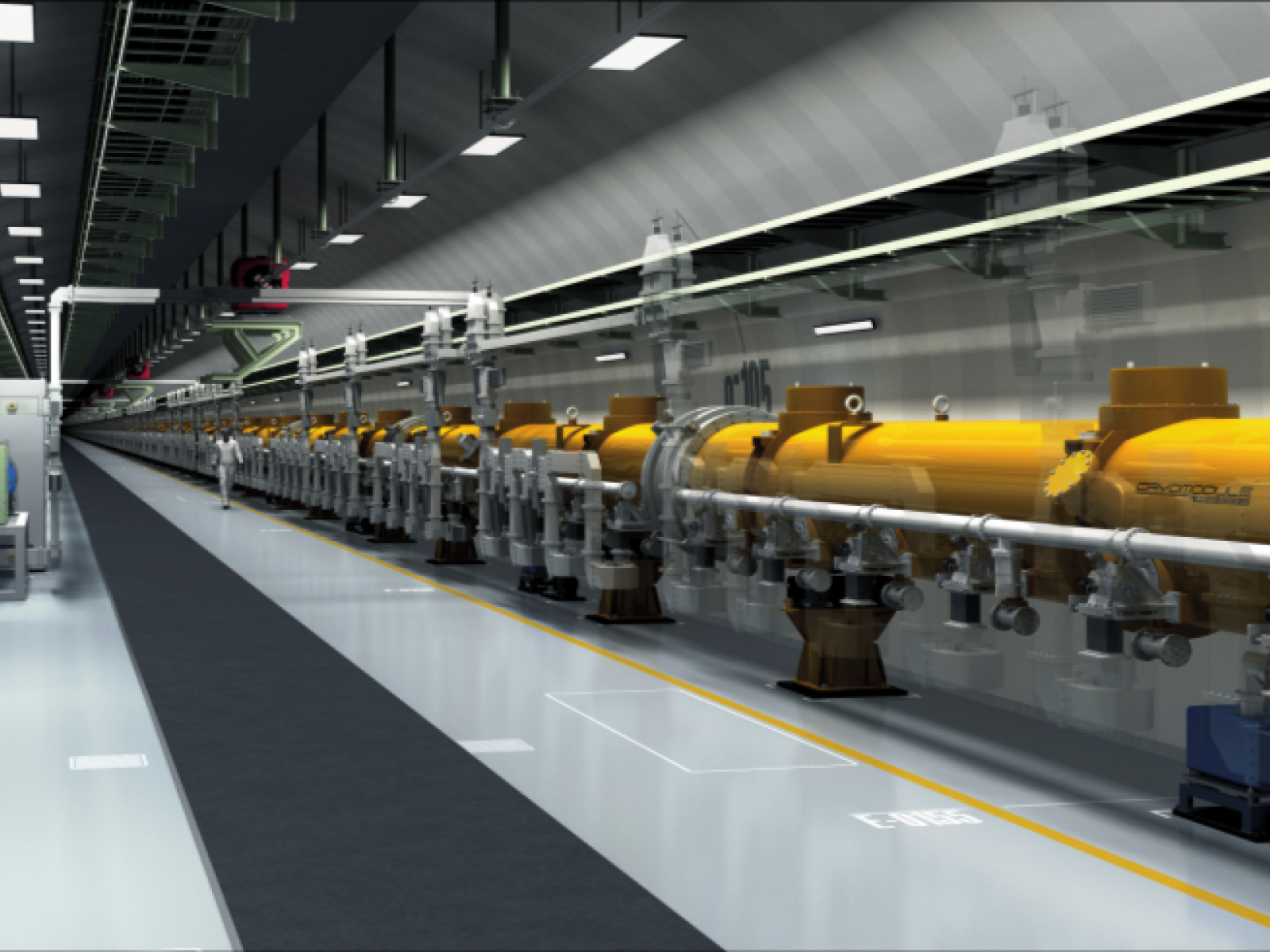}  
    \includegraphics[width=0.45\textwidth]{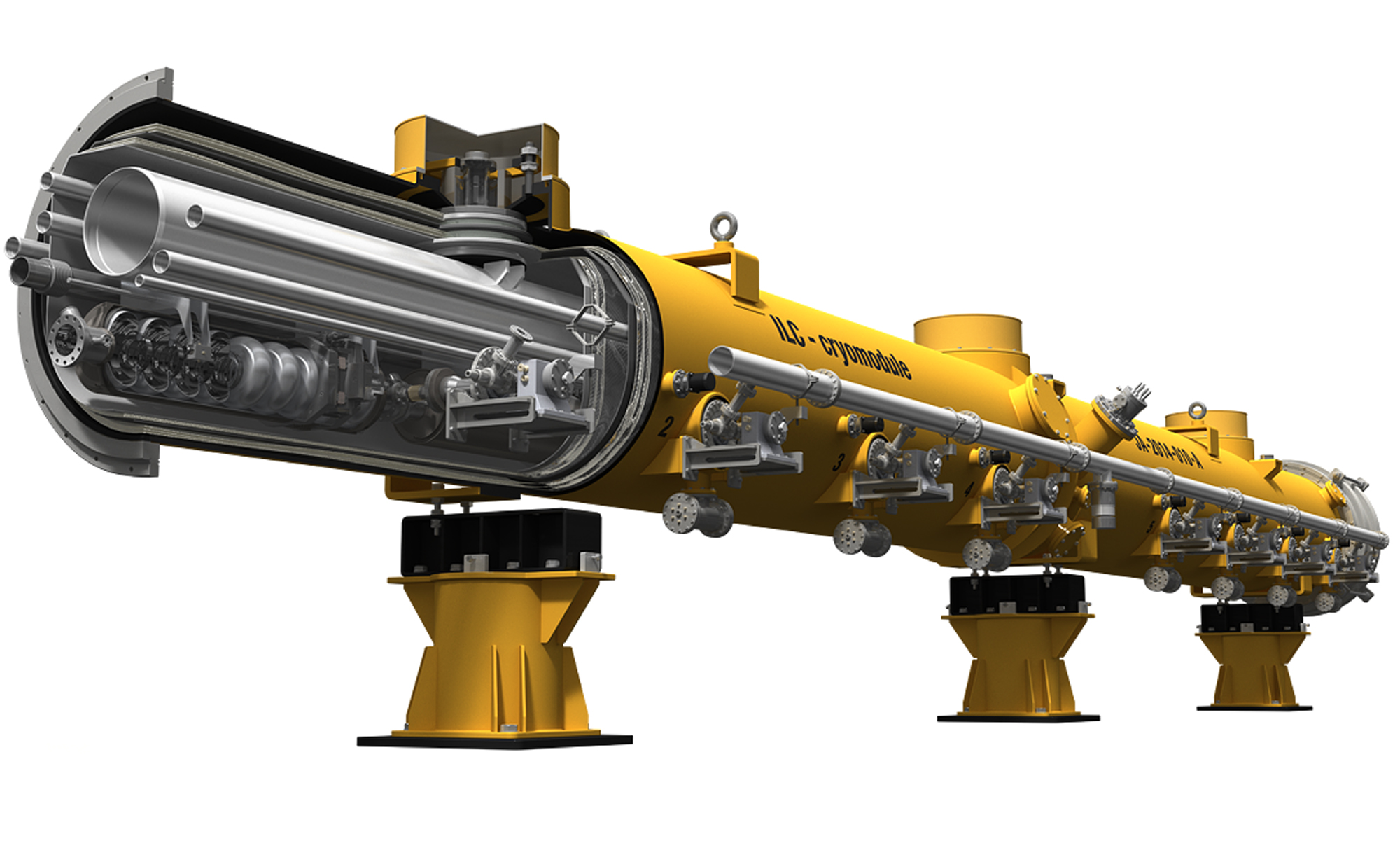}  
    \caption{Left: Rendering of the ILC main linac tunnel, with yellow cryomodules on the right. Right: Detailed view of a cryomodule. Image credit: Rey.Hori/KEK.
}
    \label{fig:ilc-rendering}
\end{figure}

For the ILC, a main focus of the accelerator LCA is the main linac, which accounts for more than \SI{50}{\%} of the physical length as well as of the cost of the whole project.
The main linac is dominated by cryomodules  (Fig.~\ref{fig:ilc-rendering}), which house the superconducting accelerating cavities made out of bulk niobium. 

\begin{figure}[ht]
    \centering
    \includegraphics[width=0.35\textwidth]{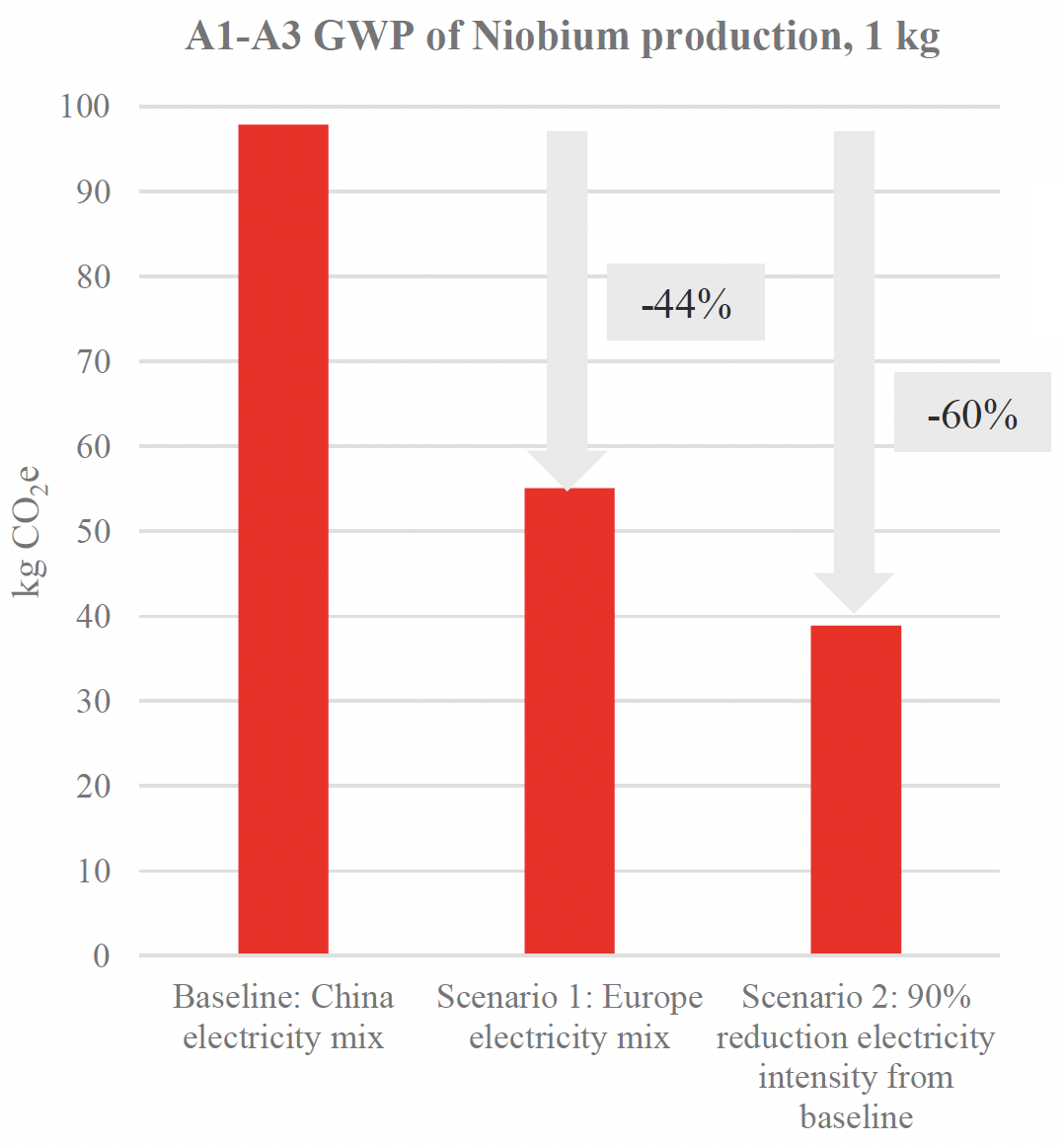}  
    \includegraphics[width=0.63\textwidth]{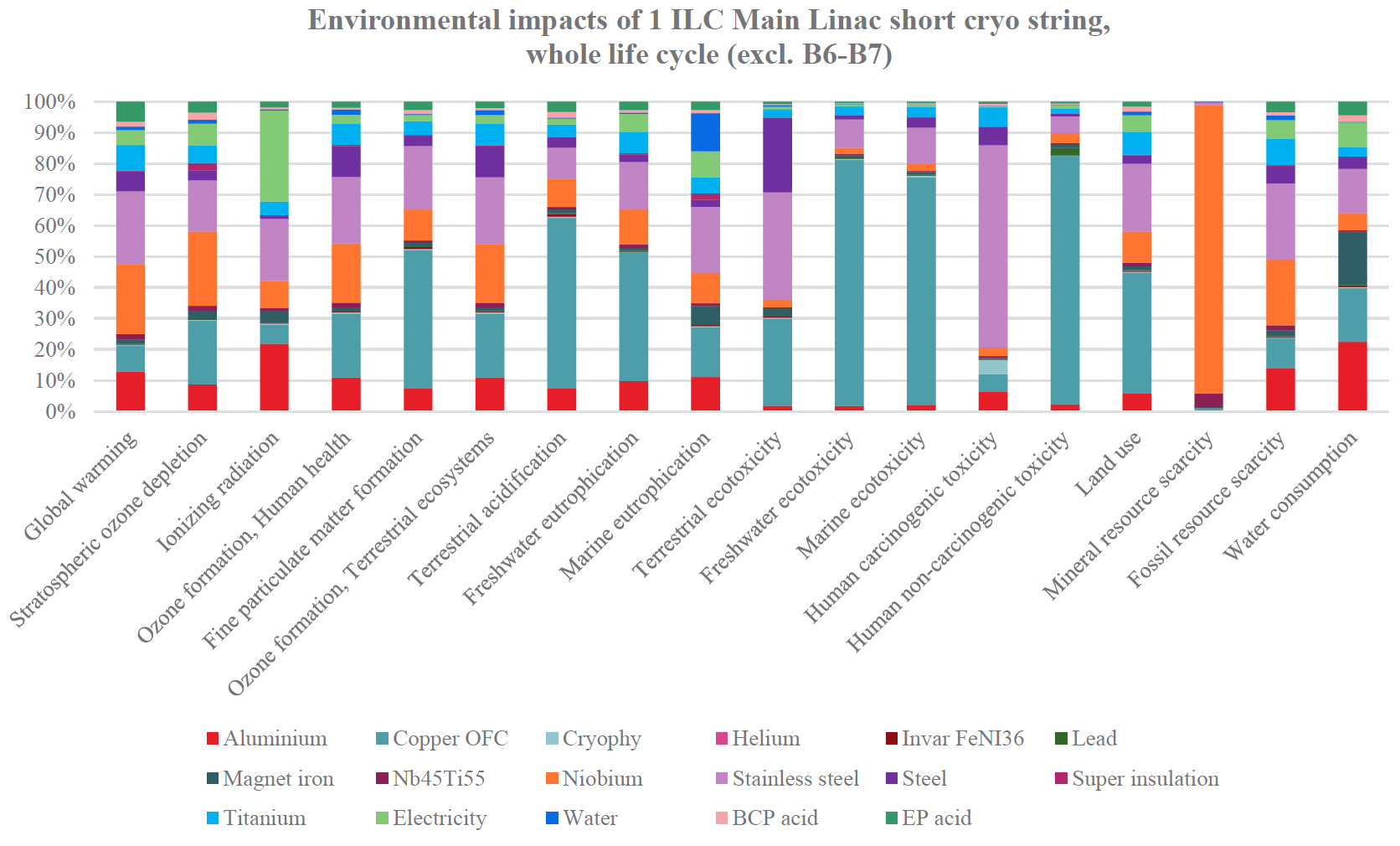}  
    \caption{Left: GWP reduction opportunities for niobium. Right: hotspot analysis for an ILC cryo-string (cryomodules, klystrons and waveguides)~\cite{clic_ilc_lca-accel_arup}.}
    \label{fig:ilc-hotspot}
\end{figure}

As a first step in the LCA it was necessary to conduct a separate LCA for high purity ($RRR>300$) niobium \cite{Kouptsidis:2000ti} as used for superconducting applications. 
Niobium is mostly used as compound in high strength steel alloys, where it is added in the form of ferro niobium (\ce{FeNb}), for which good LCA data exists \cite{Dolganova:2020aa}.
About \SI{10}{\%} of niobium is produces as niobium pentoxide (\ce{Nb2O5}), mostly for optical applications \cite{daSilvaLima:2022131327}, which is the starting point for alumino-thermic reduction into metallic (ATR) niobium.
The LCA data for ATR niobium was evaluated based on \ce{Nb2O5} data from \cite{daSilvaLima:2022131327} and aluminium data for China, based on the stoichiometric proportions, as not data on the efficiency of the process was available.
The result is a carbon intensity of \SI{24}{kg \ce{CO2}e/kg} for \SI{99.9}{\%} pure ATR niobium \cite{clic_ilc_lca-accel_arup}.
The highly pure niobium for superconducting applications, characterised by its RRR (residual resistivity ratio) value of $RRR>300$, is the result of a refining process of repeated remelting in electron beam furnaces \cite{SallesMoura:2007a}.
Taking into account the reported efficiencies of the purification process and the necessary amount of electricity (with Chinese carbon intensity factors), a carbon intensity of \SI{98}{kg \ce{CO2}e/kg} was determined \cite{clic_ilc_lca-accel_arup}, which is dominated by the emissions from electricity.
If niobium refining could be performed with electricity of \SI{0.10}{kg \ce{CO2}e/kWh}, the carbon intensity would be reduced by \SI{60}{\%} to \SI{39}{kg \ce{CO2}e/kg} (Fig.~\ref{fig:ilc-hotspot}).

To evaluate the lifecycle inventory of the ILC cryomodules, a comprehensive manufacturing boll of materials (MBOM) was set up based on the engineering model of the TESLA type IV cryomodule \cite{Adolphsen:2013kya} that forms the basis of the ILC cryomodule design, with a detailed tally of contributing parts and their materials. 
For the cavities, the production steps starting from individual niobium sheets were modeled, including an estimation of scrap material produced.
Chemicals needed for cavity treatments such as electro-polishing (EP) and buffered chemical polishing (BCP) were also taken into account, based on the estimates provided by industrial studies. 
Transport between production sites (module A2) and to the site in Japan (module A4) was included by approximately estimated transport distances and transportation means (mostly ship and lorry).
Additional components such as waveguide distributions and klystrons were also included, albeit in a coarser fashion, concentrating on the amount of materials used (modules A1-A3).
Since the main linac is such a dominant source of environmental impact, 
a hotspot analysis was performed on the cryo-string, which is the main linac building block comprising 9 cryomodules, 2 klystrons and modulators and the waveguide distributions (modulators were excluded due to lack of data), as shown in Fig.~\ref{fig:ilc-hotspot}.
The materials that contribute most to the GWP, in order of importannce, are steel and niobium, followed by aluminium, copper, titanium and BCP acid.
In other impact categories, copper -- used mostly in the klystrons -- is a key hotspot in toxicity; 
copper, steel and niobium are recurring hotspots in many categories; niobium completely dominates mineral resource-scarcity impact.

\begin{figure}[ht]
    \centering
    \includegraphics[width=0.42\textwidth]{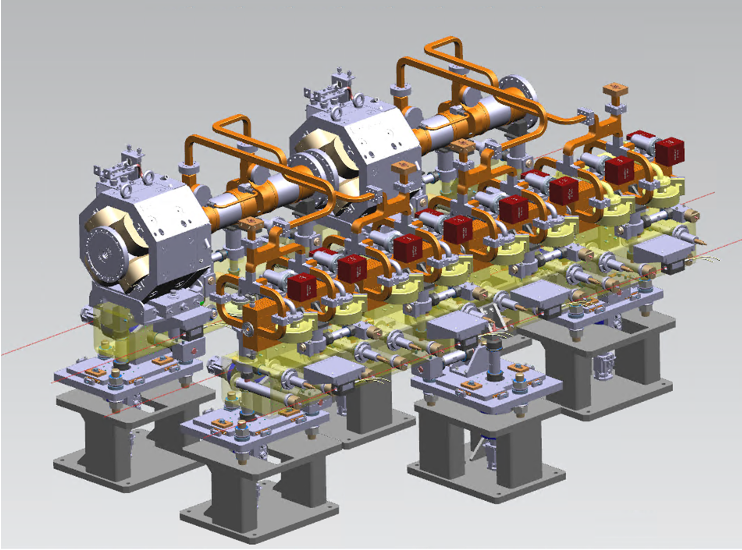}  
    \includegraphics[width=0.56\textwidth]{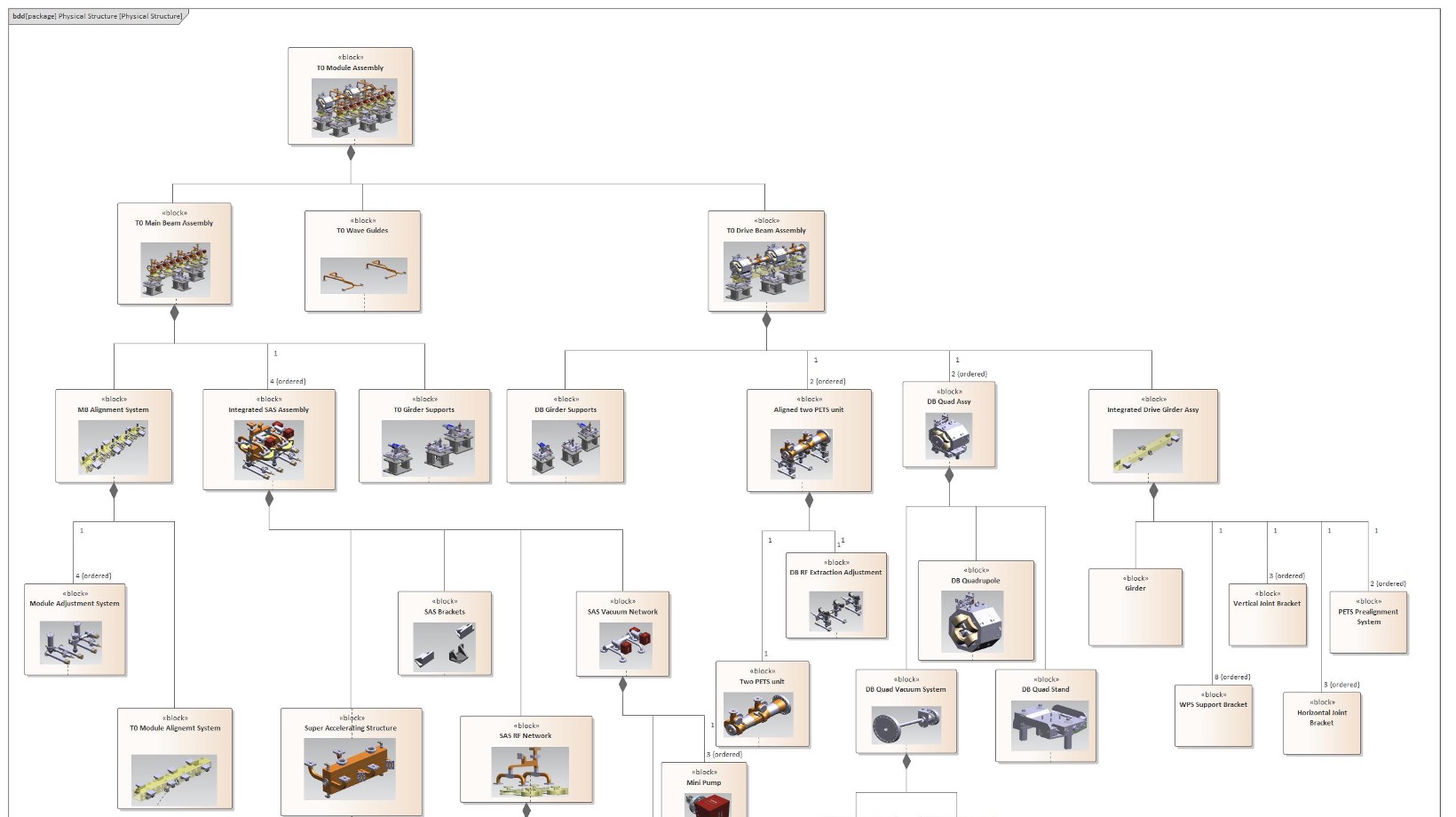}  
    \caption{Left: CAD model of the CLIC two beam module. Right: Manufacturing Bill of Materials (MBOM) of the CLIC two beam module as used for the LCI.
}
    \label{fig:clic-2-beam-module}
\end{figure}

For CLIC, the main linac for the drive-beam option was analysed in a similar fashion, focusing on the two beam acceleration module (Fig.~\ref{fig:clic-2-beam-module}).
Based on a detailed MBOM, the inventory of raw materials was collected.
Based on information about manufacturing methods, an approximation for the expected amount of scrap material was also derived. 
This is especially important for components manufactured by turning and milling solid blocks, a process that can result in 50\% or more material waste.

Significant fractions of the materials used, in particular the copper, have to be of special quality in order to fulfil the stringent requirements for vacuum and rf parts, which needs to be considered in the LCA.
For copper, highly pure oxygen-free (OFC) copper was assumed as material.

The second large contribution to material consumption in the construction of accelerators are magnets, in the case of ILC these are almost exclusively normal conducting resistive electro-magnets with iron yokes and copper coils.
For both projects, CLIC and ILC, comprehensive lists of the required magnets are available, with information on strength, aperture, current densities and coil and yoke dimensions.
For CLIC, the magnet catalogue contains preliminary designs of all the magnet types with complete cross sections and field maps, for ILC the yoke dimensions have been estimated based on the coil dimensions and aperture data.
Coils are fabricated from extruded strands of highly pure copper, with a low amount of scrap, which was neglected.
The epoxy mass used to stabilise and isolate the coil windings was excluded from the analysis.
Magnet yokes are often produced from steel laminations that are stamped out of rolled steel sheets, which produces significant amounts of scrap material. 
As an approximation, a rectangular box around the magnets was taken to calculate the gross amount of material, while the net material was taken from the actual yoke cross section. 
For certain magnet types, such as low field dipoles, other manufacturing methods such as production from solid bars as proposed for CEPC, can yield a significant optimisation potential by reducing the amount of waste material, with corresponding cost advantages.
For high field magnets, soft iron may be replaced by iron compounds rich in cobalt that have higher saturation fields.
As a benchmark, a compound made of iron and cobalt in equal mass proportions was included in the analysis (motivated by Vacoflux, a commercial product), and it was assumed that quadrupoles with high pole tip fields exceeding \SI{0.8}{T} are fabricated from such a material.
The LCA data shows that for such a compound containing \SI{50}{\%} of cobalt, the GWP impact is about six times higher (\SI{22}{kg \ce{CO2}e/kg}) than for soft iron (taken to be \SI{3.6}{kg \ce{CO2}e/kg} including the impact of fabrication).
Overall it was found that magnets contribute a very significant fraction to the accelerator's GWP impact owing to their 
mass, dominated by the yoke.

In a separate investigation, the lifecycle impact of the magnet system, comprising the magnet, power cables, power supply, water and air cooling, and the associated electricity consumption over the full lifecycle, was studied as a function of the current density in the coil, which determines the electric power loss as well as the coil and consequently the yoke dimensions.
Based on a model that parametrises these effects, the trade off between smaller coil cross sections that demand less material in the fabrication stage and larger coils that are more power efficient can be studied and optimised. 
Here, a proper definition of the functional unit is important. 
It encompasses the full system and thus takes into account that reduced ohmic losses in the coils and power cables reduce also the amount of power 
required for water and air cooling, and power supplies.
Air cooling is particularly important, as cooling \SI{1}{W} of heat going into tunnel air, especially from cables that are normally air cooled, may require \SI{1}{W} of power for air cooling.
The overall minimal GWP impact depends on the assumed operating lifetime and on the carbon intensity of materials and electricity, which means that the best (in terms of GWP) configuration actually is dependent on site specific parameters.
It was observed that this kind of optimisation typically leads to designs with lower electricity losses, reducing operating expenses at the price of increased investment costs.
This also applies to the lifetime cost optimum (considering the sum of initial investment and lifetime expenditure for electricity), indicating that a full lifecycle assessment has the potential to reduce not only the environmental but also the monetary cost of a project, addressing also the economic dimension of sustainability (see Fig.~\ref{fig:sustainability_dimensions} in Sect.~\ref{sec:account}).

In addition to the first LCA that addressed civil construction exclusively  \cite{clic_ilc_lca_arup}, for the more comprehensive LCA \cite{clic_ilc_lca-accel_arup}, tunnel services were also considered. 
Items included in the LCI were cable trays, cooling water pipes, and power supply cables, derived from engineering drawings of the main tunnel cross sections. 
The results varied between \num{16} and \SI{30}{kton \ce{CO2}e}, which adds around \SI{10}{\%} to the GWP impact of the tunnel civil construction.

Detectors were also considered in the analysis, albeit at a rather coarse level, taking into account only the net amount of material for the most massive subdetectors, which are the muon system with the magnet return yoke, the hadronic and the electromagnetic calorimeters. 
For the ILC, also material required for the reinforced concrete platforms that support the detectors was included in the LCI.
The total GWP was found to be between \num{32} for SiD and \SI{62}{kton \ce{CO2}e} for ILD.
Due to their sheer mass, the iron yokes dominate the impact by far, underlining the fact that iron or steel, despite the relatively low specific impact per \SI{}{kg} of material, are extremely important  
in the LCA results.

The inventory (LCI) stage of the study was followed by the assessment,
which used data from the ecoinvent version 3.10 database \cite{Ecoinvent} and SimaPro v. 9.6 \cite{Simapro} as LCA tool.
The impacts were evaluated in the ReCiPe 2016 Midpoint (H) categories \cite{Huijbregts:2017a,Huijbregts:2020a}, with a focus on global warming potential (GWP).
The GWP results are summarised in Tab.\ref{tab:ghg-project-CLIC} for the CLIC project and in Tab.\ref{tab:ghg-project-ILC} for ILC and LCF.

In order to evaluate the GWP reduction potential, hotspot analyses were performed to identify the materials with the largest impact.
For the accelerator and detectors, approximately \SI{70}{\%} of the GWP is caused by iron and steel, used for magnet yokes, supports, cryostats and other purposes. 
Despite the quite different acceleration technique and design, this result is very similar for ILC and CLIC \cite{List:2025aa}.
For ILC, other materials with a large overall impact are niobium (\SI{11}{\%} of the accelerator GWP), copper (\SI{10}{\%}) and aluminium (\SI{6}{\%}), 
while for CLIC, copper contributes \SI{25}{\%} of GWP, owing to its extensive use in the acceleration modules.
Based on the hotspot analysis, the study highlighted reduction opportunities for these materials, which concern mostly the raw material production pathways.
This underlines the importance of evaluating the sourcing of materials along the production chain, which requires responsible procurement policies that incentivise  producers of components to provide specific LCA data for their products and reduce impacts across the complete supply chain.

In order to evaluate which other impact category merits particular attention, the total impacts for each of the 16 ReCiPe 2016(H) midpoint categories were normalised to the global value per person in these categories, using the ReCiPe 2010 world normalisation set~\cite{ReCiPe2016:scores}.
Using weighting factors for endpoint categories would provide another approach to the question which categories are particularly important, which was not pursued.

The result of this exercise showed that toxicity, most of all human carcinogenic toxicity, were by far the most prominent category after normalisation \cite{List:2025aa}.
A subsequent hotspot analysis for toxicity revealed that it is dominated by steel production, and can be traced back to the deposition of electric arc furnace slag.
Thus, once more iron based materials were identified as a major source of concern rather than more exotic materials such as niobium or cobalt that one might have suspected.
Also here, responsible procurement is key to a reduction of adverse impacts.
On a positive note, steel and steel products are supplied by a large number of producers also in countries that do actively encourage sustainable production of goods and materials, so that significant reductions may be achievable in many cases.

\paragraph{GHG emissions from operation}

Currently, operational GHG emissions at CERN are dominated by direct gas emissions from detectors~\cite{CERN_env_report_GRI_SDG}, as discussed in section~\ref{sec:detector}.
The systematic evaluation of CERN's GHG emissions in the environmental reports has raised awareness to this fact. 
Both, CLIC and ILC aim at avoiding such emissions from detectors completely and assume that this kind of emissions can be reduced to zero.

In the absence of such direct emissions, module B6 (operational energy use) is expected to dominate operational GHG emissions.
To assess those, an estimate of the future carbon intensity of electricity at the time of accelerator operation is necessary, as discussed in section~\ref{sec:RIoperation}. 

For CLIC at CERN, a \ce{CO2} intensity of \SI{12.5}{g \ce{CO2}e/kWh} is assumed for operation in 2048 and later.
A survey of available projections for the \ce{CO2} intensity of future electricity production in France resulted in a range of \num{14} to \SI{18}{g \ce{CO2}e/kWh} for the year 2050, used  by the working group established to compare the future  accelerator projects for the 2026 Update of the European Strategy for Particle Physics~\cite{Arduini:2025a}.

For the ILC in Japan, a \ce{CO2} intensity of \SI{81}{g \ce{CO2}e/kWh} was projected for 2040 onwards, based on the availability of electricity from regenerative sources in the Tohoku region, which is better than in Japan as a whole.
It is assumed that for ILC operation, contracts are placed with electricity companies for green electricity covering \SI{90}{\%}, plus \SI{10}{\%} from Liquefied Natural Gas (LNG).

\paragraph{Overall results for CLIC and ILC}

\autoref{tab:ghg-project-CLIC} summarises the results for Greenhouse Gas emissions for CLIC, and \autoref{tab:ghg-project-ILC} for ILC and an ILC-based linear collider facility (LCF) at CERN.

\begin{table}[htbp]
 \centering 
 \caption{Data on GHG emissions for the CLIC project. Data in parenthesis are for two IPs}
 \label{tab:ghg-project-CLIC}
 \begin{tabular}{|l|c|c|}
 \hline 
   \textbf{ CLIC } & Stage 1 & Stage 2 \\
 \hline
    CoM energy [\unit{\GeV}] & 380 & 1500 \\ \hline
    Luminosity/IP $[10^{34} \unit{\cm^{-2}}\unit{\s^{-1}}~]$ & 2.3 (4.5) & 3.7\\ \hline
    Number of IPs & 1(2) & 1\\ \hline
    Operation time for physics/yr  $[10^7 \unit{s}/ \unit{yr}]$ & \multicolumn{2}{c|}{1.2} \\ \hline
    Integrated luminosity/ \unit{yr}  [1/\unit{fb}/ \unit{yr}] & 276 (524)  & 434 \\ \hline
    Host countries & 
           \multicolumn{2}{c|}{France and Switzerland } \\
 \hline 
 \multicolumn{3}{|l|}{\textbf{GHG emissions from construction, stage A1-A5}} \\
 \hline 
    Subsurface tunnels, caverns, shafts [\unit{kt}~\ce{CO2}e] & 268  & 337 \\ \hline 
    Surface sites and constructions [\unit{kt}~\ce{CO2}e] & \multicolumn{2}{c|}{118}\\ \hline 
    Accelerator (coll.) [\unit{kt}~\ce{CO2}e] & 38 & N.E.\\
 \hline 
    Accelerator (inj.) [\unit{kt}~\ce{CO2}e] & 110 & N.E. \\
     \hline
    Accelerator Services [\unit{kt}~\ce{CO2}e] & 19 & N.E.\\
 \hline 
    Detectors [\unit{kt}~\ce{CO2}e] & \multicolumn{2}{c|}{94}\\
 \hline 
    \textbf{Total [kt~\ce{CO2}e]} & 609 & \\ \hline 
    Collider tunnel length  [\unit{km}] & 11.5 & 17.6\\ \hline
    Collider tunnel diameter [\unit{m}] & \multicolumn{2}{c|}{5.6}\\ \hline
    Collider tunnel GWP / m  [\unit{t}~\ce{CO2}e/ \unit{m}] & \multicolumn{2}{c|}{8.1}\\ \hline
    Concrete GWP  [\unit{kg}~\ce{CO2}e/ kg] & \multicolumn{2}{c|}{0.16}\\
 \hline 
    Accelerator  GWP / m  [\unit{t}~\ce{CO2}e/ \unit{m}] & \multicolumn{2}{c|}{Not estimated}\\
 \hline 
 \multicolumn{3}{|l|}{\textbf{GHG emissions from operation}} \\
 \hline 
    Maximum power in operation [\unit{MW}]   & 107 (160) & 287 \\ \hline
    Electricity consumption / yr  [\unit{TWh}/ \unit{yr}] & 0.6(..) & \\ \hline
    Years of operation & 2040 & 2040 \\ \hline
    Carbon intensity of electricity [\unit{g}~\ce{CO2}e/ \unit{kWh}] & \multicolumn{2}{c|}{16} \\ \hline
    Average Scope 2 GHG emissions / yr [\unit{kt}~\ce{CO2}e] &  &\\
 \hline 

 \end{tabular}
\end{table}

\begin{table}[htbp]
 \small
 \centering
  \caption{Data on GHG emissions for the ILC project in Japan and the Linear Collider Facility (LCF) proposal for CERN~\cite{LinearCollider:2025lya}. These are baseline numbers, before application of possible \ce{CO2} reduction measures.
  The optimisation potential for tunnels, caverns and shafts is estimated to be 50\,\%, for accelerators, services and detectors it is assumed to be 25\,\%.}
 \label{tab:ghg-project-ILC}
 \scalebox{0.9}{
 \begin{tabular}{|l|c|c||c|c|}
 \hline 
   &\textbf{  ILC 250} & \textbf{ ILC 500} & \textbf{ LCF 250 LP/FP} & \textbf{ LCF 550 LP/FP} \\
 \hline
    CoM energy [\unit{\GeV}] & 250 & 500 & 250 & 550 \\ \hline
    Luminosity/IP $[10^{34} \unit{\cm^{-2}}\unit{\s^{-1}}~]$ & 1.35 / 2.7 & 3.6 & 1.35 / 2.7 & 1.9 / 3.85 \\ \hline
    Number of IPs & \multicolumn{2}{c||}{1} & \multicolumn{2}{c|}{2} \\ \hline
    Operation time for physics/yr  $[10^7 \unit{s}/ \unit{yr}]$ & \multicolumn{2}{c||}{1.6}  & \multicolumn{2}{c|}{1.2} \\ \hline
    Integrated luminosity/ \unit{yr}  [1/\unit{fb}/ \unit{yr}] & 215 / 430 & 580 & 325 / 650 & 460 / 920 \\ \hline
    Host countries & 
           \multicolumn{2}{c||}{Japan} & 
           \multicolumn{2}{c|}{France and Switzerland} \\
 \hline 
 \multicolumn{5}{|l|}{\textbf{GHG emissions from construction, stage A1-A5}} \\
 \hline 
    Subsurface tunnels, caverns, shafts [\unit{kt}~\ce{CO2}e] & 266 & 372 & \multicolumn{2}{c|}{380} \\ \hline 
    Surface sites and constructions [\unit{kt}~\ce{CO2}e] & \multicolumn{2}{c||}{Not estimated} & \multicolumn{2}{c|}{Not estimated} \\ \hline 
    Accelerator (coll.) [\unit{kt}~\ce{CO2}e] & 150 & 270 & 165 & 310 \\
 \hline 
    Accelerator (inj.) [\unit{kt}~\ce{CO2}e] & 59 / 82 & 83 & \multicolumn{2}{c|}{60 / 84} \\ 
     \hline
    Services [\unit{kt}~\ce{CO2}e] & 32 & 46 & \multicolumn{2}{c|}{46} \\
 \hline 
    Detectors [\unit{kt}~\ce{CO2}e] & \multicolumn{2}{c||}{94} & \multicolumn{2}{c|}{94}\\
 \hline 
    \textbf{Total [kt~\ce{CO2}e]} & 601 / 624 & 865 & 745 / 769 & 890 / 914 \\ \hline 
    Collider tunnel length  [\unit{km}] & 20.5 & 33.5 &\multicolumn{2}{c|}{33.5}\\ \hline
    Collider tunnel diameter [\unit{m}] & \multicolumn{2}{c||}{9.5}  & \multicolumn{2}{c|}{5.6}\\ \hline
    Collider tunnel GHG / m  [\unit{t}~\ce{CO2}e/ \unit{m}] & \multicolumn{2}{c||}{8.6} & \multicolumn{2}{c|}{8.5}\\ \hline
    Concrete GHG  [\unit{kg}~\ce{CO2}e/ kg] / [\unit{kg}~\ce{CO2}e/ \unit{m^3}] (type) & \multicolumn{2}{c||}{0.16~/~400~(C25/30)}  & \multicolumn{2}{c|}{0.16~/~400~(C25/30)} \\
 \hline 
    Main Linac accelerator GHG / m  [\unit{t}~\ce{CO2}e/ \unit{m}] & \multicolumn{2}{c||}{5.3}  & \multicolumn{2}{c|}{5.3}\\
 \hline 
 \multicolumn{5}{|l|}{\textbf{GHG emissions from operation}} \\
 \hline 
    Maximum power in operation [\unit{MW}]   & 111 / 138 & 164 & 143 / 182 & 250 / 322 \\ \hline
    Electricity consumption / yr  [\unit{TWh}/ \unit{yr}] & 0.7 / 0.9 & 1.1 & 0.8 / 1.0 & 1.4 / 1.8 \\ \hline
    Reference year of operation & \multicolumn{2}{c||}{2040}  & \multicolumn{2}{c|}{2050}\\ \hline
    Carbon intensity of electricity [\unit{g}~\ce{CO2}e/ \unit{kWh}] & \multicolumn{2}{c||}{81}  & \multicolumn{2}{c|}{16} \\ \hline
    Average Scope 2 GHG emissions / yr [\unit{kt}~\ce{CO2}e] &59 / 74 & 87 & 13 / 15 & 23 / 28  \\
 \hline 

 \end{tabular}
}
\end{table}

\clearpage

\subsubsection{Evaluation of GHG emissions for the \CCC project}
\CCC-- Cool Copper Collider -- is a concept for a linear collider based on  compact, high-gradient, normal-conducting, distributed-coupling accelerators that are operated at cryogenic temperature~\cite{vernieri2023cool,dasu2022strategy,bai2021c,nanni2023status,nanni2022c}. The design of \CCC is being optimized with a holistic approach to the main linacs, collider subsystems, and beam dynamics to deliver the required luminosity at the overall lowest cost. The \CCC design is being developed for 250 and 550~GeV center of mass operation to allow for a Higgs physics program that could also perform measurements of the Higgs self-coupling. A total facility length of 8~km is sufficient for operation at 250 and 550 GeV center of mass energy. The initial operation at 250 GeV can be upgraded to 550 GeV simply with the addition of rf power sources to the main linacs. 

A lifecycle assessment of the ILC and CLIC accelerator facilities performed by ARUP~\cite{clic_ilc_lca_arup} to evaluate their holistic GWP, has so far provided the most detailed environmental impact analysis of construction. \CCC leveraged this report to produce its own bottoms-up estimate~\cite{breidenbach2023sustainability}. In particular, the construction uses a mix of CEM1 C40 concrete and 80\% recycled steel, the GWP of concrete is taken to be 0.18 kg~CO$_2$e~/kg concrete with density 2400 kg/m$^3$~\cite{CEM1C40}, and 85\% of emissions originate from concrete and 15\% of emissions originate from steel production. Taking into account construction of the main linacs, injector linacs, damping rings, beam delivery system, and experimental hall, a conservative estimate of the total volume of construction material  is 260000 m$^3$ (consisting mostly of concrete by volume). This leads to a GWP of 133 kton CO$_2$e~ for A1-A3 components and GWP per unit length of the main linac of around 17 kton~CO$_2$e/km. Notably, this is roughly a factor two larger than the GWP/km of main tunnel construction of ILC and CLIC; this suggests that a 
detailed design will lead to further reductions. Due to the limited length and surface site construction, we estimate an additional 10\% to the GWP for surface buildings. 
This yields a final estimate of 146 kton~CO$_2$e~ for civil engineering. The power estimate for \CCC was studied and optimized \cite{ntounis2024luminosity} and is reported in the ESPPU input for \CCC \cite{ andorf2025esppuinputc3linear}.

The Linear Collider Vision (LCV) \cite{abramowicz2025linearcollidervisionfuture} calls for a Linear Collider Facility (LCF)  \cite{ LinearCollider:2025lya} with a physics reach from a Higgs Factory to the TeV-scale with $e^+e^{-}$ collisions. One of the technologies under consideration for the accelerator is a cold-copper distributed-coupling linac capable of achieving high gradient. This technology is being pursued by the \CCC collaboration to understand its applicability to future colliders and broader scientific applications. 
The construction of a LCF provides us with the opportunity to consider various pathways for future energy upgrades~\cite{abramowicz2025linearcollidervisionfuture}. These upgrades may allow for more advanced technologies presently in development to replace the original technological choices if they provide a significant advantage. A critical point is that \CCC is compatible with the injector complex and beam delivery system that will be installed in the first phase of a LCF with a superconducting rf (SRF) technology that reaches 250 GeV center of mass in 20~km, as proposed in the LCF baseline \cite{LinearCollider:2025lya,abramowicz2025linearcollidervisionfuture}. After completing the initial run of a LCF, we envision the possibility of replacing the main linac with a cold-copper distributed-coupling linac with an energy reach of up to 2~TeV in the same footprint. Scaling to 3~TeV requires a 33~km facility. Due to the minimal civil construction needed, upgrading the LCF with \CCC technology will primarily have GWP impacts from replacing the component technologies in the main linac, along with upgrades to the beam delivery system and modifications to the injector complex.

\begin{table}[htbp]
 \centering 
 \caption{Data on GHG emissions for the $C^3$ project.}
 \label{tab:ghg-project-CCC}
 \begin{tabular}{|l|c|c|}
 \hline 
   \textbf{ C$^3$ } & Stage 1 & Stage 2 \\
 \hline
    CoM energy [\unit{\GeV}] & 250 & 550 \\ \hline
    Luminosity/IP $[10^{34} \unit{\cm^{-2}}\unit{\s^{-1}}~]$ & 1.3 & 2.4\\ \hline
    Number of IPs & 1 & 1\\ \hline
    Operation time for physics/yr  $[10^7 \unit{s}/ \unit{yr}]$ & \multicolumn{2}{c|}{1.6} \\ \hline
    Integrated luminosity/ \unit{yr}  [1/\unit{fb}/ \unit{yr}] & 200  & 400 \\ \hline
    Host countries & 
           \multicolumn{2}{c|}{USA, EU } \\
 \hline 
 \multicolumn{3}{|l|}{\textbf{GHG emissions from construction, stage A1-A5}} \\
 \hline 
    Subsurface tunnels, caverns, shafts [\unit{kt}~\ce{CO2}e] & 146 & +10 \\ \hline 
    Surface sites and constructions [\unit{kt}~\ce{CO2}e] & \multicolumn{2}{c|}{Included Above}\\ \hline 
    Accelerator (coll.) [\unit{kt}~\ce{CO2}e] & \multicolumn{2}{c|}{Not estimated}\\
 \hline 
    Accelerator (inj.) [\unit{kt}~\ce{CO2}e] & \multicolumn{2}{c|}{Not estimated}\\
 \hline 
    Detectors [\unit{kt}~\ce{CO2}e] & \multicolumn{2}{c|}{Not estimated}\\
 \hline 
    \textbf{Total [kt~\ce{CO2}e]} & \multicolumn{2}{c|}{ }\\ \hline 
    Collider tunnel length  [\unit{km}] & \multicolumn{2}{c|}{ 8}\\ \hline
    Collider tunnel diameter [\unit{m}] & \multicolumn{2}{c|}{4-6.5}\\ \hline
    Collider tunnel GWP / m  [\unit{t}~\ce{CO2}e/ \unit{m}] & \multicolumn{2}{c|}{17}\\ \hline
    Concrete GWP  [\unit{kg}~\ce{CO2}e/ kg] & \multicolumn{2}{c|}{0.18 (C40)}\\
 \hline 
    Accelerator  GWP / m  [\unit{t}~\ce{CO2}e/ \unit{m}] & \multicolumn{2}{c|}{Not estimated}\\
 \hline 
 \multicolumn{3}{|l|}{\textbf{GHG emissions from operation}} \\
 \hline 
    Maximum power in operation [\unit{MW}]   & 110 & 125 \\ \hline
    Electricity consumption / yr  [\unit{TWh}/ \unit{yr}] & 0.49 & 0.56\\ \hline
    Years of operation & 10 & 10 \\ \hline
    Carbon intensity of electricity [\unit{g}~\ce{CO2}e/ \unit{kWh}] & \multicolumn{2}{c|}{20} \\ \hline
    Average Scope 2 GHG emissions / yr [\unit{kt}~\ce{CO2}e] & 9.8 & 11.2 \\
 \hline 

 \end{tabular}
\end{table}

\clearpage
\section{Mitigation and Compensation Measures}
\label{sec:mitigation}
Sustainability assessment also includes evaluating the adverse environmental impacts avoided through the implementation of mitigation strategies and compensation measures. Topics discussed in the following sections include the selection of more sustainable materials and procedures for civil engineering works, the adoption of responsible procurement practices,  the implementation of eco-designs aimed at reducing energy consumption, systems for heat recovery and redistribution, commitment to R\&D for improving accelerator energy efficiency,  and nature-based interventions.
\subsection{Better/greener materials and procedures for civil engineering works}

Particle accelerators use a wide variety of materials. These include superconducting materials like the niobium used in cavities and magnets. Copper and copper alloys are commonly used in accelerating cavities and for electrical conductors, while steel and iron provide structural support and are also used in magnets. Aluminium is sometimes used for its lightweight properties in structures and chambers. Often these materials are subject to requirements, for example that they be compatible for ultra-high vacuum materials such as stainless steel and titanium used in beam pipes. Ceramics and insulators ensure electrical insulation and withstand heat and radiation. Polymers and elastomers are used in seals and gaskets for vacuum systems. For shielding, radiation-resistant materials like lead and concrete protect against radiation. Finally, graphite and carbon-based materials are used in beam dumps and collimators to absorb or redirect particles.

While all of these materials produce emissions from production through machining, the volume of material is eclipsed by the materials that are required for the construction of the civil infrastructure. By far the most significant material by volume is concrete. Reinforced concrete is used extensively for tunnel linings, floors, and structural elements. In addition, shotcrete is sprayed 
in large quantities to stabilise the tunnel walls during excavation before the final concrete lining is installed. Steel rebar is used to reinforce concrete structures throughout the tunnel. Although steel is not the primary material by volume, a large quantity is used in combination with concrete to ensure structural integrity. Grout (often cement based) is used in significant volumes to fill gaps between the concrete lining and the surrounding rock, as well as to seal cracks and provide additional support. 

The pursuit of sustainable options for facility construction will require the use of practices that have received regulator approval, are certifiable, adopted by industry and considered extremely reliable. 
Partnering with governmental and non-governmental organisations -- such as the European Committee on Standardization~\cite{CEN_TC350} or the American Concrete Institute~\cite{ACI130} --
as well as with industrial partners on pilot projects, can help  accelerate the development and adoption of sustainable construction.

Designing tunnels to reduce the amount of concrete used can be achieved by employing several strategies that optimise the tunnel's structure and minimise material consumption while maintaining safety and durability. It is essential that these be considered early on in the design process to render them feasible.  Approaches include:
\begin{itemize}[nosep]
\item Establishing a functional and performance oriented \textbf{requirements document} that permits developing an architecture and design that responds exactly to the needs and thus will be optimised from the construction, operation and maintenance perspectives. In particular when sizing of the subsurface structures to the exact needs and defining well the required construction materials and their volumes can be important for the impact limitation. The origin of the materials, the production processes and the location of the materials production are further impactful parameters that need to be regarded. 

\item Optimising \textbf{tunnel shape}: circular tunnels are inherently strong because they evenly distribute the pressure from the surrounding ground. This reduces the need for thick concrete linings compared to flat-walled tunnels, where more material might be required to resist localised stresses.

\item Optimised \textbf{excavation methods}: using tunnel boring machines (TBM) rather than traditional drilling and blasting can create smoother, more uniform tunnel walls, reducing the need for excessive concrete lining to smooth out irregular surfaces. TBMs also allow for better control of tunnel shape and precision. Surface construction or cut-and-fill also hold the potential of significantly reduce load and allows for reduced waste material. Fully electric construction methods open up the possibility to source from renewable energy sources.

\item Detailed \textbf{geological assessment}: understanding the geology before excavation can lead to designing tunnels that rely more on natural ground stability (rock-support systems) in favourable conditions, reducing the thickness of the concrete required. In some cases, designers can use segmented lining where concrete is applied only in critical areas (for example, along the tunnel crown or areas with higher loads) and leave other sections with minimal or no concrete. This approach reduces overall material use without compromising stability. Areas that are free of concrete designers can combine composite linings, using materials like plastic membranes, steel liners, or fibre-reinforced materials. In areas where the surrounding rock is naturally strong, the use of rock bolts and a thin layer of shotcrete for stabilisation may eliminate the need for a full concrete lining. This technique allows the tunnel design to make better use of the natural ground for support, reducing concrete consumption.

\item Use of \textbf{precast concrete segments}: precast concrete segments, which are manufactured off-site, are used in place of poured-in-place concrete. These segments can be designed with thinner walls while still providing sufficient strength, leading to overall concrete savings. Their production can also be more tightly controlled, reducing material waste.
\item Fibre-Reinforced Concrete: fibre-reinforced concrete (FRC) which replaces steel with glass or synthetic fibres, can be used in tunnel linings to replace or reduce the need for traditional steel reinforcement (rebar). This material increases the strength and flexibility of the concrete, allowing for thinner linings while maintaining durability. It provides further benefits by limiting effects of corrosion. 
Furthermore, the use of FRC opens new possibilities in sustainability with reduced water quality requirements in the production of concrete. Using salt or brackish water in concrete mixes can be beneficial in specific contexts, particularly in areas where freshwater is limited. However, it requires careful consideration of potential impacts on durability and reinforcement corrosion. Proper testing, adjustments to mix design, and the use of protective measures can help mitigate some of these challenges, allowing for the effective use of alternative water sources in concrete production. 
\end{itemize}
By applying these design strategies, tunnel projects can significantly reduce concrete usage while maintaining the necessary structural integrity, ultimately making the construction process more efficient, cost-effective, and environmentally friendly.

\subsection{Responsible procurement}
\label{sec:resp_procur}

Procurement is the activity of purchasing the required materials or components according to a given set of specifications. In the past a successful procurement service would procure the required supply for the lowest possible procurement price. Shifting towards “responsible” procurement means balancing the purchasing (initial) cost with 
\begin{itemize}[nosep] 
\item the total lifecycle cost (including maintenance, operational cost, disposal cost etc.),
\item the environmental impact against a given set of environmental objectives (e.g. \ce{CO2}, radiological impact etc.),
\item social responsibility (e.g., ensuring that no child labour is involved in the manufacturing at any moment, including the extraction of ores and raw materials). 
\end{itemize}
Procurement activities in higher education institutions and research laboratories significantly contribute to the carbon footprint, especially in indirect (Scope 3) emissions. Responsible procurement plays an important role in particle accelerator projects for striving towards sustainability but also towards cost-effectiveness and ethical practices. The adoption of a set of responsible procurement policies also establishes a more favourable ground for the acceptance of new (large) scientific project from the public.
Responsible procurement best practices help to reduce the overall cost of a facility, including not only the construction but also the operation and disposal phases, sometimes at the cost of a slightly higher initial cost. For this reason, it is important to establish clear policies and objectives for the procurement actions, so to balance the construction cost and operation/dismissal costs of the facility.
In addition, it is generally recognised that adopting responsible procurement practices improves overall risk management, by fostering a more resilient supply chain and promotes ethical and social responsibility by ensuring fair labour practices and respect for human rights in the procurement of goods and services.

Environmental, Social and Governance (ESG) factors need to be taken into consideration for specific procurement actions, need to be analysed and decisions need to be justified and documented. Therefore any new Research Infrastructure shall establish a documented ESG policy and guideline to help members of the management to implement the ESG throughout the organisation. The ESG policy needs to consider technical feasibility and realistic possibilities to implement the goals with respect to financial viability in view of achieving an overall long-term sustainable research infrastructure. Such a policy and guideline needs to be be carried by top management and needs to be iteratively reviewed and updated, integrating the experience of members of the organisation in technical and administrative positions and retrieving information from potential suppliers to be able to work according to those policies and guidelines.

A responsible procurement policy will be one of the components of the overall strategy (and objectives) of a scientific institution (or a project) in terms of sustainability (or, in other words, in terms of Environment, Social and Governance aspects). 
In this sense the strategic positioning of a scientific Institution can be `defensive' (protecting the Organisation's image through risk management), `competitive' (reducing the total cost of ownership avoiding waste, and reducing energy consumption), or `offensive' (engaging in sustainable development, adopting new European strategies, and enhancing the Organisation's image to attract talent and stimulate innovation).
A strategic policy needs therefore to be established in order to drive the procurement strategy to be adopted. Independently from the positioning, the OECD provides a few axes of development, such as:
\begin{itemize}[nosep]
    \item Introducing environmental standards and best practices in technical specifications, supplier and contract selection, award criteria, and contract performance clauses.
    \item Professionalising responsible procurement to increase know-how and skills.
    \item Defining priority objectives (e.g. strategic positioning) to help procurement entities integrate sustainability criteria into their procurement procedures and stimulate commitments.
    \item Planning for responsible procurement by understanding market capacity, available technical solutions, and evaluating costs and benefits.
    \item Raising awareness on responsible procurement solutions and their benefits among the personnel technically responsible for defining the requirements and the specifications of the contracts, in order to include from the beginning environmental and social criteria.
    \item Monitoring the results of responsible procurement, for instance verifying compliance with procurement policies and standards.
\end{itemize}

The implementation of a responsible procurement policy should be based on the internationally adopted standard  ISO 20400:2017, or any equivalent standard relevant in different regions of the world. 
This  standard provides detailed guidelines on how to implement a responsible procurement framework, and does not foresee a specific certification. This means that scientific institutions can implement the guidelines which are relevant to their objectives, their status (international organisation, public institution, semi-public, etc), and the national/international legislation they are subject to. 

Concerning the use of electricity, future projects should explore the feasibility of planning for a portfolio of long-term renewable energy power purchasing agreements (PPA)~\cite{gutleber_2023_10023947}. 

For larger procurement actions the project owner should facilitate the creation of `aggregate' power purchase agreements by acting as an anchor tenant or encouraging the creation of consortia. This successfully demonstrated approach permits consortia of companies to produce the required equipment using renewable energy sources~\cite{crescenzi_2024_13166167}.

These mechanisms are valid due to the fact that electrons are fungible~\cite{EU_judgement_court_7_march_2024}. As long as a volume of electricity corresponding to a volume of purchased renewable energy is consumed from interconnected grids (national or international), spatial and temporal production and consumption patterns do not limit the capacity to lower the carbon footprint of the consumed energy and the potential to compensate for non-decarbonised energy use. Therefore, also virtual PPAs are a viable compensation approach for residual, unavoidable emissions.

Procurement actions shall include the requirement to deliver EPDs or comparable environmental indicators not only for off-the-shelf and customised off-the-shelf products but also for newly developed and integrated systems. If EPDs cannot be supplied, EPD-like reporting for the procured system based on an LCA that is disclosed to the client is an acceptable approach.

The environmental impacts of transport shall be taken into consideration in procurement actions that involve quantities of delivered products and goods which have a relevant impact on the overall environmental performance of the new project. Therefore, a cost-benefit analysis considering all benefits, negative externalities and their overall impact on cost for difference procurement scenarios is strongly recommended. The analysis should be followed by a weighting of the interests concerning the overall sustainability of the research infrastructure and should result in a well documented justification for the adopted procurement scenario.

Procurement of energy from renewable sources (RES) should be considered a priority for publicly-funded organisations such as accelerator based research infrastructures. 
Energy consumption 
of such installations permits concluding long-term power purchase agreements (PPAs). Therefore, accelerator facilities should account and forecast their energy needs in the frame of a proper management system such as ISO 50001\footnote{CERN is ISO 50001 certified since 2023, committing itself with an energy policy to continuous monitoring and energy efficiency improvement}. The first step towards this process is the definition of an energy management system with an energy policy that is anchored at organisational top level. The policy should be based on an approach that prioritises actions in the order 1) avoid, 2) reduce and 3) reuse. Procurement of energy from renewable energy sources does not need to be limited to `wire' and `physical' contracts~\cite{gutleber_2023_10023947}. Electrons are fungible and therefore any energy from renewable sources purchased on an integrated market, independent of the time of consumption, represents a valid and important contribution to the societal transition towards the use of energy from renewable sources. Typically a portfolio of PPAs ranging from actual energy supply in the form of negotiated baseload PPAs to virtual PPAs permits achieving the goals. Pooling suppliers in a sufficiently large procurement action that is particularly energy intense (e.g. construction and heavy industry dominated production) and establishing a dedicated energy supply contract for such a procurement action is an approach that has been successfully implemented already by different countries and industry consortia~\cite{crescenzi_2024_13166167}. Finally, the European Union Emission Trading System (EU ETS), the world's first and largest carbon market, is an instrument that can be leveraged to compensate the residual carbon footprint and help bring overall EU emissions down while generating revenues to finance the green transition. The system works also for Switzerland due to its linkage to the Swiss ETS since 2020.

Three issues arise for fostering the use of RES and ETS: first, additional costs occur. Therefore the funding stakeholders and the governance of a public institution need to set a policy for implementation that includes 
financial analysis, planning, and continuous monitoring of the resources involved. 
Second, energy purchase and emission trading require the availability of dedicated funds and a framework for using them. The financial instruments for this activity require expert knowledge and long-term experience to be effective, to control costs and risks. Hedging is the strategy that is typically applied and it involves trading derivatives (e.g. energy and carbon) to offset financial losses in a volatile energy market with dedicated funds. Consequently, the third issue is that funding agencies will need to allocate additional public funds 
to research infrastructures in order to plan and implement the energy transition. 

The opportunities and value of the principles of circular economy should be analysed for all procurement actions that result in a potential significant contribution to the environmental performance of the research infrastructure. 

\subsection{Energy optimisation}
\subsubsection{Electrical energy}

Concerning the optimisation of consumption of procured electrical energy, the same approach as for the optimisation of all other resources shall be used. The adoption of the hierarchical ``Avoid-Reduce-Compensate'' principle guides the iterative planning, implementation, checking and taking-action cycle that leads to a continuous improvement process.

Consequently, first of all, a comprehensive technical requirements elucidation process shall be established to document and scrutinise where and when electrical energy is needed. Following a baseline scenario, an eco-design approach can help to conceive designs and choose products that lead to reduced energy consumption. A rigorous review of the operation model that shall be guided by overall sustainability goals integrating economic, ecological and societal aspects will guide the development of different scenarios. Finally compensation measures can be explored. Some of them require re-iterating the eco-design cycle by introducing new requirements, such as for instance the integration of waste-heat recovery and supply functionality. It permits for instance offsetting fossil energy used inside and outside the project, but it comes with constraints such as additional investment and operation costs, the necessity to buffer energy, the need to dynamically adapt the operation, requirements on financial and ecological accounting, and necessity to 
establish administrative and commercial frameworks for a successful implementation. All ESG relevant parameters need to be taken into consideration and it is therefore recommended to engage experienced companies for the overall energy optimisation of the infrastructures.

For determining the carbon footprint of the consumed electricity during the design and planning phases, official national values for the selected base year of the sustainability analysis shall be used. Depending on the country, these values are made available from different sources. \autoref{ElectricityCarbonFootprint} provides an exemplary illustration only, and each research infrastructure needs to select and cite the appropriate source.

Once a specific energy supply contract is active, emission accounting shall be carried out using actual supplier provided emission information. For instance, a typical consumer-oriented electricity contract based entirely on renewable energy sources (\qty{62}{\percent} hydro, \qty{31}{\percent} wind and \qty{7}{\percent} PV) 
in France today has an actual carbon footprint of \qty{34}{\kg\ce{CO2}e\per\mega\watt\hour}\footnote{ENGIE Elec'verte contract carbon footprint: \url{https://particuliers.engie.fr/content/dam/PDF-CP/Dossier-presse-ENGIE-reduction-impact-carbone.pdf}}. This value is higher than the ADEME indicated emission factor for renewable energy sources to be used for the planning and design phases, since the consumption of electricity in the frame of an actual contract includes all emissions along the value chain and not only the production-related emissions.
\begin{table}[!ht]
\centering
\small\addtolength{\tabcolsep}{-3pt}
\caption{Energy Carbon footprint of various countries. Official sources are shown in italics.}
\label{ElectricityCarbonFootprint}
\scalebox{0.9}{
\begin{tabular}{lllccc}
\hline
\textbf{Country} & \textbf{Source} & \textbf{Energy} & \textbf{Carbon} & \textbf{Year} & \textbf{Major} \\
&&&\textbf{footprint}& &\textbf{facilities}  \\
&&&{[\unit{\kg\ce{CO2}e\per\mega\watt\hour}]}& & \\
\hline
China & Electricity Maps~\cite{china_grid_intensity} & {Unqualified  mix} & 582 & 2023 & {BEPC-II }\\
France & \textit{Ademe Base Empreinte} & {Offshore wind} & 15.6 & 2023 & \\
France & \textit{Ademe Base Empreinte} & {Unqualified mix} & 52.0 & 2022 & {LHC, ESRF,}\\
&&&&&{Soleil} \\
Germany & \textit{Umweltbundesamt (UBA)}\tablefootnote{https://www.umweltbundesamt.de/publikationen/entwicklung-der-spezifischen-treibhausgas-10} & {Unqualified mix} & 380.0 & 2023 &{BESSY-II, PETRA-III,}\\
&&&&&{EU-XFEL}\\
Italy & \textit{ISPRA} & {Unqualified mix} & 257.2 & 2022 &  \\
Japan & Electricity Maps~\cite{electricity_maps} & {Unqualified mix}&460 & 2023 & {J-PARC, SPring-8}\\
&&&&&{SuperKEKB}\\
Switzerland & 
\textit{BAFU/OFEV}\tablefootnote{https://www.bafu.admin.ch/bafu/de/home/themen/klima/fragen-antworten.html} & {Unqualified mix} & 54.7 & 2018 &\\
Switzerland & \textit{BAFU/OFEV} & {Renewable mix} & 15.7 & 2018 &\\
UK & Electricity Maps~\cite{electricity_maps} & {Unqualified mix} & 201 & 2023 & {Diamond, ISIS} \\
USA (average) &  Electricity Maps~\cite{electricity_maps} &{Unqualified mix}&410 &2023 &  \\
 \hspace{3mm}California & Electricity Maps~\cite{electricity_maps} &{Unqualified mix}& 261 &2023 & {LCLS} \\
\hspace{3mm}New York & Electricity Maps~\cite{electricity_maps} &{Unqualified mix}& 281 & 2023 & {RHIC, NSLS-II} \\
\hline
\end{tabular}
}
\end{table}

\subsubsection{Fossil energy}
Energy from fossil sources shall be avoided and where it cannot be avoided it shall be reduced. The residual quantities shall be estimated and be converted into a monetary term using the shadow price of carbon for use in an overall, integrated societal cost-benefit assessment to verify the sustainability of the project. It is valid to subtract the national carbon tax on fossil energy sources from the shadow price of carbon to avoid double accounting.

The use of fossil energy sources for heating and cooling shall be avoided. Project internal heating shall be carried out leveraging waste heat re-use and renewable energy sources whenever feasible and long-term economically viable.

The use of fossil energy sources for backup electricity supply shall be avoided. Instead, hydrogen to power a fuel-cell based system is a valid alternative in addition to short term electricity supply via battery-based energy storage systems (BESS). A cost-benefit assessment shall guide the appropriate selection of a suitable scenario among different variants and versions. Hydrogen from low-emission processes (green vs. blue and grey hydrogen) shall be preferred. Production-based footprint\footnote{IEA, Towards hydrogen definitions based on their emissions intensity, Paris, 2023, \url{https://www.iea.org/reports/towards-hydrogen-definitions-based-on-their-emissions-intensity}} of the used hydrogen should be well below \qty{3}{\kg\ce{CO2}e\per\kg\ce{H2}}.
\subsubsection{On the pathway to energy optimisation}
\label{sec:ESS-RF2.0}
 As a green-field facility, the ambition to design the European Spallation Source (ESS) as a sustainable accelerator-based infrastructure was pursued from the beginning of the project~\cite{Peggs:2013sgv,ESS_EnergyDR}.
    The ESS sustainability strategy comprises several elements (see~\cite{Sunesson:2024zux} for a brief review): i) the facility was planned to reuse and re-sell excess heat generated
    by the cooling system; 
    ii) the electricity purchase agreements with providers are based on flexible provision of electricity from renewables sources, and the infrastructure is expected to integrate purchased electricity
    with own production, with the installation of solar panels and bio-gas turbines; 
    iii) the study and realisation of energy-optimised power modulators and klystrons, with 
    an active participation to international programs 
    that target energy optimisation, technological innovation, and flexible use of energy. 

       Research Facility 2.0 (RF2.0) is a Horizon Europe project led by the Karlsruhe Institute of Technology (KIT) 
   targeting sustainable and efficient design and operation of accelerator-based research infrastructures. 
  Six large research accelerator facilities - ALBA, CERN, DESY, HZB, KIT, MAX IV --  and four small-medium enterprises (SMEs) -- Commtia, Cryoelectra, Elytt, Zaphiro -- in 
   Europe collaborate to address energy optimisation from multiple perspectives,
   including the study and analysis of individual components and systems from both,
   physics and energy engineering standpoints.
   The results of the project will be transferred in large facilities with demonstrators:
\begin{itemize}[nosep]
\item{Tuneable Permanent-Magnet (Dipoles and Quadrupoles); }
\item{Highly-efficient Solid State Amplifiers, integrating new semiconductor technologies and control systems; }
\item{Flexible High Performance Computing; }
\item{Fast grid measurement units by means of Phasor Measurement Units, 
to detect 
grid disturbances and thus take countermeasures 
to increase disturbance robustness of the research facility.} 
\end{itemize}
Together with the demonstrators, novel aspects are being investigated: i) energy-focused Digital Twins of accelerators and data centres, to support analysis and testing without the need of real hardware, and to be used as benchmark for new energy solutions; ii) artificial-based control of accelerator operations, targeting a more efficient and stable beam control leading to reduced losses; iii) optimised energy management, integrating renewables production, energy storage systems, and consumption / production forecasting as means to increase local green energy consumption.

 By the end of the project, RF2.0 will present policies, benchmarks, and roadmaps reflecting its vision for the design, implementation, and operation of energy-efficient research facilities.

\subsection{Heat recovery and supply}
\label{sec:heatrecovery}

Most of the energy used to operate the technical infrastructures and subsystems of a particle accelerator
is eventually converted into heat.
Energy used to operate accelerator magnets, amplifying RF energy, absorbing synchrotron radiation, air management systems, operating electronics and data processing equipment is almost entirely converted into low grade heat, typically below \SI{45}{\celsius}. Only rarely reaching temperatures above \SI{50}{\celsius} but still below \SI{70}{\celsius} can be reached, for instance when cooling cryogenic refrigeration system equipment and electrical transformers and substations.
This heat is typically dissipated to the ambient air via water-cooling and free-to-air cooling based systems and is thus lost.
Considering the amounts of heat that particle-accelerator based research infrastructures generate there exists an interest to explore ways to recover that heat and convert it into a valuable resource.
The use of the heat for other purposes inside and outside the project boundaries has the following socio-economic benefit potentials:

\begin{itemize}[nosep]
\item Reduction of electrical energy consumption and associated costs due to reduced cooling system operation needs.
\item Reduction of raw water consumption and associated costs due to reduced cooling system operation needs.
\item Increased cooling system lifetime and reduction of associated maintenance and repair costs due to lowered operational load on equipment.
\item Reduction of heat generation-related carbon emissions due to the avoidance of dedicated heat production.
\item Lower heat costs for consumers.
\item Opportunities for creating new economic activities in the vicinity of the heat source.
\end{itemize}

However, heat recovery and supply require also additional efforts and costs:

\begin{itemize}[nosep]
\item Additional components to recover the heat.
\item A dedicated network to transport the heat where it is needed. 
\item Potential additional components to raise the temperature of low-grade supplied heat for specific needs.
\item Short term and long term heat storage systems to assure supply stability and provide heat when needed.
\item The need to refurbish existing buildings and the need for new buildings that are equipped with heating and cooling that is able to work with the low-grade heat.
\item The need for a heat supply operator if the heat is used outside the RI boundaries.
\end{itemize}

A comprehensive and quantitative cost-benefit assessment that integrates all costs, benefits and negative externalities is needed to understand if the approach for a specific project is in principle viable and to develop the most suited heat recovery and supply scenario for that project. The LCA approach can be integrated in this process to better determine and quantify the negative externalities.

Future particle accelerator projects should first carry out a technical analysis to establish an inventory of the potential heat sources including baseline estimates for the quantities and characteristics of the heat produced.
This analysis step needs to be followed by a cartographic mapping study that identifies potential consumers of that heat inside the research infrastructure and outside its boundaries.
Next, a techno-economic study should be carried out to assess in how far heat recovery and supply is technically feasibly and economically viable.
All information must eventually be integrated in a cost-benefit assessment that yields a net present value at the end of a defined observation period to report on the overall net benefit.
The result can be compared to a counterfactual scenario in which no heat is recovered and supplied.
This approach assured that an informed decision making process is implemented.
If the system is put in place, it is strongly advised to continuously account the heat supply so to report on the actual efficiency of the approach and to be able to further optimise the waste heat recovery and supply process.

While retrofitting heat recovery and supply to existing cooling systems is technically possible, it is typically more costly than foreseeing the concept already from the onset. Depending on the existing equipment, infrastructures and environment around the particle accelerator facility, retrofitting may be less efficient, since the operation temperature of the equipment supplying the heat may not be matched to the consumer needs (e.g. too low magnet water cooling circuit temperatures, lack of data centre rack cooling infrastructure, mismatch of the waste heat characteristics with existing district heating networks or lack of presence of low-temperature district heating networks and finally lack of space and missing agreements with off-takers). It is, however, preferred over no heat recovery. Therefore an eco-design approach in which heat recovery and supply is part of the research infrastructure design and embedding the infrastructure in its socio-economic environmental context is recommended.

The following exemplary list of equipment can serve as a starting point for the consideration of integrating heat recovery and supply functionality:

\begin{itemize}[nosep]
\item Normal conducting magnets (recovery of cooling water at temperatures between \SIrange{25}{45}{\celsius}).
\item Normal conducting RF cavities (cooling water temperature between \SIrange{25}{45}{\celsius}).
\item Synchrotron radiation absorbers (cooling water temperature between \SIrange{25}{45}{\celsius}).
\item Rack mounted electronics (water cooled with with a dT of \qty{20}{\kelvin} and a temperature range between \SIrange{27}{49}{\celsius} on the outer circuit and \SIrange{39}{60}{\celsius} on the rack cooling circuit).
\item RF amplifiers of different types (e.g. solid state, klystrons, IOTs with a cooling water circuit temperature range between \SIrange{25}{35}{\celsius} with high stability and low temperature fluctuation constraint, depending on the case as tight as \SI{0.1}{\celsius}. By design, the maximum water temperature at the klystron's collector may be allowed to reach \SI{63}{\celsius}, but so far applications operating in this regime are not known.
\item Cryogenic refrigeration plants for superconducting components (e.g. magnets, RF cavities) with equipment cooling circuit water temperatures in the range of \SI{50}{\celsius} (e.g. compressors) to \SI{75}{\celsius} (e.g. oil separator).
\item Power electronics and converters (temperature range of circuit between \SIrange{30}{60}{\celsius} for directly water cooled IGBT -- Insulated Gate Bipolar Transistor -- systems as example).
\item Electrical transformation stations with water cooling based systems in the range of \SIrange{20}{70}{\celsius} (e.g. oil-based transformers).
\item Data centres~\cite{YUAN2023113777} (recovery of \SIrange{15}{20}{\celsius} air, \SIrange{40}{50}{\celsius} heat from the CRAHs and \SIrange{50}{60}{\celsius} from liquid cooling systems).
\item Ventilation and air management systems (e.g. heat from motors and air-to-air transfer) from \SI{25}{\celsius} up to about \SI{40}{\celsius}.
\end{itemize}

The initial equipment and heat load analysis will inform designers about which components generate the most heat and their characteristics (e.g., stability, temperature).

The initial equipment and heat load analysis will inform designers about which components generate the 
most 
heat and about heat-flow characteristics 
(stability, temperature).
This permits creating a hierarchy of heat producing components that can guide the heat recovery and supply concept development.
A multi-criteria analysis with different weights for the individual aspects is a suitable approach for this first step.
The analysis process should at least include the following, non-exhaustive list of aspects:

\begin{itemize}[nosep]
\item Operation temperature requirements and constraints of the equipment components to be cooled and their temperature variation tolerances.
\item Temperatures of the heat recovery potentials for the different equipment to be cooled.
\item Variability and stability of the heat generation (hourly, daily, weekly, monthly).
\item Climatic and weather conditions in the environment of the research infrastructure (for instance a particle accelerator facility and a data centre operated in the north of Europe permits the use of different heat recovery and supply technologies than in a southern European region).
\item Use cases for the recovered heat inside the research infrastructure (e.g. pre-warming of water, offices, workshops, assembly halls, guest houses).
\item Demand of industrial heat consumers outside the research infrastructure (e.g. food production and processing industries, offices, hotels, airports, shopping malls, theatres, cinemas, congress centres).
\item Demand for heating public spaces and institutions (e.g. schools and universities, hospitals, prisons, train stations)
\item Demand for heating of private spaces (e.g. apartment buildings and individual houses).
\item Demand for hot water production (the required temperature for sanitary reasons is above \SI{55}{\celsius} 
and therefore priming may require 
 water boilers and heat pumps).
\item Distances between heat production and consumers (note that distances up to \SI{10}{\kilo\meter} are well feasible for low-grade district heating systems in the \SI{50}{\celsius} range with modern pipe technology).
\item Gap analysis concerning the need for heat buffering (e.g. capacity, space, duration, technology, investment costs, operation costs).
\item Investment costs for heat recovery, buffering and supply.
\item Operation costs for heat recovery, buffering and supply.
\item Capital and operation expenditures outside the system boundary (e.g. district heating network operation host, private heat pumps and priming equipment).
\item Public co-financing possibilities.
\item Envisaged duration and observation period for the heat recovery and supply.
\item Baseline for \ce{CO2} compensation due to the possibility to avoid fossil energy sources for heating and the avoidance of primary energy for heating. Only the energy for stepping up the temperature for specific end use cases needs to be considered.
\item Definition of the interface between the heat recovery and supply system that is part of the research infrastructure and the segment that is outside the responsibility of the research infrastructure.
\item Conditions of the heat supply operator who provides the infrastructure up to the consumer and who supplies the heat with guarantees or with contractual conditions that require the consumer to generate or obtain the gap between supplied and needed heat.
\item The proposed heat supply technology (e.g. direct, indirect via a loop, indirect via heating the soil or other approaches).
\end{itemize}

Once all data are collected and the most promising heat sources that qualify for a heat recovery case have been identified and once viable heat consumers have been identified, a quantitative cost-benefit assessment shall be carried out to understand the long-term sustainability in terms of economic, environmental and societal performance. If heat recovery and supply turn out to be a viable scenario for the project, the following, non-exhaustive list of aspects to tune the heat recovery and supply scenario should also be considered:

\begin{itemize}[nosep]
\item Increase of the supplied heat by relaxing the equipment cooling requirements (e.g. water-based magnet cooling up to \SI{50}{\celsius}, increase of ambient air temperature inside the facility up to \SI{40}{\celsius} and possibly beyond).
\item Validation that mission critical systems remain within their required operation margins (e.g. increasing the cooling water temperature of klystrons or relaxing its temperature stability may lead to unacceptable performance or render the operation unfeasible).
\item Total amount of \ce{CO2} compensation potential due to the use of avoidance of fossil fuel and any primary energy, considering a credible estimate for the energy required to prime the heat for the end-use applications.
\item Optimisation of the heat supply by adjusting the operation schedule and introducing the possibility to dynamically react to heat needs.
\item Adaptation of the particle accelerator operation schedule to increase the overall socio-economic performance.
\item The potentially different energy costs for the research infrastructure operator when adjusting the operation schedule of the particle accelerator or when introducing the capability to dynamically react to both electricity supply and heat demand constraints.
\item Additional societal and economic benefits that can be generated by creating new heat consumers in the vicinity of the supplied heat (e.g. food processing industries, agricultural producers, biogas production, thermal baths and recreational installations).
\item Introduction of temporary heat buffers (daily, weekly, monthly).
\item Availability of specific public co-financing instruments and loans with specific conditions.
\item Optimisation of the interface boundary between research infrastructure, operator and heat consumers.
\end{itemize}

The recovered heat will not be consumed at all times and the consumption pattern will change. Therefore, care must be taken not to under-dimension the cooling, ventilation and evaporation systems for the research infrastructure. In case the heat is not consumed or cannot be delivered, it must be possible to cool all components reliably to assure operation for scientific research purposes as required.

Techno-economic studies and cost-benefit assessments should be carried out by 
experts 
to be sure that the entire value chain, investment and operation costs,  are included in the accounting and that the environmental externalities are properly captured and correctly converted into monetary terms.

\subsection{Energy recovery in particle accelerators}
\label{sub-sec:ERL} 

 Large-scale particle accelerators, crucial for advancing particle physics research, consume ever growing amounts of energy. As future experiments demand higher beam energies and intensities, this energy consumption becomes a significant challenge, raising the need for sustainable solutions. Energy Recovery Linacs (ERLs) offer a promising way to reduce power consumption while meeting these increasing demands.  
In conventional linear accelerators, the particle beam is discarded at high energy after completing its purpose. In contrast, an ERL recovers part of the beam’s kinetic energy by guiding the used beam back through the accelerator structures to be decelerated. During this deceleration process, the beam transfers its energy into the accelerator’s electromagnetic fields, which can then be reused to accelerate new particles. This process significantly reduces the total power needed for particle acceleration, making ERLs an energy-efficient technology.  

The basic concept of ERLs, first introduced in 1965, has evolved over time and become really viable thanks to the major advances in SRF technology within the last decades (quantified by cavity quality factors $Q_0 \geq 10^{10}$) enabling high average current operation. Recent advancements include multi-turn ERLs, where particles can pass through the accelerator multiple times for both acceleration and deceleration. This approach allows ERLs to achieve higher beam energies and efficiency with fewer components. These two aspects have paved the way to a new generation of powerful machines with relatively compact footprint and low energy consumption. The S-DALINAC team at TU Darmstadt have successfully demonstrated energy savings of up to \qty{89}{\percent} by combining energy recovery with double-turn acceleration ($E_{b, max}=34.2$~MeV, $E_{b, dump}=3.8$~MeV, no sizeable beam losses over two linac passes).  
\qty{96}{\percent} energy recovery has been demonstrated at the Cornell-BNL ERL Test Accelerator (CBETA) - an energy recovery linac (ERL) capable of accelerating electrons to \qty{150}{\MeV} in 4 linac passes ($E_{b, dump}=6$ MeV). It uses a fixed field alternating gradient return arc to return all 4 energies, from 42 to \qty{150}{\MeV}, to the linac. These two pioneering machines demonstrated the feasibility of multi-turn ERLs at low current beam operation. 
The next challenge to combine multi-turn with high current operation will be taken up by the PERLE project, initiated at the IJC Laboratory in Orsay, France [189]. PERLE aims to replicate key features of the LHeC, including a 3-turn racetrack configuration with two opposing linacs and a target beam current of \qty{20}{\mA} at \qty{500}{\MeV}. With a power range of \qty{10}{\MW}, PERLE will help address challenges related to beam dynamics, collective effects, and RF interactions, providing a platform for developing and refining ERL technology and prepare for large-scale implementation.

\subsubsection{Applications in particle colliders}  

ERLs have gained prominence due to their potential applications in high-energy physics. They are essential for future projects like the Large Hadron Electron Collider (LHeC), which aims to achieve high luminosity electron-proton collisions. The LHeC's design relies on ERL technology to meet power constraints, with a wall-plug power limit of \qty{100}{\MW}. Similarly, ERLs are being considered for the Future Circular Collider (FCC-ee) as a way to increase luminosity in specific energy ranges, such as the WW and top mass production regions.  

Another notable application is the Electron-Ion Collider (EIC), where high-current ERLs will be used to cool proton beams, improving collision efficiency. These applications illustrate that ERLs are not only energy-efficient but also capable of delivering high currents required for cutting-edge experiments.  

{\it Electron-Proton/Ion Collider LHeC~\cite{LHeC2021}}: the main guidelines for the design of the Electron Energy Recovery Linac (ERL) and the Interaction Region (IR) with the LHC are:
\begin{itemize}[nosep]
    \item Electron-hadron operation in parallel with high-luminosity hadron-hadron collisions in the LHC/HL-LHC.
    \item Centre-of-mass collision energy on the TeV scale.
    \item Power consumption of the electron accelerator below \qty{100}{\MW}.
    \item Peak luminosity approaching \qty[per-mode=power]{1e34}{\per\cm\squared\per\second}.
    \item Integrated luminosity at least two orders of magnitude higher than that achieved by HERA at DESY.
\end{itemize}

\begin{figure}[ht]
    \centering
    \includegraphics[width=0.8\textwidth]{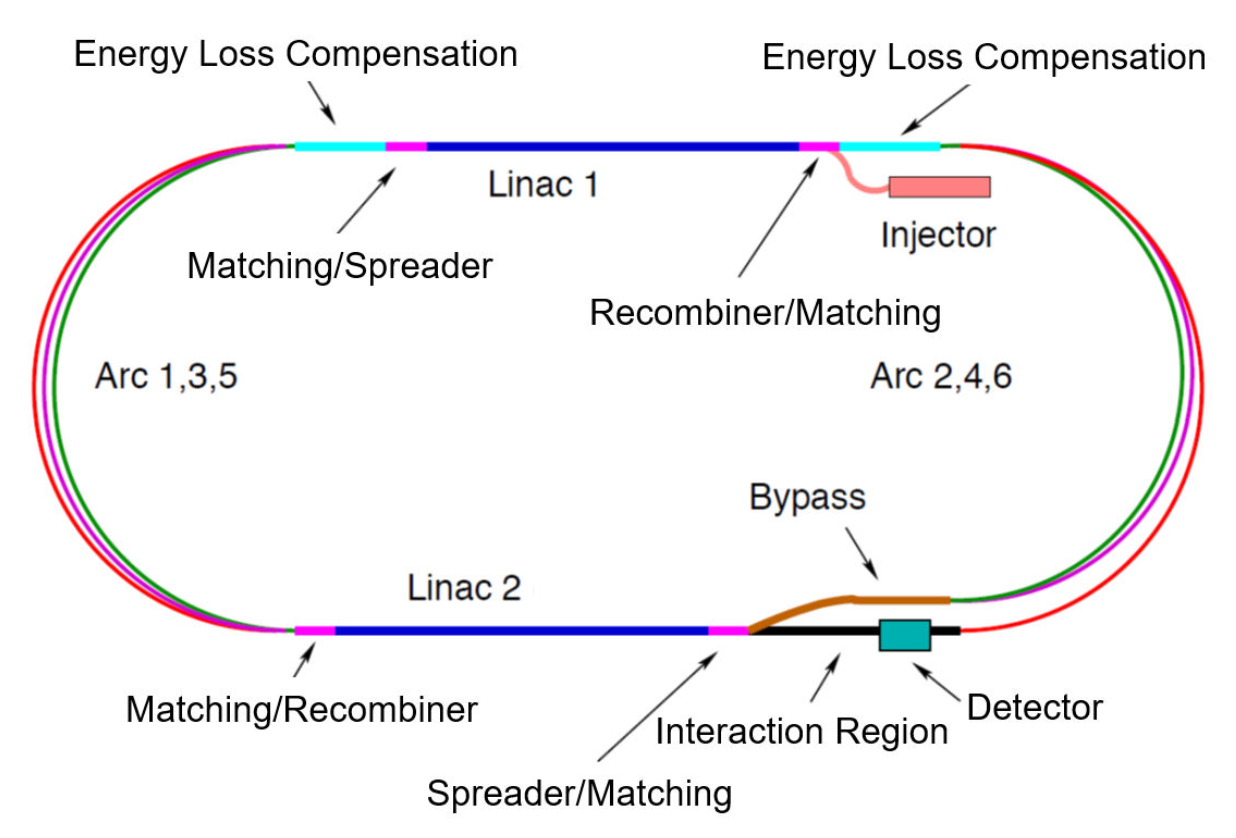} 
    \caption{Schematic of the ERL design with three accelerating and decelerating passes~\cite{LHeC2021}.}
    \label{fig:ERL}
\end{figure}

The electron energy $E_e$ chosen in the 2021 LHeC CDR update~\cite{LHeC2021} is \qty{49.2}{\GeV}. This energy could be achieved with an ERL circumference of $1/5$ of the LHC's circumference. These parameters were chosen out of the cost considerations and machine–detector performance aspects — particularly the impact of synchrotron radiation losses in the IR. The ERL design consists of two \qty{8.1}{\GeV} superconducting (SC) linacs operating in continuous wave (CW) mode, connected by at least three pairs of arcs to enable three accelerating and three decelerating passes (see \autoref{fig:ERL}). The high-energy return arc following the interaction point (IP) is designed to provide a half RF-period wavelength shift, allowing efficient beam deceleration back to the injection energy before safe disposal. SC cavities with an unloaded quality factor $Q_0 > 10^{10}$ are required to minimise the cryogenic power demands and ensure efficient operation.

The three-pass acceleration and deceleration design means that the circulating current in the linacs is six times the current colliding at the IP with the hadron beam. The ERL's high-brightness beam avoids the performance limitations caused by beam-beam interactions, which were a significant bottleneck in previous circular lepton colliders (e.g., LEP) and the LHeC’s Ring-Ring option. An operational current goal of $I_e = 20 \, \text{mA}$ has been set, corresponding to a bunch charge of \qty{500}{\pico\coulomb} at a bunch frequency of \qty{40}{\MHz}. This requires operating the superconducting RF (SRF) cavities at a very high current of \qty{120}{\mA}, yielding a virtual beam power (beam current at the IP times maximum beam energy) of \qty{1}{\GW}. Achieving stable three-turn ERL operation at these currents is a critical milestone, with validation expected through the PERLE facility~\cite{PERLE2018}.

To maximise luminosity, a small beam size at the IP is essential. The LHeC aims for peak luminosities of \qty[per-mode=power]{1e34}{\per\cm\squared\per\second} and integrated luminosities of $1 \, \text{ab}^{-1}$ over its operational lifetime. Achieving these goals requires $\beta^* < 10 \, \text{cm}$ for the colliding proton beam, consistent with optical constraints for parallel proton-proton physics during the HL-LHC era. These luminosity targets exceed those of HERA by 2-3 orders of magnitude. For comparison, HERA achieved an integrated luminosity of about 0.1 fb$^{-1}$ between 1992 and 2000 for the H1 and ZEUS experiments -- an amount the LHeC aims to collect in just one day of operation.

{\it ERL Challenges and Future Prospects}:  
despite their promise, ERLs face several technical challenges. Issues like space charge effects, multi-pass beam breakup (BBU) instabilities, and coherent synchrotron radiation (CSR) must be carefully managed. Additionally, the flexibility of ERLs to maintain long, high-charge bunches with minimal energy spread introduces further complexity in their design and operation.  

Historically, ERLs have been used in free-electron lasers (FELs) and light sources. However, their role is expanding into nuclear physics experiments, Compton backscattering sources, and strong electron cooling. As ERL technology matures, it will play a critical role in next-generation particle colliders, enabling higher luminosities and more energy-efficient operations.
 
In summary: Energy Recovery Linacs offer a transformative approach to accelerator technology by reducing power consumption while maintaining high performance. With successful demonstrations of multi-turn energy recovery and ongoing development projects like PERLE, ERLs are poised to become essential components of future particle colliders, ensuring sustainable and high-luminosity research in the years to come.

\subsection{Investment in R\&D on green technologies} 
Among all actions focused on minimising the environmental impact of future colliders, study and development of novel
technologies is the most specific to the accelerator-based RIs.
Medium- and long-term challenging development plans are in place, aiming to improve the efficiency of the RF systems, to design
advanced permanent magnets, to use high-temperature superconductors (HTSs) in high-field magnets, or to understand/demonstrate the full potentiality
of energy-recovery linacs as discussed in previous section,~\ref{sub-sec:ERL}. 
Large initiatives are in progress to coordinate and foster the programs carried out by national laboratories worldwide. 
Horizon Europe fundings support research programs such as iFAST (Innovation Fostering in Accelerator Science and Technology) and iSAS (Innovate for Sustainable Accelerator Systems) committed to the development of innovative solutions to contain the energy demand of future accelerators, EAJADE (Europe-America-Japan Accelerator Development and Exchange)  that promotes the exchange of accelerator scientists between European institutions and American and Japanese partners, and RF2.0 (Research Facility 2.0), introduced in section \ref{sec:ESS-RF2.0}, focusing on the study of energy-efficient and sustainable accelerator-based research infrastructures.

The Magnet Development Program (MDP) has been established by DOE to coordinate the R\&D activities in US laboratories. An explicit  acknowledgement of their key role in sustainability is expressed in the 2023 P5 Report  
drawn by the \emph{particle physics prioritization panel} on inputs 
from the \emph{Snowmass'21} process, presented in Annex~\ref{annex:snowmass21}.

In China, extensive efforts are being made by IHEP to sustainability studies for the Circular electron-positron Collider (CEPC)~\cite{CEPCTDRDec24}. They are focused on energy recovery plants and on the efficiency improvement of RF systems,  as summarized in the Annex~\ref{annex:CEPC}.

Significant advancements have been achieved — and are expected to continue — in key accelerator technologies, including:

\begin{itemize}[nosep] 

\item RF sources (Vacuum electron devices and solid-state amplifiers)

Future accelerators require significant beam power to achieve their target luminosities. With the majority of the beam’s acceleration provided by RF fields, the sources of these RF fields are critical elements in the power conversion chain needed for these facilities. RF sources that meet the specifications of a collider are one of the specialised components of the power conversion chain which are also limited in their broader commercial use. This is often because of the high peak and average power requirements, as well as the typically very narrow bandwidth that is needed. For most commercial applications vacuum electron devices and solid-state sources are either optimised for broader bandwidths or lower power and compactness. Therefore, while a strong industrial base exists, dedicated R\&D in optimising the sustainable performance of RF sources for colliders can have a dramatic impact. Many technological innovations could have significant impacts on these classes of devices. For example, for vacuum electron devices these include adoption of permanent or high temperature superconducting magnets for beam transport, longer lifetime or lower temperature cathodes, depressed collectors or energy recovery of spent beam, and more efficient topologies. These strategies and more are being pursued through industry, national laboratory and university collaborations and have shown significant progress in recent years. For solid state systems, optimising transistor efficiency, lifetime and power combining can also have a significant impact on overall sustainability. A critical issue to consider is the maintenance of these RF sources. For vacuum electron devices, the mean time between failures (or MTBF) can approach 100-150k hours, dramatically reducing material consumption for fabrication.

\item High-Temperature Superconductors (HTS)

The use of superconducting materials has vastly expanded the potential reach of particle colliders. Many accelerator facilities utilise superconductors for either magnets or RF cavities. However, in most cases these systems are cooled to \SIrange{2}{4}{\kelvin} resulting in significant complexity for thermal shield and limitations on cooling efficiency that result in requiring \SIrange{200}{500}{\watt} of electrical power per \qty{1}{\watt} of cooling capacity in the cryogenic environment. Higher temperature superconductors such as \ce{Nb3Sn}, \ce{MgB2} and ReBCO could increase the temperature of operation for these components and significantly reduce cooling requirements. For example, the ReBCO class of high temperature superconductors has critical temperatures in excess of \qty{90}{\kelvin}. These HTS materials can easily be cooled with liquid nitrogen based cooling systems where the electrical power consumption is \SIrange{6}{7}{\watt} per \qty{1}{\watt} of cooling at \qty{80}{\kelvin}. Developing beam screens, magnets and RF cavities with these advanced materials could have dramatic impacts on site power consumption and should be pursued vigorously.

\item Permanent Magnets

Strong magnetic fields are required in all accelerators to bend and focus beams. In most cases, these fields are provided by electromagnets. In the last few years, there has been a significant move towards permanent magnets (PMs), particularly for beamlines where fields are fixed or require very low levels of adjustment. This has been exemplified by synchrotron light source facilities, many of which are in the process of upgrading to high-brilliance, low-emittance lattices (ESRF-EBS, Diamond-II, SPring-8-II, SLS 2.0, and more). The new lattices tend to have smaller apertures and a greater packing factor: more magnets with closer spacing between them. This lends itself readily to adoption of PM-based magnets, where a simple geometric scaling does not lead to changes in field unlike electromagnets.

An example of the potential savings to be gained from switching to PM-based magnets can be found at the ESRF. The main dipoles were changed from electromagnets to fixed-field PM devices. The old normal-conducting dipoles used a total of \qty{720}{\kilo\watt}, reduced to zero for the PM version~\cite{revol-esrf-ler24}. Together with other changes to magnets, the new lattice uses \qty{7.7}{\giga\watt\hour} of energy per year, compared to \qty{16.9}{\giga\watt\hour} for the old lattice -- a reduction of \qty{54}{\percent}. There are other advantages too: PM-based magnets do not require large power supplies or current-carrying cables. The need for water cooling is eliminated, which means reduced maintenance as well as zero vibration during operation. In their raw state, PMs have a large temperature coefficient, meaning that magnetisation varies with ambient temperature (typically around \qty{-3e-4}{\per\celsius} for \ce{SmCo} and \qty{-1e-3}{\per\celsius} for \ce{NdFeB}). However, this can be overcome by using \ce{FeNi} `Thermoflux' shims in series with each PM: the magnetisation of this material varies in the opposite direction with temperature, and hence the PM temperature variation can be cancelled out around room temperature. Another concern is radiation hardness: PMs are demagnetised by prolonged exposure to radiation. Again, the effect is greater for \ce{NdFeB} than for \ce{SmCo}, and is typically worse for low-coercivity PMs. PM-based undulators have been used in light sources for many decades with negligible radiation damage, however, and usually PM-based beamline magnets have the PMs set further away from the beam where the radiation field is naturally lower.

PMs have the disadvantage of providing fixed fields, compared to electromagnets which can be readily and rapidly adjusted. Using coils for adjustment is often impractical, since the PM necessarily forms part of the magnetic circuit yet presents a relative permeability of near unity. Hence, a coil has to be set to a relatively high current to provide even a low level of adjustment. Adjustment with coils of PM-based devices is typically limited to the order of \qty{10}{\percent} before the required Ampere-turns becomes impractical compared to a pure electromagnet. There has been significant R\&D over the past few years into highly-adjustable PM-based devices~\cite{ghaith_pms_instruments3020027}, including systems which make use of moving parts to provide a high level of tuneability. In some cases, these systems achieve a factor of 10-20 between minimum and maximum field, effectively resulting in a permanent magnet that can be switched on or off~\cite{zepto_dls_ipac2022-thoysp1}.

\item SRF Coatings

The raw materials required for superconducting components in particle accelerators represent a significant impact in cost and sustainability. The use of SRF coatings could significantly reduce the requirements associated with sourcing materials such as bulk niobium. Further benefits could be simplified manufacturing or improved cooling.

\item Luminosity optimisation through lower emittance, crab cavity polarisation

Ultimately, the collider's overall design sets the beam power requirement to achieve luminosity and determine the run time of these facilities. Reduction in the phase space of the colliding particles or photons can significantly reduce power requirements or run times. Collider studies should holistically optimise performance for reducing power and infrastructure requirements, and advanced beam dynamics studies should also consider the ability to produce, accelerate and collide the lowest emittance beams possible to maximise the sustainability of the overall run plan.

\end{itemize}

\subsection{Nature-based Interventions for Carbon Removal} 

In the context of the planning towards the International Linear Collider (ILC), discussed to be located in the Kitakami region in northern Japan, the `Green ILC' concept has been developed, not least in order to comply with the Japanese policy of strict carbon neutrality by 2050. 
They comprise the use of green electricity and the integration of the research campus into local grids, the use of carbon-neutral construction elements (concrete, steel), waste heat usage (see section~\ref{sec:heatrecovery}), and nature-based interventions into the carbon usage cycle. 

`Nature-based interventions' -- also termed `Nature-based climate solutions' (NbCS) -- are ``conservation, restoration and improved management strategies (pathways) in natural and working ecosystems with the primary motivation to mitigate GHG emissions and remove \ce{CO2} from the atmosphere''~\cite{Buma:2024}. 

Nature-based climate solutions can address climate change in different ways: i) by decreasing GHG emissions related to deforestation and land use; ii) by capturing and storing \ce{CO2} from the atmosphere; iii) by enhancing the resilience of ecosystems ~\cite{Miles:2021}. 

The promise of NbCS has triggered massive interest in using them for national or regional GHG mitigation plans. For example, 104 of the 168 Paris Accord signatories included NbCS into their mitigation plans~\cite{Seddon:2020}. 
NbCS are by now considered integral for effectively countering climate change, with the highest short-term impact expected from fast maximum reductions of emissions, followed by \ce{CO2} removal actions.  

Recent studies (e.g. Ref.~\cite{Buma:2024, Miles:2021}) 
have in great detail investigated numerous pathways for their expected impact in terms of GHG emissions avoided or \ce{CO2} sequestered, and for the confidence in their feasibility or scientific basis. Results show, for example, that avoided forest lost and reforestation (for both tropical and temperate forests) rank among the highest-impact measures with the lowest uncertainties in terms of scientific confidence. 

In the Green ILC concept, the most prominent NbCS foreseen are `green carbon' and `blue carbon', or land-based and water-based carbon sinks, respectively. 

For green carbon, the assessment of \ce{CO2} absorptions by forests is calculated taking into account species, age, and management status of a given forest (given that the absorption potential comes from an increase in biomass).  A detailed analysis of the Ichinoseki city forest (Ichinoseki being the most likely place for the ILC campus) was performed in the Green ILC context: The total forest area within the city's boundaries is over \qty{66}{\hectare}, of which roughly \qty{31.5}{\hectare} are planted (not naturally grown) forests. On average, the annual \ce{CO2} absorption for this forest is of the order of \qty{4.6}{\tonne\per\hectare} (with a total of just above \qty{300}{\kilo\tonne\ce{CO2}e} removed per year). Here, the planted forest part performs better -- with about \qty{5.2}{\tonne\per\hectare}, and the potential (using e g. cedars) is even higher, up to close to \qty{9}{\tonne\per\hectare}. 

On the other hand, the annual \ce{CO2} emission of Ichinoseki city (with a population of roughly \num{110000} and some industry, e.g. concrete plant, manufacturing industries) is of the order of \qty{800}{\kilo\tonne\ce{CO2}e} -- so this particular city is over-emitting. However, several municipalities in the more northern part of the Iwate prefecture are already today carbon neutral or even absorb more than they emit. 

Blue carbon (the carbon captured by the world's oceans and coastal ecosystems) is today considered less effective and promising than green carbon (except mangrove restoration~\cite{tjendraputri:2021}) 
-- despite the fact that, according to estimates,  \qty{83}{\percent} of global carbon is circulated in oceans and coastal habitats~\cite{IUCN-BlueCarbon:2017}. 
Studies find that  a limited understanding of the basic underlying processes is hindering large-scale projects for marine CO2 removal~\cite{Boyd2024}. The Green ILC concept intends to improve and extend on existing practices. One example is the study performed in the city of Hirono, taking advantage of the unique topography of the city's marine terraces. On these terraces, in shallow waters, tidal pools are created that spread out like shelves to promote the exchange of fresh seawater using the tides, thereby nurturing algae. Most of the algae are, at some point, washed to the seafloor, where they are immobilised, permanently binding the incorporated \ce{CO2}. Various other blue carbon methods are being investigated in Japan and around the world [e.g. Ref.~\cite{NOAA-BlueCarbon:2024}]. 

In order to put the Japanese efforts described above into perspective, section~\ref{annex:EUgreenDeal} briefly summarises the approach implemented for the European Union (EU-27), with a side remark to the view taken by the United Nations.

\section{Conclusions}

Sustainability assessment of accelerator-based research infrastructures requires a long-term vision across all aspects of the project.
Early evaluations serve to identify challenges and opportunities associated with different design options. 
As the project progresses, assessments become more detailed, building on earlier studies.
The primary goal of these evaluations is to develop engineering and technical designs that address sustainability with priority and effectiveness.

This report presents a series of best practices adopted to reduce civil engineering footprint, power consumption, to redistribute heat to campus or local buildings, and integrate sustainability considerations into rules and procedures for the 
procurement of goods and services.

The study highlights three interconnected pillars of sustainability --environmental, economic, and social -- and emphasizes the need for long-term accountability. Environmental impacts arise from construction, operation, and decommissioning, with electricity consumption, civil engineering, and material production representing major contributors to greenhouse gas emissions. Life Cycle Assessments (LCA) and the use of Environmental Product Declarations (EPDs), aligned with international standards, are proposed as key tools for quantifying these impacts and informing both design decisions and policy compliance.

From a socio-economic perspective, accelerators generate substantial public value beyond scientific discovery. They act as innovation hubs, foster global collaboration, drive industrial and technological spillovers, and provide education and training that translate into long-term benefits. To capture these effects, comprehensive cost-benefit analyses are recommended, integrating both tangible and intangible contributions across the UN Sustainable Development Goals.

Mitigation and compensation measures are also essential. The report identifies pathways such as low-carbon construction practices, responsible procurement, advanced energy management systems, waste heat recovery, and investment in green technologies. Nature-based interventions and circular economy principles are proposed as complementary strategies to offset residual impacts.
The path forward requires international coordination, transparent reporting, and continuous monitoring. Establishing shared databases of materials and practices, adopting harmonized sustainability standards, and embedding sustainability criteria into all project phases will help align accelerator development with global climate and development goals.

Within the outlined reference framework, there is broad scope for effective collaboration among particle physics laboratories, focused on sharing results and coordinating efforts -- a strategy for the rational use of available resources.

In summary, sustainable accelerator infrastructures demand an integrated strategy: rigorous assessment of environmental costs, proactive socio-economic engagement, and adoption of innovative mitigation technologies. This approach will ensure that future projects not only advance fundamental science but also serve as models of responsible, forward-looking research infrastructures.

\clearpage

\appendix

\section{Snowmass process and P5 Report}
\label{annex:snowmass21}

In 2021-2023, numerous future HEP collider proposals were discussed during the US high-energy physics community strategic planning exercise, called Snowmass'21 \cite{Snowmass21}. 
 Snowmass is a U.S. particle physics community study conducted every 7-9 years, with the previous one in 2013. Organized by the American Physical Society’s divisions, it defines key scientific questions and identifies opportunities for the future of particle physics in the U.S. and globally. The \textit{Particle Physics Project Prioritization Panel (P5)}, chaired by Hitoshi Murayama (UC Berkeley), has used Snowmass input to develop a strategic 10-year plan, aligned with a 20-year global vision, by December 2023.

The \textit{Snowmass Community Summer Study Workshop} took place in Seattle from July 17-26, 2022, concluding multiple workshops and Town Hall meetings held in 2020-2022.
The workshop was divided into 10 Frontiers, including the Energy, Neutrino, and Accelerator Frontiers, and saw participation from 1,400 attendees (750 in-person). It showcased contributions from over 100 accelerator experts and featured numerous letters of interest and white papers. Discussions covered future accelerators, beam physics, advanced concepts, and key technology areas like RF cavities, magnets, and targets.

{\it Accelerators for Particle Physics}:
high-energy accelerators have driven particle and nuclear physics research for decades, continually pushing the boundaries of energy, performance, and efficiency. Recent developments include:
\begin{itemize}[leftmargin=1.5cm,nosep]
    \item \textit{PIP-II linac} and construction of \textit{LBNF/DUNE} for neutrino research.
    \item New collider proposals, including \textit{FCC-ee/CEPC} for Higgs/EW studies, and renewed interest in \textbf{muon colliders} through the \textit{International Muon Collider Collaboration} (IMCC).
    \item Shifts in focus from TeV-scale $e^+e^-$ colliders to \textbf{multi-TeV discovery machines}, signaling a shift in R\&D priorities.
\end{itemize}

{\it Future Collider Concepts}:
\begin{itemize}[leftmargin=1.5cm,nosep]
    \item A \textbf{multi-MW upgrade} of Fermilab’s accelerator complex for neutrino research is a top priority.
    \item Facilities for \textbf{axion and dark matter} searches, including the proposed \textit{PIP-II accumulator ring} (PAR), are under discussion.
    \item Growing support for \textit{FCC-ee} at CERN, alongside linear collider proposals like \textit{C³} and \textit{HELEN}.
    \item \textbf{10+ TeV muon colliders} are gaining traction, with conceptual designs expected by the end of the decade.
\end{itemize}

{\it Addressing Key Challenges and R\&D Gaps}: while the U.S. accelerator R\&D portfolio lacks collider-specific programs, the community has proposed a \textbf{national R\&D initiative} to enable U.S. participation in global efforts (e.g., FCC, IMCC) and prepare for the next P5 planning cycle (2029-2030). Focus areas include:
\begin{itemize}[leftmargin=1.5cm,nosep]
    \item \textbf{High-power targets} to support multi-MW beams for colliders and neutrino facilities.
    \item \textbf{Energy-efficient RF structures} achieving high gradients for compact linear colliders.
    \item \textbf{High-field magnets} and \textbf{advanced wakefield concepts} for next-generation energy frontier colliders.
\end{itemize}

{\it Enhancing Education and Workforce Development}: to maintain competitiveness, the community recommends:
\begin{itemize}[leftmargin=1.5cm,nosep]
    \item \textbf{Strengthening education and training programs} to attract talent, with a focus on women and underrepresented minorities.
    \item Expanding U.S. \textbf{beam test facilities} to remain competitive internationally.
    \item Enhancing \textbf{collaboration with global accelerator projects} through strategic R\&D alignment and talent recruitment initiatives.
\end{itemize}

The Snowmass'21 {\it Accelerator Frontier} has established the  Implementation Task Force (ITF) to evaluate proposed future accelerator projects for performance, technology readiness, schedule, cost, and environmental impact. Corresponding metrics have been developed for uniform comparison of proposals ranging from Higgs/EW factories to multi-TeV lepton, hadron, and $ep$ collider facilities; from those based on traditional and to advanced acceleration technologies. Ref. \cite{ITF} describes the metrics and approaches, and presents the comparative evaluations of future colliders performed by the ITF. Table \ref{tab:ITFHiggs} gives a high level summary of the ITF findings regarding the Higgs factory proposals and some 10+ pCM TeV future collider concepts: besides the c.m. energy, luminosity and facility power consumption (presented by the proponents), the table lists anticipated years of pre-project R\&D (an indicator of the readiness for construction), years to the first physics (that includes the R\&D time, pre-construction and construction - in the technically limited schedules which start at the time of the decision to proceed), and the cost range (understood as a total project cost - that includes explicit labor, etc - without contingency and the inflation escalation, in 2021 BUSD). Note, that the ITF has uniformly used several models to estimate the cost (e.g., 5- and 31-parameters) as well as known costs of existing installations and reasonably expected cost of novel equipment.  For future technologies, the cost estimates were quite conservative, and one should expect cost reductions from pre-project R\&D.

The ITF report has provided a critically important assessment of future colliders. Firstly, it has shown that there is no clear winner in all the categories among the Higgs factory proposals. Indeed, in terms of the luminosity, the leading concepts are those based on the ERL technology (CERC, ReLiC, ERLC). 
The best prepared in terms of technical readiness (years of the pre-project R\&D) are FCCee, CLIC, CEPC, and ILC (as noted above, the last two do have TDRs). ILC is the absolute leader in the category "Time to the 1st Physics" (estimated to be less than 12 years), followed by FCCee, CEPC, CLIC, and C3 
(12-18). The electric power consumption is the lowest for CERC and XCC, the next cohort includes linear $e^+e^-$ colliders ILC, CLIC, and C3. Last but not least, the lowest estimated costs are for the $\gamma \gamma$ collider XCC and $\mu \mu$ collider Higgs factory (note, that both will produce Higgs particles through $s$-channel that requires about half of the primary beams energy), while all the linear $e^+e^-$  colliders ILC, CLIC, and C3 are placed in the next cost category. 

As Table \ref{tab:ITFHiggs} indicates, all the very high energy collider proposals with 10 pCM (parton center of mass energy) TeV or above are generally more expensive than the Higgs factories, more power hungry, require significant investment in R\&D and longer time to construct. Considered were $pp$ colliders such as FCChh, SPPC and the Fermilab site-filler, advanced colliders based on the wakefield acceleration in plasma \cite{leemans:ipac2024-frxn2} and in very high frequency RF structures, and the muon colliders (at CERN \cite{Skoufaris:2885893} and at Fermilab \cite{stratakis:ipac2024-wecd3}). The latter seem to be most appealing in all the categories of the ITF comparative analysis.    

\begin{table*}[!ht]
   \begin{center}
   \caption{The {\it Snowmass'21} Implementation Task Force summary: main parameters of the Higgs factory proposals (FCCee, ILC, CLIC, $C^3$, HELEN, three ERL-based colliders, $\gamma \gamma$ and $\mu \mu$ Higgs factories), and several 10+ pCM TeV colliders (Muon Collider options, advanced wakefiled collider options, FCChh, and SPPC). Years of the pre-project R\&D indicate required effort to get to sufficient technical readiness. Estimated years to first physics are for technically limited timeline starting at the time of the decision to proceed. The total project cost range is for the single listed energy in 2021\$ (based on a parametric estimator and without escalation and contingency). All colliders listed above were assumed to be stand-alone projects, since ITF could not
assume or decide on a sequence of projects. The peak luminosity and power consumption values have not been reviewed by ITF and represent proponent inputs. (Adapted from \cite{ITF}.)} 
\label{tab:ITFHiggs}
\resizebox{1.0\textwidth}{!}{
\begin{tabular}{| l | c | c | c | c | c | c |}
\hline
\hline
 Proposal Name & CM energy & Lum./IP & Years of & Years to  & Construction & Est. operating \\ 
 & nom. (range) & @ nom. CME & pre-project & first & cost range & electric power \\  
  & [TeV] & [10$^{34}$ cm$^{-2}$s$^{-1}$] & R\&D & physics & [2021 B\$] & [MW] \\  
\hline
FCC-ee & 0.24 & 7.7 (28.9) & 0-2 & 13-18 & 12-18 & 290 \\  
 & (0.09-0.37) & & & & & \\
\hline 
CEPC & 0.24 & 8.3 (16.6) & 0-2 & 13-18 & 12-18 & 340 \\  
 & (0.09-0.37) & & & & & \\
\hline 
ILC - Higgs & 0.25 & 2.7  & 0-2 & <12 & 7-12 & 140 \\  
factory & (0.09-1) & & & & & \\
\hline 
 CLIC - Higgs & 0.38 & 2.3  & 0-2 & 13-18 & 7-12 & 110 \\  
factory & (0.09-1) & & & & & \\
 \hline 
 CCC (Cool & 0.25 & 1.3  & 3-5 & 13-18 & 7-12 & 150 \\  
Copper Collider) & (0.25-0.55) & & & & & \\
\hline 
CERC (Circular& 0.24 & 78  & 5-10 & 19-24 & 12-30 & 90 \\  
ERL Collider) & (0.09-0.6) & & & & & \\
 \hline 
ReLiC (Recycling & 0.24 & 165 (330)  & 5-10 & >25 & 7-18 & 315 \\  
Linear Collider) & (0.25-1) & & & & & \\
 \hline 
ERLC (ERL & 0.24 & 90  & 5-10 & >25 & 12-18 & 250 \\  
linear collider) & (0.25-0.5) & & & & & \\
 \hline 
XCC (FEL-based & 0.125 & 0.1  & 5-10 & 19-24 & 4-7 & 90 \\  
$\gamma \gamma$ collider) & (0.125-0.14) & & & & & \\
 \hline 
Muon Collider & 0.13 & 0.01  & >10 & 19-24 & 4-7 & 200 \\  
Higgs Factory & & & & & & \\
\hline
\hline 
LHeC & 1.2 & 1 & 0-2 ? & 13-18 & <4 & $\sim$140 \\  
 &  & & & & & \\
\hline
\hline
 Muon Collider & 10 & 20 (40)  & >10 & 19-24 & 12-18 & $\sim$300 \\  
at FNAL & (6-10) & & & & & \\
 \hline 
Muon Collider & 10 & 20 (40)  & >10 & >25 & 12-18 & $\sim$300 \\  
 & (1.5-14) & & & & & \\
 \hline 
LWFA - LC & 15 & 50  & >10 & >25 & 18-80 & $\sim$1030 \\  
(Laser-driven) & (1-15) & & & & & \\
 \hline 
PWFA - LC & 15 & 50  & >10 & >25 & 18-50 & $\sim$620  \\  
(Beam-driven) & (1-15) & & & & & \\
 \hline 
Structure WFA & 15 & 50  & >10 & >25 & 18-50 & $\sim$450  \\  
(Beam-driven) & (1-15) & & & & & \\
 \hline 
$pp$ Collider & 24 & 3.5 (7.0)  & >10 & >25 & 18-30 & $\sim$400 \\ 
at FNAL &  & & & & & \\
\hline
FCC-hh & 100 & 30 (60) & >10 & >25 & 30-50 & $\sim$560 \\  
 &  & & & & & \\
\hline 
SPPC& 125 & 13 (26)  & >10 & >25 & 30-80 & $\sim$400 \\  
 & (75-125) & & & & & \\
\hline 
\end{tabular}
}
\end{center}
\end{table*}

{\it 2023 P5 (Particle Physics Projects Prioritization Panel) Report}: 
the P5 panel has generally accepted main recommendations of the Snowmass'21 report. 
In particular, the 2023 P5 report \cite{P5} contains two recommendations specific to future colliders: \\
"{\bf Recommendation 2c}: an off-shore Higgs factory, realized in collaboration with international partners, in order to reveal the secrets of the Higgs boson. The current designs of FCC-ee and ILC meet our scientific requirements. The US should actively engage in feasibility and design studies [...]" ; and \\
"{\bf Recommendation 4a}: support vigorous R\&D toward a cost-effective 10 TeV pCM collider based on proton, muon, or possible wakefield technologies, including an evaluation of options for US siting of such a machine, with a goal of being ready to build major test facilities and demonstrator facilities within the next 10 years [...]" 

Also, there is a dedicated Chapter  {\it "6.9 Sustainability and the Environment"}: 
"...Commitment to sustainability is a high priority for particle physics activities. This includes energy and carbon management, energy efficiency and savings, and environmental impact. It concerns present and future accelerators as well as testing and computing facilities.
In the study of future accelerator projects, it is important to establish and launch at an early stage a full lifecycle sustainability effort. Because future accelerators, conceived for higher beam energies and intensities, will have higher energy demands, the promotion of energy efficient accelerator concepts and identification and development of energy-saving accelerator technology are critical. These considerations will affect the affordability of new accelerators and demonstrate the responsible role of the HEP community in society.
Accelerator technologies play a key role in sustainability. Investments in high field magnets by the DOE Magnet Development Program and NSF’s MagLab have advanced the state of the art in superconductors and magnet design to the benefit of particle physics, but also materials science, fusion energy research, and commercial development. In this context, high temperature superconducting materials, operated at temperatures higher than superfluid or liquid helium, can be significant in reducing energy consumption—for example, for future colliders including muons or FCC-hh. Accelerator structure improvements could also be significant, including higher quality factor cavities, and concepts like the Cool Copper Collider.

Other new technologies such as muon and possibly wakefield colliders have potential to improve power required for a given luminosity and reduce the size and environmental impact of future facilities. Innovation in electric power generation, management, and distribution also contribute to sustainable development and will be encouraged.
Upgrade, construction, and operation of accelerator complex 
and test facilities are part of the global effort to advance particle physics. Defining sustainable requirements on industrially procured technology, including construction, electrical, and cooling equipment, should be included in the project development.

To contribute to the global decarbonization effort, future projects will aim to reduce the CO2 footprint of their civil engineering components. Research and adoption of sustainable and eco-friendly materials will be encouraged, and this applies to gases with high global warming impact used in some types of detectors. Study of alternative solutions will be investigated and implemented, in addition to the adoption of stringent recirculation requirements when those gases are used. Reuse of materials should be encouraged to foster sustainable development and to limit the use of natural resources and raw materials, including critical minerals that are subject to potential shortages.

The international nature of particle physics activities calls for extensive travel to participate in meetings and conferences and to carry out experiments. The dramatic increase in the use of remote conferencing mandated by the COVID-19 pandemic spurred an evolution of practices and technologies that allow researchers to increase participation and inclusivity. With this new paradigm in place, laboratories should facilitate hybrid meetings and invest in upgrades to video conferencing equipment as necessary. Meeting organizers should establish protocols that ensure people connecting from remote locations can participate fully, including in discussions. At the same time, in-person meetings remain strongly valuable and should not be discouraged; they facilitate open and effective communication and build trust in international partnership. Assessing and implementing sustainability strategies require R\&D and investment of appropriate resources. The field would benefit from development of consistent metrics for sustainability of research, construction, and operations.

{\bf Area Recommendation 20}: charge HEPAP (the High Energy Physics Advisory Panel), potentially in collaboration with international partners, to conduct a dedicated study aiming at developing a sustainability strategy for particle physics.

\clearpage

\section{Sustainability researches for CEPC}
\label{annex:CEPC}
The CEPC was proposed in 2012, following the landmark discovery of the Higgs boson at the LHC. The CEPC is designed to serve as an advanced Higgs factory, operating at a beam energy of 120 GeV. Additionally, it has the capability to function across a wide energy range, from 45 GeV for Z-pole studies, 80 GeV for W boson research, and up to 180 GeV for ttbar production. The collider circumference has been optimized to 100 km, enabling high-luminosity collisions at all these energy levels. For over a decade, the CEPC research team has dedicated its efforts to physical design and key technology studies. A significant milestone was reached at the end of 2023 with the official release of the TDR, confirming that the CEPC is fully prepared for implementation.
	The CEPC TDR provides comprehensive parameters, including the electricity power requirements during operation. As shown in Table \ref{table:CePC_MW}, the high-performance Higgs mode operation, with a synchrotron radiation (SR) power load of 50 MW, requires a total electricity power of approximately 340 MW. 
The highest demand of electricity power in the system comes from the RF power source, followed by the magnet and cryogenics systems. The power needed by utilities such as cooling water, air conditioning, ventilation is also significant.
\begin{table}[htbp]
\centering
\caption{CePC total electricity power demand in Higgs mode (50 MW/beam)}
\begin{tabular}{
 | l |
  S[table-format=3.2] |
  S[table-format=2.2] |
  S[table-format=2.2] |
  S[table-format=1.2] |
  S[table-format=1.2] |
  S[table-format=2.2] |
  S[table-format=3.2] |
}
\toprule
  &  \multicolumn{6}{|c|}{  \textbf{Location and Power requirement (MW)} } &   \\
\hline
  \textbf{Systems} & \textbf{Collider} & \textbf{Booster} & \textbf{Linac} & \textbf{BTL} & \textbf{IR} & \textbf{Surface  } & \textbf{ Total} \\
   &  &  & &  &  & \textbf{Buildings  } & \textbf{(MW)} \\
\midrule
 \textbf{RF Power Source} & 161.60 & 1.73 & 14.10 & & & & 177.43 \\
\hline
  \textbf{Cryogenic System} & 9.17 & 1.77 &  &  & 0.16 &  & 11.10\\
\hline
  \textbf{Vacuum System} & 5.40 & 4.20 & 0.60 &  &  &  & 10.20\\
\hline
 \textbf{Magnet System} & 42.16 & 8.46 & 2.15 & 4.89 & 0.30 &  & 57.96\\
\hline
 \textbf{Instrumentation} & 1.30 & 0.70 & 0.20 &  &  &  & 2.20\\
\hline
 \textbf{Radiation Protection} & 0.30 &  & 0.10 &  &  &  & 0.40\\
\hline
 \textbf{Control System} & 1.00 & 0.60 & 0.20 &  &  &  & 1.80\\
\hline
 \textbf{Experimental Devices} &  &  &  &  & 4.00  &  & 4.00\\
\hline
 \textbf{Utilities} & 46.40 & 3.80 & 2.50 & 0.60 & 1.20 &  & 54.50\\
\hline
\textbf{General Services} & 7.20 & 0.30  &  & 0.20 & 0.20 & 12.00  & 19.90\\
\hline
\textbf{Total} & 274.53 & 21.26 & 20.15  & 5.69 & 5.86 & 12.00 & 339.49\\
\bottomrule
\end{tabular}
\label{table:CePC_MW}
\end{table}
The CEPC represents a substantial load on the power grid. Consequently, conducting systematic studies to reduce power consumption and ensure sustainable operation of the CEPC is a crucial topic. 
	In recent years, the CEPC host institute, the Institute of High Energy Physics (IHEP), has dedicated extensive efforts to sustainability studies for accelerators. These efforts have focused on measures such as utilizing clean energy, implementing energy and resource recovery systems, and improving the efficiency of key technologies.
\subsection{Implementing solar panels at HEPS}
\begin{figure}[ht]
    \centering
    \includegraphics[width=0.7\textwidth]{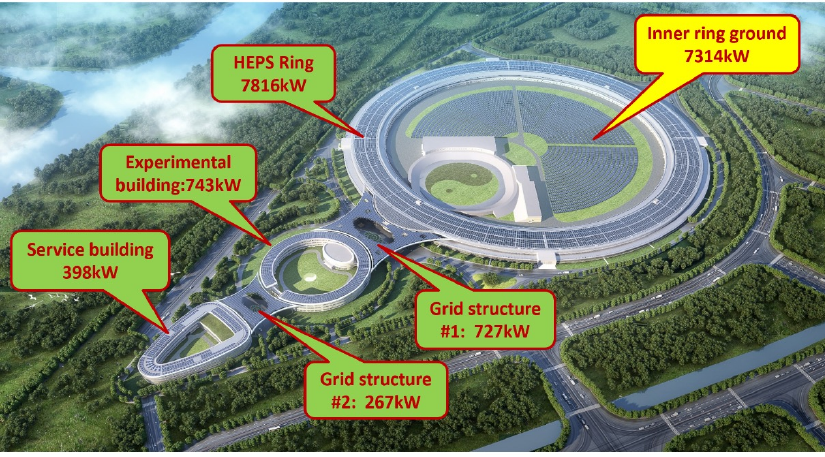}
    \caption{Solar panels at HEPS.}
    \label{fig:HEPS_solarpanels}
\end{figure}

The High Energy Photon Source (HEPS) is a fourth-generation photon source constructed by IHEP. It features a beam energy of 6 GeV and a storage ring with a circumference of 1.36 km. As a prototype for implementing solar energy and gaining experience in its efficient use, a significant number of solar panels (Fig.\ref{fig:HEPS_solarpanels}) have been installed at HEPS. 
The HEPS campus covers a total area of 153,000 m², providing an advantageous environment for solar power implementation. The roofs of various buildings at HEPS are suitable for solar panel placement, with 10 MW of solar panels already installed. Additionally, space inside the HEPS storage ring has been reserved for the future implementation of 7.3 MW of solar panels. 
The total power needed for the entire HEPS campus, including services, is approximately 38.7 MW. The facility is equipped with four 10 kV power supply stations and eleven 10/0.4 kV transformer stations. The total transformer capacity is 64 MVA. The primary target for solar energy utilization is the auxiliary facilities whose requirement on the power stability is less strict than the accelerator and beamline stations. The solar electricity connects the grid at a voltage level of 10 kV, with a total capacity of 10.1 MW in the power distribution center, where systems of transformer, AI controller, PV inverter are implemented. The theoretical power generation capacity is 11.24 GWh in the first year, with an integrated power generation of 257.69 GWh over 25 years, resulting in an average annual power generation of 10.31 GWh.
\subsection{Heat recovery at HEPS}
HEPS consumes an enormous amount of energy, which ultimately converts to heat. Additional power is consumed for energy conversion such as the cooling system. Moreover, both the accelerator and beamlines require temperature-stable environments for high-precision operation, leading to additional power consumption for air conditioning. HEPS is located in Beijing, where heating systems are required for all buildings during winter. Therefore, heat recovery is highly beneficial for HEPS.

\begin{figure}[ht]
    \centering
    \includegraphics[width=0.5\textwidth]{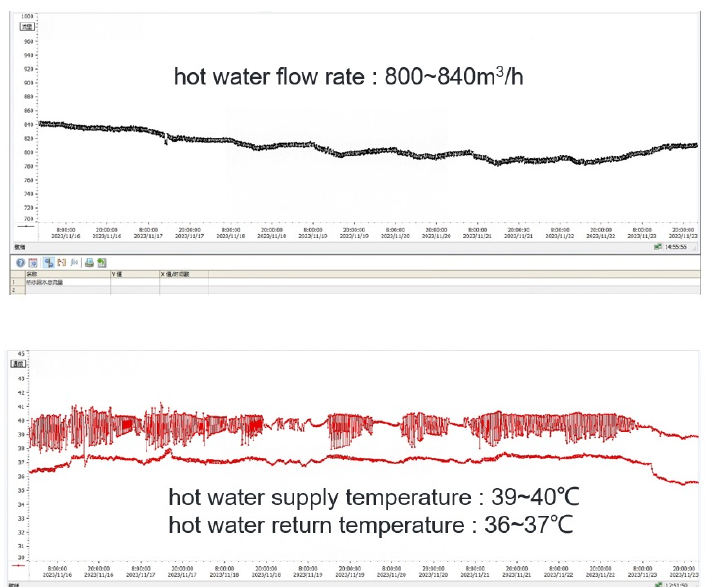}
    \caption{Key parameters of heat recovery at HEPS in the 3rd week of November 2023.}
    \label{fig:HEPS_hotwater}
\end{figure}

HEPS requires 18 MW for air conditioning and 4 MW for heating in summer, compared to 7 MW and 10 MW respectively in winter. Three sets of heat pumps are used to recover energy from the cooling water in the form of 40-42°C water. Consequently, in the winter operation scenario, 10 MW of heat power can be recovered, replacing the equivalent power from the municipal heating system. During spring and autumn, the transitional period, about 4 MW of energy is recovered from the cooling water. The entire power load for the heating system in summer is provided by the cooling water.

The system began its tentative operation in November 2023, with only the linac and booster operating. The key parameters from the 3rd week in November are depicted in Fig.\ref{fig:HEPS_hotwater}. The water flow rate varied between 800 and 840 m³/h, with the outgoing water temperature at 39-40°C and the return water temperature at 36-37°C, showing a 3°C temperature drop.
\subsection{High-efficiency klystrons}
\begin{figure}[ht]
    \centering
    \includegraphics[width=0.7\textwidth]{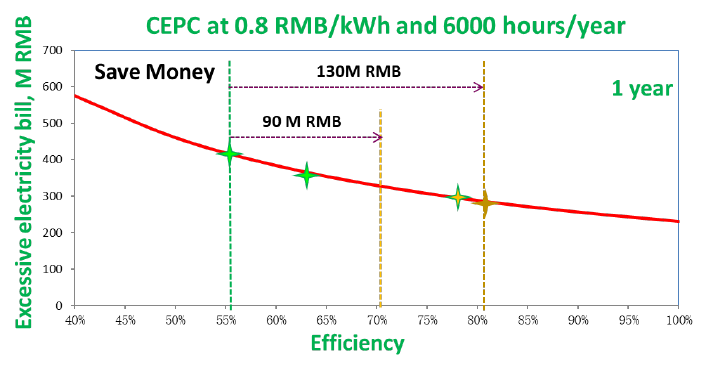}
    \caption{The savings for the CEPC operation at different klystrons efficiencies.}
    \label{fig:CePC_klyseff}
\end{figure}

The CEPC collider employs 650MHz klystrons as the RF source. The efficiency of these klystrons is crucial for the sustainable performance of the RF system, which is the major energy consumer. Therefore, the energy conversion efficiency of the klystrons is of utmost importance. Assuming an aggressive 6000 hours of operation per year and an estimated electricity cost of 0.8 RMB per kWh, improving the efficiency from the conventional 55\% to 70\% would result in annual savings of 90 million RMB, corresponding to 112 GWh. Further improving the efficiency to 80\% could save 130 million RMB and 162 GWh per year (Fig.\ref{fig:CePC_klyseff}). Consequently, CEPC aims to develop high-efficiency (80\%) klystrons for its application.
\begin{table}[h]
\centering
\caption{CePC klystrons design parameters}
\begin{tabular}{
 | l |
  c |
  c |
  c |
}
\toprule
 \textbf{Type} & \textbf{1th prototype } & \textbf{2nd prototype }  & \textbf{3rd prototype }  \\
 & & \textbf{(high-efficiency) } & \textbf{(MBK) } \\
 \hline
  \textbf{Output Power (kW)} & 800 & 800 & 800 \\
\hline  
  \textbf{Beam Voltage (kV)} & 81.5 & 110 & 52 \\
\hline  
  \textbf{Beam Current (A)} & 15.6 & 9.1 & 20.08{(2.51$\times $8)} \\
\hline  
  \textbf{Efficiency (\%)} & 65 & 78 & 80.5 \\
\hline  
\bottomrule
\end{tabular}
\label{table:CePC_klystrons}
\end{table}
To achieve this goal, three sets of klystrons are under development. The first klystron targets an efficiency of 65\%, followed by a second klystron aiming for 78\% efficiency. The third klystron explores Multi-Beam Klystron (MBK) technology with an objective of 80\% efficiency. The design parameters for these klystrons are listed in Table \ref{table:CePC_klystrons}.
Among these three prototypes, the second klystron plays the most significant role. It utilizes high voltage to enhance efficiency. The design optimizes the second and third harmonic of the cavities to achieve high efficiency.

\begin{figure}[ht]
    \includegraphics[width=0.45\textwidth]{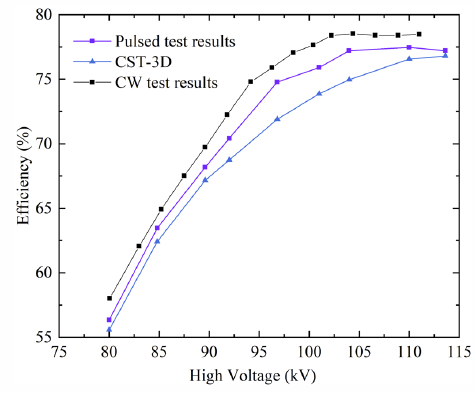}
    \hspace{4mm}
    \includegraphics[height = 0.25\textheight]{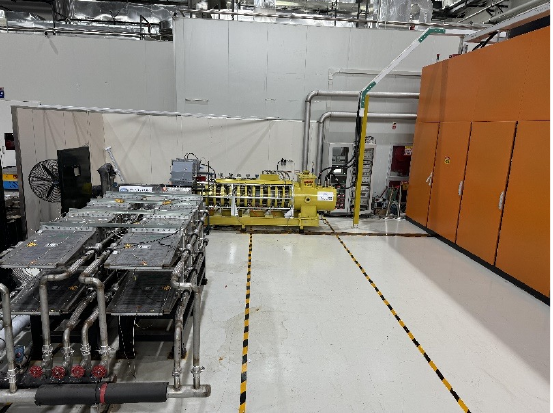}
    \caption{On the left: efficiency of the second klystron with respect to the high voltage. The dark curve indicates the test results, comparing to the blue curves of the simulations. On the right: a photo of the second klystron with its PSM power supply.}
    \label{fig:CePC_Klys2}
\end{figure}

Giving continuous efforts in testing and modification, the expected results were achieved in the summer of 2024. The efficiency slightly exceeds 78\% in the Continuous Wave (CW) operation mode at a power output of 800 kW. High voltage plays a crucial role, with the efficiency reaching saturation at approximately 105 kV. The test results and simulations are shown in fig.\ref{fig:CePC_Klys2}, left. Fig.\ref{fig:CePC_Klys2}, right shows a photo of the second klystron together with its PSM (Pulse Step Modulation) power supply. 

\subsection{High-Q superconducting RF cavities and modules}

Due to the high RF power requirements for both the booster and the collider in the CEPC, superconducting RF cavities are employed. The booster and collider utilize 1.3 GHz and 650 MHz frequencies, respectively. During the R\&D phase of the CEPC superconducting RF cavities, the mid-temperature baking technology has proved to be an effective method for enhancing the Q factor (quality factor). Since the Q factor plays a crucial role in reducing power consumption in the cryogenic system used to generate 2K liquid helium, high Q SRF cavity will help to save the electricity. 

\begin{figure}[hb]
    \centering
    \hspace{-6mm}
    \includegraphics[width=0.45\textwidth]{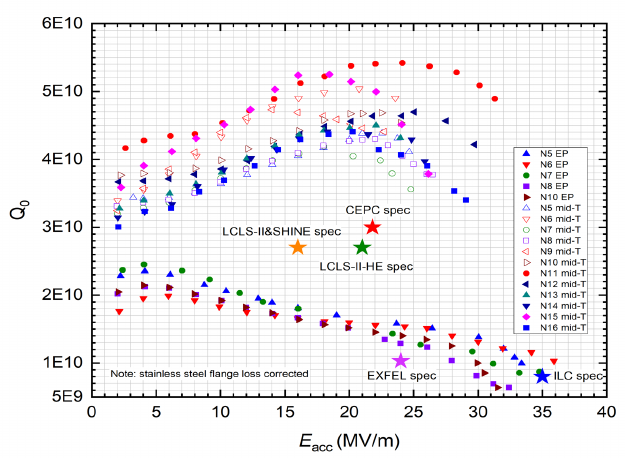}
    \hspace{2mm}
    \vspace{3mm}
    \includegraphics[height=0.20\textheight]{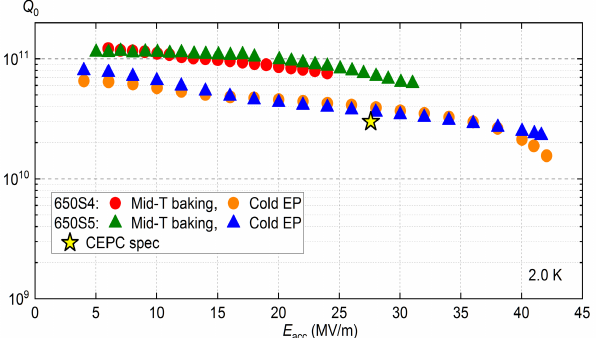}
    \caption{Q factor of the CEPC superconducting RF cavities versus acceleration gradient. On the left: the Q factor of the 9-cell 1.3 GHz cavity. On the right: the Q factor of the single cell 650MHz cavity.}
    \label{fig:CePC_Q0_Eacc}
\end{figure}

Figure \ref{fig:CePC_Q0_Eacc} illustrates the Q factor for both the 1.3 GHz (left) and 650 MHz (right) cavities. It is evident that the mid-temperature baked cavities significantly outperform the EP (Electro-Polished) cavities in terms of Q factor. Notably, all of the cavities exceed the requirements set for the CEPC.

Based on the R\&D results of the SRF cavities, the Q factor of mid-temperature baked cavities is expected to reach $3 \times 10^{10}$. This is three times higher than that of EP cavities. As a result, in the Higgs mode, the power load to the cryogenic AC power is significantly reduced from 28 MW to 17.5 MW, achieving a substantial 10 MW reduction. 

\begin{figure}[htb]
    \centering
    \includegraphics[width=0.7\textwidth]{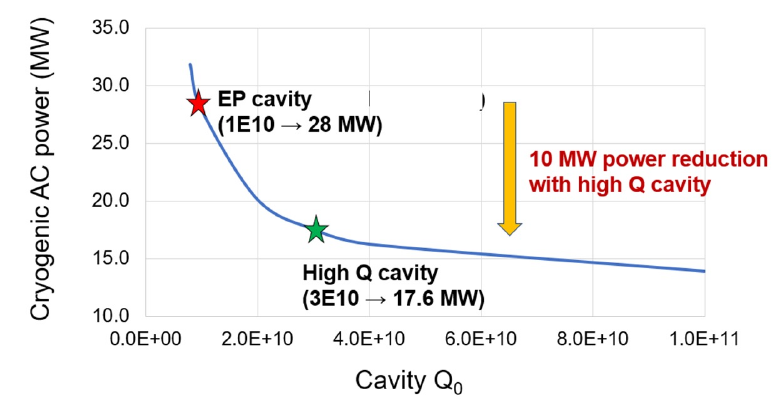}
    \caption{Cryogenic power reduction due to high Q SRF cavities.}
    \label{fig:CePC_CryoPower}
\end{figure}

Figure \ref{fig:CePC_CryoPower} compares the power reduction due to the improvement of the Q factor. Assuming 6000 hours operation per year, the high Q cavities could save 60 GWh electricity consumption. In the ttbar mode, which requires more cavities, the AC power reduction is even more significant. 

\subsection{Permanent Magnets at IHEP}

As magnets constitute the second-largest system in electricity consumption in the CEPC, it is crucial to implement measures to reduce power usage. In recent years, the application of permanent magnets in storage rings has gained popularity, especially in light sources. In China, HEPS has successfully pioneered the use of permanent magnets for all dipoles in the storage ring. Compared to electromagnetic dipoles, which require approximately 1.62 kW of excitation power, permanent magnet devices consume no energy. Moreover, electromagnetic devices need cooling, which consumes about 1.3 kW of power per unit. There are 240 dipoles in the HEPS storage ring, and consequently, 701 kW of electric power can be saved by using permanent dipoles. Typically, SR facilities operate for about 8,000 hours per year, resulting in a total annual energy savings of 5.6 GWh. This significant reduction in energy consumption not only contributes to cost-effectiveness but also aligns with sustainable practices in scientific research facilities.
Inspired by the application of permanent magnets at HEPS, researchers are currently investigating the potential implementation of permanent magnets at the CEPC. While the linac utilizes numerous quadrupoles, its energy consumption remains relatively low. The booster magnets, however, rapidly vary their field strength for beam energy ramping, making the use of permanent magnets unfeasible. In the collider magnets, the highest energy consumption is from the quadrupoles, where the field strength must be adjustable for different beam energies ranging from Z-pole to ttbar. Consequently, exploring the development of a field-switchable quadrupole holds the greatest significance.

To adjust the quadrupole field, a double magnet ring structure is designed, comprising outer and inner rings. Each ring consists of eight permanent magnet blocks arranged in a Halbach array, capable of independently generating a quadrupole field. Both rings are rotatable, and the quadrupole field strength varies with different rotation angles. When the fields generated by both rings align in the same direction, the field reaches its maximum strength. Conversely, when the fields are in opposite directions, they cancel each other out. As a result, this double-ring configuration provides an adjustable quadrupole field.
\begin{figure}[ht]
    \centering
    \includegraphics[width=0.45\textwidth]{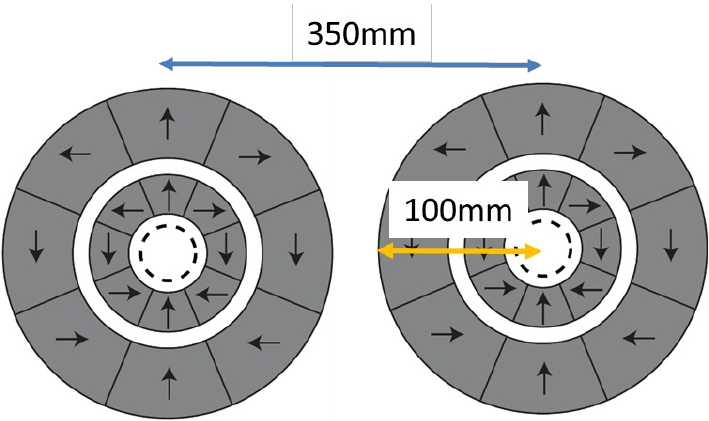}
    \caption{Schematic design of dual-aperture permanent quadrupoles  for the CEPC collider.}
    \label{fig:CePC_DualAperture_PQ}
\end{figure}
According to the CEPC requirements for quadrupoles, a dual-aperture quadrupole has been designed, as illustrated in Fig.\ref{fig:CePC_DualAperture_PQ}. The distance between the axes is 350 mm, and the outer ring radius is 100 mm. The Halbach array configuration provides excellent field shielding to its outer space, resulting in negligible field cross-talk between the two quadrupoles.
To demonstrate the feasibility of the permanent quadrupole solution for the CEPC collider, a single permanent quadrupole has been designed. 
\begin{table}[h]
\centering
\caption{Permanent quadrupole design parameters of the CePC collider}
\begin{tabular}{
 | l |
  c |
  c |
  c |
}
\toprule
 \textbf{Aperture} & $\varphi$ &  mm  & $\geq$ 66  \\
 \hline
  \textbf{Magnetic length} & $L_{eff}$ & mm & $\geq$ 200 \\
\hline  
  \textbf{Max gradient} & $G_{Max}$ & T/m & $\geq$ 10.6 \\
\hline  
  \textbf{Adjustment range} &  & T/m & 1.8 - 10.6 \\
\hline  
  \textbf{Adjustment precision} & $\Delta$ G & T/m & $\leq$ 0.002 \\
\hline  
  \textbf{Rotating precision} & $\Delta \theta $ & mrad & $\leq$ 0.08 \\
\hline  
  \textbf{Good Field region} & $R_{GFR}$ & mm & 12 \\
\hline 
  \textbf{Harmonics @$ G_{Max}$} & $B_{n}$/$B_{2} $ & $10^{-4} $ & $< $5 \\ 
\bottomrule
\end{tabular}
\label{table:CePC_Design}
\end{table}
The detailed parameters are listed in Table \ref{table:CePC_Design}. Numerical simulations conducted using OPERA and RADIA confirm that the field strength meets the CEPC requirements. Additionally, an associated engineering design for the motion system has been developed, along with a field error compensation strategy. Currently, the prototype is in the fabrication process. 

\subsection{Infrastructure development for Helium recovery}

Superconducting technologies, including superconducting magnets and Superconducting RF systems, are essential to the CEPC. The cryogenic system, which enables superconductivity, relies on liquid helium. However, helium is a non-renewable resource, and its price fluctuates significantly in the global market. To address this challenge, IHEP has implemented a comprehensive infrastructure for full helium recovery. This recovery system will be applied to the CEPC, ensuring sustainable operation with respect to helium usage.

\clearpage

\section{Research infrastructure project appraisal}
\label{annex:RI_appraisal}
\subsection{The European Green Deal}
\label{annex:EUgreenDeal}
\begin{figure*}[htbp]
\begin{center}
\includegraphics[width=0.9\textwidth, height=9cm]{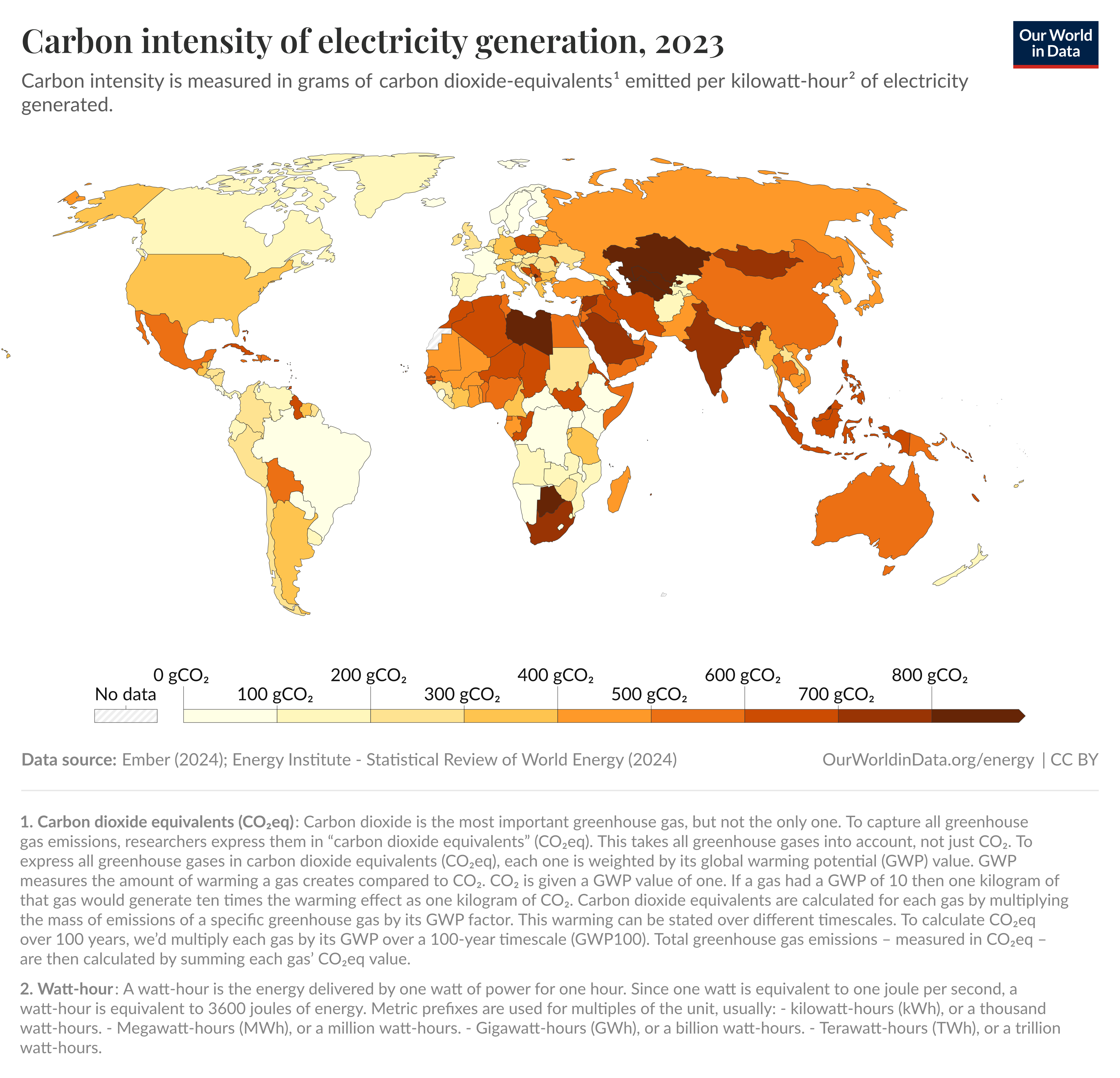}
\caption{Map of the world's electricity production \ce{CO2 }intensity~\cite{owid-electricity-mix}}. 
\label{fig:MapOfPowerSituation}
\end{center}
\end{figure*}

The map in \autoref{fig:MapOfPowerSituation} shows the \ce{CO2} intensity of electricity production across the different countries and world regions~\cite{owid-electricity-mix}. A huge spread is visible, which has its cause in the countries' level of development, natural circumstances, traditions, and policy decisions. Similarly, the strategies --- if any --- and technical solutions for fighting climate change and towards achieving carbon neutrality  can also differ significantly. 

European efforts are based on the "European Green Deal" -- founded in 2020 -- that aims at overcoming the challenges of the climate crisis by transforming the EU into a modern, resource-efficient and competitive economy~\cite{EUgreenDeal}, ensuring 
i) no net emissions of greenhouse gases by 2050; 
ii) economic growth decoupled from resource use; 
iii) no person and no place left behind.  
The goal is to be the first climate-neutral continent. In particular, the European Commission has adopted a set of proposals to make the EU's climate, energy, transport and taxation policies fit for reducing net greenhouse gas emissions by at least 55\% by 2030, compared to 1990 levels. The European Green Deal has many facets and components, and it touches upon the areas of the energy system, biodiversity and environmental protection, agriculture, transport, industry, research and innovation, and the finance sector.  The Green Deal addresses large structural topics as well as smallish or detailed elements, like the goal of 3 billion new trees to be planted in EU until 2030.

The European Green Deal can, in practice, seen as being embedded in the overarching 17 sustainable development goals (SDGs) put forward by the United Nations~\cite{UNSDG}.   

In 2021, only one year after the Green Deal, the European Climate law that ensures a climate neutral European Union by 2050, was passed~\cite{EUclimatelaw}. With the adoption of the final two proposals, the "Fit for 55 package" of measures to reduce net greenhouse gas emissions by at least 55\% by 2030, compared to 1990 levels is complete since October 2023. Since that date, the EU has legally binding climate targets covering all key sectors of the economy.

The European Climate Pact, as an accompanying element, is an initiative of the European Commission supporting the implementation of the European Green Deal~\cite{EUclimatePact}. 
It is a movement to build a greener Europe, providing a platform to work and learn together, develop solutions, and achieve real change. The Pact provides opportunities for people, communities, and organizations to participate in climate and environmental action across Europe. By pledging to the Pact, European stakeholders commit to taking concrete climate and environmental actions in a way that can be measured and/or followed up. Participating in the Pact is an opportunity for organizations to share their transition journey with their peers and collaborate with other actors towards common targets.

\begin{figure}[htbp]
\begin{center}
\includegraphics[width=0.6\linewidth]{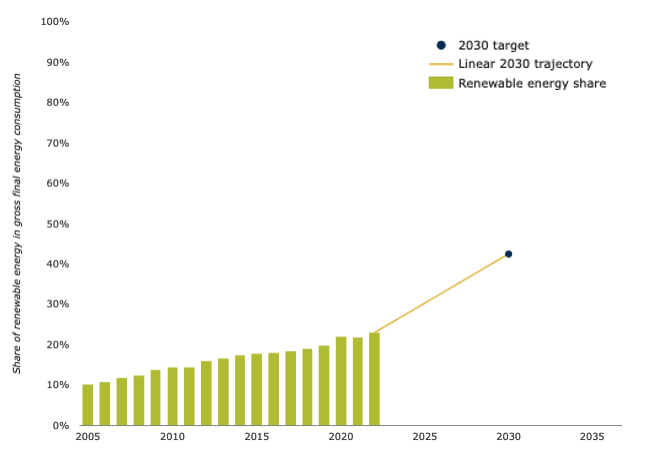}
\caption{Progress towards renewable energy source targets for EU-27~\cite{EEAReport24}.}
\label{fig:eugoals}
\end{center}
\end{figure}

Figure~\ref{fig:eugoals} shows one of the trends in the EU~\cite{EEAReport24}: the share of energy consumed in the EU during 2022 generated from renewable sources was 23\%. This increase, from a level of 21.9\% in 2021, was largely driven by a strong growth in solar power. The share is also amplified by a 2022 reduction in non-renewable energy consumption linked to high energy prices, however renewables in Europe are expected to keep growing. Meeting the new 
target of 42.5\% for 2030 will demand more than doubling the rates of renewables deployment seen over the past decade, and requires a deep transformation of the European energy system.

\begin{figure}[htbp]
\begin{center}
\includegraphics[width=0.7\linewidth]{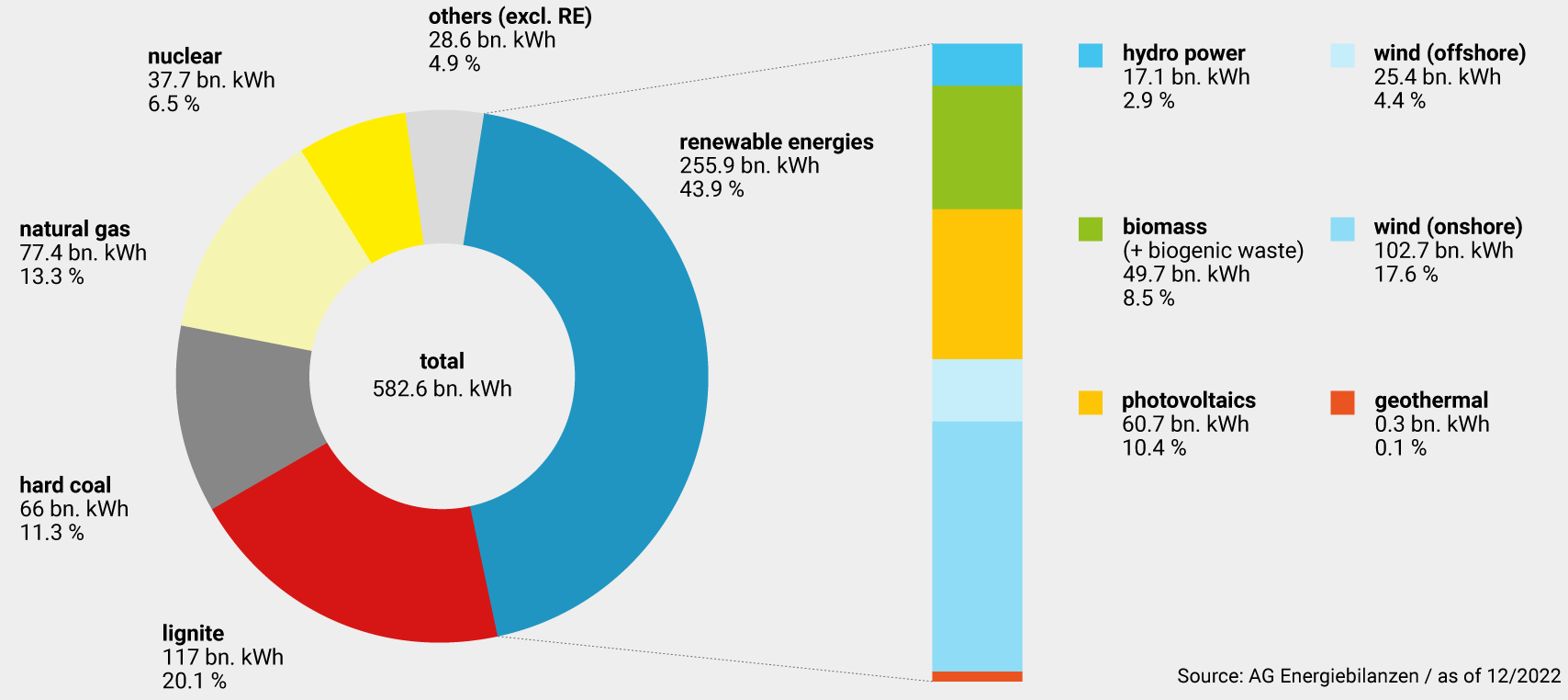}
\caption{Breakdown of electricity sources in Germany in 2022~\cite{GermanyReport22}.}
\label{fig:germany}
\end{center}
\end{figure}

The progress can be monitored, for example, via the fraction (and breakdown into sources) of sustainable electricity in the overall power mix of the EU-27 countries~\cite{GermanyReport22}. Figure~\ref{fig:germany} shows the breakdown for Germany for the year 2022. The fraction of close to 44\% sustainable electricity sources has been surpassed in 2023 by a few percent. 

\begin{figure}[htbp]
\begin{center}
\includegraphics[width=0.7\linewidth]{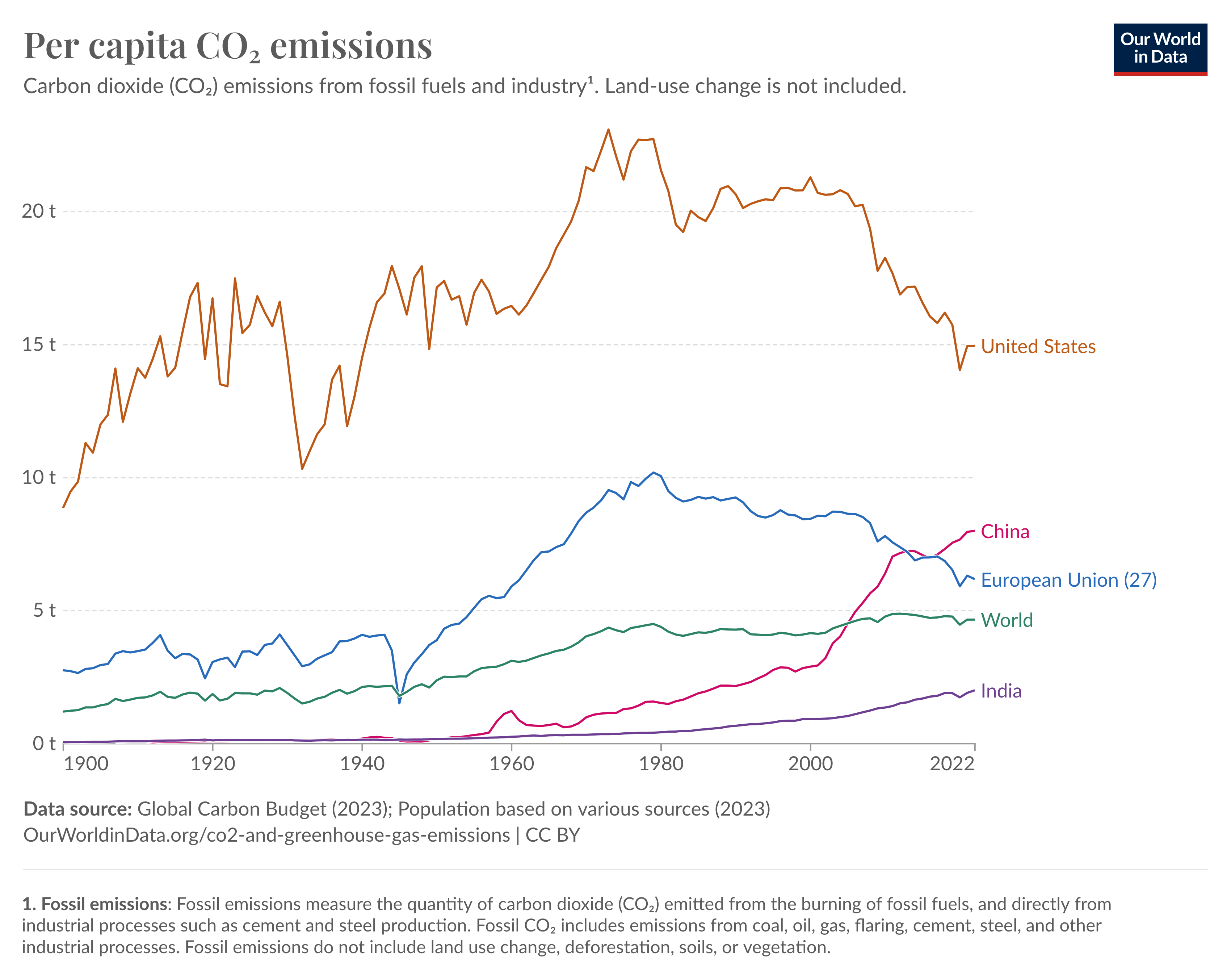}
\caption{Per capita CO2 emissions for selected world regions as function of time~\cite{owid-co2emissions-per-capita}.}
\label{fig:percapita}
\end{center}
\end{figure}

So there is movement towards climate neutrality in Europe, as can also be seen from a comparison, across selected world regions, of per capita CO2 emissions as a function of time, see fig.~\ref{fig:percapita}~\cite{owid-co2emissions-per-capita}.

\subsection{The context in the European Research Area}

Research infrastructure long-term sustainability at European Research Area (ERA\footnote{\url{https://research-and-innovation.ec.europa.eu/strategy/strategy-2020-2024/our-digital-future/european-research-area_en}}) level are primarily guided by the European Strategy Forum for Research Infrastructures (ESFRI). ESFRI was established in 2002 with the purpose of developing a European approach to Research Infrastructure policy as a key element of the emerging European Research Area. Further Research Infrastructures contribute to this effort in the frame of the EIROforum\footnote{\url{https://www.esfri.eu}} and the European Research Infrastructure Consortium (ERIC)\footnote{\url{https://research-and-innovation.ec.europa.eu/strategy/strategy-2020-2024/our-digital-future/european-research-infrastructures/eric_en}}. 

Regulation EU 2021/695 \cite{HorizonEurope} defining the Horizon Europe Framework Programme for Research and Innovation explicitly includes requirements to address global challenges including climate change and the United Nations Sustainable Development Goals (SDGs).

EU member states typically translate and integrate the strategies, policies and guidelines into their national roadmaps and plannings in addition to their existing national policies.

A robust long-term vision is the essential prerequisite in order to successfully and sustainably operate a Research Infrastructure. Therefore ESFRI issued a number of recommendations and actions \cite{ESFRI:2017}.

Sufficient funding and sustainable funding models, required across the whole lifecycle, are indispensable for a successful strategy for a new research infrastructure. Scientific excellence is the condition sine qua non. Together with adequate human resources it is crucial for the operational phase. Effective governance is another key element for ensuring long-term sustainability. Moreover, RIs should contribute to their sustainability by moving towards their own carbon neutrality.

While the existing recommendations place the explicit focus on scientific excellence, financial sustainability and societal acceptance, the call for comprehensive sustainability and impact assessment implicitly includes all environmental aspects. The recommendations also spell out that RIs should dedicate sufficient resources to periodically evaluate and communicate their socio-economic performance to different audiences. CERN's pioneering activities in this domain are explicitly mentioned by the ESFRI guide~\cite{ESFRI:2023}, encouraging national authorities to support the approach in cooperation with experts in the field.

The ESFRI policy brief \cite{ESFRI:2023} details the need to tailor the impact assessment to the specific project and recommends to assure that the "wider socio-economic impacts" are included. Following the OECD definition\footnote{\url{https://www.oecd.org/dac/evaluation/revised-evaluation-criteria-dec-2019.pdf}} these impacts comprise \textit{``the extent to which the intervention has generated or is expected to generate positive or negative, intended or unintended, higher-level effects''}. The European Commission\footnote{\url{https://ec.europa.eu/info/sites/default/files/better-regulation-guidelines-impact-assessment.pdf}} states that ``\textit{The term impact describes all the changes which are expected to happen due to the implementation and application of a given policy option/intervention [such as investment in a Research Infrastructure and its activities]. Such impacts may occur over different timescales, affect different actors and be relevant at different scales (local, regional, national and EU)}''. Ex-post assessment is needed to determine whether the intended objectives and the ex-ante estimations have actually been achieved. 

The European Commission Economic Appraisal Vademecum 2021-2027 \cite{ECVademecum} captures the general principles and provides sector application examples for impact assessment, including Research and innovation. 
It extends and complements the common provisions regulation to undertake cost-benefit analysis in line with EU regulation (EU) No 207/2015 \cite{ECCBA2015} supported by the ``European Commission Guide to Cost-Benefit Analysis of Investment Projects'' \cite{ECCBAGuide2014} . It states that for the use of European Structural and Investment Funds, a cost-benefit analysis, including an economic analysis, a financial analysis and a risk assessment is a prerequisite for the approval of a major project. Climate change adaptation and mitigation and disaster resilience are covered by this regulation. For instance the volume of the greenhouse gas (GHG) externality and the external cost of carbon are explicitly included and alignment with the EU 2050 decarbonisation objectives is required. Concerning climate adaptation, costs of measures aiming at enhancing the resilience of the project to climate change impacts that are duly justified in feasibility studies should be included in the economic analysis. The benefits of those measures, e.g. measures taken to limit the emissions of GHG or enhance the resilience to climate change and weather extremes and other natural disasters, should also be assessed and included in the economic analysis, if possible quantified, otherwise they should be properly described. An analysis of sustainable development (environmental protection, resources efficiency, climate change mitigation and adaptation, biodiversity and risk prevention) should be covered. An analysis of the options considering technical, operational, economic, environmental and social criteria for the location of the infrastructure are requested in a feasibility study. This covers also the description of the project's consistency with the applicable environmental policy. Considerations should include resource efficiency, preservation of biodiversity and ecosystem, reduction of GHG emissions, resilience to climate change impacts. The process needs to fulfil the Directive 2011/92/EU that defines the environmental impact assessment (EIA) process which ensures that projects likely to have significant effects on the environment are made subject to an assessment, prior to their authorisation. Therefore the total costs of the negative environmental impacts and their compensation have to be included. Environmental benefits can be assessed and added as well. They include for instance contributions to improve water supply and sanitation, waste management, energy capacity and stability, transport, ports (airports, seaports, inter-modal), research and innovation and broadband communication.

The appraisal process aims at assessing if a project will contribute to overall social welfare and to economic growth, taking into account benefits and costs to society. The EC \cite{ECCBAGuide2014} and UNIDO \cite{UNIDO_handbook} handbooks focus on the economic and societal topics although aspects, such as assessment of the environmental externalities are typically part of project appraisal as required per EU regulation. For instance, the shadow cost of carbon and the GHG emissions are shown in project appraisal examples. Eventually, the requirements for appraisal are defined for each project specifically by the notified body for that project. For instance, for obtaining funds from the European Investment Bank (EIB) a comprehensive appraisal study including the positive and negative environmental externalities is required. The EIB published a dedicated guide \cite{EIBCBAGuide2023} on this topic that includes among the reference for the shadow cost of carbon also comprehensive calculation examples and guidelines for capturing environmental externalities.

Other requirements emerge for instance from projects applying to the Connecting Europe Facility (CEF). The InvestEU regulation introduces climate, environmental and social sustainability as elements in the decision-making process when applying for the InvestEU Fund. The process is also a requirement in the framework of the preparatory phase of ESFRI projects. The European Bank for Reconstruction and Development (EBRD) requires project assessment with potentially relevant greenhouse gas (GHG) emissions of more than than 25 000 tonnes of \ce{CO2}e per year with respect to a baseline of 100 000 tonnes of \ce{CO2}e emissions per year. National requirements differ substantially from each other, requiring assessments for investments projects with public funding as low as 300 000 euros for instance in Lithuania.

\subsection{The context in France}

In France the "Code de l'environnement" \cite{CodeEnvironnementFrance} guides the approach to develop a sustainable project scenario in the frame of an ``environmental evaluation process'' that directly leads to the authorisation of the project. The term "environment" is to be understood in its original meaning and in a large sense: the surroundings and condition in which the project will be placed. 

\begin{figure}[ht]
    \centering
    \includegraphics[width=\textwidth]{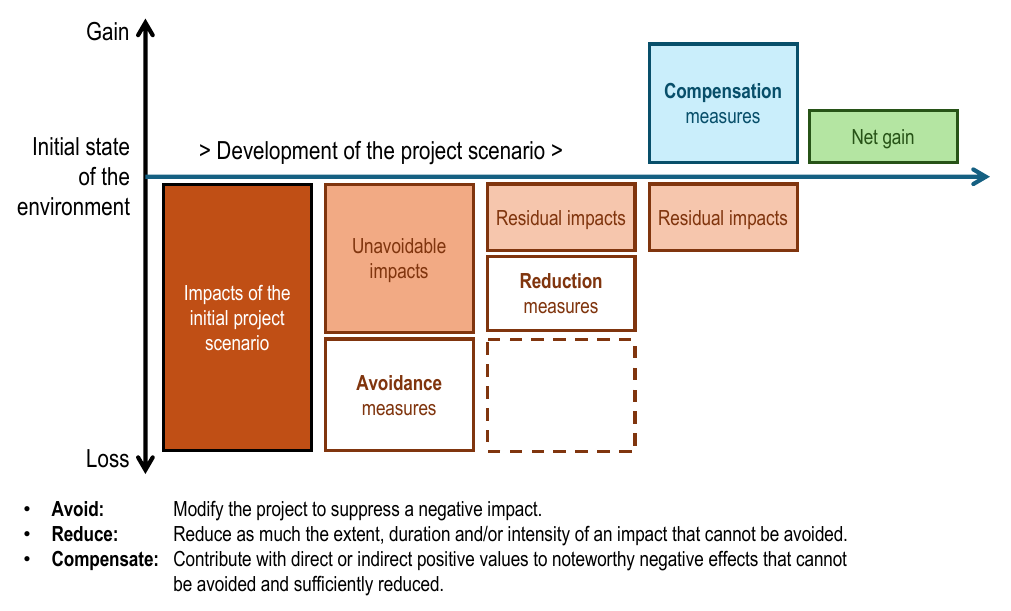}
    \caption{``Avoid-Reduce-Compensate'' approach for iterative development of a sustainable project scenario to achieve an ecological balance and, as far as possible, a net gain~\cite{ERC:2021}.}
    \label{fig:ERC}
\end{figure}

The goal of the process is to develop a feasible and sustainable project scenario following the "avoid-reduce-compensate" method (French: "éviter-réduire-compenser", ERC), which is anchored in European regulations and which is implemented in French law.\footnote{\url{https://www.legifrance.gouv.fr/codes/article_lc/LEGIARTI000038247366}} As a consequence the project scenario to be authorised aims at a net positive value (see Figure \ref{fig:ERC}).

This approach of iterative development of variants and versions that are continuously improved is also documented by the international standard ISO 14001 concerning environmental management. 
It describes the Plan-Do-Check-Act (PDCA) model as the iterative process to ensure continuous improvement (ISO 14001, Step by Step, Chapter 1). 
Other countries, for instance Switzerland, define the process with a different goal. 
There the process needs to demonstrate compliance of the scenario with all applicable, individual laws and regulations. 
It is considered best practice to follow the ERC approach with a continuous involvement of official services and the population. Consequently, for both CERN host states, the process will meet the constitutionally established, fundamental requirements applicable for large-scale publicly funded endeavours.

Considering that programme and project evaluation in France is compulsory for public investments that exceed 20 million euros and a second assessment is required for investments that exceed 100 million euros, the national assembly tracks and reports on such projects every year using a standardised information sheet approach that includes the net present value as an overall sustainability indicator and the avoided greenhouse gas emissions. 
Such information is published as an annex of the annual budget law \cite{ProjetBudget2024France}. In 2023, 13 research investments were subject to comprehensive socio-economic evaluations. 

The evaluation concerns the entire project throughout all lifecycle phases. It is understood that the level of description detail for individual phases and project segments can be initially different. The process accompanies the project throughout its entire life. Consequently, the documentation needs to be regularly updated. The process records the initial state of the environment, including nature and a variety of other topics such as urbanism, public health, population safety, economic impacts. 
The process actively identifies and assesses relevant effects on the project environment (environmental impact analysis) and gives input to the design process following the avoid, reduce and compensate steps. 
The process includes the involvement of the population for instance via an informal dialogue phase or formal consultation and public debate processes ("débat public") and public investigation ("enquête publique"). 
The last two processes are carried out with national notified bodies.

To capture the sustainability performance properly, the French government requires by law a socio-economic evaluation for each project that is funded via public investments of more than 20 million euros. The evaluation must be carried out according to the guidelines of the "Secrétariat général pour l'investissement" (SGPI)\cite{SGPI:2023}. 
Projects with a total public investment of more than 100 million euros are subject to a second expert assessment on which the SGPI provides an assessment to the French parliament and the prime minister as well as to the minister under whose topic responsibility the project falls.

The socio-economic evaluation comprises a detailed description of the project, its variants and alternatives, the key characteristics, the implementation schedule, a list of relevant socio-economic indicators, a list of indicator values showing how the project performs with respect to public policies (e.g. climate impact reduction objectives), the financing plan, the compliance with laws and regulations and a risk registry.

While until recently (around the year 2020) the socio-economic evaluation included only specific environmental aspects such as the improvement of habitats and green spaces, the ever growing importance of more global topics led to an enlargement of the covered positive impacts and negative externalities. 
The 2023 guide includes explicitly all environmental aspects that are project relevant such as \ce{CO2} equivalent emissions, noise, air pollution, water use and pollution, soil use and pollution and requires to associate them to monetary values and to equally quantify the positive contributions in monetary terms. 
For instance, the legal value of one tonne of \ce{CO2} for the socio-economic evaluation has been set to 32 euros (base year 2010) in the 2023 guideline, which correspond to about €40 in 2024. 
Despite this value, the annex that indicates the guidelines of converting non-tradeable goods and externalities into monetary terms indicates a progression of the shadow cost of carbon 250 euros per t\ce{CO2}e to 2030 and 775 euros per t\ce{CO2}e in the year 2050. This progression is in line with the EIB recommendations of the shadow cost of carbon.

The socio-economic evaluation is based on a standard cost-benefit analysis, reporting a net present value and a benefit/cost ration based on the comparison of discounted total costs and benefits over a defined observation period. The positive and negative direct effects and externalities must be taken into consideration for the entire lifecycle of the project. It is therefore recommended to capture certain relevant environmental effects via a lifecycle analysis and to convert then the result into monetary values. However, care must be taken to discount and report positive and negative contributions separately due to the fact that they may occur during different times and over different time spans.

For the social discount rate, the 2023 guidelines recommend to generally use a value of 3.2 \% for projects with a lifetime up to the year 2070. However, for very long term and low risk projects such as research infrastructures a lower social discount rate in the range of 2.5 \% can be justified.

In case of a "déclaration d'utilité publique" (DUP), which governs the authorisation process for public investment projects, the socio-economic evaluation is part of the legally required "enquête publique" files.

The socio-economic evaluation has to be regularly updated, following the iterative approach of refining the project scenario and comparing for foreseen effects against the actual effects once the project is implemented ("ex-post" analysis).

A project can be considered sustainable if both, the socio-economic and the financial net present value are positive. 
In addition, the applied social discount rate should be larger than the average weighted cost of the capital needed to finance the project. This can be verified by determining the so called "Internal Return Rate" (IRR), which is closely linked to the calculation of the net present value. It expresses at which social discount rate value the net present value of the project would become zero. For instance, if the loans for the project are granted at 3\% but the net present value becomes zero at only 5\% the condition is satisfied.

If the project does not generate revenues, i.e. in the case of a research infrastructure for fundamental scientific research, no financial net present value can be provided, the project may be justified for the society if the socio-economic net present value is positive and the Internal Return Rate is higher than the cost of the capital.
If the socio-economic net present value is negative and the financial net present value is positive, the investment may be financially viable, but it is not recommended from a societal point of view.

\subsection{The context in Germany}

In 2015, together with another 192 member states of the United Nations, Germany adopted the Agenda 2030 for Sustainable Development. The agreed goals show that the management of the wide range of ecological, social and economic tasks can only be accomplished in a joint effort. Science plays an essential role in the realisation of these goals. There are many and varied expectations: science will identify problems and challenges, develop solution proposals, show alternative paths and apply these principles to itself.
The research organisations Helmholtz Association, Fraunhofer Gesellschaft and Leibniz Association took the initiative in self-imposed responsibility and developed a "Handbook for sustainability management in non-university research organisations (LeNa)" (\url{https://www.nachhaltig-forschen.de/en/}) published in 2016. The partners developed the value of sustainability with regard to their activities for and within academic life and how it can be achieved in practice. Five areas have been defined which focus on both, science and administration. 

The framework focuses on eight criteria to be integrated: ethics, integrative approach for economic and societal systems, interdisciplinarity, user orientation, impacts on society and the environment including ex-ante or ex-post assessments, transdisciplinarity, transparency, dealing with complexity and uncertainty~\cite{LeNa-framework:2023}.

The five priorities are set according to individual needs: corporate governance, research practice with societal responsibility, human resources (management, development, equal opportunities, responsible fixed-term employment, health-preserving work conditions), need-oriented buildings and infrastructures with responsibility towards the environment and the society, supporting processes (procurement, mobility, indirect impacts on environment and society).

The German ministry of Education and Research and the Deutsche Forschungsgemeinschaft (DFG, German Research Foundation) developed their own sustainability guidelines aligned with the LeNa directives. The DFG published a “Sustainability Guide for Research Processes” focusing on ecological topics such as travelling, experiments, field trials, surveys, computing, procurement operation and use of instrumentation\cite{DFG_sustainability_guide}. The DFG also published recommendations for “Anchoring sustainability considerations in DFG funding activities”, which are recommended to be incorporated in project proposals from 2024 onwards. The DFG states that \textit{"it must be ensured that the expectations regarding contributions
to climate protection and environmental compatibility to be made by individual researchers on the one hand, and by institutions on the other, are balanced"}. While focusing on ecological sustainability, the recommendations do not ignore links to other dimensions such as social and economic sustainability. The Commission recommends\cite{DFG_sustainability_considerations} to systematically integrate sustainability considerations and raise awareness of the impact of research, but it does not encourage contributions to sustainability research.

The federal ministry for education and research follows the official “Catalog of policies for sustainability in administration”\cite{Federal_government_management_concept_for_sustainability} that are applicable for all authorities and institutions of the direct and indirect federal administration. It applies them to all publicly funded organisations including research facilities. The program comprises ten packages of measures:
\begin{enumerate}[nosep]
\item Climate-neutral federal administration by 2030,
\item Construction, renovation and operation of federal properties,
\item Mobility,
\item Procurement,
\item Events,
\item Canteens/communal catering,
\item Training for sustainable development,
\item Health,
\item Equal participation in management positions and compatibility of family/care responsibilities,
\item Diversity.
\end{enumerate}

Furthermore the ministry issued “Guidelines for preparing the draft proposal for a large research infrastructure”~\cite{BMBF:2024} in 2024 (see footnote for further information\footnote{\url{https://www.bmbf.de/bmbf/shareddocs/kurzmeldungen/de/2024/06/240617_priorisierungsverfahren_fis.html}}) for the ongoing national prioritisation process. For the first time it requires an assessment of sustainability aspects with a focus on reduction of the environmental footprint. Each proposal has to describe which “aspects are relevant for the reduction of the environmental footprint (such as emissions, resource consumption, subsequent utilisation) of the planned RI in the different phases of the entire lifecycle as well as the planned objectives and measures to address these”. In a second part it has to be outlined how “the emissions as well as the sustainability objectives and procedures, in line with a continuous monitoring process, are to be measured and controlled”, following a third part reflecting on “potential environmental and health impacts of the planned RI (such as RI impacts on the local environment and staff, on climate and land utilisation etc.)” as well as elaboration on “counteractive measures in order to prevent or, in the best possible way, to attenuate the mentioned environmental and health impacts”. 

Several German laws regarding energy and civil construction which are not used for project appraisal are relevant for research projects and need to be taken into consideration. A few examples are:
\begin{itemize}[nosep]
\item Energy Efficiency Act (Energieeffizienzgesetz EnEfG) (\url{https://www.gesetze-im-internet.de/enefg/BJNR1350B0023.html }) requires the implementation of an Energy management system (following ISO 15001 or ISO 14001), requires the publication of data on data centers, waste heat and others, requires the implementation of waste heat usage and climate neutral data centers, requires a reduction by 2\% total energy consumption per year
\item Buildings Energy Act (GebäudeEnergieGesetz GEG) (\url{https://climate-laws.org/document/buildings-energy-act-geg_ec47 }) focusing on saving energy and using renewable energies for heating and cooling in buildings; point to the role model function of the public sector; raised standards regarding new buildings and reduced the permissible annual energy demand in new buildings from 75\% to 55\%
\item Ordinance on Fees for Access to Electricity Supply Networks StromNEV §19  (Verordnung über die Entgelte für den Zugang zu Elektrizitätsversorgungsnetzen StromNEV §19) (\url{https://www.gesetze-im-internet.de/stromnev/BJNR222500005.html}): allows for several privileges for reduced fees and taxes on energy via so called "individual grid fees" in exchange for strict energy management and grid-friendly consumption. This instrument is beneficial for the environment and can save a research infrastructure the size of DESY several million euros per year.
\end{itemize}

Concerning the federal authorisation processes that include environmental impact assessments, it is foreseen to adopt the methodological convention of the German Environment Agency (UBA)\cite{UBA-MVC-2020} for the assessment of environmental costs\footnote{A comprehensive revision will become available with the publication of the Handbook on Environmental Value Factors 4.0 in 2025}. The value factors are intended for use in the public sector to assess the impacts of infrastructure investments. Particular methods and values for use in cost-benefit analysis. LCA and environmental and sustainability reporting are issued in the "Environmental Unit Cost Lists"\cite{UBA_unit_cost_list_2023}.

For instance in the domain of energy management to perform a calculation of the net present value of an energy efficiency action~\cite{UBA_energy_management}. Numerous sustainability related topics such as soil and land consumption, biodiversity and habitat loss, responsible use of resources, water, air, climate, landscape, cultural goods and material assets are part of the project authorisation and approval process that include the environmental impact assessment [\url{https://www.bmuv.de/fileadmin/Daten_BMU/Download_PDF/Umweltpruefungen/sup_leitfaden_lang_bf.pdf}]. The REACH regulation\footnote{Registration, Evaluation, Authorisation and Restriction of Chemicals, applicable to all producers, importers and users of chemical substances that may affect health and the environment.} includes the socio-economic assessment as requirement~\cite{EU_REACH,GWS_25_2019} 
to assess the economic, societal and environmental sustainability of the proposal.

\subsection{The context in Switzerland}

With the recent update of the law for the encouragement of research and innovation (LERI)~\cite{LERI:2023} and the development of a dedicated sector plan for new constructions and installations of CERN~\cite{ParlamentSchweiz:2024}, guidelines with respect to sustainability of research infrastructures become more specific.

In general, the ``avoid-reduce-compensate'' approach is required throughout the entire lifespan of the research infrastructure. It is to be accompanied by monitoring and continuous improvement measures.

The consumption of land, in particular high quality arable land that is regarded as a precious resource, needs to be avoided. Where this cannot be achieved, the residual consumption of such land that is registered in a national inventory needs in principle to be compensated. Therefore different alternatives and variants of the project need to be documented and assessed. Compensation can for instance be achieved by stripping the topsoil of the affected original land and using that soil for creating high quality arable land within an agreed geographical perimeter, for instance on wastelands or degraded soil. The authorities facilitate the process of identifying suitable compensation areas and the administrative processes.

Water bearing layers need to be protected and should not be used to provide raw water for cooling purposes. Care must be taken not to pollute water bearing layers or to mix the water of superimposed water bearing layers.

Exclusion buffers next to water streams serve the protection of habitats.

Forest clearings are generally forbidden in Switzerland. An exemption can be granted only in exceptional cases for installations that are in the interest of the public.

Measures need to be taken to manage rainwater and used water avoiding flooding and pollution of the environment. The same holds for the management of water used to extinguish fires.

Limitations for dust and particle emissions need to be respected.

Limitations for noise emissions need to be respected.

Measures need to be taken to manage waste of any kind.

The national and cantonal energy and climate protection goals are to be taken into consideration during the planning, the construction and operation of future research installations. Energy consumption is to be reduced to a minimum that is compatible with the capacity to fulfil the scientific research goals.

The emission of greenhouse gases are to be minimised within the boundaries of the needs to comply with the scientific operation requirements.

The design and implementation of technical systems has to integrate the optimisation of the energy efficiency and the recovery and use of waste heat. The available capacities and characteristics for internal and external use need to be identified and documented.

Buildings must comply with the national energy efficiency regulations.

In general, any optimisation and minimisation approach is understood to be technically feasible and economically viable. They must always be compatible with the feasibility to achieve the specified scientific research goals.

A mobility plan that supports the national and cantonal climate plans is to be developed. The parking spaces are to be optimised, considering the applicable cantonal regulations.

\subsection{The context in the UK}

There exists no legal requirement in the approach to prioritisation and decision-making for research infrastructure project selection and funding in the UK. A Cost-Benefit approach is, however, expected by the Treasury to grant public funding for investments. 'The Green Book' issued by HM Treasure and UK Government Finance Function \cite{UK_greenbook:2022} provides guidance on how to appraise policies, programmes and projects, including full sustainability analysis and environmental effects coverage.

Sustainability considerations that are part of the UK’s research and innovation infrastructure project evaluation scheme are emerging across different stages of the UK Resarch Fund’s processes. The UKRI Research and Infrastructure Programme report on opportunities includes some of them. Working closely with the UKRI Environmental Sustainability Team, the Infrastructure Fund has been increasingly incorporating this as part of its core criteria for proposal development and assessment, and it is now expected that carbon and sustainability issues will be broadly considered by Councils in their future projects\cite{UK_RI_process_evaluation_2023}. Sustainability of research infrastructures is now a part in the regular priority and landscape analysis of UK's resarch infrastructure planning\cite{UK_RI_opportunities_2023}.

\subsection{The context in the US, Canada and Australia}

CSIRO Impact Evaluation Guide \cite{CSIRO-IEG:2024} 
presents the impact evaluation
framework employed by the Commonwealth Scientific and Industrial Research
topics: why evaluations are conducted; evaluation design; evaluation methodologies;
aggregation and comparability of impacts across programmes of work; and sensitivity
primary methodology for research impact evaluation.

US Department of Energy EERE R\&D Programme Standard Impact Evaluation
Method~\cite{DOE-EERE:2014} 
is a handbook that provides guidance for impact assessments of R\&D Renewable Energy (EERE). It was developed in cooperation with the Lawrence Berkely National Laboratory to provide guidance for impact evaluations and benefit-cost analyses of R\&D programs.
It covers evaluation planning, assessing additionally, estimation of economic costs and benefits, estimation of
environmental impacts, estimation of energy security impacts, estimation of
knowledge impacts, calculating economic performance measures and sensitivity
analysis.

\subsection{Comprehensive sustainability assessment based on Cost-Benefit Analysis}
\label{annex:CBA}

\begin{figure}[ht]
    \centering
    \includegraphics[width=\textwidth]{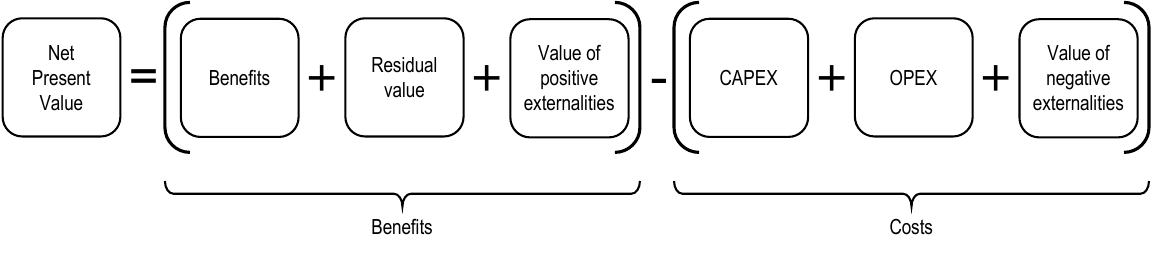}
    \caption{Expression to determine the Net Present Value (NPV) of an investment project considering economic, societal and environmental aspects.}
    \label{fig:NPV_formula}
\end{figure}

The approach is to create an inventory of negative cost and positive benefit items and convert them into monetary terms that are combined to determine a net present value (NPV) of the investment at the end of a chosen observation period (see Figure \ref{fig:NPV_formula}). The ratio of the benefits of a project relative to its costs (BCR) and the so called internal return rate (IRR) are two additional measures that can be derived from this analysis to provide valuable insight in the sustainability of the undertaking. The IRR is the discount rate that would bring the net present value of the entire project over the observation period to a zero value. Generally, an investment can be considered sustainable if the BCR is positive and if the IRR is greater than the cost of obtaining the required capital.

In line with the guidelines on cost estimation of research infrastructures \cite{StrESFRI}, the following items need to be established:
\begin{enumerate}[nosep]
\item Unit of analysis with clear scope and boundary descriptions.
\item Reference period with a start and an end date of the observation based on the expected ``useful life'' of the scientific research infrastructure (note that this period may be different from the physical life of the infrastructure and may lead to residual asset values).
\item Base year, i.e. the point in time when the quantitative estimation is made that is not necessarily coinciding with the start date. Past and future values are discounted with respect to the base year. Conversion rates of the unit of measure with respect to other relevant currencies are to be recorded for the base year.
\item Unit of measure for costs and benefits in a specific monetary currency.
\item Approach for converting in-kind contributions in monetary terms.
\item Definition and description of the counterfactual scenario.
\item Project-specific Social Discount Rate (typically between 2 and 4 \%) that is well justified, ideally approved by an economist expert advisory committee.
\item Structure of capital and operation expenditures
\item Structure of negative externalities considered
\item Structure of positive impacts (benefits) considered
\end{enumerate}

In addition to full financial costs, negative externalities must be integrated in the assessment as far as they can be reasonably identified, accounted and converted in monetary terms.

Concerning potential impacts on the climate, the methodology includes the accounting of project relevant emission factors, the estimation of the net greenhouse gas (GHG) emissions (a negative externality) and avoided emissions (positive impact or benefit) compared with a counterfactual baseline scenario. The resulting amount of generated and avoided GHG emissions in tonnes of carbon dioxide equivalent (t\ce{CO2}e) has to be converted into monetary terms using a shadow price of carbon. In line with the EC technical guidance on the climate proofing of infrastructure \cite{ECClimateProofing}, it is recommended to use as a shadow cost of carbon the values established by the European Investment Bank (EIB) as the best available evidence on the cost of meeting the goals of the Paris agreement \cite{ParisTreaty}.\footnote{The Paris Agreement is a legally binding international treaty on climate change. It was adopted by 196 Parties at the UN Climate Change Conference (COP21) in Paris, France, on 12 December 2015. It entered into force on 4 November 2016.}

Input-Output analysis \cite{IOModels} is a methodology that only captures economic linkages and serves to determine the value-added of an investment and can be used to estimate the effect on the job market and in which domains and geographical regions the economic activations take place. This instrument is regularly used by governments and at international level (e.g. EU, OECD) for empirical economic research and structural analysis and is therefore widely known. However, it is purely an economic instrument that should only be used as a complementary tool to document economic linkages, permitting the development of international in-cash and in-kind contributions, to identify project-relevant industry sector to develop targeted common activities and synergies and to understand the job market implications to develop regional specialisation policies and focused training and skilled labour mobility plans.

A complementary analysis of the public perception via a stated preference and ``Willingness-To-financially-Participate'' (WTP) approach \cite{RePEc:ucp:jaerec:doi:10.1086/691697} is recommended as was carried out recently at CERN \cite{secci_2023_7766949}. It can reveal if the full costs and cumulative impacts of a research infrastructure or scientific investment are at least justified with respect to the perception of the public.

The establishment of a risk registry, a risk assessment and an evaluation of the residual risks is important for new science missions and research infrastructure projects and programmes. It should include economic, social and environmental domains in addition to standard topics such as financial and project management related matters.

Smaller projects may choose to perform a simplified appraisal process with a limited CBA or a multi-criteria analysis (MCA). In any case, presenting several variants and versions of the project and individual segments together with their full financial costs, economic, social and environmental performance levels is considered good practice.

\subsection{Summary measures of social value}
\label{annex:social_value}

\textbf{Net Present Value (NPV)} is the discounted sum of all future benefits less the discounted sum of all future costs over the appraisal period as a whole.
To properly estimate the NPV, realistic estimates are required of the streams of benefits and costs over the appraisal period (typically around 30 years). The key to determining both these streams is knowledge of the times at which the various elements would come into play. Investment costs will typically be incurred prior to the date of opening, whilst operating costs (for example personnel and resources for operation and maintenance) and user benefits would arise after the year of opening. User benefits and operating costs and revenues can be estimated from model runs for two or more years, and the stream of benefits derived by interpolation and extrapolation between the benefits for the modelled years.

\textbf{Benefit/Cost Ratio (BCR)} is given by the ratio of the discounted sum of all future costs and benefits. The BCR is therefore a value for money measure, which indicates how much net benefit would be obtained in return for each unit of investment cost. 
Formulae for the NPV and BCR will be found in CBA textbooks, a good example of which is ref.~\cite{pearce1981Nash}.

\textbf{Internal Rate of Return (IRR)}. Whereas the NPV and BCR measures require a test discount rate to be specified, the IRR reports the average rate of return on investment costs over the appraisal period. This can be compared with the test discount rate to see whether the project yields a higher or lower return than is required to break even in social terms. Calculation of the IRR and related issues 
are discussed e.g. in Pearce's and Nash's book, ``The Social Appraisal of Projects: A Text in Cost-Benefit Analysis"~\cite{pearce1981Nash}. 

\clearpage

\section{GWP estimates of specific materials \label{annex:reference-data}}

Example reference data is provided in Table \ref{tab:LCA_materials} to exhibit potential ranges of GWP impact factors per kg of material. Data was collected from available sources in 2025 that were published within the last 20 years. Data is not guaranteed to be fully comprehensive, but is intended to be a snapshot into the discrepancies, dependencies and potential unavailability of LCAs of specific materials.

The majority of results are from cradle-to-gate-studies (of which definition of `gate' varies per study). Each material has dependencies on region, age of LCA, exact composition and extraction method of material, manufacturing methods, primary and secondary material ratios, and machining methods used, etc. Each of the listed materials need further manipulation (manufacturing, machining etc.) to be used in construction of an accelerator, which would change the LCA result quoted.
 
In addition, to our best knowledge, some specific materials did not have readily/openly available data. For example, \ce{Sm2Co17}, Niobium RRR300, and Grade 2 titanium were unavailable, but similar/related materials were available. The authors encourage the reader to share any additional information to update Table \ref{tab:LCA_materials}.  

\begin{table}[h]
\centering
\caption{Example of reference GWP data on cradle-to-factory-gate materials that are common in accelerator construction. Results are rounded to 2 significant figures.}
\label{tab:LCA_materials}
\small\addtolength{\tabcolsep}{-5pt}
\resizebox{0.90\textwidth}{!}{
\begin{tabular}{llcccl}
\toprule 
   \textbf{Material} & \textbf{Composition} & \multicolumn{1}{c}{\textbf{Density}} & \textbf{GWPs} & \textbf{References}  & \textbf{Example usage} \\
   \textbf{~} & \textbf{~} & [\unit{\tonne\per\cubic\meter}]& [\unit{\kilogram\ce{CO2}e\per\kilogram}] & \textbf{~} & \textbf{~}\\
\midrule
  \multicolumn{6}{l}{\textbf{Civil Construction}} \\
\midrule
Concrete	& Cement:fine aggregate:coarse aggregate 	& 2.30	&\textbf{0.15} - \textbf{0.24}\footnotemark[38]	&  \cite{Arup2023ConcreteEC} &	Concrete parts	 \\ 
(C30 CEM I) & (1:2:3) &&&& \\
Rebar steel	&	Fe	&	7.85	& $\textbf{1.3}$ - $\textbf{2.4}$	& \cite{Steel} &	Concrete parts	 \\ 
\midrule
  \multicolumn{6}{l}{\textbf{Magnets}} \\
\midrule
Soft Iron 	&	Fe	&	7.85	& $\textbf{1.3}$ - $\textbf{2.0}$, $\textbf{3.6}$\footnotemark[38]	& \cite{Steel, Fe_max, clic_ilc_lca-accel_arup} &	Magnet yoke, \\ 
&&&&&low/medium fields \\
Magnet iron\footnotemark[38]	&	\qty{0.49} Fe, \qty{0.49} Co, \qty{0.02} V	&	8.12	& $\textbf{22}$   & ~\cite{clic_ilc_lca-accel_arup}	&	Magnet yoke, \\
\quad(Vacoflux) &&&&&high field \\
Copper &	Cu	&	8.96	&	$\textbf{1.9}$ - $\textbf{10}$ & ~\cite{Cu_min,Cu_max}	&	Coil, cables \\  
\quad(various grades) &&&&& \\
Aluminium	&	Al&	2.70	& $\textbf{8.2}$ - $\textbf{22}$ & ~\cite{LCA-metals, Al_max}	& Coil, cables\\
Ferrite (\ce{BaFe12O19})	& \qty{0.12} Ba, \qty{0.60} Fe, \qty{0.27} O		&	5.28  &	Unavail. & -	&	Permanent magnets	\\
AlNiCo 5 	&	\qty{0.51} Fe, \qty{0.080} Al, \qty{0.14} Ni, & 7.30	 & Unavail. & -	&	Permanent magnets \\
& \qty{0.24} Co, \qty{0.03} Cu	& & & &  	\\
\ce{SmCo5}&\qty{0.36} Sm, \qty{0.64} Co	&	8.30		& \textbf{24}\footnotemark[39] - \textbf{66} &  ~\cite{SmCo5_min,SmCo5_max} 	&	Permanent magnets \\

\ce{Sm2Co17}	& \qty{0.25} Sm, \qty{0.75} Co	&	8.30 & Unavail. & -	&	Permanent magnets \\
\ce{Nd2Fe14B}	& \qty{0.27} Nd, \qty{0.72} Fe, \qty{0.01} B&	7.50	& $\textbf{18}$ - $\textbf{220}$	& \cite{Nd2Fe14B_max} &	Permanent magnets \\
Cobalt & Co & 8.83 & $\textbf{8.3}$ - $\textbf{28}$ & \cite{LCA-metals, Co_max} & Permanent magnets \\ 
\midrule
\multicolumn{6}{l}{\textbf{Vacuum System}} \\
\midrule
Stainless Steel &  304L\footnotemark[40] & 7.93 & $\textbf{1.6}$ - $\textbf{6.8}$ & \cite{StainlessSteel_min, StainlessSteel_max} &  Chamber, casing\\
Oxygen-free copper OFC\footnotemark[42] & Cu& 8.93 & $\textbf{7.6}$ & ~\cite{clic_ilc_lca-accel_arup} & chambers \\
Titanium & Ti	& 4.50 & $\textbf{8.1}$ - $\textbf{36}$ & ~\cite{LCA-metals, Al_max} & Rotor/blades\\
\midrule
\multicolumn{6}{l}{\textbf{Superconducting and Cryogenic Materials}} \\
\midrule
Niobium RRR300 \footnotemark[38]\footnotemark[41] & Nb & 8.57 & $\textbf{98}$ & ~\cite{clic_ilc_lca-accel_arup} & Cavities (high field, \\
&&&&&high purity) \\
Niobium \footnotemark[41]  &Nb	&8.57  & 5.8 - 13,  $\textbf{24}$\footnotemark[38] & ~\cite{Ni_min, LCA-metals, clic_ilc_lca-accel_arup}   & Cavities  \\
\ce{Nb45Ti55} &	\qty{0.45} Nb, \qty{0.55} Ti & 5.70 & $\textbf{38}$ & ~\cite{clic_ilc_lca-accel_arup} & Cavities \\
Titanium  Gr 2~\footnotemark[38]& Ti	& 4.50 & $\textbf{50}$ & ~\cite{clic_ilc_lca-accel_arup} & Pipes  \\
\bottomrule
\end{tabular}
}
\end{table}

\footnotetext[38]{Values used in the LCA of CLIC and ILC performed by the Arup group were the factors from Ecoinvent v.3.10 concrete with the ReCiPE2016 midpoint (H) LCIA method~\cite{clic_ilc_lca-accel_arup}.}
\footnotetext[39]{GWP is given for an electric motor with magnet mass of $m=1.26\,$kg extrapolated to 200\,000\,km.}
\footnotetext[40]{As there is a high share of alloying elements for stainless steel, only 304 stainless steel is considered in this document.}
\footnotetext[41]{Assuming production and refinement in China.}
\footnotetext[42]{From Europe.}

\printbibliography

\end{document}